\documentclass[12pt,dvipsnames]{article}
\usepackage[top=80pt,bottom=85pt,left=85pt,right=85pt]{geometry}

\pdfoutput=1
\usepackage[utf8]{inputenc}
\usepackage[T1]{fontenc} 
\usepackage[english]{babel} 
\usepackage{braket}
\usepackage[dvipsnames]{xcolor}
\usepackage{physics}
\usepackage{tabularx} 
\usepackage{ragged2e}
\usepackage{booktabs} 
\usepackage{rotating}
\usepackage{makecell,booktabs}
\usepackage{multirow} 
\usepackage{comment} 
\usepackage{amsmath}
\usepackage{cite}
\usepackage[all]{xy}

\usepackage{tikz}
\usetikzlibrary{math} 
\usetikzlibrary{3d} 
\usepackage{tikz-3dplot}
\usepackage{tikz-cd} 
\usepackage{microtype}
\usepackage{amsthm}
\usepackage{mathrsfs} 
\usepackage[strict]{changepage}
\newcommand\scalemath[2]{\scalebox{#1}{\mbox{\ensuremath{\displaystyle #2}}}}

\tikzcdset{
  cells={font=\everymath\expandafter{\the\everymath\displaystyle}},
}

\usepackage{amssymb} 
\usepackage{subcaption} 
\DeclareCaptionFormat{custom}

\usepackage[pdftex,colorlinks=true]{hyperref} 
\hypersetup{urlcolor=MidnightBlue, citecolor=red, linkcolor=MidnightBlue}

\usepackage{float} 
\usepackage{todonotes}

\usepackage[capitalise]{cleveref}

\newcolumntype{E}{>{\hfil$}p{0.65cm}<{$\hfil}}

\newcolumntype{L}{>{\hfil$}p{16cm}<{$\hfil}}

\newcolumntype{D}{>{\hfil$}p{7.4cm}<{$\hfil}}
\newcolumntype{C}{>{\hfil$}p{3cm}<{$\hfil}}
\newcolumntype{P}{>{\hfil$}p{7.7cm}<{$\hfil}}
\newcolumntype{F}{>{\hfil$}p{5.7cm}<{$\hfil}}
\newcolumntype{S}{>{\hfil$}p{1.8cm}<{$\hfil}}
\newcolumntype{R}{>{\hfil$}p{5.2cm}<{$\hfil}}
\newcolumntype{U}{>{\hfil$}p{4.2cm}<{$\hfil}}
\newcolumntype{Q}{>{\hfil$}p{6.4cm}<{$\hfil}}
\newcolumntype{T}{>{\hfil$}p{1.9cm}<{$\hfil}}
\newcolumntype{V}{>{\hfil$}p{5.8cm}<{$\hfil}}
\newcolumntype{H}{>{\hfil$}p{1.8cm}<{$\hfil}}
\newcolumntype{A}{>{\hfil$}p{6cm}<{$\hfil}}
\newcolumntype{B}{>{\hfil$}p{2cm}<{$\hfil}}

\makeatletter
\newcommand\xleftrightarrow[2][]{%
  \ext@arrow 9999{\longleftrightarrowfill@}{#1}{#2}}
\newcommand\longleftrightarrowfill@{%
  \arrowfill@\leftarrow\relbar\rightarrow}
\makeatother

\newcommand{\re}[1]{\textcolor{red}{#1}}

\numberwithin{equation}{section}

\definecolor{cambridgeblue}{rgb}{0.64, 0.76, 0.68}
\definecolor{caribbeangreen}{rgb}{0.0, 0.8, 0.6}
\definecolor{celadon}{rgb}{0.67, 0.88, 0.69}
\definecolor{champagne}{rgb}{0.97, 0.91, 0.81}
\definecolor{cream}{rgb}{1.0, 0.99, 0.82}
\definecolor{cyan(process)}{rgb}{0.0, 0.72, 0.92}
\definecolor{brilliantlavender}{rgb}{0.96, 0.73, 1.0}
\definecolor{candypink}{rgb}{0.89, 0.44, 0.48}

\usepackage{pdflscape}

\usetikzlibrary{positioning}
\usetikzlibrary{arrows}
\usetikzlibrary{decorations.pathreplacing}
\usetikzlibrary{shapes.geometric}
\usetikzlibrary{calc}
\usetikzlibrary{decorations.pathmorphing}
\usetikzlibrary{shapes.misc}

\tikzset{gaugeSU/.style={inner sep=1.7mm,draw=none,fill=yellow,minimum size=2mm,circle, draw}}
\tikzset{flavourSU/.style={draw=none,minimum size=2mm,fill=white, regular polygon,regular polygon sides=4,draw}}
\tikzset{flavour/.style={draw=none,minimum size=0.3mm,fill=white, regular polygon,regular polygon sides=4,draw}}
\tikzset{gaugeBig/.style={inner sep=1.7mm,draw=none,fill=white,minimum size=2mm,circle, draw}}
\tikzset{bd/.style={circle, draw=black, inner sep=0pt, fill=black, minimum size=2mm}}
\tikzset{wd/.style={circle, draw=black, inner sep=0pt, fill=white, minimum size=2mm}}
\tikzset{Dynkin/.style={circle, draw=black, inner sep=0pt, fill=white, minimum size=2mm}}
\tikzstyle{ligne}=[draw, very thick] 
\tikzstyle{gridline}=[draw, gray] 
\tikzset{gauge/.style={circle, draw,inner sep=2.5pt}}
\tikzset{gaugeo/.style={circle, draw,inner sep=2.5pt,fill=orange}}
\tikzset{gaugec/.style={circle, draw,inner sep=2.5pt,fill=cyan}}
\tikzset{gauger/.style={circle, draw,inner sep=2.5pt,fill=red}}
\tikzset{gaugeb/.style={circle, draw,inner sep=2.5pt,fill=blue}}
\tikzset{gaugeg/.style={circle, draw,inner sep=2.5pt,fill=green}}
\tikzset{gaugem/.style={circle, draw,inner sep=2.5pt,fill=magenta}}
\tikzset{gaugey/.style={circle, draw,inner sep=2.5pt,fill=yellow}}
\tikzset{hasse/.style={circle, fill,inner sep=2pt}}
\tikzset{shrinky/.style={circle, fill,inner sep=1pt}}
\tikzset{sized/.style={circle, draw, inner sep=1.5pt}}
\tikzset{seven/.style={circle, draw,inner sep=3pt}}

\tikzset{dotto/.style={circle, orange, draw,inner sep=1.5pt,fill=orange}}
\tikzset{dottp/.style={circle, purple, draw,inner sep=1.5pt,fill=purple}}
\tikzset{dottc/.style={circle, cyan, draw,inner sep=1.5pt,fill=cyan}}
\tikzset{dottr/.style={circle, red, draw,inner sep=1.5pt,fill=red}}
\tikzset{dottb/.style={circle, blue, draw,inner sep=1.5pt,fill=blue}}
\tikzset{dottg/.style={circle, green, draw,inner sep=1.5pt,fill=green}}
\tikzset{dottm/.style={circle, magenta, draw,inner sep=1.5pt,fill=magenta}}

\usepackage{cancel}
\newcommand{\xdownarrow}[1]{%
  {\left\downarrow\vbox to #1{}\right.\kern-\nulldelimiterspace}
}

\begin{document}

\begin{titlepage}

\phantom{wowiezowie}

\vspace{-1cm}

\begin{center}

{\Huge {\bf Remarks on the Higgs Branch}}

\bigskip

{\Huge {\bf of 5d Conformal Matter}}

\vspace{1cm}

{\Large  Mario De Marco,$^{\ast}$ Michele Del Zotto,$^{\dagger\ddagger\star}$}\\ 

\medskip

{\Large  Julius F. Grimminger,$^{\sharp}$ and Andrea Sangiovanni $^{\dagger\ddagger\star}$}\\

\vspace{1cm}

{\it
{\small

$^\dagger$ Mathematics Institute, Uppsala University, \\ Box 480, SE-75106 Uppsala, Sweden\\
\vspace{.25cm}
$^\star$ Centre for Geometry and Physics, Uppsala University, \\ Box 480, SE-75106 Uppsala, Sweden\\
\vspace{.25cm}
$^\ddagger$ Department of Physics and Astronomy, Uppsala University,\\ Box 516, SE-75120 Uppsala, Sweden\\
\vspace{.25cm}
$^{\sharp}$ Mathematical Institute, University of Oxford,\\Andrew Wiles Building, Woodstock Road, Oxford, OX2 6GG, UK\\
\vspace{.25cm}
$^{\ast}$  Physique Th\'eorique et Math\'ematique and International Solvay Institutes\\
Universit\'e Libre de Bruxelles, C.P. 231, 1050 Brussels, Belgium
}}

\vskip .5cm
{\footnotesize \tt mario.de.marco@ulb.be \hspace{1cm} michele.delzotto@math.uu.se } \\
{\footnotesize \tt    julius.grimminger@maths.ox.ac.uk \hspace{1cm} andrea.sangiovanni@math.uu.se}

\vskip 1cm
     	{\bf Abstract }
\vskip .1in

\end{center}

\noindent Among the elementary building blocks in the atomic classification of 5d SCFTs there are 5d bifundamental conformal matter theories of various kinds. In this work we study the Higgs branch of these models and of the corresponding molecules arising from their fusion. To this aim we use two complementary independent strategies. On the one hand for the type $A$ and $D$ conformal matter, we identify dual $(p,q)$ brane webs in IIB and exploit them to read off the corresponding magnetic quivers. On the other hand, we exploit circle reductions and study the resulting 4d $\mathcal N=2$ SCFTs, giving an alternative derivation of their Higgs branches which extend also to the $E$ types.

\eject

\end{titlepage}

\tableofcontents

\section{Introduction}

For a long time in the history of quantum fields it was believed that field theories in dimension greater than four were limited to systems of free fields. This was based on a Lagrangian prejudice and was proved wrong about 30 years ago with the discovery of interacting superconformal field theories (SCFTs) in 5 and 6 dimensions \cite{Witten:1995ex,Strominger:1995ac,Witten:1995em,Ganor:1996mu,Seiberg:1996qx,Seiberg1996,Morrison_1997,Douglas:1996xp}. These systems have been since then a constant source of inspiration throughout formal QFT and mathematics. An interesting pathway to study these models is to resort to various combinations of stringy techniques -- in particular string dualities and geometric engineering -- and field theoretical methods such as (twisted) compactifications. Very often a non-Lagrangian SCFT in higher dimensions can be thus related to Lagrangian field theories in lower dimensions that can be exploited to learn about features of their higher dimensional progenitor. This is the strategy we will adopt also in this work to learn about a recently introduced class of 5d SCFTs, giving the first examples of 5d Conformal Matter (CM) \cite{DeMarco:2023irn} -- 5d avatars of the 6d Conformal Matter \cite{DelZotto:2014hpa}, introduced to provide a 5d version of the atomic classification of 6d SCFTs and LSTs \cite{Heckman:2015bfa,Bhardwaj:2015oru,Bhardwaj:2019hhd}. Our main aim is to explore their Higgs branches via a concerted effort that involves studying RG flows, circle compactifications (identifying suitable 4d $\mathcal{N}=2$ systems along their extended KK Coulomb branch), as well as their dual fivebrane realizations to extract magnetic quivers. These techniques generate a coherent picture that passes several strict consistency checks operating across dimensions, and also serve to test new 5d dualities, clearly establishing the role of 5d Conformal Matter theories in the 5d SCFT landscape. 

\bigskip

After the initial spark provided by \cite{Ganor:1996mu,Seiberg:1996qx,Seiberg1996,Ganor:1996pc,Douglas:1996xp,Aharony:1997ju,Aharony:1997bh,DeWolfe:1999hj,Morrison_1997,Intriligator_1997,leung1997branes}, 5d $\mathcal{N}=1$ SCFTs have received a great deal of attention in the recent years: this renewed endeavour has fostered steadfast progress in their classification, relying on geometric approaches involving M-theory setups reduced on a non-compact canonical Calabi-Yau threefold\footnote{Oftentimes, it is assumed that a suitable complete Calabi-Yau metric exists for canonical CY3. This is a subtle assumption that requires great care -- see e.g. \cite{Acharya:2024bnt} for a recent work examining this aspect.} (CY3) \cite{Xie:2017pfl,Jefferson:2017ahm,
Jefferson:2018irk,
Apruzzi:2019opn,
Apruzzi_2020,
Closset:2020scj,
Closset:2020afy,
Closset:2021lwy,
Collinucci:2021ofd,
Collinucci:2021wty,
DeMarco:2021try,
Tian:2021cif,
Closset:2022vjj,
Collinucci:2022rii,
DeMarco:2022dgh,
Closset:2023pmc,
Mu:2023uws,
Bourget:2023wlb}, $(p,q)$-web constructions \cite{Benini:2009gi,Bergman:2013aca,Zafrir:2014ywa,Hayashi:2015zka,Hayashi:2015fsa,Bergman:2015dpa,Hayashi:2018lyv,Hayashi:2018bkd,Hayashi:2019yxj,Hayashi_2020,Bergman:2020myx},
as well as dimensional reduction from 6d SCFTs realized via F-theory on elliptic CY3 \cite{DelZotto:2017pti,Bhardwaj:2018yhy,Bhardwaj:2018vuu,Bhardwaj:2019xeg,Bhardwaj:2020gyu}. These are clearly interconnected technologies that completely agree on realms where they overlap. Eyes have been particularly set on characterizing the Coulomb and Higgs branches of the 5d SCFTs of interest: while the former can be analyzed with relatively straightforward methods,\footnote{E.g., in the context of M-theory backgrounds on CY3 $\times \mathbb{R}^{4,1}$, by performing a crepant resolution of the singularities of CY3.} in the context of geometric engineering Higgs branches receive quantum corrections due to M2-brane instantons, thus essentially hindering their understanding from an M-theory perspective. In order to overcome such difficulties, it is useful to resort to string-theoretic dualities. It turns out that 5d Higgs branches, which are typically singular hyper-K\"ahler cones, can be conveniently interpreted as Coulomb branches of suitably chosen 3d $\mathcal{N}=4$ theories called magnetic quivers. The quantum corrections to these Coulomb branches, due to monopole operators, are under control \cite{Cremonesi_2014} and are hence vastly more amenable to an explicit examination. Arguably the most successful recipe to extract magnetic quivers is via brane webs \cite{Cremonesi:2015lsa,Ferlito:2016grh,Ferlito:2017xdq,Cabrera:2018ann,Cabrera:2018jxt,Cabrera:2019izd,Bourget:2019aer,Bourget:2019rtl,Cabrera:2019dob,Grimminger:2020dmg,Bourget:2020gzi,Bourget:2020asf,Akhond:2020vhc,Bourget:2020mez,VanBeest:2020kxw}, if available. This is the case for all 5d Conformal Matter theories of types $A$ and $D$. Building on the gauge theory phases identified in \cite{DeMarco:2023irn}, in this work we give explicit $(p,q)$ web realizations for all of these systems, thus establishing further dictionaries between networks of branes and M-theory singularities. As a byproduct of this analysis, we obtain magnetic quivers for all 5d CM theories of types $A$ and $D$. But what about the 5d CM theories of type $E$? In order to determine the corresponding Higgs branches we exploit a different strategy, a circle compactification to 4d. On the one hand, this gives a nice independent consistency check on the magnetic quiver results, on the other hand this allows to obtain information about the exceptional 5d CM theories.

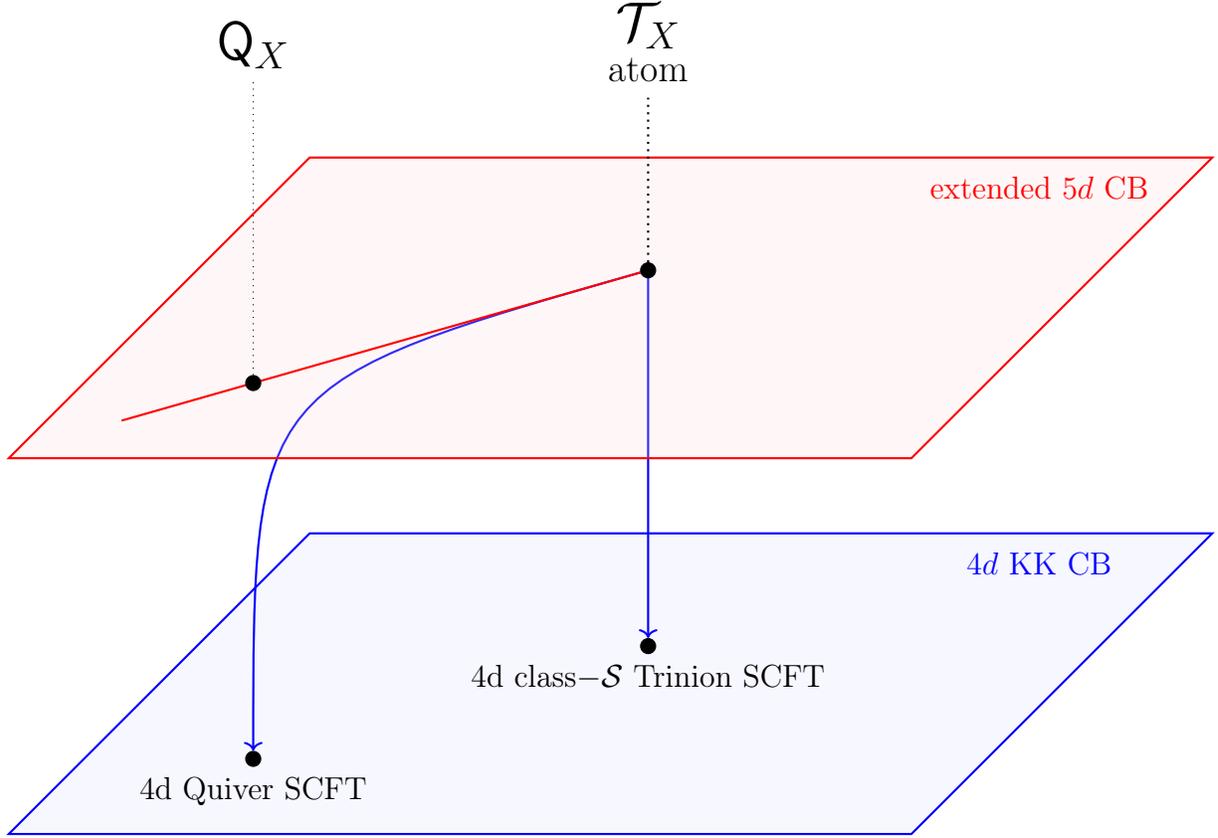
\begin{figure}
    \centering
    \begin{tikzpicture}
        \draw[color=blue,thick,fill=blue!15, fill opacity=0.2] (-6-2.5,-2.5-5)--(6-2.5,-2.5-5)--(6+1.5,1.5-5)--(-6+1.5,1.5-5)--cycle;
        \node[blue] at (6+1.5-2.3,1.5-0.4-5) {$4d$ KK CB};
        \node[bd,label=below:{4d class$-\mathcal{S}$ Trinion SCFT}] (1d) at (0,-5) {};
        \node[bd,label=below:{4d Quiver SCFT}] (2d) at (-21/4,-1.5-5) {};
        \draw[blue,->,thick] (0,0)--(1d);
        \draw[blue,->,thick] (0,0) .. controls (-21/4,-1.5) .. (2d);
        \draw[color=red,thick,fill=red!15, fill opacity=0.2] (-6-2.5,-2.5)--(6-2.5,-2.5)--(6+1.5,1.5)--(-6+1.5,1.5)--cycle;
        \node[red] at (6+1.5-2.3,1.5-0.4) {extended $5d$ CB};
        \node[bd] (1) at (0,0) {};
        \draw[red,thick] (1)--(-7,-2);
        \node[bd] (2) at (-21/4,-1.5) {};
        \node (1t) at (0,3) {\LARGE$\underset{\textnormal{atom}}{\mathcal{T}_X}$};
        \draw[dotted,thick] (1)--(1t);
        \node (2t) at (-21/4,3) {\LARGE$\mathsf{Q}_X$};
        \draw[dotted] (2)--(2t);
    \end{tikzpicture}
    \caption{Illustration of toroidal compactifications of a 5d conformal matter atom. The red shaded area is a schematic depiction of the 5d extended Coulomb branch (including mass deformations) of the SCFT $\mathcal{T}_X$. The red line emanating from the origin of the CB corresponds to the mass deformation of the SCFT to the Dynkin quiver phase. Blue arrows depict RG flows to 4 dimensions. The area shaded in blue is a schematic depiction of the 4d KK Coulomb branch. The atom $\mathcal{T}_X$ can be directly compactified to a 4d class$-\mathcal{S}$ trinion SCFT, which we may call 4d class$-\mathcal{S}$ atom. Turning on the mass deformation going to the left it can also be compactified to a 4d Lagrangian SCFT. For an atom we have $\underset{\textnormal{atom}}{\mathcal{T}_X}=\tilde{\mathsf{Q}}_X$.}
    \label{fig:InterdimensionalFlowAtom}
\end{figure}

\begin{figure}
    \centering
    \begin{tikzpicture}
        \draw[color=blue,thick,fill=blue!15, fill opacity=0.2] (-6-2.5,-2.5-5)--(6-2.5,-2.5-5)--(6+1.5,1.5-5)--(-6+1.5,1.5-5)--cycle;
        \node[blue] at (6+1.5-2.3,1.5-0.4-5) {$4d$ KK CB};
        \node[bd,label=below:{4d SCFT}] (1d) at (0,-5) {};
        \node[bd,label=below:{4d Quiver SCFT}] (2d) at (-21/4,-1.5-5) {};
        \node[bd,label=below:{4d class$-\mathcal{S}$ SCFT}] (3d) at (9/4,-1.5-5) {};
        \draw[dashed,blue,->,thick] (0,0)--(1d);
        \draw[blue,->,thick] (0,0) .. controls (-21/4,-1.5) .. (2d);
        \draw[blue,->,thick] (0,0) .. controls (9/4,-1.5) .. (3d);
        \draw[color=red,thick,fill=red!15, fill opacity=0.2] (-6-2.5,-2.5)--(6-2.5,-2.5)--(6+1.5,1.5)--(-6+1.5,1.5)--cycle;
        \node[red] at (6+1.5-2.3,1.5-0.4) {extended $5d$ CB};
        \node[bd] (1) at (0,0) {};
        \draw[red,thick] (1)--(-7,-2);
        \draw[red,thick] (1)--(3,-2);
        \node[bd] (2) at (-21/4,-1.5) {};
        \node[bd] (3) at (9/4,-1.5) {};
        \node (1t) at (0,3) {\LARGE$\underset{\textnormal{molecule}}{\mathcal{T}_X}$};
        \draw[dotted,thick] (1)--(1t);
        \node (2t) at (-21/4,3) {\LARGE$\mathsf{Q}_X$};
        \draw[dotted,thick] (2)--(2t);
        \node (3t) at (9/4,3) {\LARGE$\tilde{\mathsf{Q}}_X$};
        \draw[dotted,thick] (3)--(3t);
    \end{tikzpicture}
    \caption{Illustration of toroidal compactifications of a 5d conformal matter molecule. The red shaded area is a schematic depiction of the 5d extended Coulomb branch (including mass deformations) of the SCFT $\mathcal{T}_X$. The red line emanating from the origin of the CB to the left corresponds to the mass deformation of the SCFT to the Dynkin quiver phase. The red line emanating to the right corresponds to the mass deformation of the SCFT to the `generalized quiver' phase. Blue arrows depict RG flows to 4 dimensions. The area shaded in blue is a schematic depiction of the 4d KK Coulomb branch. The atom $\mathcal{T}_X$ cannot be directly compactified to a 4d class$-\mathcal{S}$ theory, and we don't know a good description of this theory in 4d. Turning on the mass deformation going to the left the theory can be compactified to a 4d Lagrangian SCFT. Turning on the mass deformation going to the right the theory can be compactified to a 4d class$-\mathcal{S}$ theory which is the gauging of 4d class$-\mathcal{S}$ atoms.}
    \label{fig:InterdimensionalFlowMolecule}
\end{figure}
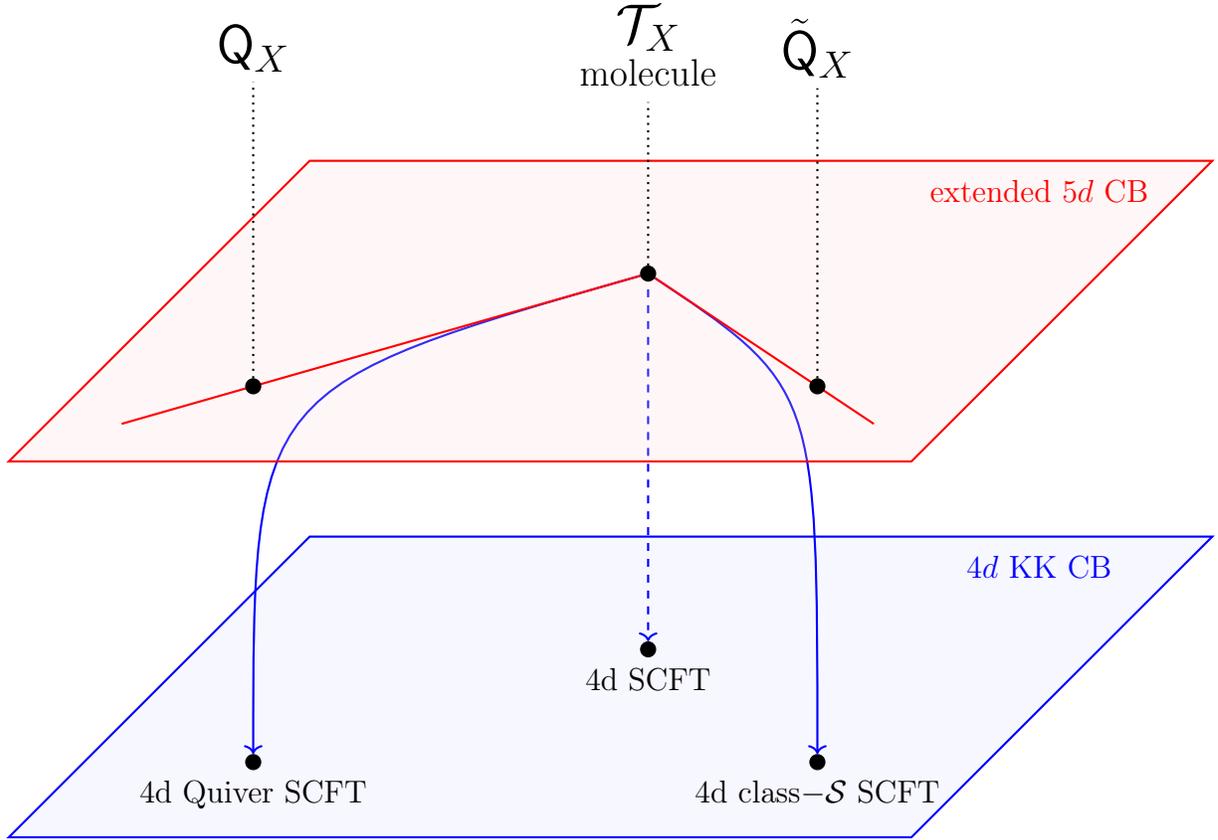

The 5d CM theories can be organized in two broad families in the atomic classification spirit: we have 5d CMs  that serve as atoms, and molecules that can be interpreted as fusions of several 5d CM systems. We denote these 5d SCFTs with $\mathcal T_X$, where $X$ is the CY3 employed to engineer them in M-theory. Fusion is a 5d process which is reminiscent of the 6d fusion \cite{Heckman:2018pqx} (a generalization of the process of conformal gauging in 4d to higher dimensional systems). We say that a 5d SCFT is a fusion of several atoms whenever along the 5d Coulomb branch of the given system, we find a non-Lagrangian gauge theory phase that can be interpreted as a generalized linear quiver that has CM atoms as edges and gauge nodes of type $\mathfrak{g}$. We denote such gauge theory phase $\widetilde{\mathsf{Q}}_X$. For a generalized bifundamental atom of type $\mathfrak{g}$, $\widetilde{\mathsf{Q}}_X = \mathcal T_X$ since by definition this theory does not have any gauge nodes of type $\mathfrak{g}$ \cite{DeMarco:2023irn}. All bifundamental CM theories also have a Lagrangian gauge theory phase, corresponding to a suitable decoration of the Dynkin quiver of type $\mathfrak{g}$ by $SU(N)$ gauge nodes, which we denote $\mathsf{Q}_X$. These two gauge theory phases, being points on the extended Coulomb branch of the same 5d SCFT, are related by a 5d duality. The behavior of 5d atoms and 5d molecules upon circle reduction is slightly different and we summarize it in Figures \ref{fig:InterdimensionalFlowAtom} and \ref{fig:InterdimensionalFlowMolecule}. In the figures we represent in red the extended 5d CB, and in blue the 4d KK Coulomb branch. Along the 4d KK Coulomb branch there are special loci, analogous to Argyres-Douglas points \cite{Argyres:1995jj}, where a 4d $\mathcal N=2$ SCFT arises. It is of particular interest to determine the point which corresponds to the circle reduction of the undeformed 5d SCFT. We denote $D^\circ_{S^1}\mathcal T_X$ the resulting 4d $\mathcal N=2$ SCFT. In the case of the circle reduction of the 5d CM atoms we consider in this paper the result is a class-$\mathcal{S}$ trinion, that we can determine uniquely thanks to the extensive work on their classification \cite{Chacaltana:2010ks,Chacaltana:2011ze,Chacaltana:2012zy,Chacaltana:2012ch,Chacaltana:2013oka,Chacaltana:2014jba,Chacaltana:2015bna,Chacaltana:2016shw,Chacaltana:2017boe,Chacaltana:2018vhp}. This identification also determines the corresponding Higgs branch, as the latter does not receive correction upon circle compactification. In the case of the circle reduction of the 5d CM molecules $D^\circ_{S^1}\mathcal T_X$ is no longer arising from regular punctures in class-$\mathcal S$ -- see Figure \ref{fig:InterdimensionalFlowMolecule}. The class-$\mathcal S$ avatar we find along the 4d KK Coulomb branch is a circle reduction of the gauge theory phase $\widetilde{\mathsf{Q}}_X$. We can still use this result to provide a conjecture about the dimension of the Higgs branch of $D^\circ_{S^1}\mathcal T_X$, which is confirmed by the magnetic quiver analysis in types $A$ and $D$. In particular, this predicts the existence of ADE families of generalized bifundamental 4d $\mathcal N=2$ SCFTs with flavor symmetry $\mathfrak{g}\times \mathfrak{g} \times F_{X}$ with $\mathfrak{g}\in ADE$ and $F_X$ an enhanced flavor symmetry we can explicitly determine that cannot be constructed via regular punctures in class-$\mathcal S$.

\bigskip

This work is organized as follows. In Section \ref{sec: atomic classification} we review the construction of 5d conformal matter atoms and molecules of \cite{DeMarco:2023irn}. In Section \ref{sec:classS} the $S^1$ reduction of 5d conformal matter atoms to $4d$ $\mathcal{N}=2$ class-$\mathcal{S}$ fixtures is discussed. In Section \ref{sec:5dAtomsBWandMQ} the brane web constructions of $A$- and $D$-type 5d conformal matter atoms are presented and their magnetic quivers are derived, testing, as a by-product, the costruction in Section \ref{sec:classS}. In Section \ref{sec:moleculesbigsec} the $S^1$ reduction of 5d conformal matter molecules is addressed; by turning on specific masses for the 5d conformal matter molecule the $S^1$ reduction leads to a $4d$ $\mathcal{N}=2$ class-$\mathcal{S}$ theory, which is provided. In Section \ref{sec:MoleculesBWandMQ} the brane web constructions of $A$- and $D$-type 5d conformal matter molecules are presented and their magnetic quivers are derived, furthermore a precise description of the mass deformation of Section \ref{sec:moleculesbigsec} is described.
Section \ref{sec:Conclusions} serves as the conclusion and provides an outlook on future work. Appendix \ref{app:QuiverPhases} summarises specific low energy quiver descriptions of 5d conformal matter atoms and molecules, which were provided in \cite{DeMarco:2023irn}. Appendix \ref{appendix A} addresses the question of identifying Higgs branch moduli of 5d a $\mathcal{N}=1$ SCFT from the geometry of a local CY3 which engineers it. In Appendix \ref{Appendix B} Higgsing is explored via flavour symmetry. In Appendix \ref{app:QuiverBWNotation} our notation for quivers and brane webs is reviewed.

\bigskip

We conclude this section by summarizing the notation used in this paper in Table \ref{tab:notation} for the convenience of our readers.

\renewcommand{\arraystretch}{1.4}
\begin{table}\centering
\begin{equation}
\begin{array}{|c|l|}
\hline
\hline
\mathcal{T}_{5d} \text{ or } \mathcal{T}_{X_{\mathfrak{g}}^{(1^{n_1},2^{n_2},3^{n_3})}}  & \makecell[l]{\text{The 5d SCFT engineered by a conformal matter}\\ \text{atom or molecule.}}\\
\hline
\mathsf{Q}_{X_{\mathfrak{g}}^{(1^{n_1},2^{n_2},3^{n_3})}} & \makecell[l]{\text{The 5d low-energy Dynkin quiver phase corresponding to}\\ \text{our preferred choice of resolution, outlined in \cref{sec:basechangeres}.}} \\
\hline
\tilde{\mathsf{Q}}_{X_{\mathfrak{g}}^{(1^{n_1},2^{n_2},3^{n_3})}} & \makecell[l]{\text{The generalized 5d conformal matter quiver that directly}\\ \text{descends to some class-$\mathcal{S}$ theory.}\\ \text{This coincides with $\mathcal{T}_{X_{\mathfrak{g}}^{(1^{n_1},2^{n_2},3^{n_3})}}$ \textit{only for atoms}.}} \\
\hline
\mathcal{T}_{4d} & \makecell[l]{\text{The 4d class-$\mathcal{S}$ theory that descends from $\tilde{\mathsf{Q}}_{X_{\mathfrak{g}}^{(1^{n_1},2^{n_2},3^{n_3})}}$}.}\\
\hline
\mathsf{MQ}(...) & \makecell[l]{\text{Magnetic quiver of `$...$' some SQFT.}}\\
\hline
\mathrm{HB}(...) & \makecell[l]{\text{Higgs branch of `$...$' some SQFT.}}\\
\hline
\mathrm{CB}_{5d}(...) & \makecell[l]{\text{5d Coulomb branch of `$...$' some 5d $\mathcal{N}=1$ SQFT.}}\\
\hline
\mathrm{CB}_{4d}(...) & \makecell[l]{\text{4d Coulomb branch of `$...$' some 4d $\mathcal{N}=2$ SQFT.}}\\
\hline
\mathrm{CB}_{3d}(...) & \makecell[l]{\text{3d Coulomb branch of `$...$' some 3d $\mathcal{N}=4$ SQFT.}}\\
\hline

\hline
\hline
\end{array}\nonumber
\end{equation}
\caption{Notation chosen for the present work.}\label{tab:notation}
\end{table}

\section{Atomic classification of 5d conformal matter theories}\label{sec: atomic classification}

In this section we summarise key features of 5d conformal matter theories, which were recently introduced in \cite{DeMarco:2023irn}.\\

\indent These are interacting 5d SCFTs, displaying at least $\mathfrak{g}\times \mathfrak{g}$ flavor symmetry, with $\mathfrak{g} \in ADE$. Further, they all admit a low-energy gauge theory phase, which can be employed to explicitly glean the UV flavor symmetry thanks to the techniques of \cite{Yonekura}. \\

\subsection{Atoms of 5d conformal matter}
\indent Concretely, 5d conformal matter bifundamental theories are obtained via M-theory geometric engineering on a specific class of threefolds with canonical non-isolated singularities. The starting point of our construction are the ADE (or Du Val) surface singularities:
\begin{equation}
P_{\mathfrak g}(x_1,x_2,x_3) = 0 \qquad \mathfrak g \in ADE,
\end{equation}
where we are using the conventional form:\\
\begin{equation}\label{ADE sing}
 \begin{cases}
P_{A_k}(x_1,x_2,x_3) = x_1^2+x_2^2-x_3^{k+1} \\
P_{D_k}(x_1,x_2,x_3) = x_1^2 +x_3x_2^2+x_3^{k-1} \\
P_{E_6}(x_1,x_2,x_3) =x_1^2+x_2^3+x_3^4 \\
P_{E_7}(x_1,x_2,x_3) = x_1^2+x_2^3+x_2x_3^3 \\
P_{E_8}(x_1,x_2,x_3) = x_1^2+x_2^3+x_3^5,
\end{cases}
\end{equation}
Replacing one of the $x_i \rightarrow uv$ we get a threefold:
\begin{equation}
\label{eq:threefoldeqbifundamental}
X^{(i)}_{\mathfrak g} \quad \colon \qquad \begin{cases}P_{\mathfrak g}(x_1,x_2,x_3) = 0 \\ x_i = uv  \end{cases}.
\end{equation}
Notice that the threefold $X^{(i)}_{\mathfrak g}$ possesses two non-compact complex lines of singularities of type $\mathfrak{g}$, on top of $u=0$ and $v=0$ respectively, intersecting at the origin $x_1 = x_2 = x_3 = u = v = 0$. The central claim of \cite{DeMarco:2023irn} can then be phrased as:\\

\indent \textit{M-theory reduced on the singularities $X^{(i)}_{\mathfrak{g}}$ geometrically engineers interacting 5d SCFTs with at least $\mathfrak{g}\times\mathfrak{g}$ flavor symmetry. We call these theories the fundamental ``atoms'' of 5d conformal matter.}\\

One can directly prove this claim by explicitly performing a crepant resolution of the singularities $X^{(i)}_{\mathfrak{g}}$: it turns out that one such resolution always exists, and that the 5d SCFT admits a low-energy quiver gauge theory phase, that can readily be inferred from the resolution procedure.
A list of the low-energy quiver gauge theories is provided in Appendix \ref{app:QuiverPhases} directly taken from \cite{DeMarco:2023irn}.
The crucial consequence of this analysis is:\\

\indent \textit{For each $\mathfrak{g}\in ADE$ there exist inequivalent atoms of 5d conformal matter, corresponding to the choice of $i$ in $X_{\mathfrak{g}}^{(i)}$}.\\

``Inequivalent'' means that they engineer different 5d SCFTs, which
have different low-energy quiver gauge theory descriptions.

Only some of the 5d conformal matter atoms are dimensional reductions of 6d conformal matter atoms. All remaining 5d CM atoms are Higgsings thereof, as we will show in \cite{AtomMoleculeHybridupcoming}.

We may represent a single ``atom'' of 5d conformal matter as a \textit{generalized quiver}, like in Figure \ref{fig:atom}, where the two $\mathfrak{g}$ flavor factors are highlighted. Notice that the edge \emph{is} an interacting 5d SCFT\footnote{In comparison a conventional edge in a quiver corresponding to a hypermultiplet may be considered a free SCFT.},
and thus it is labelled by $(\mathfrak{g},\mathfrak{g})_{(i)}$, with $i$ one among $1,2,3$. We denote these generalised quivers as $\tilde{\mathsf{Q}}_{X_{\mathfrak{g}}^{(i)}}$.

 \begin{figure}[H]
       \centering
$   \begin{array}{cc}
\makecell{\tilde{\mathsf{Q}}_{X_{\mathfrak{g}}^{(i)}} =\\ \hspace{1cm}}  &\scalemath{1}{\begin{tikzpicture}
          \draw[thick] (0.5,0.5)--(1.5,0.5)--(1.5,-0.5)--(0.5,-0.5)--cycle;
          \draw[thick, double] (1.6,0)--(2.7,0);
        \draw[thick] (2.8,0.5)--(3.8,0.5)--(3.8,-0.5)--(2.8,-0.5)--cycle;
        \node at (1,0) {\footnotesize$\mathfrak{g}$};
        \node at (3.3,0) {\footnotesize$\mathfrak{g}$};
        \node at (2.15,0.3) {\footnotesize$(\mathfrak{g},\mathfrak{g})_{(i)}$};
        \end{tikzpicture}}
        \end{array}$
        \caption{Generalised quiver representing an atom of 5d conformal matter of type $(\mathfrak{g},\mathfrak{g})_{(i)}$.}
        \label{fig:atom}
    \end{figure}
In general, the 5d conformal matter SCFT atom in Figure \ref{fig:atom} has the following properties:
\begin{itemize}
    \item The SCFT is specified by the edge $(\mathfrak{g},\mathfrak{g})_{(i)}$. One can also represent the atom via an ordinary quiver (namely, a quiver where the edges are genuine bifundamental hypermultiplets) dual to Figure \ref{fig:atom}, corresponding to the low-energy gauge theory phase arising from a specific choice of resolution (that will be reviewed in section \eqref{sec:basechangeres}) of $X_{\mathfrak{g}}^{(i)}$. 
    We denote these quivers as $\mathsf{Q}_{X_{\mathfrak{g}}^{(i)}}$.

    E.g.\ in the case $X_{E_6}^{(2)}$, one obtains the ordinary quiver of Figure \ref{E6yfig}, which is an $E_6$ Dynkin quiver.
        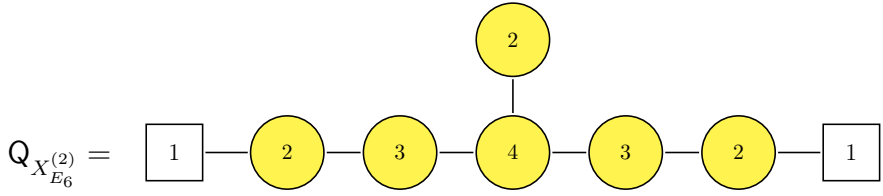
\begin{figure}[H]
    \centering
$   \begin{array}{cc}
\makecell{\mathsf{Q}_{X_{E_6}^{(2)}} =\\ \hspace{1cm}}  &  \scalebox{0.75}{
    \begin{tikzpicture}
        \draw[thick,fill=yellow!80] (0,0) circle (0.65);
        \node at (0,0) {\small$4$};
        \draw[thick] (0.7,0)--(1.3,0);
        \draw[thick,fill=yellow!80] (2,0) circle (0.65);
        \node at (2,0) {\small$3$};
        \draw[thick] (-0.7,0)--(-1.3,0);
        \draw[thick,fill=yellow!80] (-2,0) circle (0.65);
        \node at (-2,0) {\small$3$};
        \draw[thick] (0,0.7)--(0,1.3);
        \draw[thick,fill=yellow!80] (0,2) circle (0.65);
        \node at (0,2) {\small$2$};
        \draw[thick] (2.7,0)--(3.3,0);
        \draw[thick,fill=yellow!80] (4,0) circle (0.65);
        \node at (4,0) {\small$2$};
        \draw[thick] (-2.7,0)--(-3.3,0);
         \draw[thick,fill=yellow!80] (-4,0) circle (0.65);
        \node at (-4,0) {\small$2$};
        \draw[thick] (4.7,0)--(5.45,0);
             \draw[thick] (-4.7,0)--(-5.45,0);
                    \draw[thick] (5.5,-0.5)--(6.5,-0.5)--(6.5,0.5)--(5.5,0.5)--cycle ;
        \node at (6,0) {1};
               \draw[thick] (-5.5,-0.5)--(-6.5,-0.5)--(-6.5,0.5)--(-5.5,0.5)--cycle ;
        \node at (-6,0) {1};
        \end{tikzpicture}} \\
         \end{array}$
    \caption{Low-energy Dynkin quiver description for the $X_{E_6}^{(2)}$ 5d conformal matter theory. Yellow gauge nodes represent special unitary factors in the gauge group.}
    \label{E6yfig}
    \end{figure}
   
    See \cite{DeMarco:2023irn} and Appendix \ref{app:QuiverPhases} for the explicit Dynkin quiver realizations for all algebras.
    \item The flavor symmetry in the UV is\footnote{In this work we avoid discussions on the global form of the symmetry groups.}:
    \begin{equation*}
        F_{UV} = \mathfrak{g}\times \mathfrak{g} \times F_{\text{rest}},
    \end{equation*}
    with $F_{\text{rest}}$ a product of $\mathfrak{su}$ and $\mathfrak u(1)$ factors, with known rank. $F_{UV}$ can be computed employing the techniques of \cite{Yonekura} applied to the Dynkin quiver in Figure \ref{E6yfig}.
\end{itemize}

\subsection{Summary of the resolution procedure}
\label{sec:basechangeres}
\begin{figure}[H]
\centering
 \scalebox{0.9}{
    \begin{tikzpicture}
        \draw[thick] (-2.5,-3)--(2.5,3);
        \draw[thick] (-2.5,3)--(2.5,-3);

        \draw[thick,dashed,->] (-1,3.5) to [out=270, in = 40] (-1.6,2.2);
        \draw[thick,dashed,->] (1,3.5) to [out=270, in = 140] (1.6,2.2);
        \draw[thick,dashed,<-] (0.5,0)--(1.5,0);
        \draw[fill=red] (0,0) circle (0.13);
        \node at (-2.2,4.9) {$\overbrace{\hspace{3cm}}^{\text{smooth}}$};
        \draw (-3.4,3.8) circle (0.15);
        \draw[thick] (-3.2,3.8)--(-3.0,3.8);
        \draw (-2.8,3.8) circle (0.15);
        \draw[thick] (-2.6,3.8)--(-2.4,3.8);
        \draw (-2.2,3.8) circle (0.15);
        \draw[thick] (-2,3.8)--(-1.8,3.8);
        \draw (-1.6,3.8) circle (0.15);
        \draw[thick] (-1.4,3.8)--(-1.2,3.8);
        \draw (-1,3.8) circle (0.15);
        \draw[thick] (-2.2,4.0)--(-2.2,4.2);
        \draw (-2.2,4.4) circle (0.15);
        \node at (2.2,4.9) {$\overbrace{\hspace{3cm}}^{\text{smooth}}$};
        \draw (3.4,3.8) circle (0.15);
        \draw[thick] (3.2,3.8)--(3.0,3.8);
        \draw (2.8,3.8) circle (0.15);
        \draw[thick] (2.6,3.8)--(2.4,3.8);
        \draw (2.2,3.8) circle (0.15);
        \draw[thick] (2,3.8)--(1.8,3.8);
        \draw (1.6,3.8) circle (0.15);
        \draw[thick] (1.4,3.8)--(1.2,3.8);
        \draw (1,3.8) circle (0.15);
        \draw[thick] (2.2,4.0)--(2.2,4.2);
        \draw (2.2,4.4) circle (0.15);
          \node at (3.2,-0.6) {$\underbrace{\hspace{3cm}}_{\text{singular}}$};
        \draw (4.4,0) circle (0.15);
        \draw[thick] (4.2,0)--(4,0);
        \draw (3.8,0) circle (0.15);
        \draw[thick] (3.6,0)--(3.4,0);
        \draw (3.2,0) circle (0.15);
        \draw[thick] (3,0)--(2.8,0);
        \draw (2.6,0) circle (0.15);
        \draw[thick] (2.4,0)--(2.2,0);
        \draw (2,0) circle (0.15);
        \draw[thick] (3.2,0.2)--(3.2,0.4);
        \draw (3.2,0.6) circle (0.15);
        \node[rotate=49] at (-1.5,-2.2) {$\boldsymbol{u=0}$};
        \node[rotate=-49] at (1.5,-2.2) {$\boldsymbol{v=0}$};
        \end{tikzpicture}}
    \caption{Pictorial representation of $\widetilde{X}_{E_6}^{(i)}$. The red point corresponds to the origin and its preimage under $\epsilon$ consists of a collection of singular $\mathbb{P}^1$'s, arranged like a $E_6$ Dynkin diagram.}
     \label{fig:E6firstblowup}
\end{figure}
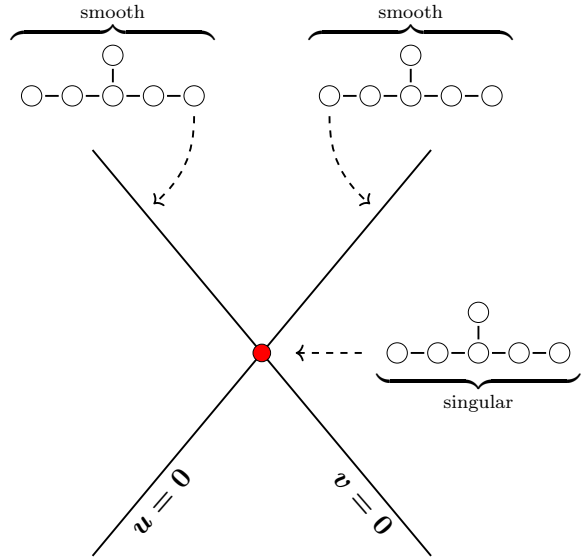

We quickly recall, for later reference, the recipe to obtain the specific resolution performed in \cite{DeMarco:2023irn}: 
\begin{enumerate}
    \item Perform the resolution of the $Y_{\mathfrak g}$ Du Val \textit{surface} singularity displayed in \eqref{ADE sing}, obtaining the following expressions: 
    \begin{equation}
        \label{eq:resfrombasechange}
        \begin{cases}
        x_{1} = x_1(a_j,b_j),  \\
        x_{2} = x_{2}(a_j,b_j), \\
        x_3 = x_3(a_j,b_j),
        \end{cases}
    \end{equation}
    where $(a_j,b_j)  \in \mathbb C^2$ form an atlas of  the \textit{resolved} Du Val surface and the functions $x_i(a_j,b_j)$ describes the blowdown map;
    \item substitute $x_i = u v$ inside \eqref{eq:resfrombasechange}, obtaining a threefold $\widetilde{X}_{\mathfrak g}^{(i)}$, that partially resolves $X_{\mathfrak g}^{(i)}$. We note that, at this point, the remaining equations $x_{k} = x_k(a_j,b_j)$, with $k \neq i$, are spurious and can be dropped. 
\end{enumerate}
This procedure fully resolves the singularities outside the origin  of \eqref{eq:threefoldeqbifundamental} and corresponds to lifting to the \textit{resolved} Du Val surface $\widetilde{Y}_{\mathfrak{g}}$ the base-change that we used to obtain $X_{\mathfrak g}^{(i)}$ from the \textit{singular} Du Val surface $Y_{\mathfrak g}$. We call  $\widetilde{X}_{\mathfrak g}^{(i)}$ the partially resolved threefold obtained in this way and $\varepsilon$ the partial resolution map 
\begin{equation}
\label{eq:partialresmap}
    \varepsilon: \quad \widetilde{X}_{\mathfrak g}^{(i)} \longrightarrow X_{\mathfrak g}^{(i)}.
\end{equation}
 The singular locus of $\widetilde{X}_{\mathfrak g}^{(i)}$ lies within the the subvariety $\varepsilon^{-1}(0)$ that is blown-down  to the origin. $\varepsilon^{-1}(0)$ is collection of $\mathbb{P}^1_i$'s, with $i = 1, \ldots, \text{rank}(\mathfrak g)$. Each of these $\mathbb P^1_i$'s supports a line of singularities of type $A_{r_i}$ with $r_i \geq 0$, from which we can read off the quivers obtained in \cite{DeMarco:2023irn}.  We pictorially represent the geometry of $\widetilde{X}_{\mathfrak g}^{(i)}$ in Figure \ref{fig:E6firstblowup}, for the $\mathfrak{g} = E_6$ example: we copied the \textit{resolved} $Y_{\mathfrak g}$ singularity along the two non-compact lines $u = 0$ and $v = 0$. Over each point of these lines we have a collection of $\text{rank}(\mathfrak g)$ $\mathbb P^1$'s that intersect each other according to the $\mathfrak g$ Dynkin diagram and that can be ``translated''\footnote{More precisely, their normal bundle contains a $\mathcal{O}(0)$ summand.} along the considered line. These two collections of ``movable'' $\mathbb P^1$'s fuse together on the $\varepsilon^{-1}(0)$, creating a singular collection of $\mathbb P^1$'s. 

A key feature of 5d conformal matter theories is that, for fixed $\mathfrak{g}$, there come physically inequivalent ``species'', labelled by $(i)$, displaying different CB dimensions and different flavor groups. Geometrically, this comes from the fact that the resolution maps $x_{i}(a_j,b_j)$ differs from $x_k(a_j,b_j)$ for $k \neq i$. 

\subsection{Molecules of 5d conformal matter}
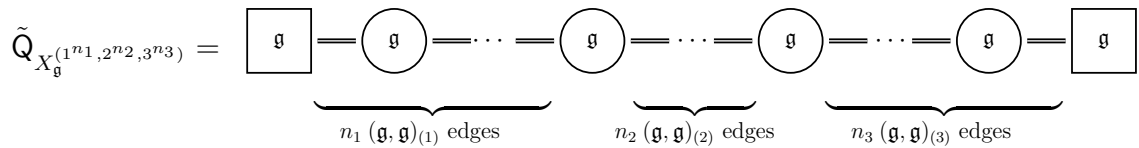
\begin{figure}[H]
       \centering
     
      $   \begin{array}{cc}
\makecell{\tilde{\mathsf{Q}}_{X_{\mathfrak{g}}^{(1^{n_1},2^{n_2},3^{n_3})}} =\\ \vspace{2cm}}  & 
\scalemath{0.85}{\begin{tikzpicture}
        \draw[thick] (0,0) circle (0.5);
        \node at (0,0) {\footnotesize$\mathfrak{g}$};
        \draw[thick, double] (0.6,0)--(1.2,0);
        \draw[thick, double] (-0.6,0)--(-1.2,0);
        \node at (0.9,0.3) {};
        \node at (-0.9,0.3) {};
          \node at (1.6,0) {$\cdots$};
          \node at (-1.6,0) {$\cdots$};
          \node at (4.7,0) {$\cdots$};
          \draw[thick, double] (1.9,0)--(2.5,0);
          \draw[thick, double] (-1.9,0)--(-2.5,0);
          \node at (2.2,0.3) {};
          \node at (-2.2,0.3) {};
          \draw[thick] (3.1,0) circle (0.5);
          \draw[thick] (-3.1,0) circle (0.5);
          \draw[thick, double] (3.7,0)--(4.3,0);
          \draw[thick, double] (-3.7,0)--(-4.3,0);
          \node at (4.0,0.3) {};
          \node at (-4.0,0.3) {};
           \draw[thick, double] (5.0,0)--(5.6,0);
           \node at (5.3,0.3) {};
           \draw[thick] (6.2,0) circle (0.5);
           \draw[thick, double] (6.8,0)--(7.4,0);
           \node at (7.1,0.3) {};
        \draw[thick] (7.5,0.5)--(8.5,0.5)--(8.5,-0.5)--(7.5,-0.5)--cycle;
        \draw[thick] (-4.4,0.5)--(-5.4,0.5)--(-5.4,-0.5)--(-4.4,-0.5)--cycle;
        \node at (3.1,0) {\footnotesize$\mathfrak{g}$};
        \node at (-3.1,0) {\footnotesize$\mathfrak{g}$};
        \node at (8,0) {\footnotesize$\mathfrak{g}$};
        \node at (-4.9,0) {\footnotesize$\mathfrak{g}$};
        \node at (6.2,0) {\footnotesize$\mathfrak{g}$};
       \node at (-2.5,-1.3) {$\underbrace{\hspace{3.7cm}}_{\scalebox{0.85}{$n_1 \hspace{0.1cm} (\mathfrak{g},\mathfrak{g})_{(1)}$ \text{edges}}}$};
       \node at (1.6,-1.3) {$\underbrace{\hspace{1.9cm}}_{\scalebox{0.85}{$n_2 \hspace{0.1cm}  (\mathfrak{g},\mathfrak{g})_{(2)}$ \text{edges}}}$};
       \node at (5.5,-1.3) {$\underbrace{\hspace{3.7cm}}_{\scalebox{0.85}{$n_3\hspace{0.1cm} (\mathfrak{g},\mathfrak{g})_{(3)}$ \text{edges}}}$};
        \end{tikzpicture}}
        \end{array}$
        
        \vspace{-1.5cm}
        
        \caption{The most general molecule of 5d conformal matter.}
        \label{fig:molecule}
    \end{figure}

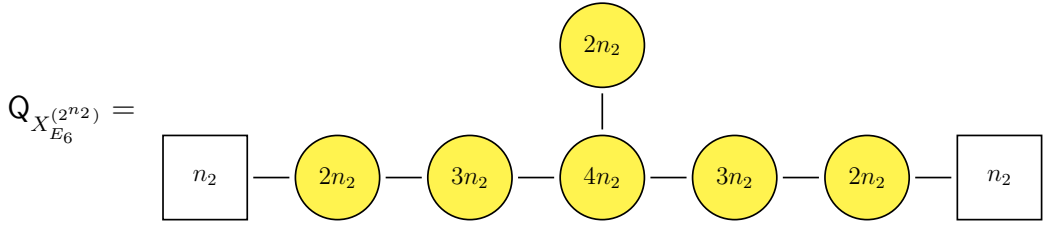
\begin{figure}[H]
\centering
  $   \begin{array}{cc}
\makecell{\mathsf{Q}_{X_{E_6}^{(2^{n_2})}} =\\ \vspace{2cm}}  & 
  \scalemath{0.8}{  \begin{tikzpicture}
        \draw[thick,fill=yellow!80] (0,0) circle (0.7);
        \node at (0,0) {$4n_2$};
        \draw[thick] (0.8,0)--(1.4,0);
        \draw[thick,fill=yellow!80] (2.2,0) circle (0.7);
        \node at (2.2,0) {$3n_2$};
        \draw[thick] (-0.8,0)--(-1.4,0);
        \draw[thick,fill=yellow!80] (-2.2,0) circle (0.7);
        \node at (-2.2,0) {$3n_2$};
        \draw[thick] (0,0.8)--(0,1.4);
        \draw[thick,fill=yellow!80] (0,2.2) circle (0.7);
        \node at (0,2.2) {$2n_2$};
        \draw[thick] (3,0)--(3.6,0);
        \draw[thick,fill=yellow!80] (4.4,0) circle (0.7);
         \draw[thick] (-3,0)--(-3.6,0);
        \node at (4.4,0) {$2n_2$};
         \draw[thick,fill=yellow!80] (-4.4,0) circle (0.7);
        \node at (-4.4,0) {$2n_2$};
         \draw[thick] (5.2,0)--(5.8,0);
            \draw[thick] (5.9,0.7)--(7.3,0.7)--(7.3,-0.7)--(5.9,-0.7)--cycle;
            \draw[thick] (-5.2,0)--(-5.8,0);
            \draw[thick] (-5.9,0.7)--(-7.3,0.7)--(-7.3,-0.7)--(-5.9,-0.7)--cycle;
        \node at (6.6,0) {$n_2$};
        \node at (-6.6,0) {$n_2$};
        \end{tikzpicture}}
        \end{array} $

        \vspace{-1.5cm}
        
    \caption{Low-energy quiver phase corresponding to the molecule with $\mathfrak{g} = E_6$, $n_1 = n_3 = 0$ and $n_2$ arbitrary.}
    \label{fig:E6molecule}
    \end{figure}

The atomic analogy for 5d conformal matter can be pushed further: atoms can be fused together to form complex molecules, that in turn correspond to well-defined 5d SCFTs. Indeed, one of the advantages of the geometric approach is that the Calabi-Yau condition furnishes an easy criterion to understand whether we can fuse a $\mathfrak g$ symmetry. The result is that, in order to preserve the Calabi-Yau condition, we can gauge together the diagonal combination of at most two flavor nodes of the 5d conformal matter atoms depicted in Figure \ref{fig:atom}, as proven in \cite{DeMarco:2023irn}. The most general molecule satisfying this criterion can be engineered from the threefold:
\begin{equation}\label{systemfourfoldgen}
  X_{\mathfrak{g}}^{(1^{n_1},2^{n_2},3^{n_3})}:\quad  \begin{cases}
        P_{\mathfrak{g}}(x_1,x_2,x_3)= 0\\
        UV = x_1^{n_1} x_2^{n_2} x_3^{n_3} \\
    \end{cases},
\end{equation}
with $n_1,n_2,n_3$ integers equal or greater than 0 (and at least one of them non-vanishing). This theory can be represented by the generalized quiver in Figure \ref{fig:molecule}, which is obtained by gluing together many of the atoms depicted in Figure \ref{fig:atom}. We remark that the edges are to be thought as carrying non-trivial gauge content, labeled by $(\mathfrak{g},\mathfrak{g})_{(i)}$.
In turn, for every threefold of the \eqref{systemfourfoldgen} kind, one can explicitly write down a Dynkin quiver with special unitary nodes corresponding to a 5d low-energy gauge theory phase. As an example, for the threefold
\begin{equation}
   X_{E_6}^{(2^{n_2})}:\quad    \begin{cases}
        P_{E_6}(x_1,x_2,x_3)= 0\\
        UV =  x_2^{n_2} \\
    \end{cases},
\end{equation}
the corresponding low-energy quiver gauge theory is depicted in Figure \ref{fig:E6molecule}.

We refer to \cite{DeMarco:2023irn} for all the other realizations.\\

In summary, given the expression \eqref{systemfourfoldgen} for the threefold canonical singularity, one needs to choose the following data, in order to specify a 5d conformal matter molecule:
\begin{itemize}
    \item the algebra $\mathfrak{g}$, that yields the $\mathfrak{g}\times\mathfrak{g}$ flavor symmetry factor;
    \item the exponents $n_1,n_2,n_3$ in \eqref{systemfourfoldgen}, that dictate how many atoms compose the molecule, and the kind of atoms that compose it\footnote{Of course, molecules can be arbitrarily long, and single atoms such as Figure \ref{fig:atom} are recovered setting all exponents $n_1,n_2,n_3$ to zero, except for one set to 1.}.
\end{itemize}
Physically, this data fixes the gauge symmetry of the corresponding 5d SCFT, as well as the rank of the flavor symmetry, which is of the form
\begin{equation}\label{total flavor}
    F_{UV} = \mathfrak{g}\times \mathfrak{g} \times F_{\text{rest}},
\end{equation}
with $F_{\text{rest}}$ depending on $n_1,n_2,n_3$.\\
\indent All in all, the aforementioned data (namely, the choice of $\mathfrak{g}$ and $n_i$) furnishes the sought-after atomic classification of the class of 5d conformal matter SCFTs introduced in \cite{DeMarco:2023irn}.

\section{5d conformal matter atoms and class \texorpdfstring{$\mathcal{S}$}{S} constructions}\label{sec:classS}

We now turn to the toroidal compactifications of 5d conformal matter atoms. It turns out that, for all the 5d conformal matter \textit{atoms} we have exhibited, there exists a 4d $\mathcal{N}=2$ descendant, based on a class-$\mathcal{S}$ construction. Such claim is substantiated explicitly by checking that the following well-known facts hold \cite{Ohmori:2015pia,Martone:2021drm}:
\begin{itemize}
    \item the number of (real-valued) $\mathcal{T}_{5d}$ CB modes must coincide with the number of (complex-valued) $\mathcal{T}_{4d}$ CB modes;
    \item the global symmetries of $\mathcal{T}_{5d}$ and $\mathcal{T}_{4d}$ must be the same\footnote{For some caveats to this claim, that will not enter our discussion, see \cite{Martone:2021drm}.};
    \item if there is no 1-form symmetry in $\mathcal{T}_{5d}$, then there is none also in $\mathcal{T}_{4d}$.
\end{itemize}
We begin, in Section \ref{sec:4dtheoriesforatoms}, finding a set of theories $\mathcal{T}_{4d}$ satisfying all the above points, by means of a class-$\mathcal{S}$ setup involving regular punctures. Such $\mathcal{T}_{4d}$ are strong candidates to be the descendants of the 5d conformal matter atomic theories $\mathcal{T}_{5d}$ via circle reduction. We will then, in Section \ref{sec:higgsbranchbigsec}, match the Higgs branch dimensions between  $\mathcal{T}_{5d}$  and  $\mathcal{T}_{4d}$, carefully taking into account the contributions of the instantonic particles and tracking them back in the geometry of the threefolds $X_{\mathfrak g}^{(i)}$.

\subsection{4d reduction of 5d conformal matter atoms}
\label{sec:4dtheoriesforatoms}

\indent The logic to extract the 4d reduction $\mathcal{T}_{4d}$ of 5d conformal matter fundamental atoms $\mathcal{T}_{5d}$ from bottom-up is straightforward and can be implemented by way of a step-by-step procedure.\\

\indent Start by picking the conformal matter theory $\mathcal{T}_{5d}$ related to the threefold $X_{\mathfrak{g}}^{(i)}$, that we have recalled in \cref{sec: atomic classification}. To access the 4d picture, consider the 6d $\mathcal{N}=(2,0)$ theory of type $\mathfrak{g}$ reduced on a sphere with three regular punctures (with a suitable topological twist). Recall that each regular puncture is related to\footnote{More precisely, the Hitchin field associated to the class $\mathcal S$ theory must have a simple pole 
\begin{equation*}
    \Phi(w) = \frac{\Phi_{-1}}{w - w_{p}} + O(1)
\end{equation*} at the puncture $w = w_{p}$, with $w$ a coordinate on the Gaiotto curve. In the conformal limit, $\Phi_{-1}$ is nilpotent, and belongs to a nilpotent orbit $\mathcal O_i$ of $\mathfrak g$.} a nilpotent orbit $\mathcal{O}_i$ of $\mathfrak{g}$, and that the Coulomb branch dimension of the 4d theory is:
\begin{equation}\label{4d CB dim}
\text{dim}\hspace{0.05cm}\text{CB}_{4d}(\mathcal{T}_{4d}) = \frac{1}{2}\sum_{i=1}^{n_{punctures}}(\text{dim}(\mathcal{O}_i))-\text{dim}(\mathfrak{g}).
\end{equation}
Moreover, the contribution of each puncture to the flavor symmetry is given by the stabilizers of the Langlands dual of $\mathcal{O}_i$, that we denote with $\mathcal{O}_i^L$ \cite{Chacaltana:2012zy}.\\
\indent Given the flavor symmetry of $\mathcal{T}_{5d}$, say (coherently with the notation of the previous sections)
\begin{equation}
    F_{UV} = \mathfrak{g}\times \mathfrak{g} \times F_{\text{rest}},
\end{equation}
the punctures are chosen according to:
\begin{itemize}
    \item Two punctures are maximal punctures related to the algebra $\mathfrak{g}$. This ensures that we obtain a $\mathfrak{g}\times\mathfrak{g}$ factor in the flavor symmetry, precisely as in the 5d parent.
    \item The third puncture $\mathcal{O}_{III}$ is chosen in such a way to match the expected CB dimension (i.e.\ the dimension of the CB of $\mathcal{T}_{5d}$). Non-trivially, the third puncture must also account for the $F_{\text{rest}}$ flavor symmetry factor of $\mathcal{T}_{5d}$.\footnote{In a follow up work \cite{AtomMoleculeHybridupcoming} we uncover an interesting relationship between the $\mathcal{O}_{III}$ orbit and the Dynkin quiver. Nevertheless, as is done in this work, it is enough to use a small set of physical data to fix the orbit.}
\end{itemize}
It is easy to check that the above criteria for the class-$\mathcal{S}$ construction uniquely constrain the choice of the three regular punctures\footnote{There are rare exceptions to this claim: e.g. for the $X_{D_7}^{(2)}$ theory we could have chosen $\mathcal{O}'_{III} = [3^2,1^8]$ instead of $\mathcal{O}_{III} = [2^6,1^2]$, keeping the CB dimension and the flavor symmetry unchanged. The orbit $\mathcal{O}'_{III}$, though, would not fit into the infinite family of Table \ref{4d table} labelled by $j$, and thus we choose to set it aside.}.\\
\indent In Table \ref{4d table} we gather all the information needed to construct the 4d $\mathcal{N}=2$ class-$\mathcal{S}$ theory, listing the threefolds $X_{\mathfrak{g}}^{(i)}$ corresponding to its 5d parent, the nilpotent orbit $\mathcal{O}_{III}$ whose numbers dictate \cite{Chacaltana:2011ze} the row lengths of the ``Hitchin partition'' Young tableaux of the third puncture, its Langlands dual $\mathcal{O}_{III}^L$ (whose numbers dictate the column heights of the ``Nahm partition'' Young tableaux of the third puncture), the rank of the 4d CB and the total 4d flavor symmetry. We employ the standard notation of \cite{collingwood1993} to label nilpotent orbits. In this notation,
\begin{itemize}
\item the Hitchin partition of a full puncture of type $A_{j}$ is denoted as $[j+1]$. The Hitchin partition of a full puncture for a $D_j$ algebra is denoted as $[2j-1,1]$;
\item the Nahm partition of a full puncture of type $A_{j}$ is denoted as $[1^{j+1}]$. The Nahm partition of a full puncture $D_j$ algebra is denoted as $[1^{2j}]$. The Young tableaux of the Nahm partition can be obtained, for $D_j$, from the Hitchin one by the ``D-collapse'' procedure \cite{Chacaltana:2011ze} (and taking care of reading the height of columns instead of the row length). The Young tableaux of the Nahm partition for the $A_j$ case coincides with the Young tableaux of the Hitchin partition (where, in the first case, we read the heights of the columns, and in the second the lengths of the rows).
\end{itemize}
Furthermore, notice that, according to \cite{Bhardwaj:2021pfz}, class-$\mathcal{S}$ theories on a sphere with only untwisted regular punctures admit no 1-form symmetries. This is in agreement with the result found in \cite{DeMarco:2023irn}: 
\begin{itemize}
    \item since all line operators are screened in $\mathcal{T}_{5d}$, there can be no electric one-form symmetry in its descendant $\mathcal{T}_{4d}$,
    \item since all magnetically charged surface operators in $\mathcal{T}_{5d}$ are screened, there must be no magnetically charged lines in  $\mathcal{T}_{4d}$.
\end{itemize}

\renewcommand{\arraystretch}{1.3}
\begin{table}\centering
\begin{equation}
\begin{array}{|c|c|c|c|c|}
\hline 
 \textbf{Singularity} & \boldsymbol{\mathcal{O}_{III}} & \boldsymbol{\mathcal{O}_{III}}^L & \textbf{rank}\hspace{0.05cm}\textbf{CB}\boldsymbol{_{\mathcal{T}_{4d}}} & \boldsymbol{F_{UV}}  \\
\hline
\hline
X_{A_{2j+1}}^{(1)} & [2^{j+1}] & [(j+1)^2] & j^2 & A_{2j+1}\times A_{2j+1}\times \mathfrak{u}(1) \\
X_{A_{2j}}^{(1)} & [2^j,1] & [j+1,j] & j(j-1) & A_{2j}\times A_{2j} \times \mathfrak{u}(1)\\
X_{A_{j}}^{(3)} & [2,1^{j-1}] & [j,1] & 0 & A_{j}\times A_{j} \times \mathfrak{u}(1)\\
X_{D_{2j+2}}^{(1)} & [3^2,2^{2j-2},1^2] & [(2j+1)^2,1^2] & j(2j+3) & D_{2j+2}\times D_{2j+2}\times \mathfrak{u}(1)^2 \\
X_{D_{2j+3}}^{(1)} & [3^2,2^{2j-2},1^4]  & [2j+3,2j+1,1^2] & j(2j+5)+1 & D_{2j+3}\times D_{2j+3} \times \mathfrak{u}(1) \\
X_{D_{2j+2}}^{(2)} & [2^{2j},1^4] & [2j+3,2j+1] & 2j^2+j-2 & D_{2j+2}\times D_{2j+2} \\
X_{D_{2j+3}}^{(2)} & [2^{2j+2},1^2] & [(2j+3)^2] & j(2j+3) & D_{2j+3}\times D_{2j+3} \times \mathfrak{u}(1) \\
X_{D_{j}}^{(3)} & [3,1^{2j-3}] & [2j-3,1^3] & j-2 & D_j\times D_j \times \mathfrak{su}(2) \\
X_{E_6}^{(1)} & A_2 & E_6(a_3) & 15 & E_6\times E_6 \\
X_{E_6}^{(2)} & 2A_1 & D_5 & 10 & E_6\times E_6 \times \mathfrak{u}(1)\\
X_{E_6}^{(3)} & A_1 & E_6(a_1) & 5  & E_6\times E_6\\
X_{E_7}^{(1)} & A_2+A_1 & E_6(a_1) & 31 &  E_7\times E_7\times \mathfrak{u}(1) \\
X_{E_7}^{(2)} & (3A_1)'' & E_6 & 20 &  E_7\times E_7 \times \mathfrak{su}(2)\\
X_{E_7}^{(3)} & A_1 & E_7(a_1) & 10 & E_7\times E_7 \\
X_{E_8}^{(1)} & A_2+A_1& E_8(a_4) & 60 & E_8\times E_8 \\
X_{E_8}^{(2)} & 2A_1 & E_8(a_2) & 38 & E_8\times E_8 \\
X_{E_8}^{(3)} & A_1 & E_8(a_1) & 21 & E_8\times E_8 \\
\hline
\end{array}\nonumber
\end{equation}
\caption{class-$\mathcal{S}$ data to construct the 4d descendants of the 5d conformal matter atoms.}\label{4d table}
\end{table}

Notice the perfect agreement between the rank of the 5d and 4d CB dimensions, as well as the full flavor symmetry.\\
\indent As a fitting conclusion to this section, we point out that a completely rigorous description of the 4d descendants of our 5d theories $X_{\mathfrak{g}}^{(i)}$ would have to go through their Type IIA description, obtained from circle reduction: the corresponding Type IIB mirror geometry precisely enforces the class-$\mathcal{S}$ construction, that we have depicted above based on consistency arguments. We leave such analysis for future work.
\subsection{5d conformal matter atoms and Higgs branch dimension}
\label{sec:higgsbranchbigsec}
In this section we compute the quaternionic Higgs branch dimension of the 5d conformal matter theories reviewed in Section \ref{sec: atomic classification}, both from the purely 5d perspective and the class-$\mathcal{S}$ 4d point of view, showing their agreement. This relies on the fact that such dimension is protected under circle compactification, and provides a further consistency check of our constructions. \\
\indent We start by computing the number of Higgs branch modes in the low-energy quiver phase of 5d conformal matter atoms of type $\mathfrak{g}$. These are the modes that are unlifted by the partial resolution \eqref{eq:partialresmap}.
We then account for extra Higgs branch directions that open up in the infinite coupling limit, due to instantonic particles becoming massless, following \cite{Yonekura}. We geometrically interpret the appearance of these instantonic particles in terms of the transformation of M2-brane ``vector-like'' modes into M2-brane ``Higgs-branch modes'' at the collision point of the two lines of Du Val singularities. \\
Finally, we compute the Higgs branch dimension of the candidate class-$\mathcal{S}$ 4d reduction of 5d conformal matter atoms, showing complete agreement with the 5d Higgs branch dimension at the UV point. \\

\indent Let us kick off our computation from the 5d perspective and examine the theories engineered by M-theory on $X_{\mathfrak{g}}^{(i)}$, whose low-energy ordinary quiver gauge theory phases were written down in\cite{DeMarco:2023irn}. We have shown an example of such quivers in Figure \ref{E6yfig}. Therefore, the dimension of the Higgs branch at finite coupling can be computed directly from the quiver, namely:
\begin{equation}
\label{eq:finitecouplingHB}
    \text{dim}_{\mathbb{H}}\text{HB}\left(\mathsf{Q}_{X_{\mathfrak{g}}^{(i)}}\right) = n_{H}-n_{V},
\end{equation}
with $n_{H}$ the number of hypermultiplets and $n_{V}$ the number of vector multiplets. We have to be careful, though, because this computation \textit{neglects} the contribution of deformations that have been obstructed by the resolution map $\epsilon$ in \eqref{eq:partialresmap}.  From the quiver perspective, this comes from the fact that \eqref{eq:finitecouplingHB} just captures the number of perturbative Higgs branch modes, and is blind to the instantonic particles becoming massless in the infinite coupling limit. These additional Higgs branch directions open up at the UV point, and can be detected by contracting the curves inflated by $\epsilon$, which are arranged as the $\mathfrak{g}$ Dynkin diagram. Indeed, the instantonic particles of the quiver are exactly labeled by the root system of $\mathfrak g$ \cite{Yonekura}, with the single-instantons being the simple roots and the multi-instantons being the other roots.\\
 \indent In order to characterize them explicitly, let us consider $\widetilde{X}_{\mathfrak g}^{(i)}$: we can zoom in on a region centered on one of the two non-compact lines of \textit{resolved} Du Val singularities of  $\widetilde{X}_{\mathfrak g}^{(i)}$, away from their intersection point. There, the threefold locally looks like $\mathbb C \times \widetilde{Y}_{\mathfrak g}$, where $\widetilde{Y}_{\mathfrak g}$ is the \textit{resolved} Du Val singularity and $\mathbb C$ is spanned, say, by $u$ (equivalently, by $v$).  We can now wrap M2-branes on the compact curves ruled over the non-compact line of resolved $\mathfrak g$ singularities. These curves exactly coincide with the $\mathbb P^1$'s resolving the (say) $u  = 0$ line of Du Val singularity (away from the $v =0$ point). These M2-branes become, in the limit of vanishing volume of the $\mathbb P^1$'s, states localized at the points where the $\mathbb P^1$'s get contracted. In this case, the M2-branes must be interpreted as 7d \textit{vector} multiplets because
\begin{itemize}
    \item the normal bundle of each $\mathbb P^1$ is $\mathcal O(0) \times \mathcal O(-2)$ and hence we have no first-order obstruction to a translation (along the $\mathcal O(0)$ direction) of the M2-brane. Mathematically, this comes from the existence of an holomorphic non-zero section of the normal bundle. Physically, this means that the corresponding M2-state is localized in 7d (rather than in 5d); 
    \item there are also no higher-order obstructions.
\end{itemize}
In general, if one of these two conditions fails to be true, the M2-brane state gets localized to 5d, and generates 5d \textit{Higgs branch} modes \cite{Witten:1996bn, DeMarco:2021try, DeMarco:2022dgh, Collinucci:2022rii, Collinucci:2021wty, Collinucci:2021ofd}.\\
\indent It is a well-known result \cite{Sen:1997kz,Sen:1997js,Acharya:2002gu} that not all the elements in the compact homology $H_{2}(Y_{\mathfrak g}, \mathbb Z)$ support an M2 BPS brane state, but, rather, the homology classes supporting an M2 state correspond to the root system of $\mathfrak g$. More precisely, the Poincar\'e duals of the $\mathbb P^1$'s resolving $Y_{\mathfrak g}$ correspond to the dual roots $\alpha_j^*$ of the $\mathfrak g$ Dynkin diagram. Furthermore, let $\alpha_j$ be the simple roots of $\mathfrak g$ and consider a two-cycle
\begin{equation}
\label{eq:genericelementh2homology}
\nu = \sum_{j=1}^{\text{rank}(\mathfrak g)} n_j [\alpha_j]
\end{equation}
with $[\alpha_j] \in H_{2}(\widetilde{Y}_{\mathfrak g},\mathbb Z)$ the homology class of the $j$-th. $\mathbb P^1$. Then, $\nu$ supports a M2 BPS state if and only if $\sum_{j=1}^{\text{rank}(\mathfrak g)} n_{j} \alpha_j$ is a root of $\mathfrak g$.

We can specialize this general picture to the partially resolved threefold $\widetilde{X}_{\mathfrak g}^{(i)}$. As we can see from Figure \ref{fig:E6firstblowup}, away from the (red) intersection point the M2-branes wrapped on the compact $\mathbb P^1$'s become 7d vector multiplets, producing the W-bosons of the $\mathfrak g \times \mathfrak g$ flavor algebra. The \textit{singular} $\mathbb P^1$'s contracted to the origin display obstructed deformation theory inside $\widetilde{X}_{\mathfrak g}^{(i)}$, and hence the corresponding M2 states are Higgs branch modes (rather than being part of a vector multiplet). This is in agreement with the 4d class-$\mathcal{S}$ picture, as we will show momentarily. In other words, at the intersection of the two lines of resolved Du Val singularities the M2-branes wrapping the  $\mathbb P^1$'s can \textit{not} be holomorphically ``translated'' along the non-compact direction, and hence they localize to 5d hypermultiplets. Indeed, as the size of the $\mathbb P^1$'s is different from zero (proportional to the square of the inverse gauge couplings) in the ordinary quiver phase, the aforementioned modes are massive and then do not contribute to \eqref{eq:finitecouplingHB}.   From a field theory perspective, the resolution \eqref{eq:partialresmap} corresponds to giving a v.e.v.\ to the Cartan elements of a diagonal factor $\mathfrak g_{\text{diag}} \subset \mathfrak g \times \mathfrak g$ 5d flavor group of M-theory on $X_{\mathfrak g}^{(i)}$. Under a minimal coupling interaction $q_{Q}\widetilde{Q} M Q$, this lifts all the 5d Higgs branch modes with charge $q_{Q}\neq 0$ under $g_{\text{diag}}$, restricting the Higgs-branch dimension. The number of these states is the number of M2-branes wrapped on the ``undeformable'' $\mathbb P^1$'s contracting on the origin, and exactly matches the number of roots of $\mathfrak g$.\\
\indent In Appendix \ref{appendix A} we exhibit an explicit example of our choice of partial resolution of a 5d singular threefold, highlighting the curves wrapped by the M2-brane states that get trapped at the origin, thus giving rise to HB modes. Such an example involves the famous 5d $T_N$ trinion theories, for a bipartite reason: it allows us to rely on toric geometry (although it is by no means a necessary choice) and, above all, shows this previously unnoticed behavior in such a well-known set of theories.\\
\indent For the purpose of computing the UV Higgs branch dimension of 5d conformal matter atoms, the previous discussion implies that \eqref{eq:finitecouplingHB} gets modified as:
\begin{equation}
\label{eq:infinitecouplingHB}
\text{dim}_{\mathbb{H}}\text{HB}\left(\mathcal{T}_{X_{\mathfrak{g}}^{(i)}}\right) = n_{H}-n_{V}+\#\text{roots}(\mathfrak{g}).
\end{equation}

\indent Let us finally make contact with the 4d class-$\mathcal{S}$ picture. The dimension of the Higgs branch for the 4d $\mathcal{N}=2$ theory can be computed according to \cite{Chacaltana:2012zy}, simply from the data of the regular punctures:
\begin{equation}
\label{eq:hbdimensionclasss}
    \text{dim}_{\mathbb{H}}\text{HB}\left(\mathcal{T}_{4d}\right) = \frac{1}{2}\sum_i\left( \text{dim}_{\mathbb{C}}(\mathfrak{g})-\text{rank}(\mathfrak{g})-\text{dim}_{\mathbb{C}}(\mathcal{O}^{L}_i) \right) + \text{rank}(\mathfrak{g}),
\end{equation}
with the choices of $\mathcal{O}^{L}_i$ of Table \eqref{4d table}.\\
\indent Let us collect the results from the 5d and 4d computation in Table \ref{tab:HB dim}, where we have listed the 5d HB dimension from the low-energy quiver phase, the 4d HB dimension, the difference between them (which corresponds to degrees of freedom that become massless in the infinite coupling limit of the 5d gauge theory, as many as the number of roots of the appropriate $\mathfrak{g}$), as well as the UV 5d HB dimension. Notice that the UV 5d HB dimension precisely agrees with the 4d HB dimension, as expected from the circle reduction.

\renewcommand{\arraystretch}{1.3}
\begin{table}[H]\centering
\begin{equation}
\scalemath{0.9}{
\begin{array}{|c|c|c|c|c|c|}
\hline 
 \textbf{Singularity} & \textbf{dim}_{\mathbb{H}}\textbf{HB}\boldsymbol{(\mathsf{Q}_{X_{\mathfrak{g}}^{(i)}})} & \textbf{dim}_{\mathbb{H}}\textbf{HB}\boldsymbol{\left(\mathcal{T}_{4d}\right)} & \textbf{dim}_{\mathbb{H}}\textbf{HB}\boldsymbol{\left(\mathcal{T}_{4d}\right)}-\textbf{dim}_{\mathbb{H}}\textbf{HB}\boldsymbol{(\mathsf{Q}_{X_{\mathfrak{g}}^{(i)}})} & \textbf{dim}_{\mathbb{H}}\textbf{HB}\boldsymbol{(\mathcal{T}_{X_{\mathfrak{g}}^{(i)}})} \\
\hline
\hline
X_{A_{2j+1}}^{(1)} & 3j+2 & 4j^2+9j+4  & 2(2j^2+3j+1) & 4j^2+9j+4  \\
X_{A_{2j}}^{(1)} & 3j & j(4j+5) &  2j(2j+1)& j(4j+5)\\
X_{A_{j}}^{(3)} & j+1 & (j+1)^2 &  j(j+1)& (j+1)^2 \\
X_{D_{2j+2}}^{(1)} & 3j+4 & 8j^2+15j+8 & 4(2j^2+3j+1) & 8j^2+15j+8 \\
X_{D_{2j+3}}^{(1)} & 3j+5 & 8j^2+23j+17 & 4 \left(2 j^2+5 j+3\right)& 8j^2+23j+17 \\
X_{D_{2j+2}}^{(2)}  & 3j+2 & 8j^2+15j+6 & 4(2j^2+3j+1) & 8j^2+15j+6 \\
X_{D_{2j+3}}^{(2)} & 3j+4 & 8j^2+23j+16 & 4 \left(2 j^2+5 j+3\right)& 8j^2+23j+16 \\
X_{D_{j}}^{(3)}  & j+2  & 2j^2-j+2 & 2j(j-1) & 2j^2-j+2\\
X_{E_6}^{(1)} & 9 & 81 & 72& 81 \\
X_{E_6}^{(2)} & 8 & 80 & 72& 80  \\
X_{E_6}^{(3)} & 7 & 79 & 72  & 79\\
X_{E_7}^{(1)} & 11 & 137 & 126 & 137\\
X_{E_7}^{(2)} & 10 & 136 & 126 & 136\\
X_{E_7}^{(3)} & 8 & 134 & 126& 134\\
X_{E_8}^{(1)} & 12 & 252 & 240 & 252 \\
X_{E_8}^{(2)} & 10 & 250 & 240& 250  \\
X_{E_8}^{(3)} & 9 & 249 & 240 & 249 \\
\hline
\end{array}\nonumber
}
\end{equation}
\caption{Higgs branch dimension from the low-energy quiver 5d phase $\mathsf{Q}_{X_{\mathfrak{g}}^{(i)}}$, the class-$\mathcal{S}$ 4d perspective $\mathcal{T}_{4d}$, and the infinite coupling 5d phase $\mathcal{T}_{X_{\mathfrak{g}}^{(i)}}$.}\label{tab:HB dim}
\end{table}

\section{5d conformal matter atoms, brane webs and magnetic quivers}
\label{sec:5dAtomsBWandMQ}

Let us now to turn to a brane web construction for 5d conformal matter atoms, which exists for $A$- and $D$-type. For the $A$-type the atoms are realised on standard brane webs of $(p,q)5$-branes. For $D$-type one needs to include an ON$5$-plane. Our conventions for brane webs are summarised in Appendix \ref{app:BWNotation}.

\subsection{A-type}

\subsubsection{$X_{A_{2j+1}}^{(1)}$}

The low energy Dynkin quiver description $\mathsf{Q}_{X_{A_{2j+1}}^{(1)}}$ of the 5d SCFT $\mathcal{T}_{X_{A_{2j+1}}^{(1)}}$ associated to the $X_{A_{2j+1}}^{(1)}$ singularity is \cite{DeMarco:2023irn}
\begin{equation}
\label{eq:A2j+1_EQ}
    \mathsf{Q}_{X_{A_{2j+1}}^{(1)}}=\raisebox{-.5\height}{\scalebox{0.6}{\begin{tikzpicture}
        \node[gaugeSU,label=below:{$1$}] (1) at (2,0) {};
        \node[gaugeSU,label=below:{$2$}] (2) at (4,0) {};
        \node (3) at (6,0) {$\cdots$};
        \node[gaugeSU,label=below:{$j$}] (4) at (8,0) {};
        \node[gaugeSU,label=below:{$j+1$}] (5) at (10,0) {};
        \node[gaugeSU,label=below:{$j$}] (6) at (12,0) {};
        \node (7) at (14,0) {$\cdots$};
        \node[gaugeSU,label=below:{$2$}] (8) at (16,0) {};
        \node[gaugeSU,label=below:{$1$}] (9) at (18,0) {};
        \node[flavourSU,label=above:{$2$}] (5f) at (10,1.5) {};
        \draw (1)--(2)--(3)--(4)--(5)--(6)--(7)--(8)--(9) (5)--(5f);
    \end{tikzpicture}}}\;.
\end{equation}
A brane web for this theory (with both flavours massive) is
\begin{equation}
    \scalebox{0.8}{\begin{tikzpicture}
        \node[seven,label=below:{\scriptsize$[0,1]7$}] (1) at (1,-1) {};
        \node[seven] (2) at (2,-1) {};
        \node[seven] (3) at (3,-1) {};
        \node[seven] (4) at (5,-1) {};
        \node[seven] (5) at (7,-1) {};
        \node[seven] (6) at (8,-1) {};
        \node[seven] (7) at (10,-1) {};
        \node[seven] (8) at (12,-1) {};
        \node[seven] (9) at (13,-1) {};
        \node[seven] (10) at (14,-1) {};

        \node[seven,label=above:{\scriptsize$[1,-1]7$}] (1f) at (-1,2) {};
        \node[seven] (2f) at (0,2) {};
        \node[seven] (3f) at (1,2) {};
        \node[seven] (4f) at (3,2) {};
        \node[seven,label=above:{\scriptsize$[0,1]7$}] (5f) at (6,2) {};
        \node[seven,label=above:{\scriptsize$[1,0]7$}] (5ff) at (5,1) {};
        \node[seven] (6ff) at (10,1) {};
        \node[seven] (6f) at (9,2) {};
        \node[seven] (7f) at (12,2) {};
        \node[seven] (8f) at (14,2) {};
        \node[seven] (9f) at (15,2) {};
        \node[seven] (10f) at (16,2) {};

        \draw (1)--(1,0)--(1f);
        \draw (2)--(2,0)--(2f);
        \draw (3)--(3,0)--(3f);
        \draw (4)--(5,0)--(4f);
        \draw (5)--(7,0)--(6,1)--(5f) (6,1)--(5ff);
        \draw (6)--(8,0)--(9,1)--(6f) (9,1)--(6ff);
        \draw (7)--(10,0)--(7f);
        \draw (8)--(12,0)--(8f);
        \draw (9)--(13,0)--(9f);
        \draw (10)--(14,0)--(10f);

        \draw (1,0)--(2,0);
        \draw[thick,double] (2,0)--(3.5,0);
        \node at (4,0) {$\cdots$};
        \draw[thick,double] (4.5,0)--(10.5,0);
        \node at (11,0) {$\cdots$};
        \draw[thick,double] (11.5,0)--(13,0);
        \draw (13,0)--(14,0);
        \node at (1.4,0.2) {\scriptsize$1$};
        \node at (2.4,0.2) {\scriptsize$2$};
        \node at (6,0.2) {\scriptsize$j$};
        \node at (7.5,0.2) {\scriptsize$j+1$};
        \node at (9,0.2) {\scriptsize$j$};
        \node at (12.6,0.2) {\scriptsize$2$};
        \node at (13.6,0.2) {\scriptsize$1$};
    \end{tikzpicture}}\;.
\end{equation}
The parallel D5 branes support the SU gauge groups, and the two D7 branes ($[1,0]7$) provide the two flavours to the central SU gauge group. The vertical position of the D7 branes corresponds to the mass of the flavours.\\

After some Hanany-Witten moves (brane creation) we obtain
\begin{equation}
    \scalebox{0.8}{\begin{tikzpicture}
        \node[seven,label=below:{\scriptsize$[0,1]7$}] (1) at (1,-1) {};
        \node[seven] (2) at (2,-1) {};
        \node[seven] (3) at (3,-1) {};
        \node[seven] (4) at (5,-1) {};
        \node[seven] (5) at (7,-1) {};
        \node[seven] (6) at (8,-1) {};
        \node[seven] (7) at (10,-1) {};
        \node[seven] (8) at (12,-1) {};
        \node[seven] (9) at (13,-1) {};
        \node[seven] (10) at (14,-1) {};

        \node[seven,label=above:{\scriptsize$[0,1]7$}] (1f) at (0,2) {};
        \node[seven] (2f) at (1,2) {};
        \node[seven] (3f) at (2,2) {};
        \node[seven] (4f) at (4,2) {};
        \node[seven,label=above:{\scriptsize$[0,1]7$}] (5f) at (6,2) {};
        \node[seven,label=above:{\scriptsize$[1,0]7$}] (5ff) at (-1,1) {};
        \node[seven] (6ff) at (16,1) {};
        \node[seven] (6f) at (9,2) {};
        \node[seven] (7f) at (11,2) {};
        \node[seven] (8f) at (13,2) {};
        \node[seven] (9f) at (14,2) {};
        \node[seven] (10f) at (15,2) {};

        \draw (1)--(1,0)--(0,1)--(1f);
        \draw (2)--(2,0)--(1,1)--(2f);
        \draw (3)--(3,0)--(2,1)--(3f);
        \draw (4)--(5,0)--(4,1)--(4f);
        \draw (5)--(7,0)--(6,1)--(5f);
        \draw (6)--(8,0)--(9,1)--(6f);
        \draw (7)--(10,0)--(11,1)--(7f);
        \draw (8)--(12,0)--(13,1)--(8f);
        \draw (9)--(13,0)--(14,1)--(9f);
        \draw (10)--(14,0)--(15,1)--(10f);

        \draw (1,0)--(2,0);
        \draw[thick,double] (2,0)--(3.5,0);
        \node at (4,0) {$\cdots$};
        \draw[thick,double] (4.5,0)--(10.5,0);
        \node at (11,0) {$\cdots$};
        \draw[thick,double] (11.5,0)--(13,0);
        \draw (13,0)--(14,0);
        \node at (1.4,0.2) {\scriptsize$1$};
        \node at (2.4,0.2) {\scriptsize$2$};
        \node at (6,0.2) {\scriptsize$j$};
        \node at (7.5,0.2) {\scriptsize$j+1$};
        \node at (9,0.2) {\scriptsize$j$};
        \node at (12.6,0.2) {\scriptsize$2$};
        \node at (13.6,0.2) {\scriptsize$1$};

        \draw (6,1)--(4,1);
        \draw[thick,double] (4,1)--(3.5,1);
        \node at (3,1) {$\cdots$};
        \draw[thick,double] (2.5,1)--(5ff);
        \draw (9,1)--(11,1);
        \draw[thick,double] (11,1)--(11.5,1);
        \node at (12,1) {$\cdots$};
        \draw[thick,double] (12.5,1)--(6ff);
        \node at (5,0.8) {\scriptsize$1$};
        \node at (1.7,0.8) {\scriptsize$j-1$};
        \node at (0.6,0.8) {\scriptsize$j$};
        \node at (-0.4,0.8) {\scriptsize$j+1$};
        \node at (10,0.8) {\scriptsize$1$};
        \node at (13.3,0.8) {\scriptsize$j-1$};
        \node at (14.4,0.8) {\scriptsize$j$};
        \node at (15.4,0.8) {\scriptsize$j+1$};
    \end{tikzpicture}}\;.
\end{equation}
Taking the masses to zero we obtain
\begin{equation}
    \scalebox{0.8}{\begin{tikzpicture}
        \node[seven] (1) at (1,-1) {};
        \node[seven] (2) at (2,-1) {};
        \node[seven] (3) at (3,-1) {};
        \node[seven] (4) at (5,-1) {};
        \node[seven] (5) at (7,-1) {};
        \node[seven] (6) at (8,-1) {};
        \node[seven] (7) at (10,-1) {};
        \node[seven] (8) at (12,-1) {};
        \node[seven] (9) at (13,-1) {};
        \node[seven] (10) at (14,-1) {};
        
        \node[seven,label=above:{\scriptsize$[0,1]7$}] (1f) at (1,1) {};
        \node[seven] (2f) at (2,1) {};
        \node[seven] (3f) at (3,1) {};
        \node[seven] (4f) at (5,1) {};
        \node[seven] (5f) at (7,1) {};
        \node[seven] (6f) at (8,1) {};
        \node[seven] (7f) at (10,1) {};
        \node[seven] (8f) at (12,1) {};
        \node[seven] (9f) at (13,1) {};
        \node[seven] (10f) at (14,1) {};
        
        \node[seven,label=left:{\scriptsize$[1,0]7$}] (5ff) at (0,0) {};
        \node[seven] (6ff) at (15,0) {};

        \draw[thick, double] (5ff)--(3.5,0) (4.5,0)--(10.5,0) (11.5,0)--(6ff);
        \node at (4,0) {$\cdots$};
        \node at (11,0) {$\cdots$};
        \node at (6,0.3) {\scriptsize$j+1$ $(1,0)5$};

        \draw (1)--(1f) (2)--(2f) (3)--(3f) (4)--(4f) (5)--(5f) (6)--(6f) (7)--(7f) (8)--(8f) (9)--(9f) (10)--(10f);

        \draw [decorate,decoration={brace,amplitude=5pt},xshift=0pt,yshift=0pt]
    (14.3,-1.3)--(0.7,-1.3) node [black,midway,yshift=-0.4cm] {$2j+2$ $[0,1]7$};
    \end{tikzpicture}}\;.
\end{equation}
From this brane web the magnetic quiver for the finite coupling electric quiver \eqref{eq:A2j+1_EQ} can be read using the prescriptions in \cite{Cabrera:2018jxt}:
\begin{equation}
    \mathsf{MQ}\left(\mathsf{Q}_{X_{A_{2j+1}}^{(1)}}\right)=\raisebox{-.5\height}{\scalebox{1}{\begin{tikzpicture}
        \node[gauge,label=below:{\scriptsize$j+1$}] (1) at (0,0) {};
        \node[gauge,label=above:{\scriptsize$1$}] (2) at (-1,1) {};
        \node[gauge,label=above:{\scriptsize$1$}] (3) at (1,1) {};
        \node at (0,1) {$\cdots$};
        \draw (2)--(1)--(3);
        \draw [decorate,decoration={brace,amplitude=5pt},xshift=0pt,yshift=0pt]
    (-1.3,1.5)--(1.3,1.5) node [black,midway,yshift=0.4cm] {\scriptsize$2j+2$};
    \end{tikzpicture}}}\;,
\end{equation}
which is the 3d mirror of \eqref{eq:A2j+1_EQ}. From the magnetic quiver it is easy to identify the Higgs branch symmetry of the finite coupling quiver \eqref{eq:A2j+1_EQ} to be U$(1)^{2j+2}$, which matches exactly the U$(1)^{2j+2}$ baryonic symmetries of said quiver. We have
\begin{equation}
    \mathrm{dim}_{\mathbb{H}}\mathrm{CB}_{3d}\left(\mathsf{MQ}\left(\mathsf{Q}_{X_{A_{2j+1}}^{(1)}}\right)\right)=3j+2=\mathrm{dim}_{\mathbb{H}}\mathrm{HB}\left(\mathsf{Q}_{X_{A_{2j+1}}^{(1)}}\right)\;,
\end{equation}
which matches Table \ref{tab:HB dim}.

Taking the couplings to infinity we obtain the brane web 
\begin{equation}
\label{eq:A2j+1_Atom_Web}
    \scalebox{0.8}{\begin{tikzpicture}
        \node[seven] (5ff) at (-2,0) {};
        \node[seven] (6ff) at (2,0) {};
        \draw[thick, double] (5ff)--(6ff);
        \node at (1,0.3) {$j+1$};

        \node[seven] (1u) at (0,2) {};
        \node[seven] (2u) at (0,3) {};
        \node (3u) at (0,4) {$\vdots$};
        \node[seven] (4u) at (0,5) {};
        \node[seven] (5u) at (0,6) {};
        \node[seven] (6u) at (0,7) {};

        \node[seven] (1d) at (0,-2) {};
        \node[seven] (2d) at (0,-3) {};
        \node (3d) at (0,-4) {$\vdots$};
        \node[seven] (4d) at (0,-5) {};
        \node[seven] (5d) at (0,-6) {};
        \node[seven] (6d) at (0,-7) {};

        \draw (5u)--(6u) (5d)--(6d);
        \draw[thick,double] (5u)--(4u)--(3u)--(2u)--(1u)--(1d)--(2d)--(3d)--(4d)--(5d);

        \node at (-0.3,6.5) {$1$};
        \node at (-0.3,5.5) {$2$};
        \node at (-0.7,2.5) {$2j+1$};
        \node at (-0.7,1) {$2j+2$};
        \node at (-0.3,-6.5) {$1$};
        \node at (-0.3,-5.5) {$2$};
        \node at (-0.7,-2.5) {$2j+1$};
    \end{tikzpicture}}\;.
\end{equation}
The corresponding GTP is
\begin{equation}
    \begin{tikzpicture}
        \node[bd] (00) at (0,0) {};
        \node[bd] (10) at (1,0) {};
        \node (20) at (2,0) {$\cdots$};
        \node[bd] (30) at (3,0) {};
        \node[bd] (40) at (4,0) {};
        \node[bd] (04) at (0,4) {};
        \node[bd] (14) at (1,4) {};
        \node (24) at (2,4) {$\cdots$};
        \node[bd] (34) at (3,4) {};
        \node[bd] (44) at (4,4) {};
        \node[wd] (01) at (0,1) {};
        \node (02) at (0,2) {$\vdots$};
        \node[wd] (03) at (0,3) {};
        \node[wd] (41) at (4,1) {};
        \node (42) at (4,2) {$\vdots$};
        \node[wd] (43) at (4,3) {};
        \draw (00)--(10)--(20)--(30)--(40)--(41)--(42)--(43)--(44)--(34)--(24)--(14)--(04)--(03)--(02)--(01)--(00);
        \draw [decorate,decoration={brace,amplitude=5pt},xshift=0pt,yshift=0pt] (4.1,-0.3)--(-0.1,-0.3) node [black,midway,yshift=-0.5cm] {\scriptsize $2j+2$ edges separated by black dots};
        \draw [decorate,decoration={brace,amplitude=5pt},xshift=0pt,yshift=0pt] (-0.3,-0.1)--(-0.3,4.1) node [black,midway,xshift=-0.5cm,rotate=90] {\scriptsize $j+1$ edges separated by white dots};
    \end{tikzpicture}\;.
\end{equation}
From this brane web we can read the magnetic quiver for the infinite coupling SCFT  $\mathcal{T}_{X_{A_{2j+1}}^{(1)}}$ of \eqref{eq:A2j+1_EQ} (i.e. the SCFT associated to the singular $X_{A_{2j+1}}^{(1)}$), which is
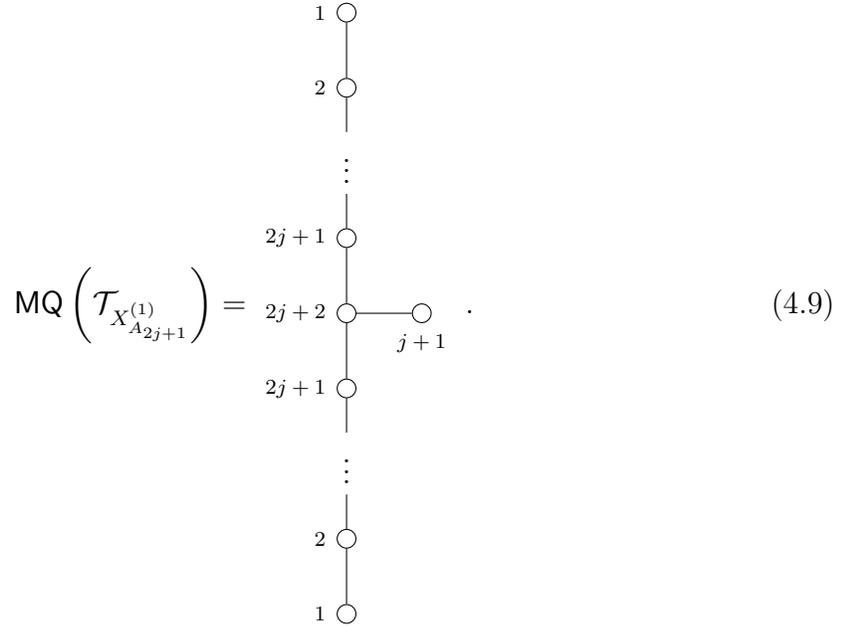
\begin{equation}
    \mathsf{MQ}\left(\mathcal{T}_{X_{A_{2j+1}}^{(1)}}\right)=\raisebox{-.5\height}{\scalebox{1}{\begin{tikzpicture}
        \node[gauge,label=below:{\scriptsize$j+1$}] (0r) at (1,0) {};
        \node[gauge,label=left:{\scriptsize$2j+2$}] (0) at (0,0) {};
        \node[gauge,label=left:{\scriptsize$2j+1$}] (1) at (0,1) {};
        \node (2) at (0,2) {$\vdots$};
        \node[gauge,label=left:{\scriptsize$2$}] (3) at (0,3) {};
        \node[gauge,label=left:{\scriptsize$1$}] (4) at (0,4) {};
        \node[gauge,label=left:{\scriptsize$2j+1$}] (-1) at (0,-1) {};
        \node (-2) at (0,-2) {$\vdots$};
        \node[gauge,label=left:{\scriptsize$2$}] (-3) at (0,-3) {};
        \node[gauge,label=left:{\scriptsize$1$}] (-4) at (0,-4) {};
        \draw (0r)--(0)--(1)--(2)--(3)--(4) (0)--(-1)--(-2)--(-3)--(-4);
    \end{tikzpicture}}}\;.
\end{equation}
From this magnetic quiver it is easy to identify the $\mathfrak{su}(2j+2)\times\mathfrak{su}(2j+2)\times\mathfrak{u}(1)$ symmetry of the SCFT, which enhances to $E_7$ for $j=1$, and which enhances to $\mathfrak{sp}(4)$ for $j=0$ where the magnetic quiver has Coulomb branch $\mathbb{H}^4$ indicating that the SCFT consists of free hypers. We have
\begin{equation}
    \makebox[\textwidth][c]{\begin{tikzpicture}
        \node at (0,0) {$\mathrm{dim}_{\mathbb{H}}\mathrm{CB}_{3d}\left(\mathsf{MQ}\left(\mathcal{T}_{X_{A_{2j+1}}^{(1)}}\right)\right)=4j^2+9j+4=(3j+2)+2(2j^2+3j+1)=\mathrm{dim}_{\mathbb{H}}\mathrm{HB}\left(\mathcal{T}_{X_{A_{2j+1}}^{(1)}}\right)\;,$};
    \end{tikzpicture}}
\end{equation}
which matches Table \ref{tab:HB dim}.\\

\paragraph{UV dual Quiver descriptions.}
As a little aside we point out two more low energy quiver descriptions of the SCFT associated to $X_{A_{2j+1}}^{(1)}$, which can be read from the brane web. In order to see this we first perform an S-duality (along with a rotation) on our brane web, and obtain
\begin{equation}
    \scalebox{0.8}{\begin{tikzpicture}[rotate=90]
        \node[seven] (5ff) at (-2,0) {};
        \node[seven] (6ff) at (2,0) {};
        \draw[thick, double] (5ff)--(6ff);
        \node at (1,0.5) {$j+1$};

        \node[seven] (1u) at (0,2) {};
        \node[seven] (2u) at (0,3) {};
        \node (3u) at (0,4) {$\cdots$};
        \node[seven] (4u) at (0,5) {};
        \node[seven] (5u) at (0,6) {};
        \node[seven] (6u) at (0,7) {};

        \node[seven] (1d) at (0,-2) {};
        \node[seven] (2d) at (0,-3) {};
        \node (3d) at (0,-4) {$\cdots$};
        \node[seven] (4d) at (0,-5) {};
        \node[seven] (5d) at (0,-6) {};
        \node[seven] (6d) at (0,-7) {};

        \draw (5u)--(6u) (5d)--(6d);
        \draw[thick,double] (5u)--(4u)--(3u)--(2u)--(1u)--(1d)--(2d)--(3d)--(4d)--(5d);

        \node at (-0.3,6.5) {$1$};
        \node at (-0.3,5.5) {$2$};
        \node at (-0.3,2.5) {\scriptsize$2j+1$};
        \node at (-0.3,1) {\scriptsize$2j+2$};
        \node at (-0.3,-6.5) {$1$};
        \node at (-0.3,-5.5) {$2$};
        \node at (-0.3,-2.5) {\scriptsize$2j+1$};
    \end{tikzpicture}}\;.
\end{equation}
After partially going on the extended Coulomb branch, and then performing some Hanany-Witten moves, we can obtain a brane web with quiver description
\begin{equation}
\label{eq:alternativeelectricphase}
    \scalebox{0.8}{\begin{tikzpicture}
        \node[flavourSU,label=below:{$2j+2$}] (0) at (0,0) {};
        \node[gaugeSU,label=below:{$2j$}] (1) at (2,0) {};
        \node[gaugeSU,label=below:{$2j-1$}] (2) at (4,0) {};
        \node (3) at (6,0) {$\cdots$};
        \node[gaugeSU,label=below:{$3$}] (4) at (8,0) {};
        \node[gaugeSU,label=below:{$2$}] (5) at (10,0) {};
        \node[flavourSU,label=below:{$2$}] (6) at (12,0) {};
        \draw (0)--(1)--(2)--(3)--(4)--(5)--(6);
    \end{tikzpicture}}\;,
\end{equation}
which has a manifest $\mathfrak{su}(2j+2)$ symmetry. At the level of the threefold, this operation amounts to performing another resolution of the singularity w.r.t. the one we presented in \Cref{sec:basechangeres}. We notice that some of the instantonic particles of the quiver \eqref{eq:A2j+1_EQ} are now local degrees of freedom for \eqref{eq:alternativeelectricphase}.

One can go to a different leaf in the extended Coulomb branch, and after again performing Hanany-Witten transitions obtain a brane web with quiver description
\begin{equation}
    \scalebox{0.8}{\begin{tikzpicture}
        \node[flavourSU,label=below:{$j+2$}] (0) at (0,0) {};
        \node[gaugeSU,label=below:{\small SU$(j+1)_{-\frac{1}{2}}$}] (1) at (2,0) {};
        \node[gaugeSU,label=below:{\small SU$(j+1)_{0}$}] (2) at (4,0) {};
        \node (3) at (6,0) {$\cdots$};
        \node[gaugeSU,label=below:{\small SU$(j+1)_{0}$}] (4) at (8,0) {};
        \node[gaugeSU,label=below:{\small SU$(j+1)_{\frac{1}{2}}$}] (5) at (10,0) {};
        \node[flavourSU,label=below:{$j+2$}] (6) at (12,0) {};
        \draw (0)--(1)--(2)--(3)--(4)--(5)--(6);
        \draw [decorate,decoration={brace,amplitude=5pt},xshift=0pt,yshift=0pt]
    (11,-1)--(1,-1) node [black,midway,yshift=-0.4cm] {\scriptsize$2j-1$};
    \end{tikzpicture}}\;,
\end{equation}
which however does not have a manifest $\mathfrak{su}(2j+2)$ symmetry.

\subsubsection{$X_{A_{2j}}^{(1)}$}

The discussion of the brane web for the 5d SCFT associated to the $X_{A_{2j}}^{(1)}$ singularity is analogous to the previous section. We only present the infinite coupling brane web and corresponding magnetic quiver.
\begin{equation}
    \textnormal{brane web: }\raisebox{-0.5\height}{\scalebox{0.8}{\begin{tikzpicture}
        \node[seven] (5ff) at (-2,0) {};
        \node[seven] (6ff) at (2,0) {};
        \draw[thick, double] (5ff)--(6ff);
        \node at (1,0.3) {$j$};

        \node[seven] (1u) at (0,2) {};
        \node[seven] (2u) at (0,3) {};
        \node (3u) at (0,4) {$\vdots$};
        \node[seven] (4u) at (0,5) {};
        \node[seven] (5u) at (0,6) {};
        \node[seven] (6u) at (0,7) {};

        \node[seven] (1d) at (0,-2) {};
        \node[seven] (2d) at (0,-3) {};
        \node (3d) at (0,-4) {$\vdots$};
        \node[seven] (4d) at (0,-5) {};
        \node[seven] (5d) at (0,-6) {};
        \node[seven] (6d) at (0,-7) {};

        \draw (5u)--(6u) (5d)--(6d);
        \draw[thick,double] (5u)--(4u)--(3u)--(2u)--(1u)--(1d)--(2d)--(3d)--(4d)--(5d);

        \node at (-0.3,6.5) {$1$};
        \node at (-0.3,5.5) {$2$};
        \node at (-0.5,2.5) {$2j$};
        \node at (-0.7,1) {$2j+1$};
        \node at (-0.3,-6.5) {$1$};
        \node at (-0.3,-5.5) {$2$};
        \node at (-0.5,-2.5) {$2j$};
    \end{tikzpicture}}}\qquad\textnormal{magnetic quiver: }\raisebox{-0.5\height}{\begin{tikzpicture}
        \node[gauge,label=below:{\scriptsize$j$}] (0r) at (1,0) {};
        \node[gauge,label=left:{\scriptsize$2j+1$}] (0) at (0,0) {};
        \node[gauge,label=left:{\scriptsize$2j$}] (1) at (0,1) {};
        \node (2) at (0,2) {$\vdots$};
        \node[gauge,label=left:{\scriptsize$2$}] (3) at (0,3) {};
        \node[gauge,label=left:{\scriptsize$1$}] (4) at (0,4) {};
        \node[gauge,label=left:{\scriptsize$2j$}] (-1) at (0,-1) {};
        \node (-2) at (0,-2) {$\vdots$};
        \node[gauge,label=left:{\scriptsize$2$}] (-3) at (0,-3) {};
        \node[gauge,label=left:{\scriptsize$1$}] (-4) at (0,-4) {};
        \draw (0r)--(0)--(1)--(2)--(3)--(4) (0)--(-1)--(-2)--(-3)--(-4);
    \end{tikzpicture}}\;.
\end{equation}
The CB symmetry of this magnetic quiver is $\mathfrak{su}(2j+1)\times\mathfrak{su}(2j+1)\times\mathfrak{u}(1)$, which enhances to $\mathfrak{su}(10)$ for $j=2$, and enhances to $\mathfrak{sp}(9)$ for $j=1$ where the magnetic quiver has Coulomb branch $\mathbb{H}^9$ indicating that the SCFT consists of free hypers. We have
\begin{equation}
    \makebox[\textwidth][c]{\begin{tikzpicture}
        \node at (0,0) {$\mathrm{dim}_{\mathbb{H}}\left(\mathrm{CB}^{3d}\left(\mathsf{MQ}\left(\mathcal{T}_{X_{A_{2j}}^{(1)}}\right)\right)\right)=4j^2+5j=\mathrm{dim}_{\mathbb{H}}\left(\mathrm{HB}\left(\mathcal{T}_{X_{A_{2j}}^{(1)}}\right)\right)\;,$};
    \end{tikzpicture}}
\end{equation}
which matches Table \ref{tab:HB dim}.

\subsection{D-type}

\subsubsection{$X_{D_{2j+2}}^{(1)}$}

A low energy quiver description of the 5d SCFT associated to the $X_{D_{2j+2}}^{(1)}$ singularity is \cite{DeMarco:2023irn}
\begin{equation}
\label{eq:D2j+2(1)_EQ}
    \mathsf{Q}_{X_{D_{2j+2}}^{(1)}}=\raisebox{-.5\height}{\scalebox{0.8}{\begin{tikzpicture}
        \node[flavourSU,label=above:{$1$}] (1uf) at (0,1) {};
        \node[gaugeSU,label=above:{$j+1$}] (1u) at (2,1) {};
        \node[flavourSU,label=below:{$1$}] (1df) at (0,-1) {};
        \node[gaugeSU,label=below:{$j+1$}] (1d) at (2,-1) {};
        \node[gaugeSU,label=below:{$2j+1$}] (2) at (4,0) {};
        \node[gaugeSU,label=below:{$2j$}] (3) at (6,0) {};
        \node (4) at (8,0) {$\cdots$};
        \node[gaugeSU,label=below:{$3$}] (5) at (10,0) {};
        \node[gaugeSU,label=below:{$2$}] (6) at (12,0) {};
        \node[flavourSU,label=below:{$1$}] (7) at (14,0) {};
        \draw (1uf)--(1u)--(2) (1df)--(1d)--(2)--(3)--(4)--(5)--(6)--(7);
    \end{tikzpicture}}}\;.
\end{equation}
The D-type bifurcation in the gauge nodes can be realised on a brane web with ON$5^-$ (drawn as a green vertical line), similar to the constructions in \cite{Hanany:1999sj}. The brane web with both flavours massive is
\begin{equation}
    \scalebox{0.8}{\begin{tikzpicture}
        \node[seven,label=below:{\scriptsize$[0,1]7$}] (4) at (6,-1) {};
        \node[seven,label=below:{\scriptsize$[0,1]7$}] (5) at (7,-1) {};
        \node[seven] (6) at (9,-1) {};
        \node[seven] (7) at (10,-1) {};
        \node[seven] (8) at (12,-1) {};
        \node[seven,label=below:{\scriptsize$[0,1]7$}] (9) at (13,-1) {};
        \node[seven,label=below:{\scriptsize$[0,1]7$}] (10) at (14,-1) {};
        
        \node[seven,label=above:{\scriptsize$[0,1]7$}] (4f) at (6,2) {};
        \node[seven,label=above:{\scriptsize$[0,1]7$}] (5f) at (8,2) {};
        \node[seven,label=above:{\scriptsize$[1,0]7$}] (5ff) at (9,1) {};
        \node[seven] (6f) at (11,2) {};
        \node[seven] (7f) at (12,2) {};
        \node[seven] (8f) at (14,2) {};
        \node[seven,label=above:{\scriptsize$[1,1]7$}] (9f) at (15,2) {};
        \node[seven,label=right:{\scriptsize$[1,1]7$}] (10f) at (17,2) {};
        \node[seven,label=right:{\scriptsize$[1,0]7$}] (10ff) at (17,1) {};

        \draw[green] (5,-2)--(5,3);
        \draw (4)--(4f);
        \draw (5)--(7,0)--(8,1)--(5f) (8,1)--(5ff);
        \draw (6)--(9,0)--(6f);
        \draw (7)--(10,0)--(7f);
        \draw (8)--(12,0)--(8f);
        \draw (9)--(13,0)--(9f);
        \draw (10)--(14,0)--(16,1)--(10f) (16,1)--(10ff);
        
        \draw[thick,double] (5,0)--(10.5,0);
        \node at (11,0) {$\cdots$};
        \draw[thick,double] (11.5,0)--(14,0);
        \node at (5.5,0.2) {\scriptsize$2j+2$};
        \node at (6.5,0.2) {\scriptsize$2j+2$};
        \node at (8,0.2) {\scriptsize$2j+1$};
        \node at (9.6,0.2) {\scriptsize$2j$};
        \node at (12.6,0.2) {\scriptsize$3$};
        \node at (13.6,0.2) {\scriptsize$2$};
    \end{tikzpicture}}\;.
\end{equation}
After some Hanany-Witten moves and taking the masses to zero we obtain
\begin{equation}
    \scalebox{0.8}{\begin{tikzpicture}
        \node[seven] (4) at (6,-1) {};
        \node[seven] (5) at (7,-1) {};
        \node[seven] (6) at (9,-1) {};
        \node[seven] (7) at (10,-1) {};
        \node[seven] (8) at (12,-1) {};
        \node[seven] (9) at (13,-1) {};
        \node[seven] (10) at (14,-1) {};
        
        \node[seven] (4f) at (6,1) {};
        \node[seven] (5f) at (7,1) {};
        \node[seven] (6f) at (9,1) {};
        \node[seven] (7f) at (10,1) {};
        \node[seven] (8f) at (12,1) {};
        \node[seven] (9f) at (13,1) {};
        \node[seven] (10f) at (14,1) {};

        \node[seven] (ff) at (15,0) {};
        \node[seven,label=right:{\scriptsize$[1,0]7$}] (fff) at (16,0) {};

        \draw[green] (5,-2)--(5,2);
        \draw (4)--(4f);
        \draw (5)--(5f);
        \draw (6)--(6f);
        \draw (7)--(7f);
        \draw (8)--(8f);
        \draw (9)--(9f);
        \draw (10)--(10f);
        
        \draw[thick,double] (5,0)--(10.5,0);
        \node at (11,0) {$\cdots$};
        \draw[thick,double] (11.5,0)--(ff);
        \draw (ff)--(fff);
        \node at (5.5,0.2) {\scriptsize$2j+2$};
        \node at (14.5,0.2) {\scriptsize$2j+2$};
        \node at (15.5,0.2) {\scriptsize$1$};

        \draw [decorate,decoration={brace,amplitude=5pt},xshift=0pt,yshift=0pt]
    (14.3,-1.3)--(5.7,-1.3) node [black,midway,yshift=-0.4cm] {$2j+2$ $[0,1]7$};
    \end{tikzpicture}}\;.
\end{equation}
The magnetic quiver for the finite coupling electric quiver \eqref{eq:D2j+2(1)_EQ} can be read from the brane web using the prescriptions in \cite{Bourget:2020gzi}:\footnote{\cite{Bourget:2020gzi} deals with brane webs in the presence of O5 orientifold branes, such a setup is reached by performing an S-duality on our brane web.}
\begin{equation}
    \mathsf{MQ}\left(\mathsf{Q}_{X_{D_{2j+2}}^{(1)}}\right)=\raisebox{-.5\height}{\scalebox{1}{\begin{tikzpicture}
        \node[gaugeb,label=below:{\scriptsize$2j+2$}] (0r) at (1,0) {};
        \node[gauge,label=below:{\scriptsize$1$}] (1r) at (2,0) {};

        \draw (0r)--(1r);
        
        \node[gauge,label=above:{\scriptsize$1$}] (0rf1) at (0,1) {};
        \node[gauge,label=above:{\scriptsize$1$}] (0rf2) at (2,1) {};
        \node at (1,1) {$\cdots$};
        \draw (0rf1)--(0r)--(0rf2);
        \draw [decorate,decoration={brace,amplitude=5pt},xshift=0pt,yshift=0pt]
    (-0.3,1.5)--(2.3,1.5) node [black,midway,yshift=0.4cm] {\scriptsize$2j+2$};

        \draw (2.5,-0.5)--(3.5,0.5);
        \node at (3.5,0) {$\mathbb{Z}_2^{(1)}$};
    \end{tikzpicture}}}\;,
\end{equation}
where $/\mathbb{Z}_2^{(1)}$ denotes gauging the $\mathbb{Z}_2$ 1-form symmetry of the quiver, which is achieved on the level of the monopole formula by considering the half-integer magnetic lattice \cite{Bourget:2020xdz}. The Coulomb branch global symmetry of the magnetic quiver is U$(1)^{2j+3}$. We have
\begin{equation}
    \mathrm{dim}_{\mathbb{H}}\mathrm{CB}_{3d}\left(\mathsf{MQ}\left(\mathsf{Q}_{X_{D_{2j+2}}^{(1)}}\right)\right)=3j+4=\mathrm{dim}_{\mathbb{H}}\mathrm{HB}\left(\mathsf{Q}_{X_{D_{2j+2}}^{(1)}}\right)\;,
\end{equation}
which matches Table \ref{tab:HB dim}.

Taking the couplings to infinity we obtain the brane web (now with both ON$5^-$ and $\widetilde{\mathrm{ON}}5^-$ planes, where $\widetilde{\mathrm{ON}}5^-$ is drawn in orange and corresponds to an ON$5^-$ with a stuck $\frac{1}{2}(0,1)5$ on top, see Appendix \ref{app:BWNotation})
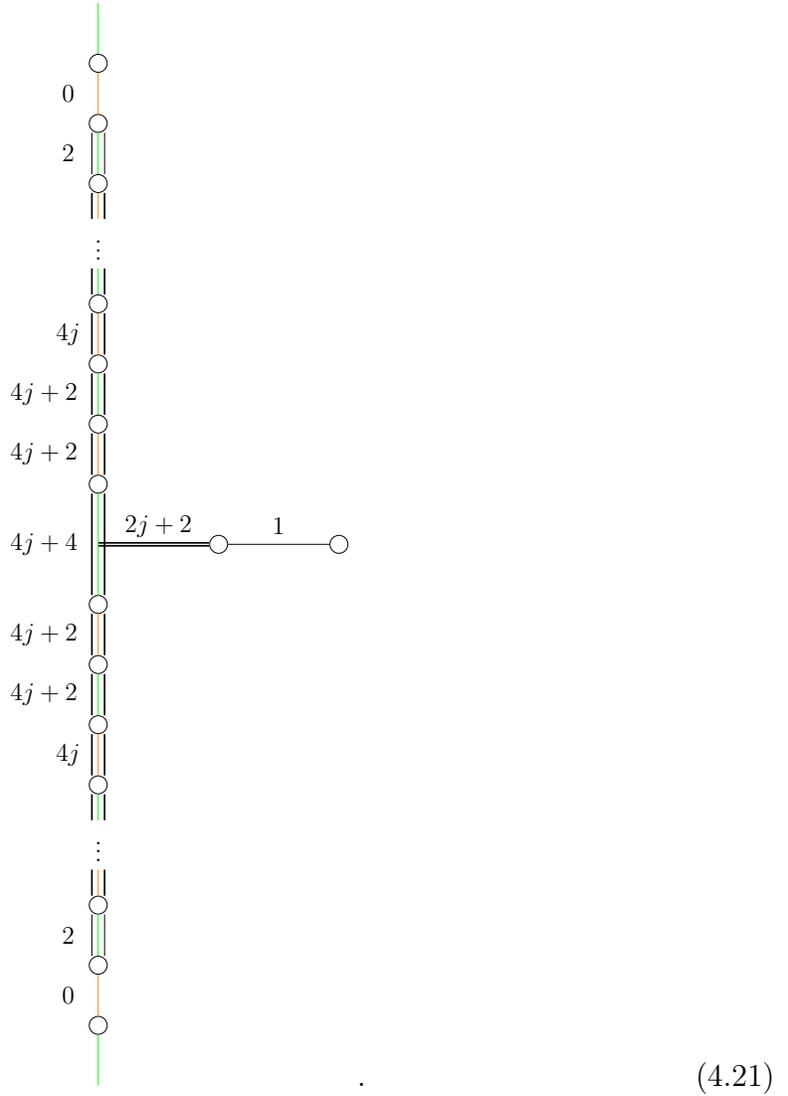
\begin{equation}
    \scalebox{0.8}{\begin{tikzpicture}
        \node[seven] (0r) at (2,0) {};
        \node[seven] (1r) at (4,0) {};

        \draw (0r)--(1r);
        \draw[thick,double] (0,0)--(0r);
        \node at (1,0.3) {$2j+2$};
        \node at (3,0.3) {$1$};

        \node[seven] (1) at (0,1) {};
        \node[seven] (2) at (0,2) {};
        \node[seven] (3) at (0,3) {};
        \node[seven] (4) at (0,4) {};
        \node (5) at (0,5) {$\vdots$};
        \node[seven] (6) at (0,6) {};
        \node[seven] (7) at (0,7) {};
        \node[seven] (8) at (0,8) {};
        
        \node[seven] (1d) at (0,-1) {};
        \node[seven] (2d) at (0,-2) {};
        \node[seven] (3d) at (0,-3) {};
        \node[seven] (4d) at (0,-4) {};
        \node (5d) at (0,-5) {$\vdots$};
        \node[seven] (6d) at (0,-6) {};
        \node[seven] (7d) at (0,-7) {};
        \node[seven] (8d) at (0,-8) {};

        \draw[transform canvas={xshift=-3pt}] (7d)--(6d) (6)--(7);
        \draw[transform canvas={xshift=3pt}] (7d)--(6d) (6)--(7);
        \draw[thick,transform canvas={xshift=-3pt}] (6d)--(5d)--(4d)--(3d)--(2d)--(1d)--(1)--(2)--(3)--(4)--(5)--(6);
        \draw[thick,transform canvas={xshift=3pt}] (6d)--(5d)--(4d)--(3d)--(2d)--(1d)--(1)--(2)--(3)--(4)--(5)--(6);

        \draw[green] (0,-9)--(8d) (7d)--(6d) (5d)--(4d) (3d)--(2d) (1d)--(1) (2)--(3) (4)--(5) (6)--(7) (8)--(0,9);
        \draw[orange] (8d)--(7d) (6d)--(5d) (4d)--(3d) (2d)--(1d) (1)--(2) (3)--(4) (5)--(6) (7)--(8);

        \node at (-0.9,0) {$4j+4$};
        \node at (-0.9,1.5) {$4j+2$};
        \node at (-0.9,2.5) {$4j+2$};
        \node at (-0.5,3.5) {$4j$};
        \node at (-0.5,6.5) {$2$};
        \node at (-0.5,7.5) {$0$};
        \node at (-0.9,-1.5) {$4j+2$};
        \node at (-0.9,-2.5) {$4j+2$};
        \node at (-0.5,-3.5) {$4j$};
        \node at (-0.5,-6.5) {$2$};
        \node at (-0.5,-7.5) {$0$};
    \end{tikzpicture}}\;.
\end{equation}
The numbers next to the brane segments on top of ON$5$ planes indicate the number of $\frac{1}{2}(0,1)5$-branes not accounting for the stuck $\frac{1}{2}(0,1)5$ involved in the $\widetilde{\mathrm{ON}}5^-$. From this brane web we can read the unitary-orthosymplectic magnetic quiver\footnote{For notation of quivers and types of gauge nodes, see Appendix \ref{app:QuiverNotation}.}
\begin{equation}
    \mathsf{MQ}\left(\mathcal{T}_{X_{D_{2j+2}}^{(1)}}\right)=\raisebox{-.5\height}{\scalebox{1}{\begin{tikzpicture}
        \draw (2.5,-0.5)--(3.5,0.5);
        \node at (3.5,0) {$\mathbb{Z}_2^{(1)}$};
        \node[gaugeb,label=below:{\scriptsize$2j+2$}] (0r) at (1,0) {};
        \node[gauge,label=below:{\scriptsize$1$}] (1r) at (2,0) {};
        \node[gauger,label=left:{\scriptsize$4j+4$}] (0) at (0,0) {};
        \node[gaugeb,label=left:{\scriptsize$4j+2$}] (1) at (0,1) {};
        \node[gauger,label=left:{\scriptsize$4j+2$}] (2) at (0,2) {};
        \node[gaugeb,label=left:{\scriptsize$4j$}] (3) at (0,3) {};
        \node (4) at (0,4) {$\vdots$};
        \node[gauger,label=left:{\scriptsize$2$}] (5) at (0,5) {};
        \node[gaugeb,label=left:{\scriptsize$0$}] (6) at (0,6) {};
        \node[gaugeb,label=left:{\scriptsize$4j+2$}] (-1) at (0,-1) {};
        \node[gauger,label=left:{\scriptsize$4j+2$}] (-2) at (0,-2) {};
        \node[gaugeb,label=left:{\scriptsize$4j$}] (-3) at (0,-3) {};
        \node (-4) at (0,-4) {$\vdots$};
        \node[gauger,label=left:{\scriptsize$2$}] (-5) at (0,-5) {};
        \node[gaugeb,label=left:{\scriptsize$0$}] (-6) at (0,-6) {};
        \draw (1r)--(0r)--(0)--(1)--(2)--(3)--(4)--(5)--(6) (0)--(-1)--(-2)--(-3)--(-4)--(-5)--(-6);
    \end{tikzpicture}}}\;,
\end{equation}
from which we can identify the $\mathfrak{so}(4j+4)\times\mathfrak{so}(4j+4)\times \mathfrak{u}(1)^2$ global symmetry of the SCFT, which exists for $j\geq1$. We have
\begin{equation}
    \makebox[\textwidth][c]{\begin{tikzpicture}
        \node at (0,0) {$\mathrm{dim}_{\mathbb{H}}\mathrm{CB}_{3d}\left(\mathsf{MQ}\left(\mathcal{T}_{X_{D_{2j+2}}^{(1)}}\right)\right)=8j^2+15j+8=(3j+4)+4(2j^2+3j+1)=\mathrm{dim}_{\mathbb{H}}\mathrm{HB}\left(\mathcal{T}_{X_{D_{2j+2}}^{(1)}}\right)\;,$};
    \end{tikzpicture}}
\end{equation}
which matches Table \ref{tab:HB dim}.

\subsubsection{$X_{D_{2j+3}}^{(1)}$}

The discussion of the brane web for the 5d SCFT associated to the $X_{D_{2j+3}}^{(1)}$ singularity is analogous to the previous section. We only present the infinite coupling brane web and corresponding magnetic quiver:
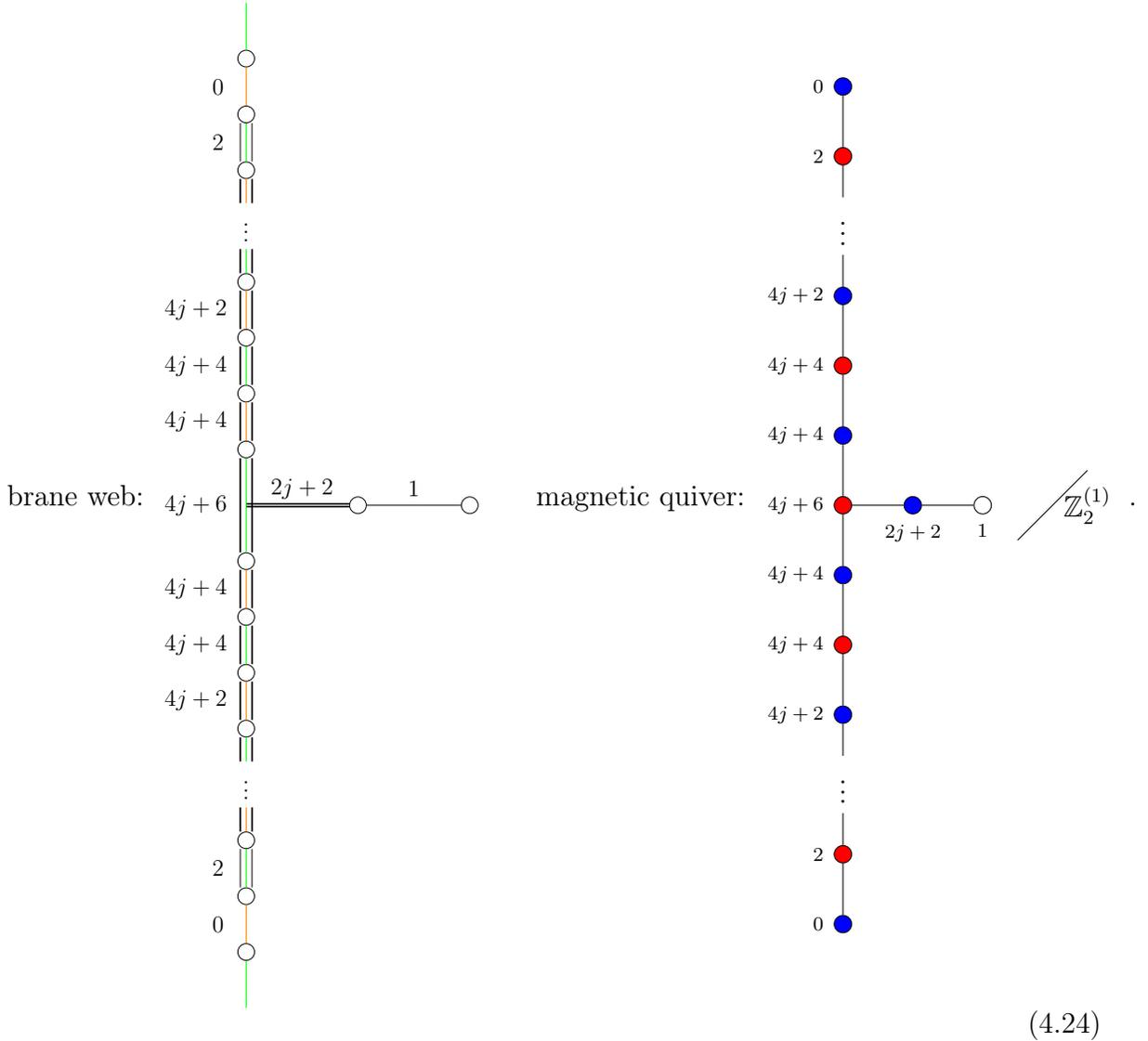
\begin{equation}
    \textnormal{brane web: }\raisebox{-0.5\height}{\scalebox{0.8}{\begin{tikzpicture}
        \node[seven] (0r) at (2,0) {};
        \node[seven] (1r) at (4,0) {};

        \draw (0r)--(1r);
        \draw[thick,double] (0,0)--(0r);
        \node at (1,0.3) {$2j+2$};
        \node at (3,0.3) {$1$};

        \node[seven] (1) at (0,1) {};
        \node[seven] (2) at (0,2) {};
        \node[seven] (3) at (0,3) {};
        \node[seven] (4) at (0,4) {};
        \node (5) at (0,5) {$\vdots$};
        \node[seven] (6) at (0,6) {};
        \node[seven] (7) at (0,7) {};
        \node[seven] (8) at (0,8) {};
        
        \node[seven] (1d) at (0,-1) {};
        \node[seven] (2d) at (0,-2) {};
        \node[seven] (3d) at (0,-3) {};
        \node[seven] (4d) at (0,-4) {};
        \node (5d) at (0,-5) {$\vdots$};
        \node[seven] (6d) at (0,-6) {};
        \node[seven] (7d) at (0,-7) {};
        \node[seven] (8d) at (0,-8) {};

        \draw[transform canvas={xshift=-3pt}] (7d)--(6d) (6)--(7);
        \draw[transform canvas={xshift=3pt}] (7d)--(6d) (6)--(7);
        \draw[thick,transform canvas={xshift=-3pt}] (6d)--(5d)--(4d)--(3d)--(2d)--(1d)--(1)--(2)--(3)--(4)--(5)--(6);
        \draw[thick,transform canvas={xshift=3pt}] (6d)--(5d)--(4d)--(3d)--(2d)--(1d)--(1)--(2)--(3)--(4)--(5)--(6);

        \draw[green] (0,-9)--(8d) (7d)--(6d) (5d)--(4d) (3d)--(2d) (1d)--(1) (2)--(3) (4)--(5) (6)--(7) (8)--(0,9);
        \draw[orange] (8d)--(7d) (6d)--(5d) (4d)--(3d) (2d)--(1d) (1)--(2) (3)--(4) (5)--(6) (7)--(8);

        \node at (-0.9,0) {$4j+6$};
        \node at (-0.9,1.5) {$4j+4$};
        \node at (-0.9,2.5) {$4j+4$};
        \node at (-0.9,3.5) {$4j+2$};
        \node at (-0.5,6.5) {$2$};
        \node at (-0.5,7.5) {$0$};
        \node at (-0.9,-1.5) {$4j+4$};
        \node at (-0.9,-2.5) {$4j+4$};
        \node at (-0.9,-3.5) {$4j+2$};
        \node at (-0.5,-6.5) {$2$};
        \node at (-0.5,-7.5) {$0$};
    \end{tikzpicture}}}\qquad\textnormal{magnetic quiver: }\raisebox{-0.5\height}{\begin{tikzpicture}
        \draw (2.5,-0.5)--(3.5,0.5);
        \node at (3.5,0) {$\mathbb{Z}_2^{(1)}$};
        \node[gaugeb,label=below:{\scriptsize$2j+2$}] (0r) at (1,0) {};
        \node[gauge,label=below:{\scriptsize$1$}] (1r) at (2,0) {};
        \node[gauger,label=left:{\scriptsize$4j+6$}] (0) at (0,0) {};
        \node[gaugeb,label=left:{\scriptsize$4j+4$}] (1) at (0,1) {};
        \node[gauger,label=left:{\scriptsize$4j+4$}] (2) at (0,2) {};
        \node[gaugeb,label=left:{\scriptsize$4j+2$}] (3) at (0,3) {};
        \node (4) at (0,4) {$\vdots$};
        \node[gauger,label=left:{\scriptsize$2$}] (5) at (0,5) {};
        \node[gaugeb,label=left:{\scriptsize$0$}] (6) at (0,6) {};
        \node[gaugeb,label=left:{\scriptsize$4j+4$}] (-1) at (0,-1) {};
        \node[gauger,label=left:{\scriptsize$4j+4$}] (-2) at (0,-2) {};
        \node[gaugeb,label=left:{\scriptsize$4j+2$}] (-3) at (0,-3) {};
        \node (-4) at (0,-4) {$\vdots$};
        \node[gauger,label=left:{\scriptsize$2$}] (-5) at (0,-5) {};
        \node[gaugeb,label=left:{\scriptsize$0$}] (-6) at (0,-6) {};
        \draw (1r)--(0r)--(0)--(1)--(2)--(3)--(4)--(5)--(6) (0)--(-1)--(-2)--(-3)--(-4)--(-5)--(-6);
    \end{tikzpicture}}\;.
\end{equation}
We note that, again, in the brane-web many D5 end on the same D7, and hence there exists a dual description, for the 5d conformal matter of type $X_{A_{2j}}^{(1)}$, in terms of a toric diagram with white dots. Again, this means that we can think to the procedure  of passing from the aforementioned white-dots toric diagram to the singularity $X_{A_{2j}}^{(1)}$ as the geometric analogous of an Hanany-Witten move \cite{Bourget:2023wlb}.\\

The Coulomb branch global symmetry of the magnetic quiver is $\mathfrak{so}(4j+6)\times\mathfrak{so}(4j+6)\times \mathfrak{u}(1)$ for $j\geq1$, and $E_7$ for $j=0$. We have 
\begin{equation}
    \makebox[\textwidth][c]{\begin{tikzpicture}
        \node at (0,0) {$\mathrm{dim}_{\mathbb{H}}\mathrm{CB}_{3d}\left(\mathsf{MQ}\left(\mathcal{T}_{X_{D_{2j+3}}^{(1)}}\right)\right)=8j^2+23j+17=\mathrm{dim}_{\mathbb{H}}\mathrm{HB}\left(\mathcal{T}_{X_{D_{2j+3}}^{(1)}}\right)\;,$};
    \end{tikzpicture}}
\end{equation}
which matches Table \ref{tab:HB dim}.

The magnetic quiver for $\mathcal{T}_{X_{D_{3}}^{(1)}}$ is IR dual to the magnetic quiver for $\mathcal{T}_{X_{A_3}^{(1)}}$ \cite{Bourget:2020xdz}, as is expected since $\mathfrak{su}(4)=\mathfrak{so}(6)$.

\subsubsection{$X_{D_{2j+2}}^{(2)}$}
The discussion of the brane web for the 5d SCFT associated to the $X_{D_{2j+2}}^{(2)}$ singularity is analogous to the previous section. We only present the infinite coupling brane web and corresponding magnetic quiver:
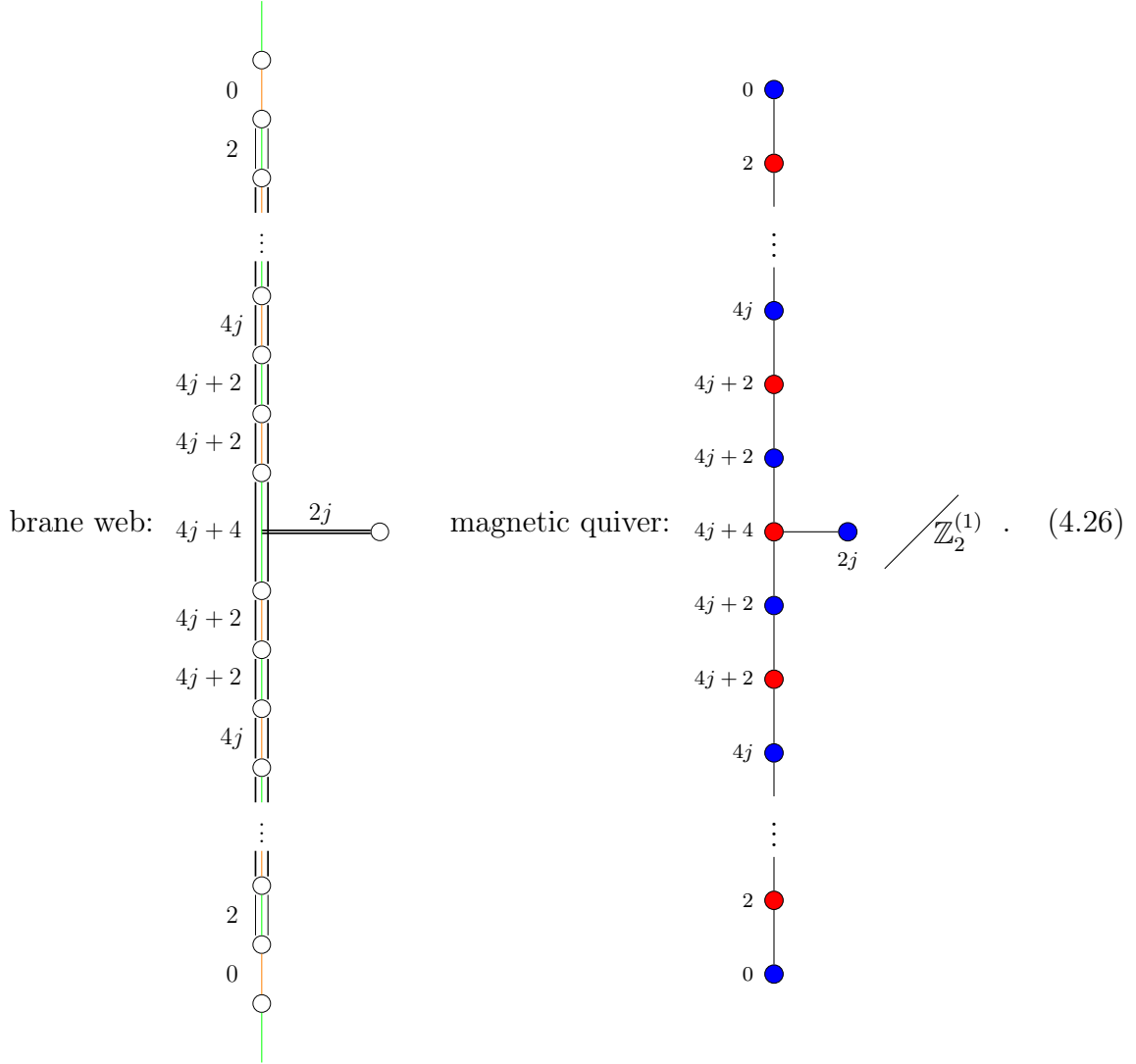
\begin{equation}
    \textnormal{brane web: }\raisebox{-0.5\height}{\scalebox{0.8}{\begin{tikzpicture}
        \node[seven] (0r) at (2,0) {};
        
        \draw[thick,double] (0,0)--(0r);
        \node at (1,0.3) {$2j$};

        \node[seven] (1) at (0,1) {};
        \node[seven] (2) at (0,2) {};
        \node[seven] (3) at (0,3) {};
        \node[seven] (4) at (0,4) {};
        \node (5) at (0,5) {$\vdots$};
        \node[seven] (6) at (0,6) {};
        \node[seven] (7) at (0,7) {};
        \node[seven] (8) at (0,8) {};
        
        \node[seven] (1d) at (0,-1) {};
        \node[seven] (2d) at (0,-2) {};
        \node[seven] (3d) at (0,-3) {};
        \node[seven] (4d) at (0,-4) {};
        \node (5d) at (0,-5) {$\vdots$};
        \node[seven] (6d) at (0,-6) {};
        \node[seven] (7d) at (0,-7) {};
        \node[seven] (8d) at (0,-8) {};

        \draw[transform canvas={xshift=-3pt}] (7d)--(6d) (6)--(7);
        \draw[transform canvas={xshift=3pt}] (7d)--(6d) (6)--(7);
        \draw[thick,transform canvas={xshift=-3pt}] (6d)--(5d)--(4d)--(3d)--(2d)--(1d)--(1)--(2)--(3)--(4)--(5)--(6);
        \draw[thick,transform canvas={xshift=3pt}] (6d)--(5d)--(4d)--(3d)--(2d)--(1d)--(1)--(2)--(3)--(4)--(5)--(6);

        \draw[green] (0,-9)--(8d) (7d)--(6d) (5d)--(4d) (3d)--(2d) (1d)--(1) (2)--(3) (4)--(5) (6)--(7) (8)--(0,9);
        \draw[orange] (8d)--(7d) (6d)--(5d) (4d)--(3d) (2d)--(1d) (1)--(2) (3)--(4) (5)--(6) (7)--(8);

        \node at (-0.9,0) {$4j+4$};
        \node at (-0.9,1.5) {$4j+2$};
        \node at (-0.9,2.5) {$4j+2$};
        \node at (-0.5,3.5) {$4j$};
        \node at (-0.5,6.5) {$2$};
        \node at (-0.5,7.5) {$0$};
        \node at (-0.9,-1.5) {$4j+2$};
        \node at (-0.9,-2.5) {$4j+2$};
        \node at (-0.5,-3.5) {$4j$};
        \node at (-0.5,-6.5) {$2$};
        \node at (-0.5,-7.5) {$0$};
    \end{tikzpicture}}}\qquad\textnormal{magnetic quiver: }\raisebox{-0.5\height}{\begin{tikzpicture}
        \draw (1.5,-0.5)--(2.5,0.5);
        \node at (2.5,0) {$\mathbb{Z}_2^{(1)}$};
        \node[gaugeb,label=below:{\scriptsize$2j$}] (0r) at (1,0) {};
        \node[gauger,label=left:{\scriptsize$4j+4$}] (0) at (0,0) {};
        \node[gaugeb,label=left:{\scriptsize$4j+2$}] (1) at (0,1) {};
        \node[gauger,label=left:{\scriptsize$4j+2$}] (2) at (0,2) {};
        \node[gaugeb,label=left:{\scriptsize$4j$}] (3) at (0,3) {};
        \node (4) at (0,4) {$\vdots$};
        \node[gauger,label=left:{\scriptsize$2$}] (5) at (0,5) {};
        \node[gaugeb,label=left:{\scriptsize$0$}] (6) at (0,6) {};
        \node[gaugeb,label=left:{\scriptsize$4j+2$}] (-1) at (0,-1) {};
        \node[gauger,label=left:{\scriptsize$4j+2$}] (-2) at (0,-2) {};
        \node[gaugeb,label=left:{\scriptsize$4j$}] (-3) at (0,-3) {};
        \node (-4) at (0,-4) {$\vdots$};
        \node[gauger,label=left:{\scriptsize$2$}] (-5) at (0,-5) {};
        \node[gaugeb,label=left:{\scriptsize$0$}] (-6) at (0,-6) {};
        \draw (0r)--(0)--(1)--(2)--(3)--(4)--(5)--(6) (0)--(-1)--(-2)--(-3)--(-4)--(-5)--(-6);
    \end{tikzpicture}}\;.
\end{equation}
The Coulomb branch global symmetry of the magnetic quiver is $\mathfrak{so}(4j+4)\times\mathfrak{so}(4j+4)$ for $j>1$, and $E_8$ for $j=1$. We have 
\begin{equation}
    \makebox[\textwidth][c]{\begin{tikzpicture}
        \node at (0,0) {$\mathrm{dim}_{\mathbb{H}}\mathrm{CB}_{3d}\left(\mathsf{MQ}\left(\mathcal{T}_{X_{D_{2j+2}}^{(2)}}\right)\right)=8j^2+15j+6=\mathrm{dim}_{\mathbb{H}}\mathrm{HB}\left(\mathcal{T}_{X_{D_{2j+2}}^{(2)}}\right)\;,$};
    \end{tikzpicture}}
\end{equation}
which matches Table \ref{tab:HB dim}.

\subsubsection{$X_{D_{2j+3}}^{(2)}$}
The discussion of the brane web for the 5d SCFT associated to the $X_{D_{2j+3}}^{(2)}$ singularity is analogous to the previous section. We only present the infinite coupling brane web and corresponding magnetic quiver:
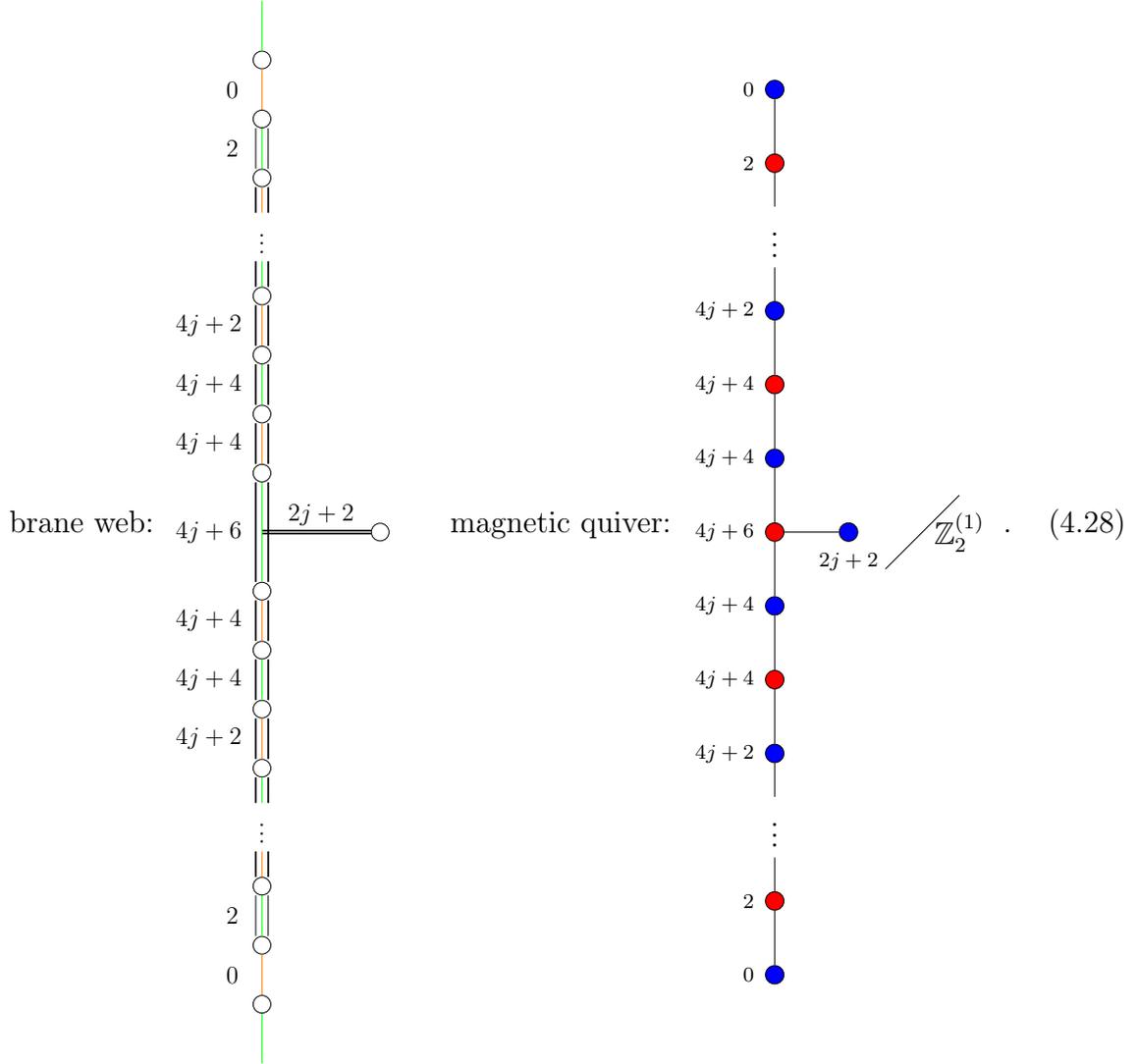
\begin{equation}
    \textnormal{brane web: }\raisebox{-0.5\height}{\scalebox{0.8}{\begin{tikzpicture}
        \node[seven] (0r) at (2,0) {};
        
        \draw[thick,double] (0,0)--(0r);
        \node at (1,0.3) {$2j+2$};

        \node[seven] (1) at (0,1) {};
        \node[seven] (2) at (0,2) {};
        \node[seven] (3) at (0,3) {};
        \node[seven] (4) at (0,4) {};
        \node (5) at (0,5) {$\vdots$};
        \node[seven] (6) at (0,6) {};
        \node[seven] (7) at (0,7) {};
        \node[seven] (8) at (0,8) {};
        
        \node[seven] (1d) at (0,-1) {};
        \node[seven] (2d) at (0,-2) {};
        \node[seven] (3d) at (0,-3) {};
        \node[seven] (4d) at (0,-4) {};
        \node (5d) at (0,-5) {$\vdots$};
        \node[seven] (6d) at (0,-6) {};
        \node[seven] (7d) at (0,-7) {};
        \node[seven] (8d) at (0,-8) {};

        \draw[transform canvas={xshift=-3pt}] (7d)--(6d) (6)--(7);
        \draw[transform canvas={xshift=3pt}] (7d)--(6d) (6)--(7);
        \draw[thick,transform canvas={xshift=-3pt}] (6d)--(5d)--(4d)--(3d)--(2d)--(1d)--(1)--(2)--(3)--(4)--(5)--(6);
        \draw[thick,transform canvas={xshift=3pt}] (6d)--(5d)--(4d)--(3d)--(2d)--(1d)--(1)--(2)--(3)--(4)--(5)--(6);

        \draw[green] (0,-9)--(8d) (7d)--(6d) (5d)--(4d) (3d)--(2d) (1d)--(1) (2)--(3) (4)--(5) (6)--(7) (8)--(0,9);
        \draw[orange] (8d)--(7d) (6d)--(5d) (4d)--(3d) (2d)--(1d) (1)--(2) (3)--(4) (5)--(6) (7)--(8);

        \node at (-0.9,0) {$4j+6$};
        \node at (-0.9,1.5) {$4j+4$};
        \node at (-0.9,2.5) {$4j+4$};
        \node at (-0.9,3.5) {$4j+2$};
        \node at (-0.5,6.5) {$2$};
        \node at (-0.5,7.5) {$0$};
        \node at (-0.9,-1.5) {$4j+4$};
        \node at (-0.9,-2.5) {$4j+4$};
        \node at (-0.9,-3.5) {$4j+2$};
        \node at (-0.5,-6.5) {$2$};
        \node at (-0.5,-7.5) {$0$};
    \end{tikzpicture}}}\qquad\textnormal{magnetic quiver: }\raisebox{-0.5\height}{\begin{tikzpicture}
        \draw (1.5,-0.5)--(2.5,0.5);
        \node at (2.5,0) {$\mathbb{Z}_2^{(1)}$};
        \node[gaugeb,label=below:{\scriptsize$2j+2$}] (0r) at (1,0) {};
        \node[gauger,label=left:{\scriptsize$4j+6$}] (0) at (0,0) {};
        \node[gaugeb,label=left:{\scriptsize$4j+4$}] (1) at (0,1) {};
        \node[gauger,label=left:{\scriptsize$4j+4$}] (2) at (0,2) {};
        \node[gaugeb,label=left:{\scriptsize$4j+2$}] (3) at (0,3) {};
        \node (4) at (0,4) {$\vdots$};
        \node[gauger,label=left:{\scriptsize$2$}] (5) at (0,5) {};
        \node[gaugeb,label=left:{\scriptsize$0$}] (6) at (0,6) {};
        \node[gaugeb,label=left:{\scriptsize$4j+4$}] (-1) at (0,-1) {};
        \node[gauger,label=left:{\scriptsize$4j+4$}] (-2) at (0,-2) {};
        \node[gaugeb,label=left:{\scriptsize$4j+2$}] (-3) at (0,-3) {};
        \node (-4) at (0,-4) {$\vdots$};
        \node[gauger,label=left:{\scriptsize$2$}] (-5) at (0,-5) {};
        \node[gaugeb,label=left:{\scriptsize$0$}] (-6) at (0,-6) {};
        \draw (0r)--(0)--(1)--(2)--(3)--(4)--(5)--(6) (0)--(-1)--(-2)--(-3)--(-4)--(-5)--(-6);
    \end{tikzpicture}}\;.
\end{equation}
The Coulomb branch global symmetry of the magnetic quiver is $\mathfrak{so}(4j+6)\times\mathfrak{so}(4j+6)\times\mathfrak{u}(1)$ for $j\geq1$. For $j=1$ the magnetic quiver has Coulomb branch $\mathbb{H}^{16}$ indicating that the SCFT consists of free hypers. We have 
\begin{equation}
    \makebox[\textwidth][c]{\begin{tikzpicture}
        \node at (0,0) {$\mathrm{dim}_{\mathbb{H}}\mathrm{CB}_{3d}\left(\mathsf{MQ}\left(\mathcal{T}_{X_{D_{2j+3}}^{(2)}}\right)\right)=8j^2+23j+16=\mathrm{dim}_{\mathbb{H}}\mathrm{HB}\left(\mathcal{T}_{X_{D_{2j+3}}^{(2)}}\right)\;,$};
    \end{tikzpicture}}
\end{equation}
which matches Table \ref{tab:HB dim}.

\subsubsection{$X_{D_{j}}^{(3)}$}
The discussion of the brane web for the 5d SCFT associated to the $X_{D_{2}}^{(3)}$ singularity is analogous to the previous section. We only present the infinite coupling brane web and corresponding magnetic quiver:
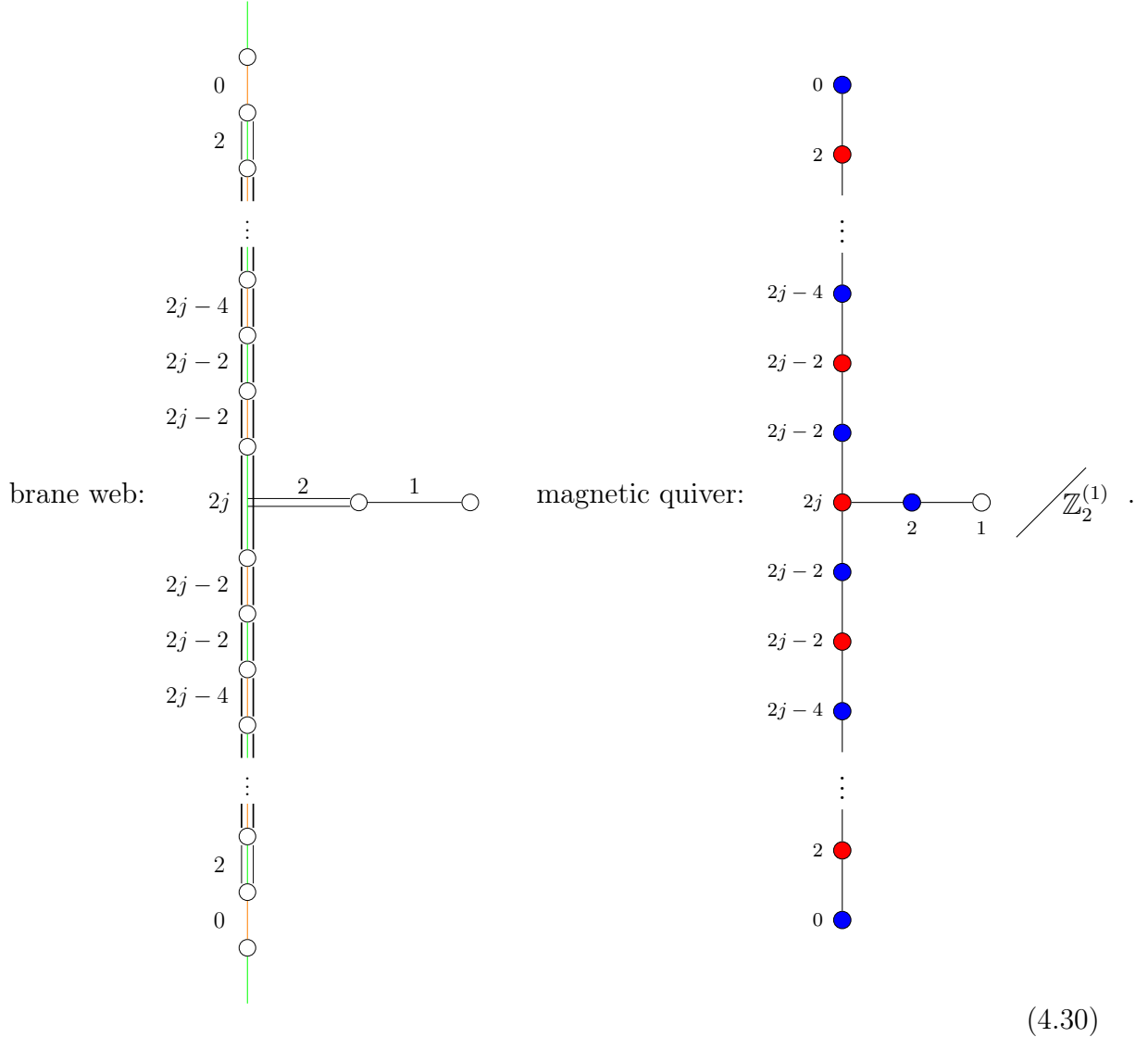
\begin{equation}
    \textnormal{brane web: }\raisebox{-0.5\height}{\scalebox{0.8}{\begin{tikzpicture}
        \node[seven] (0r) at (2,0) {};
        \node[seven] (1r) at (4,0) {};

        \draw (0r)--(1r);
        \draw[transform canvas={yshift=-2pt}] (0,0)--(0r);
        \draw[transform canvas={yshift=2pt}] (0,0)--(0r);
        \node at (1,0.3) {$2$};
        \node at (3,0.3) {$1$};

        \node[seven] (1) at (0,1) {};
        \node[seven] (2) at (0,2) {};
        \node[seven] (3) at (0,3) {};
        \node[seven] (4) at (0,4) {};
        \node (5) at (0,5) {$\vdots$};
        \node[seven] (6) at (0,6) {};
        \node[seven] (7) at (0,7) {};
        \node[seven] (8) at (0,8) {};
        
        \node[seven] (1d) at (0,-1) {};
        \node[seven] (2d) at (0,-2) {};
        \node[seven] (3d) at (0,-3) {};
        \node[seven] (4d) at (0,-4) {};
        \node (5d) at (0,-5) {$\vdots$};
        \node[seven] (6d) at (0,-6) {};
        \node[seven] (7d) at (0,-7) {};
        \node[seven] (8d) at (0,-8) {};

        \draw[transform canvas={xshift=-3pt}] (7d)--(6d) (6)--(7);
        \draw[transform canvas={xshift=3pt}] (7d)--(6d) (6)--(7);
        \draw[thick,transform canvas={xshift=-3pt}] (6d)--(5d)--(4d)--(3d)--(2d)--(1d)--(1)--(2)--(3)--(4)--(5)--(6);
        \draw[thick,transform canvas={xshift=3pt}] (6d)--(5d)--(4d)--(3d)--(2d)--(1d)--(1)--(2)--(3)--(4)--(5)--(6);

        \draw[green] (0,-9)--(8d) (7d)--(6d) (5d)--(4d) (3d)--(2d) (1d)--(1) (2)--(3) (4)--(5) (6)--(7) (8)--(0,9);
        \draw[orange] (8d)--(7d) (6d)--(5d) (4d)--(3d) (2d)--(1d) (1)--(2) (3)--(4) (5)--(6) (7)--(8);

        \node at (-0.5,0) {$2j$};
        \node at (-0.9,1.5) {$2j-2$};
        \node at (-0.9,2.5) {$2j-2$};
        \node at (-0.9,3.5) {$2j-4$};
        \node at (-0.5,6.5) {$2$};
        \node at (-0.5,7.5) {$0$};
        \node at (-0.9,-1.5) {$2j-2$};
        \node at (-0.9,-2.5) {$2j-2$};
        \node at (-0.9,-3.5) {$2j-4$};
        \node at (-0.5,-6.5) {$2$};
        \node at (-0.5,-7.5) {$0$};
    \end{tikzpicture}}}\qquad\textnormal{magnetic quiver: }\raisebox{-0.5\height}{\begin{tikzpicture}
        \draw (2.5,-0.5)--(3.5,0.5);
        \node at (3.5,0) {$\mathbb{Z}_2^{(1)}$};
        \node[gaugeb,label=below:{\scriptsize$2$}] (0r) at (1,0) {};
        \node[gauge,label=below:{\scriptsize$1$}] (1r) at (2,0) {};
        \node[gauger,label=left:{\scriptsize$2j$}] (0) at (0,0) {};
        \node[gaugeb,label=left:{\scriptsize$2j-2$}] (1) at (0,1) {};
        \node[gauger,label=left:{\scriptsize$2j-2$}] (2) at (0,2) {};
        \node[gaugeb,label=left:{\scriptsize$2j-4$}] (3) at (0,3) {};
        \node (4) at (0,4) {$\vdots$};
        \node[gauger,label=left:{\scriptsize$2$}] (5) at (0,5) {};
        \node[gaugeb,label=left:{\scriptsize$0$}] (6) at (0,6) {};
        \node[gaugeb,label=left:{\scriptsize$2j-2$}] (-1) at (0,-1) {};
        \node[gauger,label=left:{\scriptsize$2j-2$}] (-2) at (0,-2) {};
        \node[gaugeb,label=left:{\scriptsize$2j-4$}] (-3) at (0,-3) {};
        \node (-4) at (0,-4) {$\vdots$};
        \node[gauger,label=left:{\scriptsize$2$}] (-5) at (0,-5) {};
        \node[gaugeb,label=left:{\scriptsize$0$}] (-6) at (0,-6) {};
        \draw (1r)--(0r)--(0)--(1)--(2)--(3)--(4)--(5)--(6) (0)--(-1)--(-2)--(-3)--(-4)--(-5)--(-6);
    \end{tikzpicture}}\;.
\end{equation}
The Coulomb branch global symmetry of the magnetic quiver is $\mathfrak{so}(4j)\times\mathfrak{su}(2)$\footnote{Notice that the global symmetry always enhances from the $\mathfrak{g}\times \mathfrak{g} \times F_{\text{rest}}$ type in this case.} for $j\geq3$, and $E_7$ for $j=3$ \cite{Bourget:2020xdz}. We have 
\begin{equation}
    \makebox[\textwidth][c]{\begin{tikzpicture}
        \node at (0,0) {$\mathrm{dim}_{\mathbb{H}}\mathrm{CB}_{3d}\left(\mathsf{MQ}\left(\mathcal{T}_{X_{D_{j}}^{(3)}}\right)\right)=2j^2-j+2=\mathrm{dim}_{\mathbb{H}}\mathrm{HB}\left(\mathcal{T}_{X_{D_{j}}^{(3)}}\right)\;,$};
    \end{tikzpicture}}
\end{equation}
which matches Table \ref{tab:HB dim}.

\subsection{Relation to class-$\mathcal{S}$ construction}
We observe that all A- and D-type 5d conformal matter atoms can be constructed as worldvolume theories of 5-brane webs. The magnetic quivers read from these brane webs match exactly those read from the class-$\mathcal{S}$ construction (from puncture data). This further confirms that the identified toroidal reduction is the correct one.

\section{5d conformal matter molecules and class \texorpdfstring{$\mathcal{S}$}{S}}
\label{sec:moleculesbigsec}
In this section we naturally extend the construction presented in \cref{sec:classS} and determine the 4d descendants of the 5d SCFTs encoded by the complete intersections \eqref{systemfourfoldgen}, that correspond to 5d conformal matter molecules (or, equivalently, to 5d generalized linear quivers). The 5d molecules in general admit a flavor symmetry of the form:
\begin{equation}
    F_{UV} = \mathfrak{g}\times \mathfrak{g} \times F_{\text{rest}}.
\end{equation}
It turns out that on a specific point of the extended CB of the 5d SCFT, corresponding to turning on some masses for $F_{\text{rest}}$, its circle reduction lends itself to a 4d $\mathcal{N}=2$ class-$\mathcal{S}$ interpretation. Hence, we match the CB dimension across dimensional reduction in \cref{sec:CBdimensionmolecules}, and we relate the dimension of the HB and the flavor symmetry of the 5d theory to the ones of its 4d descendant in \cref{sec:HBdimensionmolecules}. In Section \eqref{sec:MoleculesBWandMQ} we then compute the brane-webs and magnetic quivers for the 5d conformal matter molecules.

\subsection{4d reduction of 5d conformal matter molecules}
\label{sec:CBdimensionmolecules}
\begin{figure}[H]
\centering
\begin{subfigure}[c]{.9\textwidth}
\centering
$  \begin{array}{cc}
\makecell{\tilde{\mathsf{Q}}_{X_{\mathfrak{g}}^{(i)}} =\\ \vspace{0.5cm}} & 
  \scalemath{1}{  \begin{tikzpicture}
        \draw[thick] (0,0.7)--(1.4,0.7)--(1.4,-0.7)--(0,-0.7)--cycle;
        \draw[thick,double] (1.5,0)--(2.9,0);
        \draw[thick] (3,0.7)--(4.4,0.7)--(4.4,-0.7)--(3,-0.7)--cycle;
        \node at (0.7,0) {$\mathfrak{g}$};
        \node at (3.7,0) {$\mathfrak{g}$};
        \node at (2.2,0.3) {$(\mathfrak{g},\mathfrak{g})_{(i)}$};
        \end{tikzpicture}}
            \end{array}$
    \caption*{}
    \end{subfigure}
    \vspace{-0.5cm}
    
 \begin{subfigure}[c]{0.9\textwidth}
    \centering
    \scalemath{1}{   \begin{tikzpicture}
   \hspace{1.5cm} \node at (6.2,0) {$\displaystyle\left\downarrow\vphantom{\int_A^B}\right.$};
    \node at (7.8,0) {\text{circle reduction}};
    \end{tikzpicture}}
    \end{subfigure}
       \begin{subfigure}[c]{.9\textwidth}
       \centering
       \vspace{0.8cm}
\scalemath{1}{
         \begin{tikzpicture}
        \draw[thick] (2.2,0) circle (1.5);
        \node at (2.8,-0.2) {\small$\mathcal{O}_{\text{max}}$};
        \node at (3.4,-0.2) {$\times$};
         \node at (1.6,-0.2) {\small$\mathcal{O}_{\text{max}}$};
          \node at (1.0,-0.2) {$\times$};
          \node at (2.2,0.8) {\small$\mathcal{O}_{III}^{(i)}$};
           \node at (2.2,1.2) {$\times$};
        \end{tikzpicture}}
    \end{subfigure}
    \caption{The conformal matter 5d theory of type $X_{\mathfrak{g}}^{(i)}$ and its 4d descendant realized as a class-$\mathcal{S}$ setup with three regular punctures. This corresponds to the vertical blue arrow in Figure \ref{fig:InterdimensionalFlowAtom}.}
     \label{fig: class S duality}
\end{figure}
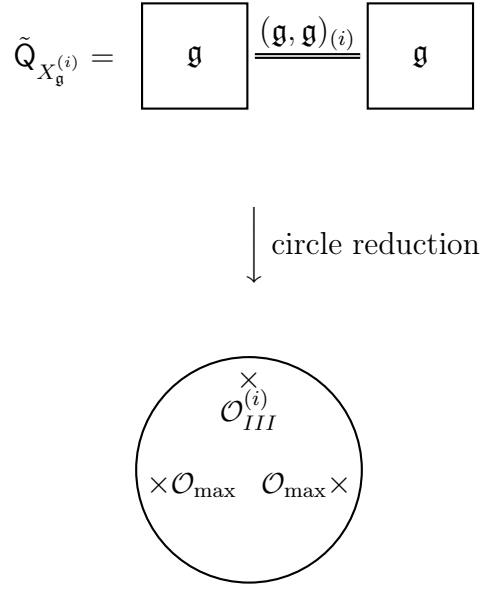

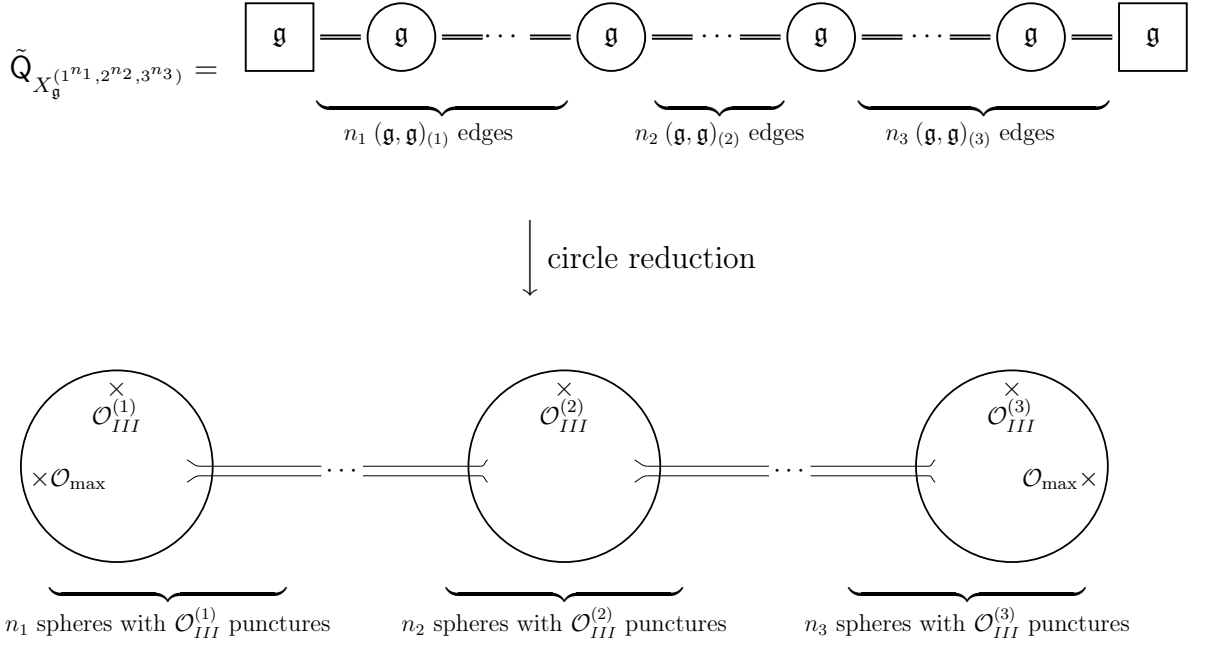
\begin{figure}[H]
\centering
\begin{subfigure}[c]{.9\textwidth}
\centering
$   \begin{array}{cc}
\makecell{\tilde{\mathsf{Q}}_{X_{\mathfrak{g}}^{(1^{n_1},2^{n_2},3^{n_3})}} =\\ \vspace{1.5cm}} &
  \scalemath{0.9}{ 
    \begin{tikzpicture}
        \draw[thick] (0,0) circle (0.5);
        \node at (0,0) {$\mathfrak{g}$};
        \draw[thick,double] (0.6,0)--(1.2,0);
        \draw[thick,double] (-0.6,0)--(-1.2,0);
        \node at (0.9,0.3) {};
        \node at (-0.9,0.3) {};
          \node at (1.6,0) {$\cdots$};
          \node at (-1.6,0) {$\cdots$};
          \node at (4.7,0) {$\cdots$};
          \draw[thick,double] (1.9,0)--(2.5,0);
          \draw[thick,double] (-1.9,0)--(-2.5,0);
          \node at (2.2,0.3) {};
          \node at (-2.2,0.3) {};
          \draw[thick] (3.1,0) circle (0.5);
          \draw[thick] (-3.1,0) circle (0.5);
          \draw[thick,double] (3.7,0)--(4.3,0);
          \draw[thick,double] (-3.7,0)--(-4.3,0);
          \node at (4.0,0.3) {};
          \node at (-4.0,0.3) {};
           \draw[thick,double] (5.0,0)--(5.6,0);
           \node at (5.3,0.3) {};
           \draw[thick] (6.2,0) circle (0.5);
           \draw[thick,double] (6.8,0)--(7.4,0);
           \node at (7.1,0.3) {};
        \draw[thick] (7.5,0.5)--(8.5,0.5)--(8.5,-0.5)--(7.5,-0.5)--cycle;
        \draw[thick] (-4.4,0.5)--(-5.4,0.5)--(-5.4,-0.5)--(-4.4,-0.5)--cycle;
        \node at (3.1,0) {$\mathfrak{g}$};
        \node at (-3.1,0) {$\mathfrak{g}$};
        \node at (8,0) {$\mathfrak{g}$};
        \node at (-4.9,0) {$\mathfrak{g}$};
        \node at (6.2,0) {$\mathfrak{g}$};
       \node at (-2.5,-1.3) {$\underbrace{\hspace{3.7cm}}_{\scalebox{0.85}{$n_1\hspace{0.1cm}(\mathfrak{g},\mathfrak{g})_{(1)}$ \text{edges}}}$};
       \node at (1.6,-1.3) {$\underbrace{\hspace{1.9cm}}_{\scalebox{0.85}{$n_2\hspace{0.1cm} (\mathfrak{g},\mathfrak{g})_{(2)}$ \text{edges}}}$};
       \node at (5.5,-1.3) {$\underbrace{\hspace{3.7cm}}_{\scalebox{0.85}{$n_3\hspace{0.1cm}(\mathfrak{g},\mathfrak{g})_{(3)}$ \text{edges}}}$};
        \end{tikzpicture}}
        \end{array} $
    \caption*{}
    \end{subfigure}
    
\vspace{-1.5cm}
    
 \begin{subfigure}[c]{0.9\textwidth}
    \centering
    \scalemath{1}{   \begin{tikzpicture}
   \hspace{1.5cm} \node at (6.2,0) {$\displaystyle\left\downarrow\vphantom{\int_A^B}\right.$};
    \node at (7.8,0) {\text{circle reduction}};
    \end{tikzpicture}}
    \end{subfigure}
       \begin{subfigure}[c]{.9\textwidth}
       \centering
       \vspace{0.8cm}
\scalemath{0.85}{
\begin{tikzpicture}
        \draw[thick] (2.2,0) circle (1.5);
         \node at (1.6,-0.2) {\small$\mathcal{O}_{\text{max}}$};
          \node at (1.0,-0.2) {$\times$};
          \node at (2.2,0.8) {\small$\mathcal{O}_{III}^{(1)}$};
           \node at (2.2,1.2) {$\times$};
        \draw (3.5,0)--(5.4,0);
        \draw (3.5,-0.15)--(5.4,-0.15);
        \node at (5.75,-0.075) {$\cdots$};
        \draw (6.05,0)--(7.9,0);
        \draw (6.05,-0.15)--(7.9,-0.15);
        \draw (3.5,0) to [out=180,in=-30] (3.3,0.1);
        \draw (3.5,-0.15) to [out=180,in=30] (3.3,-0.25);
        \draw (7.9,0) to [out=0,in=210] (8,0.1);
        \draw (7.9,-0.15) to [out=0,in=150] (8,-0.25);
        \draw[thick] (9.2,0) circle (1.5);

          \node at (9.2,0.8) {\small$\mathcal{O}_{III}^{(2)}$};
           \node at (9.2,1.2) {$\times$};
        \draw (10.5,0)--(12.4,0);
        \draw (10.5,-0.15)--(12.4,-0.15);
        \node at (12.75,-0.075) {$\cdots$};
        \draw (13.05,0)--(14.9,0);
        \draw (13.05,-0.15)--(14.9,-0.15);
        \draw (10.5,0) to [out=180,in=-30] (10.3,0.1);
        \draw (10.5,-0.15) to [out=180,in=30] (10.3,-0.25);
        \draw (14.9,0) to [out=0,in=210] (15,0.1);
        \draw (14.9,-0.15) to [out=0,in=150] (15,-0.25);
        \draw[thick] (16.2,0) circle (1.5);
         \node at (16.8,-0.2) {\small$\mathcal{O}_{\text{max}}$};
          \node at (17.4,-0.2) {$\times$};
          \node at (16.2,0.8) {\small$\mathcal{O}_{III}^{(3)}$};
           \node at (16.2,1.2) {$\times$};
            \node at (3,-2.3) {$\underbrace{\hspace{3.7cm}}_{\scalebox{0.9}{$n_1 \text{ spheres with } \mathcal{O}_{III}^{(1)}$ \text{punctures}}}$};
       \node at (9.2,-2.3) {$\underbrace{\hspace{3.7cm}}_{\scalebox{0.9}{$n_2 \text{ spheres with } \mathcal{O}_{III}^{(2)}$ \text{punctures}}}$};
       \node at (15.5,-2.3) {$\underbrace{\hspace{3.7cm}}_{\scalebox{0.9}{$n_3 \text{ spheres with } \mathcal{O}_{III}^{(3)}$ \text{punctures}}}$};
        \end{tikzpicture}}
    \end{subfigure}
    \caption{The generalized quiver with 5d conformal matter edges and its 4d descendant realized as a class-$\mathcal{S}$ setup with regular punctures. This corresponds to the blue arrow going to the bottom right in Figure \ref{fig:InterdimensionalFlowMolecule}.}
     \label{fig: generalized class S duality}
\end{figure}

The procedure to find the 4d descendant of a 5d generalized linear quiver (molecule) is straightforward: we start from the fundamental 5d conformal matter atoms of Table \ref{4d table}, that specify the descendant of the 5d theory $X_{\mathfrak{g}}^{(i)}$ by means of a class-$\mathcal{S}$ setup. The latter is given by the 6d $\mathcal{N}=(2,0)$ theory of type $\mathfrak{g}$ compactified on a sphere with three appropriately chosen regular punctures. As previously remarked, there is a one to one correspondence between $X_{\mathfrak{g}}^{(i)}$ and the choice of the nilpotent orbit associated to the third puncture $\mathcal{O}_{III}^{(i)}$ (where we have explicitly highlighted the dependence on $i$, that can be read off from Table \ref{4d table}).\\
\indent Pictorially, we can represent the 5d theory and its 4d descendant as in Figure \ref{fig: class S duality}.
Notice that the sphere corresponds to an ``edge'' in the generalized 5d quiver. Consequently, in order to construct the 4d descendants of the generalized 5d quiver theories $\tilde{\mathsf{Q}}_{X_{\mathfrak{g}}^{(1^{n_1},2^{n_2},3^{n_3})}}$ of Figure \ref{fig: generalized class S duality} it suffices to consider as many class-$\mathcal{S}$ spheres as the number of edges of the 5d generalized quiver. Of course, we must choose the third puncture of each sphere in such a way to match the corresponding 5d conformal matter type $(\mathfrak{g},\mathfrak{g})_{(i)}$ of the corresponding edge. The gluing happens along the maximal punctures: all in all, we are left with a single sphere with two maximal punctures of type $\mathfrak{g}$, and as many non-maximal punctures as the number of edges, accounting for the extra flavor symmetry. Intuitively, we can represent this correspondence as in Figure \ref{fig: generalized class S duality}. Crucially, notice that the generalized quiver $\tilde{\mathsf{Q}}_{X_{\mathfrak{g}}^{(1^{n_1},2^{n_2},3^{n_3})}}$ does \textit{not} correspond to the same extended CB phase as the quivers specified by our preferred resolution procedure of Section \ref{sec:basechangeres}. This will impact on the relation between the Higgs branch dimension of the 5d SCFT and the class-$\mathcal{S}$ reduction corresponding to the choice of resolution that yields $\tilde{\mathsf{Q}}_{X_{\mathfrak{g}}^{(1^{n_1},2^{n_2},3^{n_3})}}$. We will extensively comment on this in Section \ref{sec:HBdimensionmolecules}.
It is easy to see that the ranks of the 5d and 4d Coulomb branch match: in the 5d quiver, we have $n_1,n_2,n_3$ edges of type $(\mathfrak{g},\mathfrak{g})_{(1)},(\mathfrak{g},\mathfrak{g})_{(2)},(\mathfrak{g},\mathfrak{g})_{(3)}$, respectively, each contributing with $r_1,r_2,r_3$ to the total rank of the Coulomb branch. Each gauge node further adds rank$(\mathfrak{g})$ to such rank. Hence we obtain:
\begin{equation}\label{5d CB dim}
    \text{rank}(\text{CB})_{5d} = n_1 r_1+n_2 r_2+n_3r_3 +(n_1+n_2+n_3-1)\text{rank}(\mathfrak{g}).
\end{equation}
From the 4d side, we have to consider the contribution of two maximal regular punctures of type $\mathfrak{g}$, as well as that of $n_1,n_2,n_3$ regular punctures of type $\mathcal{O}_{III}^{(1)},\mathcal{O}_{III}^{(2)},\mathcal{O}_{III}^{(3)}$, respectively. As a consequence of the choices exhibited in Table \ref{4d table}, it is immediate to check that the dimensions $d_i$ of the $\mathcal{O}_{III}^{(i)}$ orbits are such that:
\begin{equation}
    d_{i} = 2(r_{i}+\text{rank}(\mathfrak{g})) \quad\quad \text{for }i = 1,2,3.
\end{equation}
Then, employing \eqref{4d CB dim}, we can compute the CB dimension in the 4d class-$\mathcal{S}$ theory:
\begin{equation}
\begin{split}
    \mathrm{rank}\hspace{0.05cm}\mathrm{CB}_{4d}\left(\mathcal{T}_{4d}\right) &= \frac{1}{2}\left(2\text{dim}(\mathcal{O}_{\text{max}})+n_1\text{dim} (\mathcal{O}_{III}^{(1)})+n_2\text{dim}(\mathcal{O}_{III}^{(2)})+n_3\text{dim}(\mathcal{O}_{III}^{(3)})\right)-\text{dim}(\mathfrak{g}) = \\
    & =  n_1 r_1+n_2 r_2+n_3 r_3 +(n_1+n_2+n_3-1)\text{rank}(\mathfrak{g}),
    \end{split}
\end{equation}
that precisely agrees with \eqref{5d CB dim}.\\
\indent As regards the flavor symmetry, we expect the 5d theory to possess at least the flavor symmetry \eqref{total flavor}, that we repeat here for convenience, sticking to the same conventions:
\begin{equation}
\label{total flavor 2}
    F_{UV} = \mathfrak{g}\times \mathfrak{g} \times F_{\text{rest}}.
\end{equation}
\indent From the 4d perspective, notice that the two maximal punctures give rise to the expected $\mathfrak{g}\times\mathfrak{g}$ factor, and that the remaining punctures should take care of the leftover flavor symmetry. A quick tracking of the factors shows that the 4d picture does not account for the rank of the global symmetry $F_{\text{rest}}$ in the general case. 
We explain this symmetry breaking effect in the following section, where we also comment on the matching between the 4d and 5d HB dimensions.

\subsection{5d conformal matter molecules and Higgs branch dimension}
\label{sec:HBdimensionmolecules}
In this section, we are going to relate the Higgs branch dimension, as well as the flavor symmetries, of the 5d conformal matter molecules with those of their 4d class $\mathcal S$ descendants presented in Figure \ref{fig: generalized class S duality}. We add more details on the matching of the flavor symmetry in Appendix \ref{Appendix B}.

Let us start by noticing that, due to \eqref{eq:hbdimensionclasss}, the dimension of the HB for the candidate class $\mathcal S$ theory depicted in Figure \ref{fig:molecule} scales linearly with the number of junctions of the generalized quiver of the corresponding 5d theory. More precisely, for the 4d theory corresponding to a quiver with $n_1$ junctions of type $i =1$, $n_2$ of type $i=2$ and $n_3$ of type $i=3$, we have
\begin{eqnarray}
\label{eq:4dmoleculehbdimension}
\text{dim}_{\mathbb H}\text{HB}\left(\mathcal{T}_{4d}\right) &=&
\text{dim}(\mathfrak g)  - \text{rank}(\mathfrak g)  \nonumber \\ && \qquad +\frac{1}{2}\sum_{i} n_i\left( \text{dim}_{\mathbb{C}}(\mathfrak{g})-\text{rank}(\mathfrak{g})-\text{dim}_{\mathbb{C}}(\mathcal{O}^{L}_i) \right) + \text{rank}(\mathfrak g) \nonumber \\
&=& \text{dim}(\mathfrak g) + \frac{1}{2}\sum_{i} n_i\left( \text{dim}_{\mathbb{C}}(\mathfrak{g})-\text{rank}(\mathfrak{g})-\text{dim}_{\mathbb{C}}(\mathcal{O}^{L}_i) \right), 
\end{eqnarray}
where in the first step we singled out the contribution $\text{dim}(\mathfrak g)  - \text{rank}(\mathfrak g)$ coming from the full punctures, and the sum over $i$ accounts for the other punctures.\\
\indent We can now compute the HB dimension of the 5d theory. As explained in \cite{DeMarco:2023irn}, M-theory on \eqref{systemfourfoldgen} admits a mass deformation to an ordinary quiver gauge theory phase. The gauge nodes of the quiver are of type $\mathfrak{su}(N_k)$, with $k = 1,...,\text{rank}(\mathfrak g)$ and intersect each other according to the $\mathfrak g$ Dynkin diagram. The quiver might also display some flavor nodes, again of type $\mathfrak{su}$ or of type $\mathfrak{u}(1)$. In particular, the ranks of the quiver gauge nodes scale linearly with the $n_i$ \cite{DeMarco:2023irn} and hence the HB dimension of the quiver, $n_{H}-n_{V}$, scales \textit{quadratically} with the $n_i$: 
\begin{equation}
\label{eq:hbdimension5dfull}
    \text{dim}_{\mathbb H}\text{HB}\left(\mathcal{T}_{5d}\right) = n_{H}-n_{V}+\underbrace{\text{dim}(\mathfrak g) - \text{rank}(\mathfrak g)}_{\textit{instantonic contribution}},
\end{equation}
where we have separated, on the r.h.s., the HB contributions that can be counted at finite coupling from the quiver ($n_{H}-n_V$) from those coming from the instantonic particles \cite{Yonekura} (or, equivalently, from the M2-brane states described in  \cref{sec:higgsbranchbigsec}). We listed the $\text{dim}_{\mathbb H}(\text{HB})_{5d}^{UV}$ for all the conformal matter molecules in \cref{table:hbmodesfullmolecules}, where we colored in red the explicit expressions for $n_{H}-n_{V}$. It is immediate to see that \eqref{eq:hbdimension5dfull} and \eqref{eq:4dmoleculehbdimension} can not be matched. The resolution of the puzzle, though, is straightforward and can be traced back to our choices of partial resolutions for the molecule geometries.\\
\indent Indeed, we constructed the 4d class $\mathcal S$ as the circle reduction of the generalized quiver in Figure \ref{fig: generalized class S duality}. The generalized quiver phase is itself a partially resolved phase (different from the resolution presented in \cref{sec:basechangeres}) of M-theory on \eqref{systemfourfoldgen}. Consequently, the generalized quiver is a valid description away from the origin of the 5d extended Coulomb branch of the SCFT, and holds only when the compact curves supporting the compact lines of $\mathfrak g$ Du Val singularities have a non-zero volume. It then follows that the M2-brane states\footnote{It is important not to confuse these M2 states with the ones appearing in \cref{sec:higgsbranchbigsec}, as they are associated to different resolutions of the threefold singularity.} wrapped on the inflated compact curves have a non-vanishing BPS mass, and then some HB directions are lifted. We can count the unlifted HB dimensions as follows: the generalized linear quiver in Figure \ref{fig: generalized class S duality} corresponds to a \textit{partial resolution} of \eqref{systemfourfoldgen} displaying $n_1 + n_2 + n_3 - 1$ compact lines of $\mathfrak g$ singularities intersecting according to the $A_{n_1+n_2+n_3-1}$ Dynkin diagram. Furthermore, we have two non-compact lines of $\mathfrak g$ singularities intersecting the compact lines corresponding to the leftmost and to the rightmost nodes of the $A_{n_1+n_2+n_3-1}$ Dynkin diagram. In total, we have $n_1 + n_2 + n_3$ intersection points between these lines where the unlifted HB modes localize. As these intersection points are separated, we can count their contributions one by one using the third\footnote{This is equivalent to adding the contributions from the 5d instantonic particles to the second column of \cref{tab:HB dim}.} column of \cref{tab:HB dim}. The final HB dimension is obtained by subtracting the contributions coming from the $\mathfrak g$ gauge nodes, obtaining:
\begin{eqnarray}
\label{eq:hb5dobstructed}
\text{dim}_{\mathbb H}\text{HB}\left(\tilde{\mathsf{Q}}_{X_{\mathfrak{g}}^{(1^{n_1},2^{n_2},3^{n_3})}}\right) &=& \sum_{i} n_i \text{dim}_{\mathbb H}\text{HB}\left(Q_{X_{\mathfrak{g}}^{(i)}}\right) - (N-1)\text{dim} (\mathfrak g) \nonumber \\
 &=&
N (\text{dim}(\mathfrak g)-\text{rank}(\mathfrak g))+\frac{1}{2}\sum_{i} n_i\left( \text{dim}_{\mathbb{C}}(\mathfrak{g})-\text{rank}(\mathfrak{g}) -\text{dim}_{\mathbb{C}}(\mathcal{O}^{L}_i) \right) \nonumber \\
&& + N \text{rank}(\mathfrak g)- (N-1) \text{dim}(\mathfrak g) \nonumber \\
&=& \text{dim}(\mathfrak g) + \frac{1}{2}\sum_{i} n_i\left( \text{dim}_{\mathbb{C}}(\mathfrak{g})-\text{rank}(\mathfrak{g}) -\text{dim}_{\mathbb{C}}(\mathcal{O}^{L}_i) \right) \nonumber \\
&=& \text{dim}_{\mathbb H}\text{HB}\left(\mathcal{T}_{4d}\right),
\end{eqnarray}
where $N = n_1 + n_2 + n_3$ and, in the second step, we singled out the contribution $N (\text{dim}(\mathfrak g)-\text{rank}(\mathfrak g))$ of the maximal regular punctures.\\
\indent We can use this result to understand the apparent mismatch between the flavor symmetries of the 5d SCFT associated to \eqref{systemfourfoldgen} and those of the 4d class $\mathcal S$ corresponding theory. Indeed, in order to obtain $\mathcal T_{4d}$, we gave mass to some 5d HB modes, and hence we expect to get a tinier 4d flavor group after dimensional reduction. The final flavor group is expected to be the commutant, inside the initial flavor group, of the elements of the flavor algebra corresponding to the mass parameters. Stated differently, the class $\mathcal S$ flavor group is expected to coincide with the flavor group of the mass-deformed phase of the 5d SCFT corresponding to the generalized linear quiver, that, from a 5d perspective, gets enhanced at the UV fixed point to \eqref{total flavor} only when all the compact $\mathbb P^1$'s are contracted. In analogy to the instanton particles in the quiver gauge theory setup, the M2 branes wrapping the resolved $\mathbb P^1$ enlarge, after the contraction, both the flavor symmetry of the theory and the Higgs branch dimensions. More specifically, we notice that, since the generalized linear quiver still displays two non-compact lines of \textit{singular} $\mathfrak g$ singularities, the mass parameters that we turned on to reach the generalized linear quiver phase must come from the $F_{\text{rest}}$ factor of the 5d flavor group. Consequently, the 4d punctures capture just the commutants, inside $F_{\text{rest}}$, of the mass parameters that we turned on. Indeed, we can check this statement as follows: a mass deformation for $F_{rest}$ is a background field along $\mathfrak h' \subset \mathfrak h \subset F_{\text{rest}}$, with $\mathfrak h$ the Cartan subalgebra. This breaks the $F_{\text{rest}}$ to the commutant of $\mathfrak h'$ (excluding $\mathfrak h'$ itself). We hence expect 
\begin{equation}
    \text{rank}(G_{III, 4d}) = \text{rank}(F_{\text{rest}}) - N.
\end{equation}
We check this formula for the most general molecule, highlighting the precise breaking pattern of the flavor symmetry, in \cref{Appendix B}.

To conclude, we note that \eqref{eq:hb5dobstructed} is an independent check of the fact that the candidate 4d class $\mathcal S$ theories are exactly those in Figure \ref{fig: class S duality} and Figure \ref{fig: generalized class S duality}, validating our construction. In Table \ref{table:hbmodesfullmolecules} we summarize the \textit{full} UV quaternionic Higgs Branch dimensions for all molecules of 5d conformal matter. In other words, given the defining threefold \eqref{systemfourfoldgen}, the HB dimension depends on the non-negative integers $n_1,n_2,n_3$ as in the column on the right of \cref{table:hbmodesfullmolecules}.

\renewcommand{\arraystretch}{1.5}
\begin{table}[H]\centering
\begin{equation}
\scalemath{1}{
\begin{array}{|c|c|}
\hline 
 \textbf{Molecule class} & \textbf{dim}_{\mathbb{H}}\textbf{HB}\left(\mathcal{T}_{5d}\right) \\
\hline
\hline
A_{2j+1} &  n_1^2 (j+1)+2 j+1+\underbrace{ 2 (1 + j) (1 + 2 j)}_{\text{obstructed}}\\
A_{2j} & \left(n_1^2+2\right) j +\underbrace{2 j (1 + 2 j)}_{\text{obstructed}}\\
D_{2j+2} & \scalemath{0.85}{(j+2) n_1^2+n_1 \left(2 j n_2+n_2+4
   n_3\right)+j \left(n_2^2+2\right)+2 n_3
   \left(n_2+n_3\right)+2 +\underbrace{ 4 (1 + j) (1 + 2 j)}_{\text{obstructed}}} \\
D_{2j+3} & \scalemath{0.85}{(j+2) n_1^2+2 n_1 \left((j+1) n_2+2
   n_3\right)+(j+1) n_2^2+2 j+2 n_3^2+2
   n_2 n_3+3 + \underbrace{4 (1 + j) (3 + 2 j)}_{\text{obstructed}}}\\
E_6  & 3 n_1^2+4 n_1 n_2+3 n_1 n_3+2 n_2^2+2 n_2 n_3+n_3^2+6+\underbrace{72}_{\text{obstructed}} \\
E_7 & 4 n_1^2+6 n_1 n_2+3 n_1 n_3+3 n_2^2+2 n_2 n_3+n_3^2+7+\underbrace{126}_{\text{obstructed}}\\
E_8  &  4 n_1^2+5 n_1 n_2+3 n_1 n_3+2 n_2^2+2 n_2 n_3+n_3^2+8+\underbrace{240}_{\text{obstructed}}\\
\hline
\end{array}\nonumber
}
\end{equation}
\caption{UV Higgs branch dimension of 5d conformal matter molecules. We highlighted with brackets the HB modes that are obstructed after the partial resolution of \cref{sec:basechangeres}. Note that, for the $\mathfrak{g} = A$ case, in \cite{DeMarco:2023irn} we have just one type of 5d conformal matter atom, and hence just $n_1$ appears in the HB dimension formula.}\label{table:hbmodesfullmolecules}
\end{table}

\section{5d conformal matter molecules, brane webs and magnetic quivers}
\label{sec:MoleculesBWandMQ}

In this Section we explore the brane-web perspective on 5d conformal matter molecules. That is, we now turn to gauging a diagonal symmetry of two 5d conformal matter atoms of the same algebra directly in 5d, and then going to the UV fixed point, i.e.\ a 5d conformal matter molecule.

We first discuss the A-type conformal matter brane systems in some detail, and then present the D-type brane systems which behave completely analogously.

\subsection{A-type}

\subsubsection{$X_{A_{2j+1}}^{(1^n)}$}

\paragraph{$X_{A_{2j+1}}^{(1^2)}$.}

Consider gauging the diagonal $SU(2j+2)$ symmetry of two $(A_{2j+1},A_{2j+1})_{(1)}$ 5d conformal matter atoms:
\begin{equation}
\label{eq:A2j+1_molecule_gauging}
    \begin{tikzpicture}
        \node at (0,0) {$\scalebox{0.5}{\begin{tikzpicture}[rotate=90]
        \draw[thick] (0,0.7)--(1.4,0.7)--(1.4,-0.7)--(0,-0.7)--cycle;
        \draw[thick] (1.5,0)--(2.9,0);
        \draw[thick] (3,0.7)--(4.4,0.7)--(4.4,-0.7)--(3,-0.7)--cycle;
        \node at (0.7,0) {$A_{2j+1}$};
        \node at (3.7,0) {$A_{2j+1}$};
        \node at (2.2,1.5) {$(A_{2j+1},A_{2j+1})_{(1)}$};
        \end{tikzpicture}}$};
        \node at (1.5,-1.6) {$\scalebox{0.5}{\begin{tikzpicture}[rotate=90]
        \draw[thick] (0,0.7)--(1.4,0.7)--(1.4,-0.7)--(0,-0.7)--cycle;
        \draw[thick] (1.5,0)--(2.9,0);
        \draw[thick] (3,0.7)--(4.4,0.7)--(4.4,-0.7)--(3,-0.7)--cycle;
        \node at (0.7,0) {$A_{2j+1}$};
        \node at (3.7,0) {$A_{2j+1}$};
        \node at (2.2,1.5) {$(A_{2j+1},A_{2j+1})_{(1)}$};
        \end{tikzpicture}}$};
        \draw[->,thick] (3,-0.8)--(5,-0.8);
        \node at (4,-0.5) {\small glue};
        \node at (4,-1.1) {\small boxes};
        \node at (6.5,-0.8) {$\scalebox{0.5}{\begin{tikzpicture}[rotate=90]
        \draw[thick] (0,0.7)--(1.4,0.7)--(1.4,-0.7)--(0,-0.7)--cycle;
        \draw[thick,double] (1.5,0)--(2.7,0);
        \draw[thick] (3.7,0) circle (1);
        \node at (3.7,0) {\small$SU(2j+2)$};
        \node at (3.3,-1.8) {\small$g_{SU(2j+2)}$};
        \draw[thick,double] (4.7,0)--(5.9,0);
        \draw[thick] (6,0.7)--(7.4,0.7)--(7.4,-0.7)--(6,-0.7)--cycle;
        \node at (0.7,0) {$A_{2j+1}$};
        \node at (6.7,0) {$A_{2j+1}$};
        \node at (2.2,1.5) {$(A_{2j+1},A_{2j+1})_{(1)}$};
        \node at (5.2,1.5) {$(A_{2j+1},A_{2j+1})_{(1)}$};
        \end{tikzpicture}}$};
    \end{tikzpicture}\;,
\end{equation}
where $g_{SU(2j+2)}$ is the coupling of the gauged diagonal $SU(2j+2)$ symmetry.

The brane web for the the $(A_{2j+1},A_{2j+1})_{(1)}$ 5d conformal matter atom is given in \ref{eq:A2j+1_Atom_Web}. The two $SU(2j+2)$ symmetries of the SCFT can be seen from the $[0,1]7$ branes which the (vertical) $(0,1)5$ branes end on.

The diagonal $SU(2j+2)$ gauging can be done on the level of the brane web by `gluing' the two webs corresponding to the two atoms together in the following way:
\begin{equation}
\label{eq:A2j+1_BW_gluing}
    \begin{tikzpicture}
        \node at (0,0) {$\scalebox{0.5}{\begin{tikzpicture}
        \node[seven] (5ff) at (-2,0) {};
        \node[seven] (6ff) at (2,0) {};
        \draw[thick, double] (5ff)--(6ff);
        \node at (1,0.3) {$j+1$};

        \node[seven] (1u) at (0,2) {};
        \node[seven] (2u) at (0,3) {};
        \node (3u) at (0,4) {$\vdots$};
        \node[seven] (4u) at (0,5) {};
        \node[seven] (5u) at (0,6) {};
        \node[seven] (6u) at (0,7) {};

        \node[seven] (1d) at (0,-2) {};
        \node[seven] (2d) at (0,-3) {};
        \node (3d) at (0,-4) {$\vdots$};
        \node[seven] (4d) at (0,-5) {};
        \node[seven] (5d) at (0,-6) {};
        \node[seven] (6d) at (0,-7) {};

        \draw (5u)--(6u) (5d)--(6d);
        \draw[thick,double] (5u)--(4u)--(3u)--(2u)--(1u)--(1d)--(2d)--(3d)--(4d)--(5d);

        \node at (-0.3,6.5) {$1$};
        \node at (-0.3,5.5) {$2$};
        \node at (-0.7,2.5) {$2j+1$};
        \node at (-0.7,1) {$2j+2$};
        \node at (-0.3,-6.5) {$1$};
        \node at (-0.3,-5.5) {$2$};
        \node at (-0.7,-2.5) {$2j+1$};
        \draw (1,-7.5)--(1,-1.5)--(-1.4,-1.5)--(-1.4,-7.5)--(1,-7.5);
        \node at (0,-7.8) {$su(2j+2)$};
    \end{tikzpicture}}$};
    \node at (2,-4.5) {$\scalebox{0.5}{\begin{tikzpicture}
        \node[seven] (5ff) at (-2,0) {};
        \node[seven] (6ff) at (2,0) {};
        \draw[thick, double] (5ff)--(6ff);
        \node at (1,0.3) {$j+1$};

        \node[seven] (1u) at (0,2) {};
        \node[seven] (2u) at (0,3) {};
        \node (3u) at (0,4) {$\vdots$};
        \node[seven] (4u) at (0,5) {};
        \node[seven] (5u) at (0,6) {};
        \node[seven] (6u) at (0,7) {};

        \node[seven] (1d) at (0,-2) {};
        \node[seven] (2d) at (0,-3) {};
        \node (3d) at (0,-4) {$\vdots$};
        \node[seven] (4d) at (0,-5) {};
        \node[seven] (5d) at (0,-6) {};
        \node[seven] (6d) at (0,-7) {};

        \draw (5u)--(6u) (5d)--(6d);
        \draw[thick,double] (5u)--(4u)--(3u)--(2u)--(1u)--(1d)--(2d)--(3d)--(4d)--(5d);

        \node at (-0.3,6.5) {$1$};
        \node at (-0.3,5.5) {$2$};
        \node at (-0.7,2.5) {$2j+1$};
        \node at (-0.7,1) {$2j+2$};
        \node at (-0.3,-6.5) {$1$};
        \node at (-0.3,-5.5) {$2$};
        \node at (-0.7,-2.5) {$2j+1$};
        \draw (1,7.5)--(1,1.5)--(-1.4,1.5)--(-1.4,7.5)--(1,7.5);
        \node at (0,7.8) {$su(2j+2)$};
    \end{tikzpicture}}$};
    \draw[->,thick] (3.5,-2.25)--(5,-2.25);
    \node at (4.25,-2) {\small glue};
    \node at (4.25,-2.5) {\small boxes};
    \node at (7.5,-2.25) {$\scalebox{0.6}{\begin{tikzpicture}
        \node[seven] (5ff) at (-2,0) {};
        \node[seven] (6ff) at (2,0) {};
        \draw[thick, double] (5ff)--(6ff);
        \node at (1,0.3) {$j+1$};

        \node[seven] (5fff) at (-2,-2) {};
        \node[seven] (6fff) at (2,-2) {};
        \draw[thick, double] (5fff)--(6fff);
        \node at (1,-2.3) {$j+1$};

        \node[seven] (1u) at (0,2) {};
        \node[seven] (2u) at (0,3) {};
        \node (3u) at (0,4) {$\vdots$};
        \node[seven] (4u) at (0,5) {};
        \node[seven] (5u) at (0,6) {};
        \node[seven] (6u) at (0,7) {};

        \node[seven] (1d) at (0,-4) {};
        \node[seven] (2d) at (0,-5) {};
        \node (3d) at (0,-6) {$\vdots$};
        \node[seven] (4d) at (0,-7) {};
        \node[seven] (5d) at (0,-8) {};
        \node[seven] (6d) at (0,-9) {};

        \draw (5u)--(6u) (5d)--(6d);
        \draw[thick,double] (5u)--(4u)--(3u)--(2u)--(1u)--(1d)--(2d)--(3d)--(4d)--(5d);

        \node at (-0.3,6.5) {$1$};
        \node at (-0.3,5.5) {$2$};
        \node at (-0.7,2.5) {$2j+1$};
        \node at (-0.7,1) {$2j+2$};
        \node at (-0.7,-1) {$2j+2$};
        \node at (-0.3,-8.5) {$1$};
        \node at (-0.3,-7.5) {$2$};
        \node at (-0.7,-4.5) {$2j+1$};
        \draw (2.5,0)--(2.5,-2) (2.3,0)--(2.7,0) (2.3,-2)--(2.7,-2);
        \node at (3.6,-1) {$\propto\frac{1}{g^2_{SU(2j+2)}}$};
    \end{tikzpicture}}$};
    \end{tikzpicture}\;.
\end{equation}
Note that this is the brane web for
\begin{equation}
    \mathsf{Q}_{X_{A_{2j+1}}^{(1^2)}}=\raisebox{-.5\height}{\scalebox{0.6}{\begin{tikzpicture}
        \node[gaugeSU,label=below:{$2$}] (1) at (2,0) {};
        \node[gaugeSU,label=below:{$4$}] (2) at (4,0) {};
        \node (3) at (6,0) {$\cdots$};
        \node[gaugeSU,label=below:{$2j$}] (4) at (8,0) {};
        \node[gaugeSU,label=below:{$2j+2$}] (5) at (10,0) {};
        \node[gaugeSU,label=below:{$2j$}] (6) at (12,0) {};
        \node (7) at (14,0) {$\cdots$};
        \node[gaugeSU,label=below:{$4$}] (8) at (16,0) {};
        \node[gaugeSU,label=below:{$2$}] (9) at (18,0) {};
        \node[flavourSU,label=above:{$4$}] (5f) at (10,1.5) {};
        \draw (1)--(2)--(3)--(4)--(5)--(6)--(7)--(8)--(9) (5)--(5f);
    \end{tikzpicture}}}\;,
\end{equation}
after taking all the couplings to infinity, but giving a mass $m$ to two of the four flavours leaving the other two massless, where $m=\frac{1}{g^2_{SU(2j+2)}}$. This $m$ corresponds to the mass deformation corresponding to the red line going to the right in Figure \ref{fig:InterdimensionalFlowMolecule}.

On the level of magnetic quivers this is realised by gauging a diagonal Coulomb branch symmetry (see for example \cite{Benini:2010uu} or \cite[Appendix B]{Dancer:2024lra} how these are implemented):
\begin{equation}
\label{eq:A2j+1_MQ_quotient}
    \begin{tikzpicture}
        \node at (0,0) {$\scalebox{0.6}{\begin{tikzpicture}
        \node[gauge,label=below:{\scriptsize$j+1$}] (0r) at (1,0) {};
        \node[gauge,label=left:{\scriptsize$2j+2$}] (0) at (0,0) {};
        \node[gauge,label=left:{\scriptsize$2j+1$}] (1) at (0,1) {};
        \node (2) at (0,2) {$\vdots$};
        \node[gauge,label=left:{\scriptsize$2$}] (3) at (0,3) {};
        \node[gauge,label=left:{\scriptsize$1$}] (4) at (0,4) {};
        \node[gauge,label=left:{\scriptsize$2j+1$}] (-1) at (0,-1) {};
        \node (-2) at (0,-2) {$\vdots$};
        \node[gauge,label=left:{\scriptsize$2$}] (-3) at (0,-3) {};
        \node[gauge,label=left:{\scriptsize$1$}] (-4) at (0,-4) {};
        \draw (0r)--(0)--(1)--(2)--(3)--(4) (0)--(-1)--(-2)--(-3)--(-4);
        \draw (-1.5,-4.5)--(1,-4.5)--(1,-0.6)--(-1.5,-0.6)--(-1.5,-4.5);
        \node at (0,-4.8) {$su(2j+2)$};
    \end{tikzpicture}}$};
        \node at (2,-3) {$\scalebox{0.6}{\begin{tikzpicture}
        \node[gauge,label=below:{\scriptsize$j+1$}] (0r) at (1,0) {};
        \node[gauge,label=left:{\scriptsize$2j+2$}] (0) at (0,0) {};
        \node[gauge,label=left:{\scriptsize$2j+1$}] (1) at (0,1) {};
        \node (2) at (0,2) {$\vdots$};
        \node[gauge,label=left:{\scriptsize$2$}] (3) at (0,3) {};
        \node[gauge,label=left:{\scriptsize$1$}] (4) at (0,4) {};
        \node[gauge,label=left:{\scriptsize$2j+1$}] (-1) at (0,-1) {};
        \node (-2) at (0,-2) {$\vdots$};
        \node[gauge,label=left:{\scriptsize$2$}] (-3) at (0,-3) {};
        \node[gauge,label=left:{\scriptsize$1$}] (-4) at (0,-4) {};
        \draw (0r)--(0)--(1)--(2)--(3)--(4) (0)--(-1)--(-2)--(-3)--(-4);
        \draw (-1.5,4.5)--(1,4.5)--(1,0.6)--(-1.5,0.6)--(-1.5,4.5);
        \node at (0,4.8) {$su(2j+2)$};
    \end{tikzpicture}}$};
    \draw[->,thick] (3.5,-1.5)--(5.5,-1.5);
    \node at (4.5,-1.2) {\small gauge diag};
    \node at (4.5,-1.8) {\small CB symm};
        \node at (7,-1.5) {$\scalebox{0.8}{\begin{tikzpicture}
        \node[gauge,label=above:{\scriptsize$j+1$}] (0r) at (1,0.5) {};
        \node[gauge,label=below:{\scriptsize$j+1$}] (0rr) at (1,-0.5) {};
        \node[gauge,label=left:{\scriptsize$2j+2$}] (0) at (0,0) {};
        \node[gauge,label=left:{\scriptsize$2j+1$}] (1) at (0,1) {};
        \node (2) at (0,2) {$\vdots$};
        \node[gauge,label=left:{\scriptsize$2$}] (3) at (0,3) {};
        \node[gauge,label=left:{\scriptsize$1$}] (4) at (0,4) {};
        \node[gauge,label=left:{\scriptsize$2j+1$}] (-1) at (0,-1) {};
        \node (-2) at (0,-2) {$\vdots$};
        \node[gauge,label=left:{\scriptsize$2$}] (-3) at (0,-3) {};
        \node[gauge,label=left:{\scriptsize$1$}] (-4) at (0,-4) {};
        \draw (0r)--(0)--(1)--(2)--(3)--(4) (0rr)--(0)--(-1)--(-2)--(-3)--(-4);
    \end{tikzpicture}}$};
    \end{tikzpicture}\;,
\end{equation}
where the quiver on the RHS of \eqref{eq:A2j+1_MQ_quotient} can also be read straight from the brane web on the RHS of \eqref{eq:A2j+1_BW_gluing}.

The Higgs branch of the RHS of \eqref{eq:A2j+1_molecule_gauging} (at finite $g_{SU(2j+2)}$) is obtained as a hyper-K\"ahler quotient of the product of the Higgs branches on the LHS of \eqref{eq:A2j+1_molecule_gauging} by $SU(2j+2)$:
\begin{equation}
    \mathrm{HB}\left(\mathrm{RHS}\eqref{eq:A2j+1_molecule_gauging}\right)=\left(\mathrm{HB}\left(\mathrm{LHS1}\eqref{eq:A2j+1_molecule_gauging}\right)\times\mathrm{HB}\left(\mathrm{LHS2}\eqref{eq:A2j+1_molecule_gauging}\right)\right)///SU(2j+2)\;,
\end{equation}
with dimension
\begin{equation}
    \mathrm{dim}_{\mathbb{H}}\left(\mathrm{HB}\left(\mathrm{RHS}\eqref{eq:A2j+1_molecule_gauging}\right)\right)=\mathrm{dim}_{\mathbb{H}}\left(\mathrm{HB}\left(\mathrm{LHS1}\eqref{eq:A2j+1_molecule_gauging}\right)\right)+\mathrm{dim}_{\mathbb{H}}\left(\mathrm{HB}\left(\mathrm{LHS2}\eqref{eq:A2j+1_molecule_gauging}\right)\right)-(2j+2)^2+1\;.
\end{equation}

\paragraph{Infinite coupling.} The SCFT engineered by $X_{A_{2j+1}}^{(1^2)}$ is realised by taking $\frac{1}{g^2_{SU(2j+2)}}\rightarrow 0$ in the RHS brane web of \eqref{eq:A2j+1_BW_gluing}:
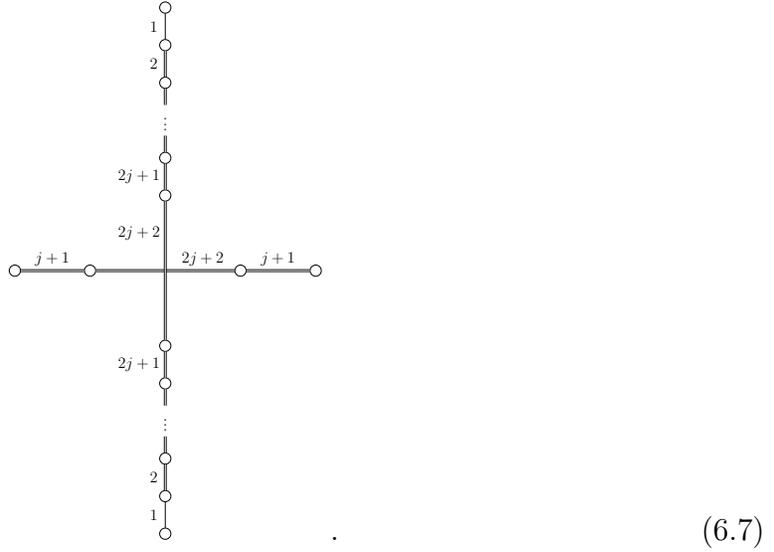
\begin{equation}
    \scalebox{0.5}{\begin{tikzpicture}
        \node[seven] (5ff) at (-2,0) {};
        \node[seven] (6ff) at (2,0) {};
        \node[seven] (5fff) at (-4,0) {};
        \node[seven] (6fff) at (4,0) {};
        \draw[thick, double] (5fff)--(5ff)--(6ff)--(6fff);
        \node at (1,0.3) {$2j+2$};
        \node at (3,0.3) {$j+1$};
        \node at (-3,0.3) {$j+1$};

        \node[seven] (1u) at (0,2) {};
        \node[seven] (2u) at (0,3) {};
        \node (3u) at (0,4) {$\vdots$};
        \node[seven] (4u) at (0,5) {};
        \node[seven] (5u) at (0,6) {};
        \node[seven] (6u) at (0,7) {};

        \node[seven] (1d) at (0,-2) {};
        \node[seven] (2d) at (0,-3) {};
        \node (3d) at (0,-4) {$\vdots$};
        \node[seven] (4d) at (0,-5) {};
        \node[seven] (5d) at (0,-6) {};
        \node[seven] (6d) at (0,-7) {};

        \draw (5u)--(6u) (5d)--(6d);
        \draw[thick,double] (5u)--(4u)--(3u)--(2u)--(1u)--(1d)--(2d)--(3d)--(4d)--(5d);

        \node at (-0.3,6.5) {$1$};
        \node at (-0.3,5.5) {$2$};
        \node at (-0.7,2.5) {$2j+1$};
        \node at (-0.7,1) {$2j+2$};
        \node at (-0.3,-6.5) {$1$};
        \node at (-0.3,-5.5) {$2$};
        \node at (-0.7,-2.5) {$2j+1$};
    \end{tikzpicture}}\;.
\end{equation}
The magnetic quiver read from this web is
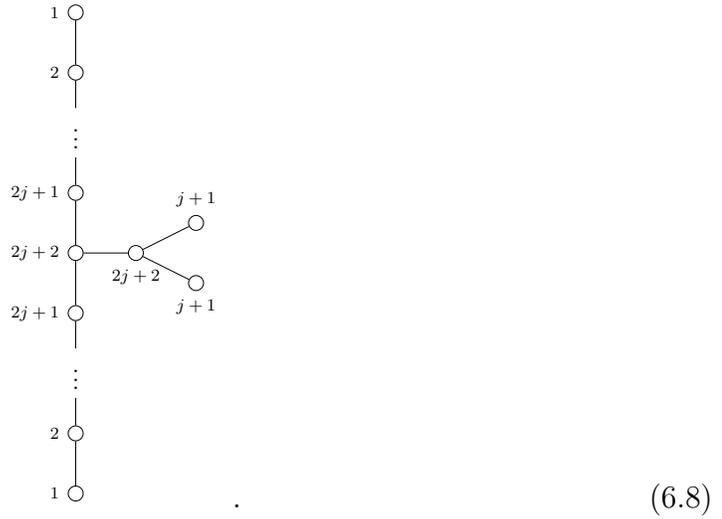
\begin{equation}
    \scalebox{0.8}{\begin{tikzpicture}
        \node[gauge,label=below:{\scriptsize$2j+2$}] (0r) at (1,0) {};
        \node[gauge,label=above:{\scriptsize$j+1$}] (1r) at (2,0.5) {};
        \node[gauge,label=below:{\scriptsize$j+1$}] (-1r) at (2,-0.5) {};
        \node[gauge,label=left:{\scriptsize$2j+2$}] (0) at (0,0) {};
        \node[gauge,label=left:{\scriptsize$2j+1$}] (1) at (0,1) {};
        \node (2) at (0,2) {$\vdots$};
        \node[gauge,label=left:{\scriptsize$2$}] (3) at (0,3) {};
        \node[gauge,label=left:{\scriptsize$1$}] (4) at (0,4) {};
        \node[gauge,label=left:{\scriptsize$2j+1$}] (-1) at (0,-1) {};
        \node (-2) at (0,-2) {$\vdots$};
        \node[gauge,label=left:{\scriptsize$2$}] (-3) at (0,-3) {};
        \node[gauge,label=left:{\scriptsize$1$}] (-4) at (0,-4) {};
        \draw (1r)--(0r)--(0)--(1)--(2)--(3)--(4) (-1r)--(0r) (0)--(-1)--(-2)--(-3)--(-4);
    \end{tikzpicture}}\;.
\end{equation}

\paragraph{$X_{A_{2j+1}}^{(1^n)}$.}
Now consider gauging $n$ $(A_{2j+1},A_{2j+1})_{(1)}$ 5d conformal matter atoms, in order to obtain a generalised linear quiver. The corresponding brane web and magnetic quiver are:
\begin{equation}   
    \begin{tikzpicture}
        \node at (0,0) {brane web:};
        \node at (3,0) {$\scalebox{0.6}{\begin{tikzpicture}
            \node[seven] (5ff) at (-2,0) {};
            \node[seven] (6ff) at (2,0) {};
            \draw[thick, double] (5ff)--(6ff);
            \node at (1,0.3) {$j+1$};
    
            \node[seven] (5fff) at (-2,-2) {};
            \node[seven] (6fff) at (2,-2) {};
            \draw[thick, double] (5fff)--(6fff);
            \node at (1,-2.3) {$j+1$};
    
            \node[seven] (1u) at (0,2) {};
            \node[seven] (2u) at (0,3) {};
            \node (3u) at (0,4) {$\vdots$};
            \node[seven] (4u) at (0,5) {};
            \node[seven] (5u) at (0,6) {};
            \node[seven] (6u) at (0,7) {};
    
            \node[seven] (1d) at (0,-4) {};
            \node[seven] (2d) at (0,-5) {};
            \node (3d) at (0,-6) {$\vdots$};
            \node[seven] (4d) at (0,-7) {};
            \node[seven] (5d) at (0,-8) {};
            \node[seven] (6d) at (0,-9) {};
    
            \node (1) at (0,-1) {$\vdots$};
            \node at (1.5,-1) {$\vdots$};
    
            \draw (5u)--(6u) (5d)--(6d);
            \draw[thick,double] (5u)--(4u)--(3u)--(2u)--(1u)--(1)--(1d)--(2d)--(3d)--(4d)--(5d);
    
            \node at (-0.3,6.5) {$1$};
            \node at (-0.3,5.5) {$2$};
            \node at (-0.7,2.5) {$2j+1$};
            \node at (-0.7,1) {$2j+2$};
            \node at (-0.3,-8.5) {$1$};
            \node at (-0.3,-7.5) {$2$};
            \node at (-0.7,-4.5) {$2j+1$};
    
            \draw [decorate,decoration={brace,amplitude=5pt},xshift=0pt,yshift=0pt]
        (2.3,0.3)--(2.3,-2.3) node [black,midway,xshift=0.4cm] {\large$n$};
        \end{tikzpicture}}$};
        \node at (7,0) {magnetic quiver:};
        \node at (10,0) {$\scalebox{0.8}{\begin{tikzpicture}
            \node[gauge,label=above:{\scriptsize$j+1$}] (0r) at (1,0.5) {};
            \node[gauge,label=below:{\scriptsize$j+1$}] (0rr) at (1,-0.5) {};
            \node at (1,0) {$\vdots$};
            \node[gauge,label=left:{\scriptsize$2j+2$}] (0) at (0,0) {};
            \node[gauge,label=left:{\scriptsize$2j+1$}] (1) at (0,1) {};
            \node (2) at (0,2) {$\vdots$};
            \node[gauge,label=left:{\scriptsize$2$}] (3) at (0,3) {};
            \node[gauge,label=left:{\scriptsize$1$}] (4) at (0,4) {};
            \node[gauge,label=left:{\scriptsize$2j+1$}] (-1) at (0,-1) {};
            \node (-2) at (0,-2) {$\vdots$};
            \node[gauge,label=left:{\scriptsize$2$}] (-3) at (0,-3) {};
            \node[gauge,label=left:{\scriptsize$1$}] (-4) at (0,-4) {};
            \draw (0r)--(0)--(1)--(2)--(3)--(4) (0rr)--(0)--(-1)--(-2)--(-3)--(-4);
            \draw [decorate,decoration={brace,amplitude=5pt},xshift=0pt,yshift=0pt]
            (1.3,0.7)--(1.3,-0.7) node [black,midway,xshift=0.4cm] {\small$n$};
        \end{tikzpicture}}$};
    \end{tikzpicture}\;.
\end{equation}

\paragraph{Infinite coupling.}
Taking all the couplings of the generalised quiver to infinity realises the SCFT associated to the $X_{A_{2j+1}}^{(1^n)}$ singularity. The corresponding brane web is:
\begin{equation}
    \scalebox{0.5}{\begin{tikzpicture}
        \node[seven] (0l1) at (-2,0) {};
        \node[seven] (0l2) at (-4,0) {};
        \node (0l3) at (-5,0) {$\cdots$};
        \node[seven] (0l4) at (-6,0) {};
        \node[seven] (0l5) at (-8,0) {};
        \node[seven] (0r1) at (2,0) {};
        \node[seven] (0r2) at (4,0) {};
        \node (0r3) at (5,0) {$\cdots$};
        \node[seven] (0r4) at (6,0) {};
        \node[seven] (0r5) at (8,0) {};
        \draw[thick, double] (0l5)--(0l4)--(0l3)--(0l2)--(0l1)--(0r1)--(0r2)--(0r3)--(0r4)--(0r5);
        \node at (1,0.3) {\small$n(j+1)$};
        \node[rotate=45] at (3,1) {\small$(n-1)(j+1)$};
        \node at (7,0.3) {$1(j+1)$};
        \node[rotate=45] at (-3,1) {\small$(n-1)(j+1)$};
        \node at (-7,0.3) {$1(j+1)$};

        \node[seven] (1u) at (0,2) {};
        \node[seven] (2u) at (0,3) {};
        \node (3u) at (0,4) {$\vdots$};
        \node[seven] (4u) at (0,5) {};
        \node[seven] (5u) at (0,6) {};
        \node[seven] (6u) at (0,7) {};

        \node[seven] (1d) at (0,-2) {};
        \node[seven] (2d) at (0,-3) {};
        \node (3d) at (0,-4) {$\vdots$};
        \node[seven] (4d) at (0,-5) {};
        \node[seven] (5d) at (0,-6) {};
        \node[seven] (6d) at (0,-7) {};

        \draw (5u)--(6u) (5d)--(6d);
        \draw[thick,double] (5u)--(4u)--(3u)--(2u)--(1u)--(1d)--(2d)--(3d)--(4d)--(5d);

        \node at (-0.3,6.5) {$1$};
        \node at (-0.3,5.5) {$2$};
        \node at (-0.7,2.5) {$2j+1$};
        \node at (-0.7,1) {$2j+2$};
        \node at (-0.3,-6.5) {$1$};
        \node at (-0.3,-5.5) {$2$};
        \node at (-0.7,-2.5) {$2j+1$};
    \end{tikzpicture}}
\end{equation}
from which we can read the magnetic quiver
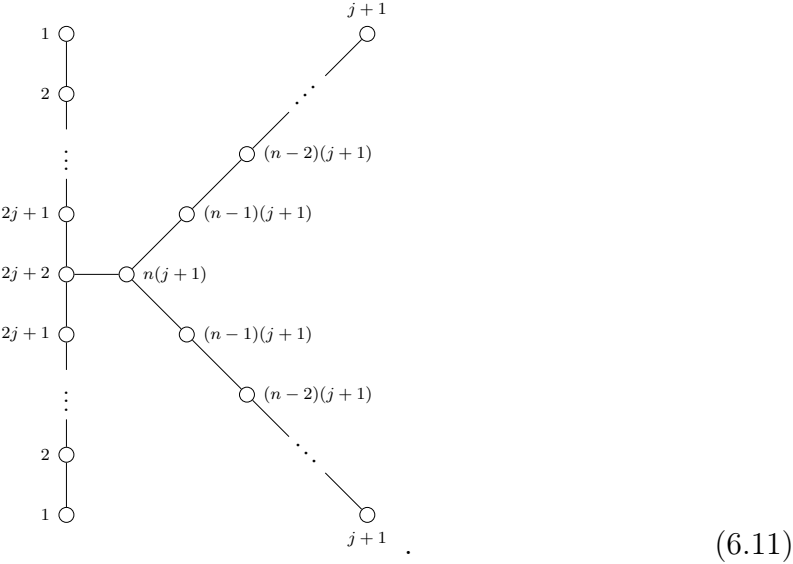
\begin{equation}
    \scalebox{0.8}{\begin{tikzpicture}
        \node[gauge,label=right:{\scriptsize$n(j+1)$}] (0r) at (1,0) {};
        \node[gauge,label=right:{\scriptsize$(n-1)(j+1)$}] (1r1) at (2,1) {};
        \node[gauge,label=right:{\scriptsize$(n-2)(j+1)$}] (1r2) at (3,2) {};
        \node[rotate=45] (1r3) at (4,3) {$\cdots$};
        \node[gauge,label=above:{\scriptsize$j+1$}] (1r4) at (5,4) {};
        \node[gauge,label=right:{\scriptsize$(n-1)(j+1)$}] (1l1) at (2,-1) {};
        \node[gauge,label=right:{\scriptsize$(n-2)(j+1)$}] (1l2) at (3,-2) {};
        \node[rotate=-45] (1l3) at (4,-3) {$\cdots$};
        \node[gauge,label=below:{\scriptsize$j+1$}] (1l4) at (5,-4) {};
        \node[gauge,label=left:{\scriptsize$2j+2$}] (0) at (0,0) {};
        \node[gauge,label=left:{\scriptsize$2j+1$}] (1) at (0,1) {};
        \node (2) at (0,2) {$\vdots$};
        \node[gauge,label=left:{\scriptsize$2$}] (3) at (0,3) {};
        \node[gauge,label=left:{\scriptsize$1$}] (4) at (0,4) {};
        \node[gauge,label=left:{\scriptsize$2j+1$}] (-1) at (0,-1) {};
        \node (-2) at (0,-2) {$\vdots$};
        \node[gauge,label=left:{\scriptsize$2$}] (-3) at (0,-3) {};
        \node[gauge,label=left:{\scriptsize$1$}] (-4) at (0,-4) {};
        \draw (1r4)--(1r3)--(1r2)--(1r1)--(0r)--(0)--(1)--(2)--(3)--(4) (1l4)--(1l3)--(1l2)--(1l1)--(0r) (0)--(-1)--(-2)--(-3)--(-4);
    \end{tikzpicture}}\;.
\end{equation}
We notice that the quiver is not star-shaped, which would be the case for a magnetic quiver for a class-S theory with regular punctures \cite{Benini:2010uu}, but rather it has two `centers' stemming from the stack of NS5 branes and D5 branes in the middle of the brane web respectively. This suggests that the reduction to 4d is not a class-$\mathcal{S}$ theory with regular punctures.\footnote{This probably suggests a 4d-mechanism analogous to the one of \cite{Ohmori:2015pua,Ohmori:2015pia}}
\subsubsection{$X_{A_{2j}}^{(1^n)}$}
Now consider gauging $n$ $(A_{2j},A_{2j})_{(1)}$ 5d conformal matter atoms, in order to obtain a generalised linear quiver. This case is completely analogous to the previous one. The corresponding brane web and magnetic quiver are:
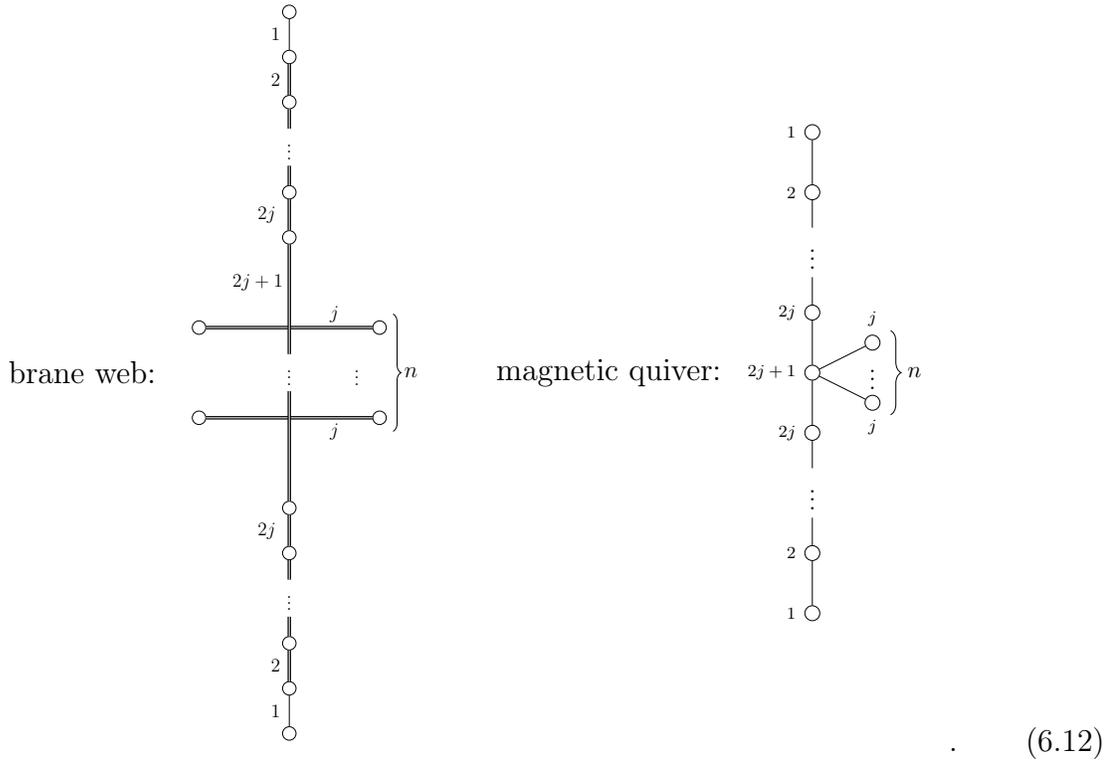
\begin{equation}   
    \begin{tikzpicture}
        \node at (0,0) {brane web:};
        \node at (3,0) {$\scalebox{0.6}{\begin{tikzpicture}
            \node[seven] (5ff) at (-2,0) {};
            \node[seven] (6ff) at (2,0) {};
            \draw[thick, double] (5ff)--(6ff);
            \node at (1,0.3) {$j$};
    
            \node[seven] (5fff) at (-2,-2) {};
            \node[seven] (6fff) at (2,-2) {};
            \draw[thick, double] (5fff)--(6fff);
            \node at (1,-2.3) {$j$};
    
            \node[seven] (1u) at (0,2) {};
            \node[seven] (2u) at (0,3) {};
            \node (3u) at (0,4) {$\vdots$};
            \node[seven] (4u) at (0,5) {};
            \node[seven] (5u) at (0,6) {};
            \node[seven] (6u) at (0,7) {};
    
            \node[seven] (1d) at (0,-4) {};
            \node[seven] (2d) at (0,-5) {};
            \node (3d) at (0,-6) {$\vdots$};
            \node[seven] (4d) at (0,-7) {};
            \node[seven] (5d) at (0,-8) {};
            \node[seven] (6d) at (0,-9) {};
    
            \node (1) at (0,-1) {$\vdots$};
            \node at (1.5,-1) {$\vdots$};
    
            \draw (5u)--(6u) (5d)--(6d);
            \draw[thick,double] (5u)--(4u)--(3u)--(2u)--(1u)--(1)--(1d)--(2d)--(3d)--(4d)--(5d);
    
            \node at (-0.3,6.5) {$1$};
            \node at (-0.3,5.5) {$2$};
            \node at (-0.5,2.5) {$2j$};
            \node at (-0.7,1) {$2j+1$};
            \node at (-0.3,-8.5) {$1$};
            \node at (-0.3,-7.5) {$2$};
            \node at (-0.5,-4.5) {$2j$};
    
            \draw [decorate,decoration={brace,amplitude=5pt},xshift=0pt,yshift=0pt]
        (2.3,0.3)--(2.3,-2.3) node [black,midway,xshift=0.4cm] {\large$n$};
        \end{tikzpicture}}$};
        \node at (7,0) {magnetic quiver:};
        \node at (10,0) {$\scalebox{0.8}{\begin{tikzpicture}
            \node[gauge,label=above:{\scriptsize$j$}] (0r) at (1,0.5) {};
            \node[gauge,label=below:{\scriptsize$j$}] (0rr) at (1,-0.5) {};
            \node at (1,0) {$\vdots$};
            \node[gauge,label=left:{\scriptsize$2j+1$}] (0) at (0,0) {};
            \node[gauge,label=left:{\scriptsize$2j$}] (1) at (0,1) {};
            \node (2) at (0,2) {$\vdots$};
            \node[gauge,label=left:{\scriptsize$2$}] (3) at (0,3) {};
            \node[gauge,label=left:{\scriptsize$1$}] (4) at (0,4) {};
            \node[gauge,label=left:{\scriptsize$2j$}] (-1) at (0,-1) {};
            \node (-2) at (0,-2) {$\vdots$};
            \node[gauge,label=left:{\scriptsize$2$}] (-3) at (0,-3) {};
            \node[gauge,label=left:{\scriptsize$1$}] (-4) at (0,-4) {};
            \draw (0r)--(0)--(1)--(2)--(3)--(4) (0rr)--(0)--(-1)--(-2)--(-3)--(-4);
            \draw [decorate,decoration={brace,amplitude=5pt},xshift=0pt,yshift=0pt]
            (1.3,0.7)--(1.3,-0.7) node [black,midway,xshift=0.4cm] {\small$n$};
        \end{tikzpicture}}$};
    \end{tikzpicture}\;.
\end{equation}

\paragraph{Infinite coupling.}
Taking all the couplings of the generalised quiver to infinity realises the SCFT associated to the $X_{A_{2j}}^{(1^n)}$ singularity. The corresponding brane web is:
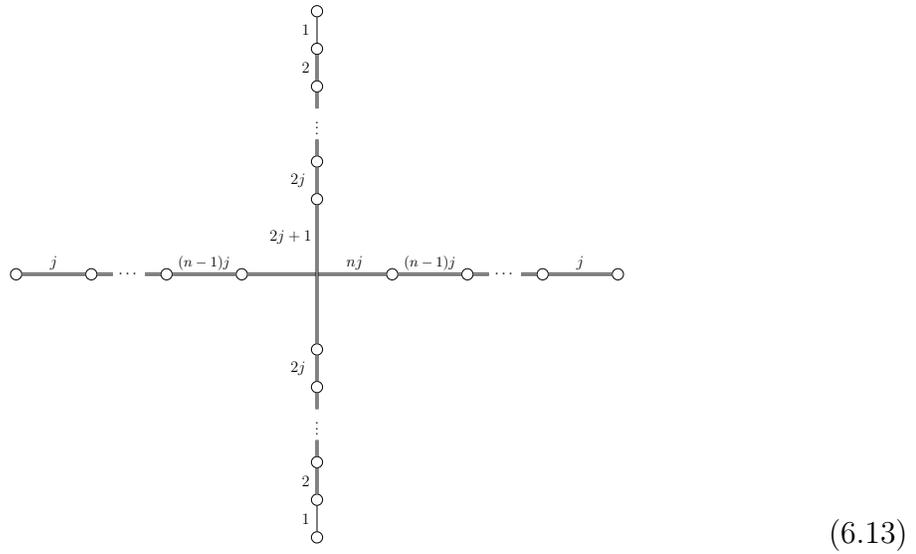
\begin{equation}
    \scalebox{0.5}{\begin{tikzpicture}
        \node[seven] (0l1) at (-2,0) {};
        \node[seven] (0l2) at (-4,0) {};
        \node (0l3) at (-5,0) {$\cdots$};
        \node[seven] (0l4) at (-6,0) {};
        \node[seven] (0l5) at (-8,0) {};
        \node[seven] (0r1) at (2,0) {};
        \node[seven] (0r2) at (4,0) {};
        \node (0r3) at (5,0) {$\cdots$};
        \node[seven] (0r4) at (6,0) {};
        \node[seven] (0r5) at (8,0) {};
        \draw[thick, double] (0l5)--(0l4)--(0l3)--(0l2)--(0l1)--(0r1)--(0r2)--(0r3)--(0r4)--(0r5);
        \node at (1,0.3) {$n j$};
        \node at (3,0.3) {\small$(n-1)j$};
        \node at (7,0.3) {$j$};
        \node at (-3,0.3) {\small$(n-1)j$};
        \node at (-7,0.3) {$j$};

        \node[seven] (1u) at (0,2) {};
        \node[seven] (2u) at (0,3) {};
        \node (3u) at (0,4) {$\vdots$};
        \node[seven] (4u) at (0,5) {};
        \node[seven] (5u) at (0,6) {};
        \node[seven] (6u) at (0,7) {};

        \node[seven] (1d) at (0,-2) {};
        \node[seven] (2d) at (0,-3) {};
        \node (3d) at (0,-4) {$\vdots$};
        \node[seven] (4d) at (0,-5) {};
        \node[seven] (5d) at (0,-6) {};
        \node[seven] (6d) at (0,-7) {};

        \draw (5u)--(6u) (5d)--(6d);
        \draw[thick,double] (5u)--(4u)--(3u)--(2u)--(1u)--(1d)--(2d)--(3d)--(4d)--(5d);

        \node at (-0.3,6.5) {$1$};
        \node at (-0.3,5.5) {$2$};
        \node at (-0.5,2.5) {$2j$};
        \node at (-0.7,1) {$2j+1$};
        \node at (-0.3,-6.5) {$1$};
        \node at (-0.3,-5.5) {$2$};
        \node at (-0.5,-2.5) {$2j$};
    \end{tikzpicture}}
\end{equation}
from which we can read the magnetic quiver
\begin{equation}
    \scalebox{0.8}{\begin{tikzpicture}
        \node[gauge,label=right:{\scriptsize$nj$}] (0r) at (1,0) {};
        \node[gauge,label=above:{\scriptsize$(n-1)j$},rotate=10] (1r1) at (2,1) {};
        \node[gauge,label=above:{\scriptsize$(n-2)j$},rotate=10] (1r2) at (3,2) {};
        \node[rotate=45] (1r3) at (4,3) {$\cdots$};
        \node[gauge,label=above:{\scriptsize$j$}] (1r4) at (5,4) {};
        \node[gauge,label=below:{\scriptsize$(n-1)j$},rotate=-10] (1l1) at (2,-1) {};
        \node[gauge,label=below:{\scriptsize$(n-2)j$},rotate=-10] (1l2) at (3,-2) {};
        \node[rotate=-45] (1l3) at (4,-3) {$\cdots$};
        \node[gauge,label=below:{\scriptsize$j$}] (1l4) at (5,-4) {};
        \node[gauge,label=left:{\scriptsize$2j+1$}] (0) at (0,0) {};
        \node[gauge,label=left:{\scriptsize$2j$}] (1) at (0,1) {};
        \node (2) at (0,2) {$\vdots$};
        \node[gauge,label=left:{\scriptsize$2$}] (3) at (0,3) {};
        \node[gauge,label=left:{\scriptsize$1$}] (4) at (0,4) {};
        \node[gauge,label=left:{\scriptsize$2j$}] (-1) at (0,-1) {};
        \node (-2) at (0,-2) {$\vdots$};
        \node[gauge,label=left:{\scriptsize$2$}] (-3) at (0,-3) {};
        \node[gauge,label=left:{\scriptsize$1$}] (-4) at (0,-4) {};
        \draw (1r4)--(1r3)--(1r2)--(1r1)--(0r)--(0)--(1)--(2)--(3)--(4) (1l4)--(1l3)--(1l2)--(1l1)--(0r) (0)--(-1)--(-2)--(-3)--(-4);
    \end{tikzpicture}}\;.
\end{equation}

\subsection{D-type}

\subsubsection{$X_{D_{2j+2}}^{(1^{n_1},2^{n_2},3^{n_3})}$}
Let us now turn to $D_{2j+2}$ conformal matter molecules. We start with building generalised quivers from $(D_{2j+2},D_{2j+2})_{i\in\{1,2,3\}}$ conformal matter atoms by gauging diagonal $SO(4j+4)$ symmetries:
\begin{equation}
    \begin{array}{cc}
\makecell{\tilde{\mathsf{Q}}_{X_{D_{2j+2}}^{(1^{n_1},2^{n_2},3^{n_3})}} =\\ \vspace{1.5cm}} &  \begin{tikzpicture}
        \draw[thick] (0,0) circle (0.5);
        \node at (0,0) {\scalebox{0.4}{$SO(4j+4)$}};
        \draw[thick,double] (0.6,0)--(1.2,0);
        \draw[thick,double] (-0.6,0)--(-1.2,0);
        \node at (0.9,0.3) {};
        \node at (-0.9,0.3) {};
          \node at (1.6,0) {$\cdots$};
          \node at (-1.6,0) {$\cdots$};
          \node at (4.7,0) {$\cdots$};
          \draw[thick,double] (1.9,0)--(2.5,0);
          \draw[thick,double] (-1.9,0)--(-2.5,0);
          \node at (2.2,0.3) {};
          \node at (-2.2,0.3) {};
          \draw[thick] (3.1,0) circle (0.5);
          \draw[thick] (-3.1,0) circle (0.5);
          \draw[thick,double] (3.7,0)--(4.3,0);
          \draw[thick,double] (-3.7,0)--(-4.3,0);
          \node at (4.0,0.3) {};
          \node at (-4.0,0.3) {};
           \draw[thick,double] (5.0,0)--(5.6,0);
           \node at (5.3,0.3) {};
           \draw[thick] (6.2,0) circle (0.5);
           \draw[thick,double] (6.8,0)--(7.4,0);
           \node at (7.1,0.3) {};
        \draw[thick] (7.5,0.5)--(8.5,0.5)--(8.5,-0.5)--(7.5,-0.5)--cycle;
        \draw[thick] (-4.4,0.5)--(-5.4,0.5)--(-5.4,-0.5)--(-4.4,-0.5)--cycle;
        \node at (3.1,0) {\scalebox{0.4}{$SO(4j+4)$}};
        \node at (-3.1,0) {\scalebox{0.4}{$SO(4j+4)$}};
        \node at (8,0) {\scalebox{0.8}{$D_{2j+2}$}};
        \node at (-4.9,0) {\scalebox{0.8}{$D_{2j+2}$}};
        \node at (6.2,0) {\scalebox{0.4}{$SO(4j+4)$}};
       \node at (-2.5,-1.3) {$\underbrace{\hspace{3.7cm}}_{\scalebox{0.7}{$n_1\hspace{0.1cm}(D_{2j+2},D_{2j+2})_{(1)}$ \text{edges}}}$};
       \node at (1.6,-1.3) {$\underbrace{\hspace{1.9cm}}_{\scalebox{0.7}{$n_2\hspace{0.1cm} (D_{2j+2},D_{2j+2})_{(2)}$ \text{edges}}}$};
       \node at (5.5,-1.3) {$\underbrace{\hspace{3.7cm}}_{\scalebox{0.7}{$n_3\hspace{0.1cm}(D_{2j+2},D_{2j+2})_{(3)}$ \text{edges}}}$};
    \end{tikzpicture}\;.
    \vspace{-1.5cm}
    
    \end{array}
\end{equation}
The infinite coupling SCFT will not depend on the order in which the atoms are gauge together, and neither will the finite coupling (of the $SO(4j+4)$s) Higgs branch, therefore we fix the order above.

The brane web is
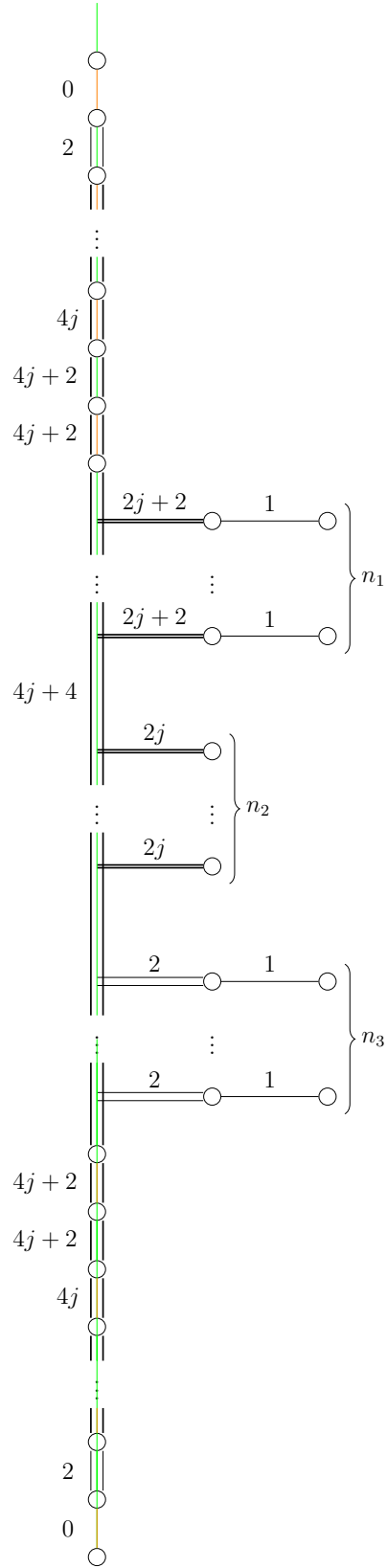
\begin{equation}
    \scalebox{0.8}{\begin{tikzpicture}
        \node[seven] (r11) at (2,0) {};
        \node[seven] (r12) at (4,0) {};
        \node at (2,-1) {$\vdots$};
        \node[seven] (r21) at (2,-2) {};
        \node[seven] (r22) at (4,-2) {};

        \draw (r11)--(r12);
        \draw[thick,double] (0,0)--(r11);
        \node at (1,0.3) {$2j+2$};
        \node at (3,0.3) {$1$};

        \draw (r21)--(r22);
        \draw[thick,double] (0,-2)--(r21);
        \node at (1,-1.7) {$2j+2$};
        \node at (3,-1.7) {$1$};

        \node[seven] (r31) at (2,-4) {};
        \node at (2,-5) {$\vdots$};
        \node[seven] (r41) at (2,-6) {};

        \draw[thick,double] (0,-4)--(r31);
        \node at (1,-3.7) {$2j$};

        \draw[thick,double] (0,-6)--(r41);
        \node at (1,-5.7) {$2j$};
        
        \node[seven] (r51) at (2,-8) {};
        \node[seven] (r52) at (4,-8) {};
        \node at (2,-9) {$\vdots$};
        \node[seven] (r61) at (2,-10) {};
        \node[seven] (r62) at (4,-10) {};

        \draw (r51)--(r52);
        \draw[transform canvas={yshift=-2pt}] (0,-8)--(r51);
        \draw[transform canvas={yshift=2pt}] (0,-8)--(r51);
        \node at (1,-7.7) {$2$};
        \node at (3,-7.7) {$1$};

        \draw (r61)--(r62);
        \draw[transform canvas={yshift=-2pt}] (0,-10)--(r61);
        \draw[transform canvas={yshift=2pt}] (0,-10)--(r61);
        \node at (1,-9.7) {$2$};
        \node at (3,-9.7) {$1$};

        \node[seven] (1) at (0,1) {};
        \node[seven] (2) at (0,2) {};
        \node[seven] (3) at (0,3) {};
        \node[seven] (4) at (0,4) {};
        \node (5) at (0,5) {$\vdots$};
        \node[seven] (6) at (0,6) {};
        \node[seven] (7) at (0,7) {};
        \node[seven] (8) at (0,8) {};

        \node (01) at (0,-1) {$\vdots$};
        \node (02) at (0,-5) {$\vdots$};
        \node (03) at (0,-9) {$\vdots$};
        
        \node[seven] (1d) at (0,-11) {};
        \node[seven] (2d) at (0,-12) {};
        \node[seven] (3d) at (0,-13) {};
        \node[seven] (4d) at (0,-14) {};
        \node (5d) at (0,-15) {$\vdots$};
        \node[seven] (6d) at (0,-16) {};
        \node[seven] (7d) at (0,-17) {};
        \node[seven] (8d) at (0,-18) {};

        \draw[transform canvas={xshift=-3pt}] (7d)--(6d) (6)--(7);
        \draw[transform canvas={xshift=3pt}] (7d)--(6d) (6)--(7);
        \draw[thick,transform canvas={xshift=-3pt}] (6d)--(5d)--(4d)--(3d)--(2d)--(1d)--(03)--(02)--(01)--(1)--(2)--(3)--(4)--(5)--(6);
        \draw[thick,transform canvas={xshift=3pt}] (6d)--(5d)--(4d)--(3d)--(2d)--(1d)--(03)--(02)--(01)--(1)--(2)--(3)--(4)--(5)--(6);

        \draw[green] (0,-9)--(8d) (7d)--(6d) (5d)--(4d) (3d)--(2d) (1d)--(03)--(02)--(01)--(1) (2)--(3) (4)--(5) (6)--(7) (8)--(0,9);
        \draw[orange] (8d)--(7d) (6d)--(5d) (4d)--(3d) (2d)--(1d) (1)--(2) (3)--(4) (5)--(6) (7)--(8);

        \node at (-0.9,-3) {$4j+4$};
        \node at (-0.9,1.5) {$4j+2$};
        \node at (-0.9,2.5) {$4j+2$};
        \node at (-0.5,3.5) {$4j$};
        \node at (-0.5,6.5) {$2$};
        \node at (-0.5,7.5) {$0$};
        \node at (-0.9,-11.5) {$4j+2$};
        \node at (-0.9,-12.5) {$4j+2$};
        \node at (-0.5,-13.5) {$4j$};
        \node at (-0.5,-16.5) {$2$};
        \node at (-0.5,-17.5) {$0$};

        \draw [decorate,decoration={brace,amplitude=5pt},xshift=0pt,yshift=0pt]
        (4.3,0.3)--(4.3,-2.3) node [black,midway,xshift=0.5cm] {$n_1$};

        \draw [decorate,decoration={brace,amplitude=5pt},xshift=0pt,yshift=0pt]
        (2.3,0.3-4)--(2.3,-2.3-4) node [black,midway,xshift=0.5cm] {$n_2$};

        \draw [decorate,decoration={brace,amplitude=5pt},xshift=0pt,yshift=0pt]
        (4.3,0.3-8)--(4.3,-2.3-8) node [black,midway,xshift=0.5cm] {$n_3$};
    \end{tikzpicture}}
\end{equation}
and the magnetic quiver is
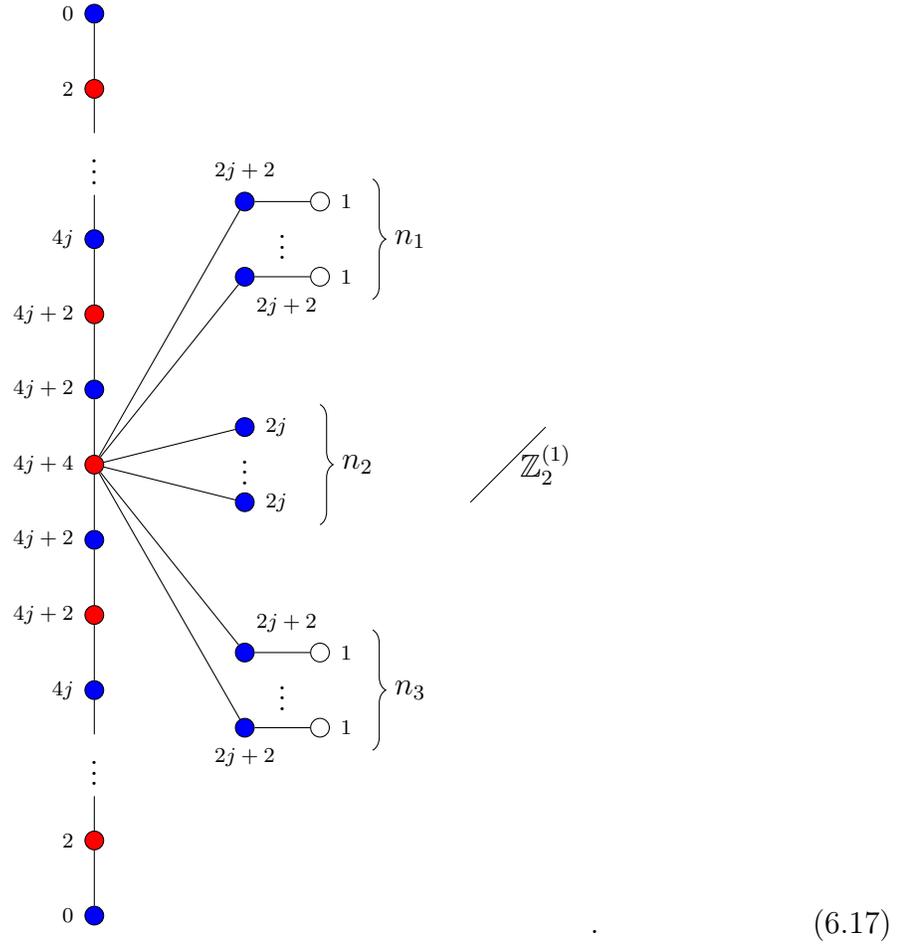
\begin{equation}
    \begin{tikzpicture}
        \draw (5,-0.5)--(6,0.5);
        \node at (6,0) {$\mathbb{Z}_2^{(1)}$};
        
        \node[gaugeb,label=above:{\scriptsize$2j+2$}] (r11) at (2,3.5) {};
        \node[gauge,label=right:{\scriptsize$1$}] (r12) at (3,3.5) {};
        \node at (2.5,3) {$\vdots$};
        \node[gaugeb,label=below:{\scriptsize$2j+2$},rotate=5] (r21) at (2,2.5) {};
        \node[gauge,label=right:{\scriptsize$1$}] (r22) at (3,2.5) {};
        
        \node[gaugeb,label=right:{\scriptsize$2j$}] (r31) at (2,0.5) {};
        \node at (2,0) {$\vdots$};
        \node[gaugeb,label=right:{\scriptsize$2j$}] (r41) at (2,-0.5) {};
        
        \node[gaugeb,label=above:{\scriptsize$2j+2$},rotate=-5] (r51) at (2,-2.5) {};
        \node[gauge,label=right:{\scriptsize$1$}] (r52) at (3,-2.5) {};
        \node at (2.5,-3) {$\vdots$};
        \node[gaugeb,label=below:{\scriptsize$2j+2$}] (r61) at (2,-3.5) {};
        \node[gauge,label=right:{\scriptsize$1$}] (r62) at (3,-3.5) {};
        
        \node[gauger,label=left:{\scriptsize$4j+4$}] (0) at (0,0) {};
        \node[gaugeb,label=left:{\scriptsize$4j+2$}] (1) at (0,1) {};
        \node[gauger,label=left:{\scriptsize$4j+2$}] (2) at (0,2) {};
        \node[gaugeb,label=left:{\scriptsize$4j$}] (3) at (0,3) {};
        \node (4) at (0,4) {$\vdots$};
        \node[gauger,label=left:{\scriptsize$2$}] (5) at (0,5) {};
        \node[gaugeb,label=left:{\scriptsize$0$}] (6) at (0,6) {};
        \node[gaugeb,label=left:{\scriptsize$4j+2$}] (-1) at (0,-1) {};
        \node[gauger,label=left:{\scriptsize$4j+2$}] (-2) at (0,-2) {};
        \node[gaugeb,label=left:{\scriptsize$4j$}] (-3) at (0,-3) {};
        \node (-4) at (0,-4) {$\vdots$};
        \node[gauger,label=left:{\scriptsize$2$}] (-5) at (0,-5) {};
        \node[gaugeb,label=left:{\scriptsize$0$}] (-6) at (0,-6) {};
        \draw (0)--(1)--(2)--(3)--(4)--(5)--(6) (0)--(-1)--(-2)--(-3)--(-4)--(-5)--(-6) (0)--(r11)--(r12) (0)--(r21)--(r22) (0)--(r31) (0)--(r41) (0)--(r51)--(r52) (0)--(r61)--(r62);

        \draw [decorate,decoration={brace,amplitude=5pt},xshift=0pt,yshift=0pt] (3.7,3.5+0.3)--(3.7,2.5-0.3) node [black,midway,xshift=0.5cm] {$n_1$};
        
        \draw [decorate,decoration={brace,amplitude=5pt},xshift=0pt,yshift=0pt] (3,0.5+0.3)--(3,-0.5-0.3) node [black,midway,xshift=0.5cm] {$n_2$};

        \draw [decorate,decoration={brace,amplitude=5pt},xshift=0pt,yshift=0pt] (3.7,-2.5+0.3)--(3.7,-3.5-0.3) node [black,midway,xshift=0.5cm] {$n_3$};
    \end{tikzpicture}\;.
\end{equation}

\paragraph{Infinite coupling.}
Taking all the couplings of the generalised quiver to infinity realises the SCFT associated to the $X_{D_{2j+2}}^{(1^{n_1},2^{n_2},3^{n_3})}$ singularity.

\begin{landscape}
The corresponding brane web is:
\begin{equation}
    \scalebox{0.75}{\begin{tikzpicture}
        \def\x{0.9};
        \node[seven] (r1) at (2*\x,0) {};
        \node[seven] (r2) at (4*\x,0) {};
        \node (r3) at (5*\x,0) {$\cdots$};
        \node[seven] (r4) at (6*\x,0) {};
        \node[seven] (r5) at (8*\x,0) {};
        \node[seven] (r6) at (10*\x,0) {};
        \node[seven] (r7) at (12*\x,0) {};
        \node (r8) at (13*\x,0) {$\cdots$};
        \node[seven] (r9) at (14*\x,0) {};
        \node[seven] (r10) at (16*\x,0) {};
        \node[seven] (r11) at (18*\x,0) {};
        \node[seven] (r12) at (20*\x,0) {};
        \node (r13) at (21*\x,0) {$\cdots$};
        \node[seven] (r14) at (22*\x,0) {};
        \node[seven] (r15) at (24*\x,0) {};
        \node[seven] (r16) at (26*\x,0) {};
        \node[seven] (r17) at (28*\x,0) {};
        \node (r18) at (29*\x,0) {$\cdots$};
        \node[seven] (r19) at (30*\x,0) {};
        \node[seven] (r20) at (32*\x,0) {};

        \draw[thick,double] (0,0)--(r1)--(r2)--(r3)--(r4)--(r5)--(r6)--(r7)--(r8)--(r9)--(r10)--(r11)--(r12)--(r13)--(r14)--(r15)--(r16)--(r17)--(r18)--(r19);
        \draw (r19)--(r20);
        \node[anchor=west,rotate=90] at (1*\x,0.2) {$n_1(2j+2)+n_2(2j)+n_3(2)$};
        \node[anchor=west,rotate=90] at (3*\x,0.2) {$(n_1-1)(2j+2)+n_2(2j)+n_3(2)+1$};
        \node[anchor=west,rotate=90] at (7*\x,0.2) {$1(2j+2)+n_2(2j)+n_3(2)+(n_1-1)$};
        \node[anchor=west,rotate=90] at (9*\x,0.2) {$n_2(2j)+n_3(2)+n_1$};
        \node[anchor=west,rotate=90] at (11*\x,0.2) {$(n_2-1)(2j)+n_3(2)+n_1$};
        \node[anchor=west,rotate=90] at (15*\x,0.2) {$1(2j)+n_3(2)+n_1$};
        \node[anchor=west,rotate=90] at (17*\x,0.2) {$n_3(2)+n_1$};
        \node[anchor=west,rotate=90] at (19*\x,0.2) {$(n_3-1)(2)+n_1+1$};
        \node[anchor=west,rotate=90] at (23*\x,0.2) {$1(2)+n_1+(n_3-1)$};
        \node[anchor=west,rotate=90] at (25*\x,0.2) {$n_1+n_3$};
        \node[anchor=west,rotate=90] at (27*\x,0.2) {$n_1+n_3-1$};
        \node[anchor=west,rotate=90] at (31*\x,0.2) {$1$};

        \node[seven] (1) at (0,1) {};
        \node[seven] (2) at (0,2) {};
        \node[seven] (3) at (0,3) {};
        \node[seven] (4) at (0,4) {};
        \node (5) at (0,5) {$\vdots$};
        \node[seven] (6) at (0,6) {};
        \node[seven] (7) at (0,7) {};
        \node[seven] (8) at (0,8) {};
        
        \node[seven] (1d) at (0,-1) {};
        \node[seven] (2d) at (0,-2) {};
        \node[seven] (3d) at (0,-3) {};
        \node[seven] (4d) at (0,-4) {};
        \node (5d) at (0,-5) {$\vdots$};
        \node[seven] (6d) at (0,-6) {};
        \node[seven] (7d) at (0,-7) {};
        \node[seven] (8d) at (0,-8) {};

        \draw[transform canvas={xshift=-3pt}] (7d)--(6d) (6)--(7);
        \draw[transform canvas={xshift=3pt}] (7d)--(6d) (6)--(7);
        \draw[thick,transform canvas={xshift=-3pt}] (6d)--(5d)--(4d)--(3d)--(2d)--(1d)--(1)--(2)--(3)--(4)--(5)--(6);
        \draw[thick,transform canvas={xshift=3pt}] (6d)--(5d)--(4d)--(3d)--(2d)--(1d)--(1)--(2)--(3)--(4)--(5)--(6);

        \draw[green] (0,-9)--(8d) (7d)--(6d) (5d)--(4d) (3d)--(2d) (1d)--(1) (2)--(3) (4)--(5) (6)--(7) (8)--(0,9);
        \draw[orange] (8d)--(7d) (6d)--(5d) (4d)--(3d) (2d)--(1d) (1)--(2) (3)--(4) (5)--(6) (7)--(8);

        \node at (-0.9,0) {$4j+4$};
        \node at (-0.9,1.5) {$4j+2$};
        \node at (-0.9,2.5) {$4j+2$};
        \node at (-0.5,3.5) {$4j$};
        \node at (-0.5,6.5) {$2$};
        \node at (-0.5,7.5) {$0$};
        \node at (-0.9,-1.5) {$4j+2$};
        \node at (-0.9,-2.5) {$4j+2$};
        \node at (-0.5,-3.5) {$4j$};
        \node at (-0.5,-6.5) {$2$};
        \node at (-0.5,-7.5) {$0$};
    \end{tikzpicture}}\;,
\end{equation}
from which we can read the magnetic quiver:
\begin{equation}
    \begin{tikzpicture}
        \def\x{0.6};
        
        \draw (33*\x,-0.5)--(34*\x,0.5);
        \node at (34.5*\x,0) {$\mathbb{Z}_2^{(1)}$};

        \node[gaugeb] (r1) at (2*\x,0) {};
        \node[gauge] (r2) at (4*\x,0) {};
        \node (r3) at (6*\x,0) {$\cdots$};
        \node[gauge] (r4) at (8*\x,0) {};
        \node[gauge] (r5) at (10*\x,0) {};
        \node[gauge] (r6) at (12*\x,0) {};
        \node (r7) at (14*\x,0) {$\cdots$};
        \node[gauge] (r8) at (16*\x,0) {};
        \node[gauge] (r9) at (18*\x,0) {};
        \node[gauge] (r10) at (20*\x,0) {};
        \node (r11) at (22*\x,0) {$\cdots$};
        \node[gauge] (r12) at (24*\x,0) {};
        \node[gauge] (r13) at (26*\x,0) {};
        \node[gauge] (r14) at (28*\x,0) {};
        \node (r15) at (30*\x,0) {$\cdots$};
        \node[gauge] (r16) at (32*\x,0) {};

        \node[anchor=west,rotate=90] at (2*\x,0.2) {\scriptsize$n_1(2j+2)+n_2(2j)+n_3(2)$};
        \node[anchor=west,rotate=90] at (4*\x,0.2) {\scriptsize$(n_1-1)(2j+2)+n_2(2j)+n_3(2)+1$};
        \node[anchor=west,rotate=90] at (8*\x,0.2) {\scriptsize$1(2j+2)+n_2(2j)+n_3(2)+(n_1-1)$};
        \node[anchor=west,rotate=90] at (10*\x,0.2) {\scriptsize$n_2(2j)+n_3(2)+n_1$};
        \node[anchor=west,rotate=90] at (12*\x,0.2) {\scriptsize$(n_2-1)(2j)+n_3(2)+n_1$};
        \node[anchor=west,rotate=90] at (16*\x,0.2) {\scriptsize$1(2j)+n_3(2)+n_1$};
        \node[anchor=west,rotate=90] at (18*\x,0.2) {\scriptsize$n_3(2)+n_1$};
        \node[anchor=west,rotate=90] at (20*\x,0.2) {\scriptsize$(n_3-1)(2)+n_1+1$};
        \node[anchor=west,rotate=90] at (24*\x,0.2) {\scriptsize$1(2)+n_1+(n_3-1)$};
        \node[anchor=west,rotate=90] at (26*\x,0.2) {\scriptsize$n_1+n_3$};
        \node[anchor=west,rotate=90] at (28*\x,0.2) {\scriptsize$n_1+n_3-1$};
        \node[anchor=west,rotate=90] at (32*\x,0.2) {\scriptsize$1$};
        
        \node[gauger,label=left:{\scriptsize$4j+4$}] (0) at (0,0) {};
        \node[gaugeb,label=left:{\scriptsize$4j+2$}] (1) at (0,1) {};
        \node[gauger,label=left:{\scriptsize$4j+2$}] (2) at (0,2) {};
        \node[gaugeb,label=left:{\scriptsize$4j$}] (3) at (0,3) {};
        \node (4) at (0,4) {$\vdots$};
        \node[gauger,label=left:{\scriptsize$2$}] (5) at (0,5) {};
        \node[gaugeb,label=left:{\scriptsize$0$}] (6) at (0,6) {};
        \node[gaugeb,label=left:{\scriptsize$4j+2$}] (-1) at (0,-1) {};
        \node[gauger,label=left:{\scriptsize$4j+2$}] (-2) at (0,-2) {};
        \node[gaugeb,label=left:{\scriptsize$4j$}] (-3) at (0,-3) {};
        \node (-4) at (0,-4) {$\vdots$};
        \node[gauger,label=left:{\scriptsize$2$}] (-5) at (0,-5) {};
        \node[gaugeb,label=left:{\scriptsize$0$}] (-6) at (0,-6) {};
        \draw (0)--(1)--(2)--(3)--(4)--(5)--(6) (0)--(-1)--(-2)--(-3)--(-4)--(-5)--(-6) (0)--(r1)--(r2)--(r3)--(r4)--(r5)--(r6)--(r7)--(r8)--(r9)--(r10)--(r11)--(r12)--(r13)--(r14)--(r15)--(r16);
    \end{tikzpicture}
\end{equation}
\end{landscape}

\subsubsection{$X_{D_{2j+3}}^{(1^{n_1},2^{n_2},3^{n_3})}$}
Let us now turn to $D_{2j+3}$ conformal matter molecules. We start with building generalised quivers from $(D_{2j+3},D_{2j+3})_{i\in\{1,2,3\}}$ conformal matter atoms by gauging diagonal $SO(4j+6)$ symmetries:
\begin{equation*}
   \begin{array}{cc}
\makecell{\tilde{\mathsf{Q}}_{X_{D_{2j+3}}^{(1^{n_1},2^{n_2},3^{n_3})}} =\\ \vspace{2cm}}  &
    \begin{tikzpicture}
        \draw[thick] (0,0) circle (0.5);
        \node at (0,0) {\scalebox{0.4}{$SO(4j+6)$}};
        \draw[thick,double] (0.6,0)--(1.2,0);
        \draw[thick,double] (-0.6,0)--(-1.2,0);
        \node at (0.9,0.3) {};
        \node at (-0.9,0.3) {};
          \node at (1.6,0) {$\cdots$};
          \node at (-1.6,0) {$\cdots$};
          \node at (4.7,0) {$\cdots$};
          \draw[thick,double] (1.9,0)--(2.5,0);
          \draw[thick,double] (-1.9,0)--(-2.5,0);
          \node at (2.2,0.3) {};
          \node at (-2.2,0.3) {};
          \draw[thick] (3.1,0) circle (0.5);
          \draw[thick] (-3.1,0) circle (0.5);
          \draw[thick,double] (3.7,0)--(4.3,0);
          \draw[thick,double] (-3.7,0)--(-4.3,0);
          \node at (4.0,0.3) {};
          \node at (-4.0,0.3) {};
           \draw[thick,double] (5.0,0)--(5.6,0);
           \node at (5.3,0.3) {};
           \draw[thick] (6.2,0) circle (0.5);
           \draw[thick,double] (6.8,0)--(7.4,0);
           \node at (7.1,0.3) {};
        \draw[thick] (7.5,0.5)--(8.5,0.5)--(8.5,-0.5)--(7.5,-0.5)--cycle;
        \draw[thick] (-4.4,0.5)--(-5.4,0.5)--(-5.4,-0.5)--(-4.4,-0.5)--cycle;
        \node at (3.1,0) {\scalebox{0.4}{$SO(4j+6)$}};
        \node at (-3.1,0) {\scalebox{0.4}{$SO(4j+6)$}};
        \node at (8,0) {\scalebox{0.8}{$D_{2j+3}$}};
        \node at (-4.9,0) {\scalebox{0.8}{$D_{2j+3}$}};
        \node at (6.2,0) {\scalebox{0.4}{$SO(4j+6)$}};
       \node at (-2.5,-1.3) {$\underbrace{\hspace{3.7cm}}_{\scalebox{0.7}{$n_1\hspace{0.1cm}(D_{2j+3},D_{2j+3})_{(1)}$ \text{edges}}}$};
       \node at (1.6,-1.3) {$\underbrace{\hspace{1.9cm}}_{\scalebox{0.7}{$n_2\hspace{0.1cm} (D_{2j+3},D_{2j+3})_{(2)}$ \text{edges}}}$};
       \node at (5.5,-1.3) {$\underbrace{\hspace{3.7cm}}_{\scalebox{0.7}{$n_3\hspace{0.1cm}(D_{2j+3},D_{2j+3})_{(3)}$ \text{edges}}}$};
    \end{tikzpicture}\;.
    \end{array}
    \vspace{-1.5cm}
\end{equation*}
As before the infinite coupling SCFT will not depend on the order in which the atoms are gauged together \cite{DeMarco:2023irn}, and neither will the finite coupling (of the $SO(4j+6)$s) Higgs branch. Therefore we may fix the order above.

The brane web is
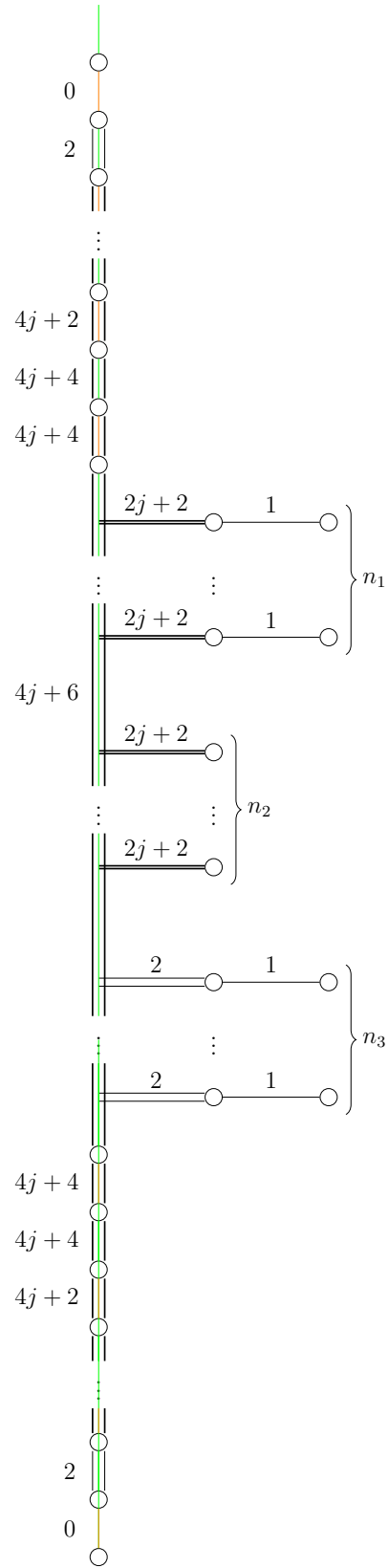
\begin{equation}
    \scalebox{0.8}{\begin{tikzpicture}
        \node[seven] (r11) at (2,0) {};
        \node[seven] (r12) at (4,0) {};
        \node at (2,-1) {$\vdots$};
        \node[seven] (r21) at (2,-2) {};
        \node[seven] (r22) at (4,-2) {};

        \draw (r11)--(r12);
        \draw[thick,double] (0,0)--(r11);
        \node at (1,0.3) {$2j+2$};
        \node at (3,0.3) {$1$};

        \draw (r21)--(r22);
        \draw[thick,double] (0,-2)--(r21);
        \node at (1,-1.7) {$2j+2$};
        \node at (3,-1.7) {$1$};

        \node[seven] (r31) at (2,-4) {};
        \node at (2,-5) {$\vdots$};
        \node[seven] (r41) at (2,-6) {};

        \draw[thick,double] (0,-4)--(r31);
        \node at (1,-3.7) {$2j+2$};

        \draw[thick,double] (0,-6)--(r41);
        \node at (1,-5.7) {$2j+2$};
        
        \node[seven] (r51) at (2,-8) {};
        \node[seven] (r52) at (4,-8) {};
        \node at (2,-9) {$\vdots$};
        \node[seven] (r61) at (2,-10) {};
        \node[seven] (r62) at (4,-10) {};

        \draw (r51)--(r52);
        \draw[transform canvas={yshift=-2pt}] (0,-8)--(r51);
        \draw[transform canvas={yshift=2pt}] (0,-8)--(r51);
        \node at (1,-7.7) {$2$};
        \node at (3,-7.7) {$1$};

        \draw (r61)--(r62);
        \draw[transform canvas={yshift=-2pt}] (0,-10)--(r61);
        \draw[transform canvas={yshift=2pt}] (0,-10)--(r61);
        \node at (1,-9.7) {$2$};
        \node at (3,-9.7) {$1$};

        \node[seven] (1) at (0,1) {};
        \node[seven] (2) at (0,2) {};
        \node[seven] (3) at (0,3) {};
        \node[seven] (4) at (0,4) {};
        \node (5) at (0,5) {$\vdots$};
        \node[seven] (6) at (0,6) {};
        \node[seven] (7) at (0,7) {};
        \node[seven] (8) at (0,8) {};

        \node (01) at (0,-1) {$\vdots$};
        \node (02) at (0,-5) {$\vdots$};
        \node (03) at (0,-9) {$\vdots$};
        
        \node[seven] (1d) at (0,-11) {};
        \node[seven] (2d) at (0,-12) {};
        \node[seven] (3d) at (0,-13) {};
        \node[seven] (4d) at (0,-14) {};
        \node (5d) at (0,-15) {$\vdots$};
        \node[seven] (6d) at (0,-16) {};
        \node[seven] (7d) at (0,-17) {};
        \node[seven] (8d) at (0,-18) {};

        \draw[transform canvas={xshift=-3pt}] (7d)--(6d) (6)--(7);
        \draw[transform canvas={xshift=3pt}] (7d)--(6d) (6)--(7);
        \draw[thick,transform canvas={xshift=-3pt}] (6d)--(5d)--(4d)--(3d)--(2d)--(1d)--(03)--(02)--(01)--(1)--(2)--(3)--(4)--(5)--(6);
        \draw[thick,transform canvas={xshift=3pt}] (6d)--(5d)--(4d)--(3d)--(2d)--(1d)--(03)--(02)--(01)--(1)--(2)--(3)--(4)--(5)--(6);

        \draw[green] (0,-9)--(8d) (7d)--(6d) (5d)--(4d) (3d)--(2d) (1d)--(03)--(02)--(01)--(1) (2)--(3) (4)--(5) (6)--(7) (8)--(0,9);
        \draw[orange] (8d)--(7d) (6d)--(5d) (4d)--(3d) (2d)--(1d) (1)--(2) (3)--(4) (5)--(6) (7)--(8);

        \node at (-0.9,-3) {$4j+6$};
        \node at (-0.9,1.5) {$4j+4$};
        \node at (-0.9,2.5) {$4j+4$};
        \node at (-0.9,3.5) {$4j+2$};
        \node at (-0.5,6.5) {$2$};
        \node at (-0.5,7.5) {$0$};
        \node at (-0.9,-11.5) {$4j+4$};
        \node at (-0.9,-12.5) {$4j+4$};
        \node at (-0.9,-13.5) {$4j+2$};
        \node at (-0.5,-16.5) {$2$};
        \node at (-0.5,-17.5) {$0$};

        \draw [decorate,decoration={brace,amplitude=5pt},xshift=0pt,yshift=0pt]
        (4.3,0.3)--(4.3,-2.3) node [black,midway,xshift=0.5cm] {$n_1$};

        \draw [decorate,decoration={brace,amplitude=5pt},xshift=0pt,yshift=0pt]
        (2.3,0.3-4)--(2.3,-2.3-4) node [black,midway,xshift=0.5cm] {$n_2$};

        \draw [decorate,decoration={brace,amplitude=5pt},xshift=0pt,yshift=0pt]
        (4.3,0.3-8)--(4.3,-2.3-8) node [black,midway,xshift=0.5cm] {$n_3$};
    \end{tikzpicture}}
\end{equation}
and the magnetic quiver is
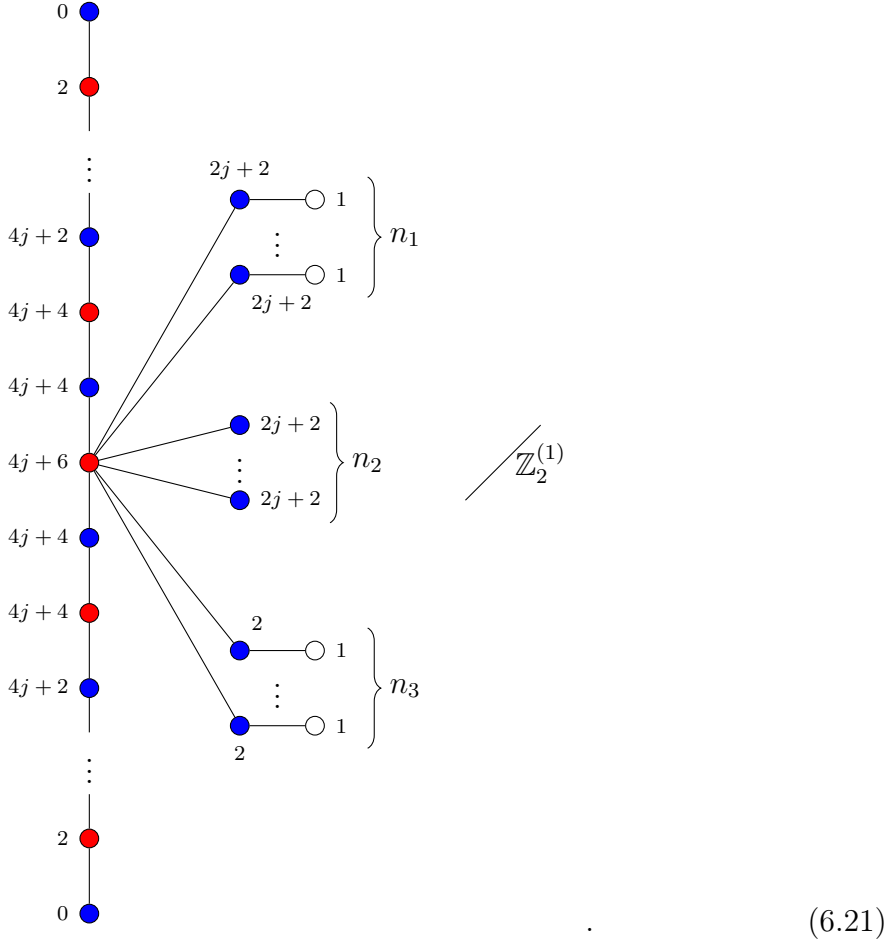
\begin{equation}
    \begin{tikzpicture}
        \draw (5,-0.5)--(6,0.5);
        \node at (6,0) {$\mathbb{Z}_2^{(1)}$};
        
        \node[gaugeb,label=above:{\scriptsize$2j+2$}] (r11) at (2,3.5) {};
        \node[gauge,label=right:{\scriptsize$1$}] (r12) at (3,3.5) {};
        \node at (2.5,3) {$\vdots$};
        \node[gaugeb,label=below:{\scriptsize$2j+2$},rotate=5] (r21) at (2,2.5) {};
        \node[gauge,label=right:{\scriptsize$1$}] (r22) at (3,2.5) {};
        
        \node[gaugeb,label=right:{\scriptsize$2j+2$}] (r31) at (2,0.5) {};
        \node at (2,0) {$\vdots$};
        \node[gaugeb,label=right:{\scriptsize$2j+2$}] (r41) at (2,-0.5) {};
        
        \node[gaugeb,label=above:{\scriptsize$2$},rotate=-5] (r51) at (2,-2.5) {};
        \node[gauge,label=right:{\scriptsize$1$}] (r52) at (3,-2.5) {};
        \node at (2.5,-3) {$\vdots$};
        \node[gaugeb,label=below:{\scriptsize$2$}] (r61) at (2,-3.5) {};
        \node[gauge,label=right:{\scriptsize$1$}] (r62) at (3,-3.5) {};
        
        \node[gauger,label=left:{\scriptsize$4j+6$}] (0) at (0,0) {};
        \node[gaugeb,label=left:{\scriptsize$4j+4$}] (1) at (0,1) {};
        \node[gauger,label=left:{\scriptsize$4j+4$}] (2) at (0,2) {};
        \node[gaugeb,label=left:{\scriptsize$4j+2$}] (3) at (0,3) {};
        \node (4) at (0,4) {$\vdots$};
        \node[gauger,label=left:{\scriptsize$2$}] (5) at (0,5) {};
        \node[gaugeb,label=left:{\scriptsize$0$}] (6) at (0,6) {};
        \node[gaugeb,label=left:{\scriptsize$4j+4$}] (-1) at (0,-1) {};
        \node[gauger,label=left:{\scriptsize$4j+4$}] (-2) at (0,-2) {};
        \node[gaugeb,label=left:{\scriptsize$4j+2$}] (-3) at (0,-3) {};
        \node (-4) at (0,-4) {$\vdots$};
        \node[gauger,label=left:{\scriptsize$2$}] (-5) at (0,-5) {};
        \node[gaugeb,label=left:{\scriptsize$0$}] (-6) at (0,-6) {};
        \draw (0)--(1)--(2)--(3)--(4)--(5)--(6) (0)--(-1)--(-2)--(-3)--(-4)--(-5)--(-6) (0)--(r11)--(r12) (0)--(r21)--(r22) (0)--(r31) (0)--(r41) (0)--(r51)--(r52) (0)--(r61)--(r62);

        \draw [decorate,decoration={brace,amplitude=5pt},xshift=0pt,yshift=0pt] (3.7,3.5+0.3)--(3.7,2.5-0.3) node [black,midway,xshift=0.5cm] {$n_1$};
        
        \draw [decorate,decoration={brace,amplitude=5pt},xshift=0pt,yshift=0pt] (3.2,0.5+0.3)--(3.2,-0.5-0.3) node [black,midway,xshift=0.5cm] {$n_2$};

        \draw [decorate,decoration={brace,amplitude=5pt},xshift=0pt,yshift=0pt] (3.7,-2.5+0.3)--(3.7,-3.5-0.3) node [black,midway,xshift=0.5cm] {$n_3$};
    \end{tikzpicture}\;.
\end{equation}

\paragraph{Infinite coupling.}
Taking all the couplings of the generalised quiver to infinity realises the SCFT associated to the $X_{D_{2j+3}}^{(1^{n_1},2^{n_2},3^{n_3})}$ singularity.

\begin{landscape}
The corresponding brane web is:
\begin{equation}
    \scalebox{0.75}{\begin{tikzpicture}
        \def\x{0.9};
        \node[seven] (r1) at (2*\x,0) {};
        \node[seven] (r2) at (4*\x,0) {};
        \node (r3) at (5*\x,0) {$\cdots$};
        \node[seven] (r4) at (6*\x,0) {};
        \node[seven] (r5) at (8*\x,0) {};
        \node[seven] (r6) at (10*\x,0) {};
        \node[seven] (r7) at (12*\x,0) {};
        \node (r8) at (13*\x,0) {$\cdots$};
        \node[seven] (r9) at (14*\x,0) {};
        \node[seven] (r10) at (16*\x,0) {};
        \node[seven] (r11) at (18*\x,0) {};
        \node[seven] (r12) at (20*\x,0) {};
        \node (r13) at (21*\x,0) {$\cdots$};
        \node[seven] (r14) at (22*\x,0) {};
        \node[seven] (r15) at (24*\x,0) {};
        \node[seven] (r16) at (26*\x,0) {};
        \node[seven] (r17) at (28*\x,0) {};
        \node (r18) at (29*\x,0) {$\cdots$};
        \node[seven] (r19) at (30*\x,0) {};
        \node[seven] (r20) at (32*\x,0) {};

        \draw[thick,double] (0,0)--(r1)--(r2)--(r3)--(r4)--(r5)--(r6)--(r7)--(r8)--(r9)--(r10)--(r11)--(r12)--(r13)--(r14)--(r15)--(r16)--(r17)--(r18)--(r19);
        \draw (r19)--(r20);
        \node[anchor=west,rotate=90] at (1*\x,0.2) {$n_1(2j+2)+n_2(2j+2)+n_3(2)$};
        \node[anchor=west,rotate=90] at (3*\x,0.2) {$(n_1-1)(2j+2)+n_2(2j+2)+n_3(2)+1$};
        \node[anchor=west,rotate=90] at (7*\x,0.2) {$1(2j+2)+n_2(2j+2)+n_3(2)+(n_1-1)$};
        \node[anchor=west,rotate=90] at (9*\x,0.2) {$n_2(2j+2)+n_3(2)+n_1$};
        \node[anchor=west,rotate=90] at (11*\x,0.2) {$(n_2-1)(2j+2)+n_3(2)+n_1$};
        \node[anchor=west,rotate=90] at (15*\x,0.2) {$1(2j+2)+n_3(2)+n_1$};
        \node[anchor=west,rotate=90] at (17*\x,0.2) {$n_3(2)+n_1$};
        \node[anchor=west,rotate=90] at (19*\x,0.2) {$(n_3-1)(2)+n_1+1$};
        \node[anchor=west,rotate=90] at (23*\x,0.2) {$1(2)+n_1+(n_3-1)$};
        \node[anchor=west,rotate=90] at (25*\x,0.2) {$n_1+n_3$};
        \node[anchor=west,rotate=90] at (27*\x,0.2) {$n_1+n_3-1$};
        \node[anchor=west,rotate=90] at (31*\x,0.2) {$1$};

        \node[seven] (1) at (0,1) {};
        \node[seven] (2) at (0,2) {};
        \node[seven] (3) at (0,3) {};
        \node[seven] (4) at (0,4) {};
        \node (5) at (0,5) {$\vdots$};
        \node[seven] (6) at (0,6) {};
        \node[seven] (7) at (0,7) {};
        \node[seven] (8) at (0,8) {};
        
        \node[seven] (1d) at (0,-1) {};
        \node[seven] (2d) at (0,-2) {};
        \node[seven] (3d) at (0,-3) {};
        \node[seven] (4d) at (0,-4) {};
        \node (5d) at (0,-5) {$\vdots$};
        \node[seven] (6d) at (0,-6) {};
        \node[seven] (7d) at (0,-7) {};
        \node[seven] (8d) at (0,-8) {};

        \draw[transform canvas={xshift=-3pt}] (7d)--(6d) (6)--(7);
        \draw[transform canvas={xshift=3pt}] (7d)--(6d) (6)--(7);
        \draw[thick,transform canvas={xshift=-3pt}] (6d)--(5d)--(4d)--(3d)--(2d)--(1d)--(1)--(2)--(3)--(4)--(5)--(6);
        \draw[thick,transform canvas={xshift=3pt}] (6d)--(5d)--(4d)--(3d)--(2d)--(1d)--(1)--(2)--(3)--(4)--(5)--(6);

        \draw[green] (0,-9)--(8d) (7d)--(6d) (5d)--(4d) (3d)--(2d) (1d)--(1) (2)--(3) (4)--(5) (6)--(7) (8)--(0,9);
        \draw[orange] (8d)--(7d) (6d)--(5d) (4d)--(3d) (2d)--(1d) (1)--(2) (3)--(4) (5)--(6) (7)--(8);

        \node at (-0.9,0) {$4j+6$};
        \node at (-0.9,1.5) {$4j+4$};
        \node at (-0.9,2.5) {$4j+4$};
        \node at (-0.9,3.5) {$4j+2$};
        \node at (-0.5,6.5) {$2$};
        \node at (-0.5,7.5) {$0$};
        \node at (-0.9,-1.5) {$4j+4$};
        \node at (-0.9,-2.5) {$4j+4$};
        \node at (-0.9,-3.5) {$4j+2$};
        \node at (-0.5,-6.5) {$2$};
        \node at (-0.5,-7.5) {$0$};
    \end{tikzpicture}}\;,
\end{equation}
from which we can read the magnetic quiver:
\begin{equation}
    \begin{tikzpicture}
        \def\x{0.6};
        
        \draw (33*\x,-0.5)--(34*\x,0.5);
        \node at (34.5*\x,0) {$\mathbb{Z}_2^{(1)}$};

        \node[gaugeb] (r1) at (2*\x,0) {};
        \node[gauge] (r2) at (4*\x,0) {};
        \node (r3) at (6*\x,0) {$\cdots$};
        \node[gauge] (r4) at (8*\x,0) {};
        \node[gauge] (r5) at (10*\x,0) {};
        \node[gauge] (r6) at (12*\x,0) {};
        \node (r7) at (14*\x,0) {$\cdots$};
        \node[gauge] (r8) at (16*\x,0) {};
        \node[gauge] (r9) at (18*\x,0) {};
        \node[gauge] (r10) at (20*\x,0) {};
        \node (r11) at (22*\x,0) {$\cdots$};
        \node[gauge] (r12) at (24*\x,0) {};
        \node[gauge] (r13) at (26*\x,0) {};
        \node[gauge] (r14) at (28*\x,0) {};
        \node (r15) at (30*\x,0) {$\cdots$};
        \node[gauge] (r16) at (32*\x,0) {};

        \node[anchor=west,rotate=90] at (2*\x,0.2) {\scriptsize$n_1(2j+2)+n_2(2j+2)+n_3(2)$};
        \node[anchor=west,rotate=90] at (4*\x,0.2) {\scriptsize$(n_1-1)(2j+2)+n_2(2j+2)+n_3(2)+1$};
        \node[anchor=west,rotate=90] at (8*\x,0.2) {\scriptsize$1(2j+2)+n_2(2j+2)+n_3(2)+(n_1-1)$};
        \node[anchor=west,rotate=90] at (10*\x,0.2) {\scriptsize$n_2(2j+2)+n_3(2)+n_1$};
        \node[anchor=west,rotate=90] at (12*\x,0.2) {\scriptsize$(n_2-1)(2j+2)+n_3(2)+n_1$};
        \node[anchor=west,rotate=90] at (16*\x,0.2) {\scriptsize$1(2j+2)+n_3(2)+n_1$};
        \node[anchor=west,rotate=90] at (18*\x,0.2) {\scriptsize$n_3(2)+n_1$};
        \node[anchor=west,rotate=90] at (20*\x,0.2) {\scriptsize$(n_3-1)(2)+n_1+1$};
        \node[anchor=west,rotate=90] at (24*\x,0.2) {\scriptsize$1(2)+n_1+(n_3-1)$};
        \node[anchor=west,rotate=90] at (26*\x,0.2) {\scriptsize$n_1+n_3$};
        \node[anchor=west,rotate=90] at (28*\x,0.2) {\scriptsize$n_1+n_3-1$};
        \node[anchor=west,rotate=90] at (32*\x,0.2) {\scriptsize$1$};
        
        \node[gauger,label=left:{\scriptsize$4j+6$}] (0) at (0,0) {};
        \node[gaugeb,label=left:{\scriptsize$4j+4$}] (1) at (0,1) {};
        \node[gauger,label=left:{\scriptsize$4j+4$}] (2) at (0,2) {};
        \node[gaugeb,label=left:{\scriptsize$4j+2$}] (3) at (0,3) {};
        \node (4) at (0,4) {$\vdots$};
        \node[gauger,label=left:{\scriptsize$2$}] (5) at (0,5) {};
        \node[gaugeb,label=left:{\scriptsize$0$}] (6) at (0,6) {};
        \node[gaugeb,label=left:{\scriptsize$4j+4$}] (-1) at (0,-1) {};
        \node[gauger,label=left:{\scriptsize$4j+4$}] (-2) at (0,-2) {};
        \node[gaugeb,label=left:{\scriptsize$4j+2$}] (-3) at (0,-3) {};
        \node (-4) at (0,-4) {$\vdots$};
        \node[gauger,label=left:{\scriptsize$2$}] (-5) at (0,-5) {};
        \node[gaugeb,label=left:{\scriptsize$0$}] (-6) at (0,-6) {};
        \draw (0)--(1)--(2)--(3)--(4)--(5)--(6) (0)--(-1)--(-2)--(-3)--(-4)--(-5)--(-6) (0)--(r1)--(r2)--(r3)--(r4)--(r5)--(r6)--(r7)--(r8)--(r9)--(r10)--(r11)--(r12)--(r13)--(r14)--(r15)--(r16);
    \end{tikzpicture}
\end{equation}
\end{landscape}

\subsection{Relation to class$-\mathcal{S}$}
We observe that all A- and D-type 5d conformal matter molecules can be constructed as worldvolume theories of 5-brane webs. The magnetic quivers read from these brane webs are not of the type obtained from class-$\mathcal{S}$ theories with regular punctures, which would be star-shaped. The deformations of conformal matter theories to generalized quivers $\tilde{\mathsf{Q}}$ produces magnetic quivers which match exactly those read from the class-$\mathcal{S}$ construction (from puncture data). This further confirms that the identified toroidal reduction is the correct one.

\section{Outlook}
\label{sec:Conclusions}
In this work, we further characterised the 5d Conformal Matter SCFTs introduced in \cite{DeMarco:2023irn}. We specifically analysed three interconnected problems: the dimensional reduction of the aformentioned SCFTs, their $(p,q)$-web brane constructions and their magnetic quivers.

This work is situated at the intersection of a variety of tightly-related points of view on the geometric engineering landscape, and naturally paves the way for a plethora of possible developments:
\begin{enumerate}
    \item One may ask, whether the 5d CM atoms (and molecules built thereof) introduced in \cite{DeMarco:2023irn}, and studied in the present paper, are all the 5d SCFTs with at least $\mathfrak{g}\times\mathfrak{g}$ symmetry and a Dynkin quiver phase. In the upcoming paper \cite{AtomMoleculeHybridupcoming} we answer this question by completing the classification and matching with corresponding CY singularities. Most of the results of this paper extend to the more general class, but we find also some new exotic hybrid 5d CM systems (new atoms that can be Higgsed to molecules).
    \item The Higgs branch phase diagram, or Hasse diagram, of a 5d theory can be obtained from its brane web construction via magnetic quiver technology. On the geometric engineering side complex structure deformations achieve the same Higgsing. We address how to match such deformations with the the Hasse diagram for specific theories in a further upcoming work \cite{HiggsRGupcoming}. The Higgs branch RG-flows between atoms are summarised in Figure \ref{fig:RGflowAtoms}.
    \item While first steps in reading magnetic quivers from geometry were made in \cite{Closset:2020scj,Closset:2020afy,Closset:2021lwy}, in general, it is still not clear how to precisely rephrase some aspects of the magnetic quiver analysis in the language of geometric engineering.  Indeed, if on the one hand the geometric engineering allows in principle to study theories and Higgs branches that do not admit a (either electric or magnetic) quiver descriptions, it is less clear how to analyze fine aspects of the Higgs branches from a geometric perspective.
    \item Our study leads to the prediction of the existence of new ADE families of 4d $\mathcal N=2$ SCFTs with high rank and flavor symmetry at least $\mathfrak{g} \times \mathfrak{g}$. These models do not admit a Lagrangian nor a class-$\mathcal{S}$ realisation with regular punctures, which we predict from our magnetic quiver analysis (the magnetic quiver of these SCFTs is not of class-$\mathcal{S}$ type with regular punctures). We leave for the future the problem of characterizing these 4d SCFTs (a subset of which have been investigated by \cite{Ohmori:2015pia}), in particular to determine their CB dimensions, their conformal central charges and whether their flavor symmetries are gaugeable in 4d (which we expect to be the case).
\end{enumerate}
    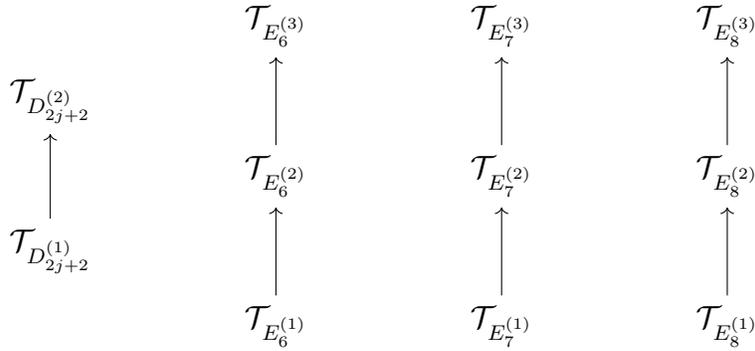
\begin{figure}
        \centering
        \begin{tikzpicture}
            \node (D1) at (0,-1) {$\mathcal{T}_{D_{2j+2}^{(1)}}$};
            \node (D2) at (0,1) {$\mathcal{T}_{D_{2j+2}^{(2)}}$};
            \draw[->] (D1)--(D2);
            \node (E61) at (3,-2) {$\mathcal{T}_{E_{6}^{(1)}}$};
            \node (E62) at (3,0) {$\mathcal{T}_{E_{6}^{(2)}}$};
            \node (E63) at (3,2) {$\mathcal{T}_{E_{6}^{(3)}}$};
            \draw[->] (E61)--(E62);
            \draw[->] (E62)--(E63);
            \node (E71) at (6,-2) {$\mathcal{T}_{E_{7}^{(1)}}$};
            \node (E72) at (6,0) {$\mathcal{T}_{E_{7}^{(2)}}$};
            \node (E73) at (6,2) {$\mathcal{T}_{E_{7}^{(3)}}$};
            \draw[->] (E71)--(E72);
            \draw[->] (E72)--(E73);
            \node (E81) at (9,-2) {$\mathcal{T}_{E_{8}^{(1)}}$};
            \node (E82) at (9,0) {$\mathcal{T}_{E_{8}^{(2)}}$};
            \node (E83) at (9,2) {$\mathcal{T}_{E_{8}^{(3)}}$};
            \draw[->] (E81)--(E82);
            \draw[->] (E82)--(E83);
        \end{tikzpicture}
        \caption{Higgsings between conformal matter atoms. (Many more phases exists in the Higgs branch of the conformal matter atom at the bottom of each flow, we omit them from this Figure.)}
        \label{fig:RGflowAtoms}
    \end{figure} 

\newpage
\section{Acknowledgement}
We would like to thank Sergio Benvenuti, Antoine Bourget, Amihay Hanany, and Sakura Sch\"afer-Nameki for discussions.

JFG is supported by the EPSRC Open Fellowship (Sch\"afer-Nameki) EP/X01276X/1. JFG and AS thank the ``Simons Collaboration on Special Holonomy in Geometry, Analysis and Physics'' for support during the initial stage of this project. The work of AS is funded by the VR Centre for Geometry and Physics (VR grant No. 2022-06593). The work of MDZ is supported by the European Research Council (ERC) under the European Union’s Horizon 2020 research and innovation program (grant agreement No. 851931), the Simons Foundation Grant \#888984 (Simons Collaboration on Global Categorical Symmetries) and by the VR project grant No. 2023-05590. MDZ also acknowledges the VR Centre for Geometry and Physics (VR grant No. 2022-06593). 

\appendix

\section{Quiver Phases for 5d Conformal Atoms and Molecules}
\label{app:QuiverPhases}

In this appendix we collect the low energy quiver descriptions of 5d conformal matter atoms (and molecules) which were provided in \cite{DeMarco:2023irn}. It is important to remember that 5d SCFTs may have many different low energy Lagrangians (called UV duals), and the quivers provided here are very specific deformations of the 5d SCFTs in question.
\begin{table}[h!]
    \centering
        \begin{tabular}{lcc}
            $\mathsf{Q}_{X_{A_{2j+1}}^{(1)}}$ & $=$ & $\raisebox{-.5\height}{\scalebox{0.6}{\begin{tikzpicture}
                \node[gaugeSU,label=below:{$1$}] (1) at (2,0) {};
                \node[gaugeSU,label=below:{$2$}] (2) at (4,0) {};
                \node (3) at (6,0) {$\cdots$};
                \node[gaugeSU,label=below:{$j$}] (4) at (8,0) {};
                \node[gaugeSU,label=below:{$j+1$}] (5) at (10,0) {};
                \node[gaugeSU,label=below:{$j$}] (6) at (12,0) {};
                \node (7) at (14,0) {$\cdots$};
                \node[gaugeSU,label=below:{$2$}] (8) at (16,0) {};
                \node[gaugeSU,label=below:{$1$}] (9) at (18,0) {};
                \node[flavourSU,label=above:{$2$}] (5f) at (10,1.5) {};
                \draw (1)--(2)--(3)--(4)--(5)--(6)--(7)--(8)--(9) (5)--(5f);
            \end{tikzpicture}}}$ \\
            $\mathsf{Q}_{X_{A_{2j}}^{(1)}}$ & $=$ & $\raisebox{-.5\height}{\scalebox{0.6}{\begin{tikzpicture}
                \node[gaugeSU,label=below:{$1$}] (1) at (2,0) {};
                \node[gaugeSU,label=below:{$2$}] (2) at (4,0) {};
                \node (3) at (6,0) {$\cdots$};
                \node[gaugeSU,label=below:{$j-1$}] (4) at (8,0) {};
                \node[gaugeSU,label=below:{$j$}] (5) at (10,0) {};
                \node[gaugeSU,label=below:{$j$}] (6) at (12,0) {};
                \node[gaugeSU,label=below:{$j-1$}] (7) at (14,0) {};
                \node (8) at (16,0) {$\cdots$};
                \node[gaugeSU,label=below:{$2$}] (9) at (18,0) {};
                \node[gaugeSU,label=below:{$1$}] (10) at (20,0) {};
                \node[flavourSU,label=above:{$1$}] (5f) at (10,1.5) {};
                \node[flavourSU,label=above:{$1$}] (6f) at (12,1.5) {};
                \draw (1)--(2)--(3)--(4)--(5)--(6)--(7)--(8)--(9)--(10) (5)--(5f) (6)--(6f);
            \end{tikzpicture}}}$ \\
            $\mathsf{Q}_{X_{D_{2j+2}}^{(1)}}$ & $=$ & $\raisebox{-.5\height}{\scalebox{0.6}{\begin{tikzpicture}
                \node at (-1,0) {};
                \node at (15,0) {};
                \node[flavourSU,label=above:{$1$}] (1uf) at (0,1) {};
                \node[gaugeSU,label=above:{$j+1$}] (1u) at (2,1) {};
                \node[flavourSU,label=below:{$1$}] (1df) at (0,-1) {};
                \node[gaugeSU,label=below:{$j+1$}] (1d) at (2,-1) {};
                \node[gaugeSU,label=below:{$2j+1$}] (2) at (4,0) {};
                \node[gaugeSU,label=below:{$2j$}] (3) at (6,0) {};
                \node (4) at (8,0) {$\cdots$};
                \node[gaugeSU,label=below:{$3$}] (5) at (10,0) {};
                \node[gaugeSU,label=below:{$2$}] (6) at (12,0) {};
                \node[flavourSU,label=below:{$1$}] (7) at (14,0) {};
                \draw (1uf)--(1u)--(2) (1df)--(1d)--(2)--(3)--(4)--(5)--(6)--(7);
            \end{tikzpicture}}}$ \\
            $\mathsf{Q}_{X_{D_{2j+3}}^{(1)}}$ & $=$ & $\raisebox{-.5\height}{\scalebox{0.6}{\begin{tikzpicture}
                \node at (-1,0) {};
                \node at (15,0) {};
                \node[gaugeSU,label=above:{$j+1$}] (1u) at (2,1) {};
                \node[gaugeSU,label=below:{$j+1$}] (1d) at (2,-1) {};
                \node[gaugeSU,label=below:{$2j+2$}] (2) at (4,0) {};
                \node[flavourSU,label=above:{$1$}] (2u) at (4,1.5) {};
                \node[gaugeSU,label=below:{$2j+1$}] (3) at (6,0) {};
                \node (4) at (8,0) {$\cdots$};
                \node[gaugeSU,label=below:{$3$}] (5) at (10,0) {};
                \node[gaugeSU,label=below:{$2$}] (6) at (12,0) {};
                \node[flavourSU,label=below:{$1$}] (7) at (14,0) {};
                \draw (1u)--(2)--(2u) (1d)--(2)--(3)--(4)--(5)--(6)--(7);
            \end{tikzpicture}}}$ \\
            $\mathsf{Q}_{X_{D_{2j+2}}^{(2)}}$ & $=$ & $\raisebox{-.5\height}{\scalebox{0.6}{\begin{tikzpicture}
                \node at (-1,0) {};
                \node at (15,0) {};
                \node[gaugeSU,label=above:{$j$}] (1u) at (2,1) {};
                \node[gaugeSU,label=below:{$j$}] (1d) at (2,-1) {};
                \node[gaugeSU,label=below:{$2j$}] (2) at (4,0) {};
                \node[flavourSU,label=above:{$1$}] (2u) at (4,1.5) {};
                \node[gaugeSU,label=below:{$2j-1$}] (3) at (6,0) {};
                \node (4) at (8,0) {$\cdots$};
                \node[gaugeSU,label=below:{$2$}] (5) at (10,0) {};
                \node[gaugeSU,label=below:{$1$}] (6) at (12,0) {};
                \draw (1u)--(2)--(2u) (1d)--(2)--(3)--(4)--(5)--(6);
            \end{tikzpicture}}}$ \\
            $\mathsf{Q}_{X_{D_{2j+3}}^{(2)}}$ & $=$ & $\raisebox{-.5\height}{\scalebox{0.6}{\begin{tikzpicture}
                \node at (-1,0) {};
                \node at (15,0) {};
                \node[flavourSU,label=above:{$1$}] (1uf) at (0,1) {};
                \node[gaugeSU,label=above:{$j+1$}] (1u) at (2,1) {};
                \node[flavourSU,label=below:{$1$}] (1df) at (0,-1) {};
                \node[gaugeSU,label=below:{$j+1$}] (1d) at (2,-1) {};
                \node[gaugeSU,label=below:{$2j+1$}] (2) at (4,0) {};
                \node[gaugeSU,label=below:{$2j$}] (3) at (6,0) {};
                \node (4) at (8,0) {$\cdots$};
                \node[gaugeSU,label=below:{$2$}] (5) at (10,0) {};
                \node[gaugeSU,label=below:{$1$}] (6) at (12,0) {};
                \draw (1uf)--(1u)--(2) (1df)--(1d)--(2)--(3)--(4)--(5)--(6);
            \end{tikzpicture}}}$ \\
            $\mathsf{Q}_{X_{D_{j}}^{(3)}}$ & $=$ & $\raisebox{-.5\height}{\scalebox{0.6}{\begin{tikzpicture}
                \node at (-1,0) {};
                \node at (15,0) {};
                \node[gaugeSU,label=above:{$1$}] (1u) at (2,1) {};
                \node[gaugeSU,label=below:{$1$}] (1d) at (2,-1) {};
                \node[gaugeSU,label=below:{$2$}] (2) at (4,0) {};
                \node[gaugeSU,label=below:{$2$}] (3) at (6,0) {};
                \node (4) at (8,0) {$\cdots$};
                \node[gaugeSU,label=below:{$2$}] (5) at (10,0) {};
                \node[gaugeSU,label=below:{$2$}] (6) at (12,0) {};
                \node[flavourSU,label=below:{$2$}] (7) at (14,0) {};
                \draw (1u)--(2) (1d)--(2)--(3)--(4)--(5)--(6)--(7);
            \end{tikzpicture}}}$
        \end{tabular}
    \caption{Specific low energy quiver deformations of A/D-type $5d$ conformal matter atoms \cite{DeMarco:2023irn}.}
    \label{tab:A/D-Atoms_EQs}
\end{table}

\begin{table}[h!]
    \centering
        \begin{tabular}{lcc}    
            $\mathsf{Q}_{X_{E_6^{(1)}}}$ & $=$ & $\raisebox{-.5\height}{\scalebox{0.6}{\begin{tikzpicture}
                \node at (-1,0) {};
                \node at (17,0) {};
                \node[gaugeSU,label=below:{$2$}] (1) at (2,0) {};
                \node[gaugeSU,label=below:{$4$}] (2) at (4,0) {};
                \node[gaugeSU,label=below:{$6$}] (3) at (6,0) {};
                \node[gaugeSU,label=below:{$4$}] (4) at (8,0) {};
                \node[gaugeSU,label=below:{$2$}] (5) at (10,0) {};
                \node[gaugeSU,label=right:{$3$}] (3u) at (6,2) {};
                \node[flavourSU,label=left:{$1$}] (3f) at (4,2) {};
                \draw (1)--(2)--(3)--(4)--(5) (3f)--(3)--(3u);
            \end{tikzpicture}}}$ \\
            $\mathsf{Q}_{X_{E_6^{(2)}}}$ & $=$ & $\raisebox{-.5\height}{\scalebox{0.6}{\begin{tikzpicture}
                \node at (-1,0) {};
                \node at (17,0) {};
                \node[gaugeSU,label=below:{$2$}] (1) at (2,0) {};
                \node[gaugeSU,label=below:{$3$}] (2) at (4,0) {};
                \node[gaugeSU,label=below:{$4$}] (3) at (6,0) {};
                \node[gaugeSU,label=below:{$3$}] (4) at (8,0) {};
                \node[gaugeSU,label=below:{$2$}] (5) at (10,0) {};
                \node[gaugeSU,label=right:{$2$}] (3u) at (6,2) {};
                \node[flavourSU,label=below:{$1$}] (1f) at (0,0) {};
                \node[flavourSU,label=below:{$1$}] (5f) at (12,0) {};
                \draw (1f)--(1)--(2)--(3)--(4)--(5)--(5f) (3)--(3u);
            \end{tikzpicture}}}$ \\
            $\mathsf{Q}_{X_{E_6^{(3)}}}$ & $=$ & $\raisebox{-.5\height}{\scalebox{0.6}{\begin{tikzpicture}
                \node at (-1,0) {};
                \node at (17,0) {};
                \node[gaugeSU,label=below:{$1$}] (1) at (2,0) {};
                \node[gaugeSU,label=below:{$2$}] (2) at (4,0) {};
                \node[gaugeSU,label=below:{$3$}] (3) at (6,0) {};
                \node[gaugeSU,label=below:{$2$}] (4) at (8,0) {};
                \node[gaugeSU,label=below:{$1$}] (5) at (10,0) {};
                \node[gaugeSU,label=right:{$2$}] (3u) at (6,2) {};
                \node[flavourSU,label=right:{$1$}] (3uf) at (6,4) {};
                \draw (1)--(2)--(3)--(4)--(5) (3)--(3u)--(3uf);
            \end{tikzpicture}}}$ \\
            $\mathsf{Q}_{X_{E_7^{(1)}}}$ & $=$ & $\raisebox{-.5\height}{\scalebox{0.6}{\begin{tikzpicture}
                \node at (-1,0) {};
                \node at (17,0) {};
                \node[gaugeSU,label=below:{$3$}] (1) at (2,0) {};
                \node[gaugeSU,label=below:{$6$}] (2) at (4,0) {};
                \node[gaugeSU,label=below:{$9$}] (3) at (6,0) {};
                \node[gaugeSU,label=below:{$7$}] (4) at (8,0) {};
                \node[gaugeSU,label=below:{$5$}] (5) at (10,0) {};
                \node[gaugeSU,label=below:{$3$}] (6) at (12,0) {};
                \node[gaugeSU,label=right:{$5$}] (3u) at (6,2) {};
                \node[flavourSU,label=right:{$1$}] (3uf) at (6,4) {};
                \node[flavourSU,label=below:{$1$}] (6f) at (14,0) {};
                \draw (1)--(2)--(3)--(4)--(5)--(6)--(6f) (3)--(3u)--(3uf);
            \end{tikzpicture}}}$ \\
            $\mathsf{Q}_{X_{E_7^{(2)}}}$ & $=$ & $\raisebox{-.5\height}{\scalebox{0.6}{\begin{tikzpicture}
                \node at (-1,0) {};
                \node at (17,0) {};
                \node[gaugeSU,label=below:{$2$}] (1) at (2,0) {};
                \node[gaugeSU,label=below:{$4$}] (2) at (4,0) {};
                \node[gaugeSU,label=below:{$6$}] (3) at (6,0) {};
                \node[gaugeSU,label=below:{$5$}] (4) at (8,0) {};
                \node[gaugeSU,label=below:{$4$}] (5) at (10,0) {};
                \node[gaugeSU,label=below:{$3$}] (6) at (12,0) {};
                \node[gaugeSU,label=right:{$3$}] (3u) at (6,2) {};
                \node[flavourSU,label=below:{$2$}] (6f) at (14,0) {};
                \draw (1)--(2)--(3)--(4)--(5)--(6)--(6f) (3)--(3u);
            \end{tikzpicture}}}$ \\
            $\mathsf{Q}_{X_{E_7^{(3)}}}$ & $=$ & $\raisebox{-.5\height}{\scalebox{0.6}{\begin{tikzpicture}
                \node at (-1,0) {};
                \node at (17,0) {};
                \node[gaugeSU,label=below:{$2$}] (1) at (2,0) {};
                \node[gaugeSU,label=below:{$3$}] (2) at (4,0) {};
                \node[gaugeSU,label=below:{$4$}] (3) at (6,0) {};
                \node[gaugeSU,label=below:{$3$}] (4) at (8,0) {};
                \node[gaugeSU,label=below:{$2$}] (5) at (10,0) {};
                \node[gaugeSU,label=below:{$1$}] (6) at (12,0) {};
                \node[gaugeSU,label=right:{$2$}] (3u) at (6,2) {};
                \node[flavourSU,label=below:{$1$}] (1f) at (0,0) {};
                \draw (1f)--(1)--(2)--(3)--(4)--(5)--(6) (3)--(3u);
            \end{tikzpicture}}}$ \\
            $\mathsf{Q}_{X_{E_8^{(1)}}}$ & $=$ & $\raisebox{-.5\height}{\scalebox{0.6}{\begin{tikzpicture}
                \node at (-1,0) {};
                \node at (17,0) {};
                \node[gaugeSU,label=below:{$5$}] (1) at (2,0) {};
                \node[gaugeSU,label=below:{$10$}] (2) at (4,0) {};
                \node[gaugeSU,label=below:{$15$}] (3) at (6,0) {};
                \node[gaugeSU,label=below:{$12$}] (4) at (8,0) {};
                \node[gaugeSU,label=below:{$9$}] (5) at (10,0) {};
                \node[gaugeSU,label=below:{$6$}] (6) at (12,0) {};
                \node[gaugeSU,label=below:{$3$}] (7) at (14,0) {};
                \node[gaugeSU,label=right:{$8$}] (3u) at (6,2) {};
                \node[flavourSU,label=right:{$1$}] (3uf) at (6,4) {};
                \draw (1)--(2)--(3)--(4)--(5)--(6)--(7) (3)--(3u)--(3uf);
            \end{tikzpicture}}}$ \\
            $\mathsf{Q}_{X_{E_8^{(2)}}}$ & $=$ & $\raisebox{-.5\height}{\scalebox{0.6}{\begin{tikzpicture}
                \node at (-1,0) {};
                \node at (17,0) {};
                \node[gaugeSU,label=below:{$4$}] (1) at (2,0) {};
                \node[gaugeSU,label=below:{$7$}] (2) at (4,0) {};
                \node[gaugeSU,label=below:{$10$}] (3) at (6,0) {};
                \node[gaugeSU,label=below:{$8$}] (4) at (8,0) {};
                \node[gaugeSU,label=below:{$6$}] (5) at (10,0) {};
                \node[gaugeSU,label=below:{$4$}] (6) at (12,0) {};
                \node[gaugeSU,label=below:{$2$}] (7) at (14,0) {};
                \node[gaugeSU,label=right:{$5$}] (3u) at (6,2) {};
                \node[flavourSU,label=below:{$1$}] (1f) at (0,0) {};
                \draw (1f)--(1)--(2)--(3)--(4)--(5)--(6)--(7) (3)--(3u);
            \end{tikzpicture}}}$ \\
            $\mathsf{Q}_{X_{E_8^{(3)}}}$ & $=$ & $\raisebox{-.5\height}{\scalebox{0.6}{\begin{tikzpicture}
                \node at (-1,0) {};
                \node at (17,0) {};
                \node[gaugeSU,label=below:{$2$}] (1) at (2,0) {};
                \node[gaugeSU,label=below:{$4$}] (2) at (4,0) {};
                \node[gaugeSU,label=below:{$6$}] (3) at (6,0) {};
                \node[gaugeSU,label=below:{$5$}] (4) at (8,0) {};
                \node[gaugeSU,label=below:{$4$}] (5) at (10,0) {};
                \node[gaugeSU,label=below:{$3$}] (6) at (12,0) {};
                \node[gaugeSU,label=below:{$2$}] (7) at (14,0) {};
                \node[gaugeSU,label=right:{$3$}] (3u) at (6,2) {};
                \node[flavourSU,label=below:{$1$}] (7f) at (16,0) {};
                \draw (1)--(2)--(3)--(4)--(5)--(6)--(7)--(7f) (3)--(3u);
            \end{tikzpicture}}}$
        \end{tabular}
    \caption{Specific low energy quiver deformations of E-type $5d$ conformal matter atoms \cite{DeMarco:2023irn}.}
    \label{tab:E-Atoms_EQs}
\end{table}

\begin{table}[h!]
    \centering
        \begin{tabular}{lcc}
            $\mathsf{Q}_{X_{A_{2j+1}}^{(1^{n_1})}}$ & $=$ & $\raisebox{-.5\height}{\scalebox{0.6}{\begin{tikzpicture}
                \node[gaugeSU,label=below:{$n_1$}] (1) at (2,0) {};
                \node[gaugeSU,label=below:{$2n_1$}] (2) at (4,0) {};
                \node (3) at (6,0) {$\cdots$};
                \node[gaugeSU,label=below:{$jn_1$}] (4) at (8,0) {};
                \node[gaugeSU,label=below:{$(j+1)n_1$}] (5) at (10,0) {};
                \node[gaugeSU,label=below:{$jn_1$}] (6) at (12,0) {};
                \node (7) at (14,0) {$\cdots$};
                \node[gaugeSU,label=below:{$2n_1$}] (8) at (16,0) {};
                \node[gaugeSU,label=below:{$n_1$}] (9) at (18,0) {};
                \node[flavourSU,label=above:{$2n_1$}] (5f) at (10,1.5) {};
                \draw (1)--(2)--(3)--(4)--(5)--(6)--(7)--(8)--(9) (5)--(5f);
            \end{tikzpicture}}}$ \\
            $\mathsf{Q}_{X_{A_{2j}}^{(1^{n_1})}}$ & $=$ & $\raisebox{-.5\height}{\scalebox{0.6}{\begin{tikzpicture}
                \node[gaugeSU,label=below:{$n_1$}] (1) at (2,0) {};
                \node[gaugeSU,label=below:{$2n_1$}] (2) at (4,0) {};
                \node (3) at (6,0) {$\cdots$};
                \node[gaugeSU,label=below:{$(j-1)n_1$}] (4) at (8,0) {};
                \node[gaugeSU,label=below:{$jn_1$}] (5) at (10,0) {};
                \node[gaugeSU,label=below:{$jn_1$}] (6) at (12,0) {};
                \node[gaugeSU,label=below:{$(j-1)n_1$}] (7) at (14,0) {};
                \node (8) at (16,0) {$\cdots$};
                \node[gaugeSU,label=below:{$2n_1$}] (9) at (18,0) {};
                \node[gaugeSU,label=below:{$n_1$}] (10) at (20,0) {};
                \node[flavourSU,label=above:{$n_1$}] (5f) at (10,1.5) {};
                \node[flavourSU,label=above:{$n_1$}] (6f) at (12,1.5) {};
                \draw (1)--(2)--(3)--(4)--(5)--(6)--(7)--(8)--(9)--(10) (5)--(5f) (6)--(6f);
            \end{tikzpicture}}}$ \\
            $\mathsf{Q}_{X_{D_{2j+2}}^{(1^{n_1},2^{n_2},3^{n_3})}}$ & $=$ & $\raisebox{-.5\height}{\scalebox{0.6}{\begin{tikzpicture}
                \node[flavourSU,label=above:{$n_1$}] (1uf) at (0,1) {};
                \node[gaugeSU,label=above:{\rotatebox{90}{$(j+1)n_1+jn_2+n_3$}}] (1u) at (2,1) {};
                \node[flavourSU,label=below:{$n_1$}] (1df) at (0,-1) {};
                \node[gaugeSU,label=below:{\rotatebox{90}{$(j+1)n_1+jn_2+n_3$}}] (1d) at (2,-1) {};
                \node[gaugeSU,label=below:{\rotatebox{90}{$(2j+1)n_1+2jn_2+2n_3$}}] (2) at (4,0) {};
                \node[flavourSU,label=above:{$n_2$}] (2u) at (4,1.5) {};
                \node[gaugeSU,label=below:{\rotatebox{90}{$2jn_1+(2j-1)n_2+2n_3$}}] (3) at (6,0) {};
                \node (4) at (8,0) {$\cdots$};
                \node[gaugeSU,label=below:{\rotatebox{90}{$3n_1+2n_2+2n_3$}}] (5) at (10,0) {};
                \node[gaugeSU,label=below:{\rotatebox{90}{$2n_1+1n_2+2n_3$}}] (6) at (12,0) {};
                \node[flavourSU,label=below:{$n_1+2n_3$}] (7) at (14,0) {};
                \draw (1uf)--(1u)--(2)--(2u) (1df)--(1d)--(2)--(3)--(4)--(5)--(6)--(7);
            \end{tikzpicture}}}$ \\ \\
            $\mathsf{Q}_{X_{D_{2j+3}}^{(1^{n_1},2^{n_2},3^{n_3})}}$ & $=$ & $\raisebox{-.5\height}{\scalebox{0.6}{\begin{tikzpicture}
                \node[flavourSU,label=above:{$n_2$}] (1uf) at (0,1) {};
                \node[gaugeSU,label=above:{\rotatebox{90}{$(j+1)n_1+(j+1)n_2+n_3$}}] (1u) at (2,1) {};
                \node[flavourSU,label=below:{$n_2$}] (1df) at (0,-1) {};
                \node[gaugeSU,label=below:{\rotatebox{90}{$(j+1)n_1+(j+1)n_2+n_3$}}] (1d) at (2,-1) {};
                \node[gaugeSU,label=below:{\rotatebox{90}{$(2j+2)n_1+(2j+1)n_2+2n_3$}}] (2) at (4,0) {};
                \node[flavourSU,label=above:{$n_1$}] (2u) at (4,1.5) {};
                \node[gaugeSU,label=below:{\rotatebox{90}{$(2j+1)n_1+2jn_2+2n_3$}}] (3) at (6,0) {};
                \node (4) at (8,0) {$\cdots$};
                \node[gaugeSU,label=below:{\rotatebox{90}{$3n_1+2n_2+2n_3$}}] (5) at (10,0) {};
                \node[gaugeSU,label=below:{\rotatebox{90}{$2n_1+1n_2+2n_3$}}] (6) at (12,0) {};
                \node[flavourSU,label=below:{$n_1+2n_3$}] (7) at (14,0) {};
                \draw (1uf)--(1u)--(2)--(2u) (1df)--(1d)--(2)--(3)--(4)--(5)--(6)--(7);
            \end{tikzpicture}}}$
        \end{tabular}
    \caption{Specific low energy quiver deformations of A/D-type $5d$ conformal matter molecules \cite{DeMarco:2023irn}.}
    \label{tab:A/D-Molecules_EQs}
\end{table}

\begin{table}[h!]
    \centering
        \begin{tabular}{lcc}   
            $\mathsf{Q}_{X_{E_6^{(1^{n_1},2^{n_2},3^{n_3})}}}$ & $=$ & $\raisebox{-.5\height}{\scalebox{0.6}{\begin{tikzpicture}
                \node at (-1,0) {};
                \node at (17,0) {};
                \node[gaugeSU,label=below:{\rotatebox{90}{$2n_1+2n_2+n_3$}}] (1) at (2,0) {};
                \node[gaugeSU,label=below:{\rotatebox{90}{$4n_1+3n_2+2n_3$}}] (2) at (4,0) {};
                \node[gaugeSU,label=below:{\rotatebox{90}{$6n_1+4n_2+3n_3$}}] (3) at (6,0) {};
                \node[gaugeSU,label=below:{\rotatebox{90}{$4n_1+3n_2+2n_3$}}] (4) at (8,0) {};
                \node[gaugeSU,label=below:{\rotatebox{90}{$2n_1+2n_2+n_3$}}] (5) at (10,0) {};
                \node[gaugeSU,label=right:{$3n_1+2n_2+2n_3$}] (3u) at (6,2) {};
                \node[flavourSU,label=left:{$1n_1$}] (3f) at (4,2) {};
                \node[flavourSU,label=below:{$1n_2$}] (1f) at (0,0) {};
                \node[flavourSU,label=below:{$1n_2$}] (5f) at (12,0) {};
                \node[flavourSU,label=right:{$1n_3$}] (3uf) at (6,4) {};
                \draw (1f)--(1)--(2)--(3)--(4)--(5)--(5f) (3f)--(3)--(3u)--(3uf);
            \end{tikzpicture}}}$ \\ \\
            $\mathsf{Q}_{X_{E_7^{(1^{n_1},2^{n_2},3^{n_3})}}}$ & $=$ & $\raisebox{-.5\height}{\scalebox{0.6}{\begin{tikzpicture}
                \node at (-1,0) {};
                \node at (17,0) {};
                \node[gaugeSU,label=below:{\rotatebox{90}{$3n_1+2n_2+2n_3$}}] (1) at (2,0) {};
                \node[gaugeSU,label=below:{\rotatebox{90}{$6n_1+4n_2+3n_3$}}] (2) at (4,0) {};
                \node[gaugeSU,label=below:{\rotatebox{90}{$9n_1+6n_2+4n_3$}}] (3) at (6,0) {};
                \node[gaugeSU,label=below:{\rotatebox{90}{$7n_1+5n_2+3n_3$}}] (4) at (8,0) {};
                \node[gaugeSU,label=below:{\rotatebox{90}{$5n_1+4n_2+2n_3$}}] (5) at (10,0) {};
                \node[gaugeSU,label=below:{\rotatebox{90}{$3n_1+3n_2+n_3$}}] (6) at (12,0) {};
                \node[gaugeSU,label=right:{$5n_1+2n_3$}] (3u) at (6,2) {};
                \node[flavourSU,label=right:{$n_1$}] (3uf) at (6,4) {};
                \node[flavourSU,label=below:{$n_1+2n_2$}] (6f) at (14,0) {};
                \node[flavourSU,label=below:{$n_3$}] (1f) at (0,0) {};
                \draw (1f)--(1)--(2)--(3)--(4)--(5)--(6)--(6f) (3)--(3u)--(3uf);
            \end{tikzpicture}}}$ \\ \\
            $\mathsf{Q}_{X_{E_8^{(1^{n_1},2^{n_2},3^{n_3})}}}$ & $=$ & $\raisebox{-.5\height}{\scalebox{0.6}{\begin{tikzpicture}
                \node at (-1,0) {};
                \node at (17,0) {};
                \node[gaugeSU,label=below:{\rotatebox{90}{$5n_1+4n_2+2n_3$}}] (1) at (2,0) {};
                \node[gaugeSU,label=below:{\rotatebox{90}{$10n_1+7n_2+4n_3$}}] (2) at (4,0) {};
                \node[gaugeSU,label=below:{\rotatebox{90}{$15n_1+10n_2+6n_3$}}] (3) at (6,0) {};
                \node[gaugeSU,label=below:{\rotatebox{90}{$12n_1+8n_2+5n_3$}}] (4) at (8,0) {};
                \node[gaugeSU,label=below:{\rotatebox{90}{$9n_1+6n_2+4n_3$}}] (5) at (10,0) {};
                \node[gaugeSU,label=below:{\rotatebox{90}{$6n_1+4n_2+3n_3$}}] (6) at (12,0) {};
                \node[gaugeSU,label=below:{\rotatebox{90}{$3n_1+2n_2+2n_3$}}] (7) at (14,0) {};
                \node[gaugeSU,label=right:{$8n_1+5n_2+3n_3$}] (3u) at (6,2) {};
                \node[flavourSU,label=right:{$n_1$}] (3uf) at (6,4) {};
                \node[flavourSU,label=below:{$n_2$}] (1f) at (0,0) {};
                \node[flavourSU,label=below:{$n_3$}] (7f) at (16,0) {};
                \draw (1f)--(1)--(2)--(3)--(4)--(5)--(6)--(7)--(7f) (3)--(3u)--(3uf);
            \end{tikzpicture}}}$
        \end{tabular}
    \caption{Specific low energy quiver deformations of E-type $5d$ conformal matter molecules \cite{DeMarco:2023irn}.}
    \label{tab:E-Molecules_EQs}
\end{table}

\clearpage
\pagebreak
\section{Partial resolution of trinion theories}\label{appendix A}

\indent In this appendix we test the following statement:\\

\textit{Consider a 5d conformal matter singular threefold of the form \eqref{systemfourfoldgen}, related to the algebra $\mathfrak{g}$, and perform the partial resolution of \cref{sec:basechangeres}: then, the M2-brane states wrapped on the singular $\mathbb{P}^1$'s trapped at the origin give rise to $\text{dim}(\mathfrak{g})-\text{rank}(\mathfrak{g})$ HB modes for the corresponding 5d SCFT.}\\

We do so by examining a 5d SCFT that has been vastly studied in the literature, namely the $\mathbb C^3/(\mathbb Z_{m} \times \mathbb Z_{m})$ ``trinion'' singularity. This is completely equivalent to considering 5d conformal matter theories. Indeed, we can employ the same resolution technique that we used to resolve the 5d conformal matter threefolds also to obtain a partial resolution of $\mathbb C^3/(\mathbb Z_{m} \times \mathbb Z_{m})$. M-theory on $\mathbb C^3/(\mathbb Z_{m} \times \mathbb Z_{m})$ is known to engineer a 5d trinions of type $\mathfrak{su}(m)$: a 5d SCFT with a flavor group at least of type $\mathfrak{su}(m) \times \mathfrak{su}(m) \times \mathfrak{su}(m)$. This can be geometrically guessed by the fact that the threefold displays three non-compact lines of $A_{m-1}$ Du Val singularities. Performing a triangulation of the type called ``allowed'' in \cite{Closset_2019} and then taking the IIA limit \cite{Closset_2019} we get a gauge theory phase described by the ascending quiver in Figure \ref{fig:trinionquiver}:
       \begin{figure}[H]
    \centering
     $   \begin{array}{cc}
\makecell{\mathsf{Q}_{\mathbb C^3/(\mathbb Z_{m} \times \mathbb Z_{m})} =\\ \vspace{0.5cm}}  &
    \scalebox{1}{
    \begin{tikzpicture}
        \node at (0,0) {$\ldots$};
        \draw[thick] (-0.7,0)--(-1.3,0);
        \draw[thick] (0.7,0)--(1.3,0);
        \draw[thick] (-2.7,0)--(-3.4,0);
        \draw[thick] (2,0) circle (0.65);
        \node at (2,0) {\small$1$};
        \draw[thick] (-2,0) circle (0.65);
        \node at (-2,0) {\small$m-1$};
        \draw[thick] (-2,0) circle (0.65);
        \node at (-2,0) {\small$m-1$};
        \node at (-4.1,0) {\small$m$};
        \draw[thick] (-3.45,-0.65)--(-4.75,-0.65)--(-4.75,0.65)--(-3.45,0.65)--cycle;
        \end{tikzpicture}}
           \end{array}$
            \vspace{-0.5cm}
            
    \caption{Quiver description of the low-energy gauge theory limit of M-theory on $\mathbb C^3/(\mathbb Z_m \times \mathbb Z_m)$. A round (resp. squared) node labeled with $j$ corresponds to a  $\mathfrak{su}(j)$ gauge (resp. flavor) node.}
    \label{fig:trinionquiver}
    \end{figure}
We can again count the number of Higgs branch modes from the quiver, after tuning to zero the mass parameters of the $\mathfrak{su}(m)$ leftmost node of Figure \ref{fig:trinionquiver}. Tuning to zero these mass parameters corresponds to going to the partially resolved phase depicted in Figure \ref{fig:partialresgaugecouplingtrinion}, and the Higgs branch modes of the quiver correspond to complex deformations of the threefold that are not obstructed by the partial resolution.

\begin{figure}[H]
\centering
    \scalebox{1.}{
    \begin{tikzpicture}
        \draw (0,4)--(1,3)--(2,2)--(3,1)--(4,0)--(0,0)--(0,4);
        \draw[dashed] (1,0)--(1,3);
        \draw[dashed] (2,0)--(2,2);
        \draw[dashed]  (3,0)--(3,1);
        \filldraw[blue] (0,0) circle (2pt);
        \filldraw[red] (0,1) circle (2pt);
        \filldraw[red] (0,2) circle (2pt);
        \filldraw[red] (0,3) circle (2pt);
        \filldraw[blue] (0,4) circle (2pt);
        \filldraw (1,0) circle (2pt);
        \filldraw[red] (1,1) circle (2pt);
        \filldraw[red] (1,2) circle (2pt);
        \filldraw (1,3) circle (2pt);
        \filldraw (2,0) circle (2pt);
        \filldraw[red] (2,1) circle (2pt);
        \filldraw (2,2) circle (2pt);
        \filldraw (3,0) circle (2pt);
        \filldraw (3,1) circle (2pt);
        \filldraw[blue] (4,0) circle (2pt);
    \end{tikzpicture}}
    \caption{Toric diagram of the partial resolution of the $\mathbb C^3/(\mathbb Z_4 \times \mathbb Z_4)$ trinion that corresponds the quiver in Figure \ref{fig:trinionquiver} with all the zero-modes tuned to zero (keeping the gauge couplings finite). The red dots correspond to shrunk divisors, the black and the blue ones to inflated divisors. The black (resp. blue) divisors are non-compact and correspond to half-cigar bundles over compact (resp. non-compact) rational curves.}
    \label{fig:partialresgaugecouplingtrinion}
    \end{figure}
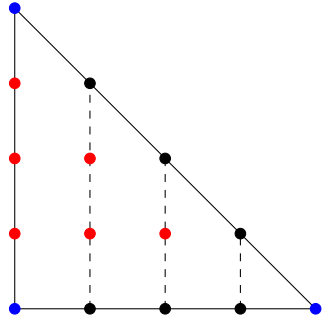
\begin{figure}[H]
\centering
    \scalebox{1.}{
    \begin{tikzpicture}
        \draw (0,4)--(1,3)--(2,2)--(3,1)--(4,0)--(0,0)--(0,4);
        \draw[red] (1,0)--(1,1);
        \draw (1,1)--(1,3);
        \draw (0,3)--(3,0);
        \draw (0,2)--(2,0);
        \draw (0,1)--(1,0);
        \draw[red]  (2,0)--(2,1);
        \draw  (2,1)--(2,2);
        \draw[red]  (3,0)--(3,1);
        \draw  (0,3)--(1,3);
        \draw  (0,2)--(2,2);
        \draw  (0,1)--(3,1);
        \filldraw (0,0) circle (2pt);
        \filldraw(0,1) circle (2pt);
        \filldraw (0,2) circle (2pt);
        \filldraw (0,3) circle (2pt);
        \filldraw (0,4) circle (2pt);
        \filldraw (1,0) circle (2pt);
        \filldraw (1,1) circle (2pt);
        \filldraw(1,2) circle (2pt);
        \filldraw (1,3) circle (2pt);
        \filldraw (2,0) circle (2pt);
        \filldraw (2,1) circle (2pt);
        \filldraw (2,2) circle (2pt);
        \filldraw (3,0) circle (2pt);
        \filldraw (3,1) circle (2pt);
        \filldraw (4,0) circle (2pt);
    \end{tikzpicture}}
    \caption{Toric diagram of the full resolution of the $\mathbb C^3/(\mathbb Z_4 \times \mathbb Z_4)$ trinion. The red edges correspond to the zero-section of the various Hirzebruch surfaces resolving the singularity and are naturally identified with the intersection with one of the three non-compact lines of $A_3$ singularities colliding at the origin of $\mathbb C^3/(\mathbb Z_4 \times \mathbb Z_4)$.}
    \label{fig:fullresgaugecouplingtrinion}
    \end{figure}
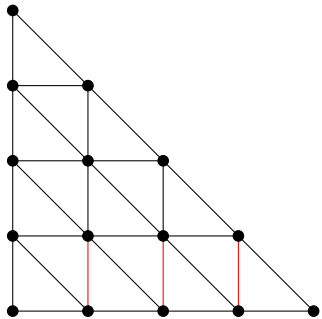

From the quiver, we obtain the following data 
\begin{equation}
    n_{V} = \frac{1}{6} (m-2) (m-1) (2 m+3), \quad n_{H} =\frac{1}{3} m \left(m^2-1\right),
\end{equation}
and hence 
\begin{equation}
\label{eq:unobstrinion5d}
    \text{dim}_{\mathbb{H}}\text{HB}\left(\mathsf{Q}_{\mathbb C^3/(\mathbb Z_{m} \times \mathbb Z_{m})}\right) = n_{V}-n_H = \frac{1}{2} \left(m^2+m-2\right). 
\end{equation}
$\text{dim}_{\mathbb{H}}\text{HB}\left(\mathsf{Q}_{\mathbb C^3/(\mathbb Z_{m} \times \mathbb Z_{m})}\right) $ corresponds to the number of complex deformations of the threefold that can still be performed when, from a generic triangulation respecting the structure of $\mathbb C^*$-fibration over the resolved $A_{m-1}$ singularity (called ``allowed'' in \cite{Closset_2019}), we contract, as in Figure \ref{fig:partialresgaugecouplingtrinion}, the volumes of the fibral curves inside the ruled surfaces inflated by the triangulation. We depicted this procedure graphically in Figure \ref{fig:pqwebtrinion}, using the dual brane web, with the contracted curves being the oblique edges of the uppermost part of Figure \ref{fig:pqwebtrinion}.\\
\indent We can now relate \eqref{eq:unobstrinion5d} with  the dimension of the 4d Higgs branch. In the case of M-theory on $\mathbb C^3/(\mathbb Z_m \times \mathbb Z_m)$, the corresponding 4d class-$\mathcal{S}$ theory is obtained from the 6d $(2,0)$ $\mathfrak{su}(m)$ theory, with three full punctures. One can use the methods of \cref{sec:classS} to find the number of 4d Higgs branch modes:
\begin{equation}
\text{dim}_{\mathbb{H}}\text{HB}\left(\mathcal{T}_{4d}\right) =  \frac{1}{2} (m-1) (3 m+2).
\end{equation}
We then can eventually perform our check: we expect that, at the collision point, the M2-brane states supported on the $\mathbb P^1$'s of the non-compact lines get ``transformed'' into 5d Higgs branch modes. This number $n_{\text{missing}}$ should then coincide with the mismatch between $\text{dim}_{\mathbb{H}}\text{HB}(\mathcal{T}_{4d})$ and $\text{dim}_{\mathbb{H}}\text{HB}\left(\mathsf{Q}_{\mathbb C^3/(\mathbb Z_{m} \times \mathbb Z_{m})}\right) $, as the aforementioned $\mathbb P^1$'s are inflated in the partial resolution of Figure \ref{fig:partialresgaugecouplingtrinion}, and the corresponding M2-brane states are massive.  In fact, we can check that
\begin{equation}
    \label{eq:checkequation}
    n_{\text{missing}} = \text{dim}_{\mathbb{H}}\text{HB}\left(\mathcal{T}_{4d}\right) -\text{dim}_{\mathbb{H}}\text{HB}\left(\mathsf{Q}_{\mathbb C^3/(\mathbb Z_{m} \times \mathbb Z_{m})}\right) =m^2-m=\textnormal{dim}(\mathfrak{su}(m))-\textnormal{rank}(\mathfrak{su}(m)),
\end{equation}
as expected by our general argument. Indeed, one can complete the resolution in Figure \ref{fig:partialresgaugecouplingtrinion} to the full resolution depicted in Figure \ref{fig:fullresgaugecouplingtrinion}. In this full resolution, the vertical lines of Figure \ref{fig:partialresgaugecouplingtrinion} have normal bundle $\mathcal O(-1) \times \mathcal O(-1)$. This can be shown by noticing that, if we contract them, we find a subdiagram of the toric diagram that is equivalent, up to $Sl(2,\mathbb Z)$ moves, to the conifold one.  The red vertical lines are the intersection of the non-compact $\mathbb P^1$-bundles resolving the non-compact line of $A_{m-1}$ singularities corresponding to the horizontal external edge of the toric diagram in Figure \ref{fig:fullresgaugecouplingtrinion} with the Hirzebruch surfaces associated to the gauge nodes of Figure \ref{fig:trinionquiver}. We then conclude that the M2 states wrapping the $\mathbb P^1$'s of the aforementioned non-compact line get trapped at the intersection point (as they have now a normal bundle without sections), and get transformed to 5d HB modes.\\

\indent We can finally exhibit an alternative perspective considering the $(p,q)$-web picture dual to the toric diagram, shown in Figure \ref{fig: partial resolution}. From the image at the top, corresponding to the complete resolution of the singularity, one can readily see that the M2-brane modes trapped at the origin wrap the base $\mathbb{P}^1$'s of the red and blue divisors, as well as the green curve\footnote{Of course, a rotational $\mathbb{Z}_3$ symmetry is allowed in the choice of the base $\mathbb{P}^1$'s, in agreement with the symmetry of the brane web.}. In the picture, these are nothing but horizontal segments, that have normal bundle $\mathcal{O}(-1)\oplus \mathcal{O}(-1)$, and thus are rigid. Shrinking the fibers produces the image at the center of Figure \ref{fig: partial resolution}, where the base $\mathbb{P}^1$'s are evident. The HB modes wrap precisely these singular $\mathbb{P}^1$'s. Notice that this partial resolution has explicitly broken the $\mathbb{Z}_3$ rotation symmetry of the fully resolved phase: in fact, the set of non-compact divisors labelled by I, II and III has been shrunk to a non-compact line supporting a $A_3$ singularity.

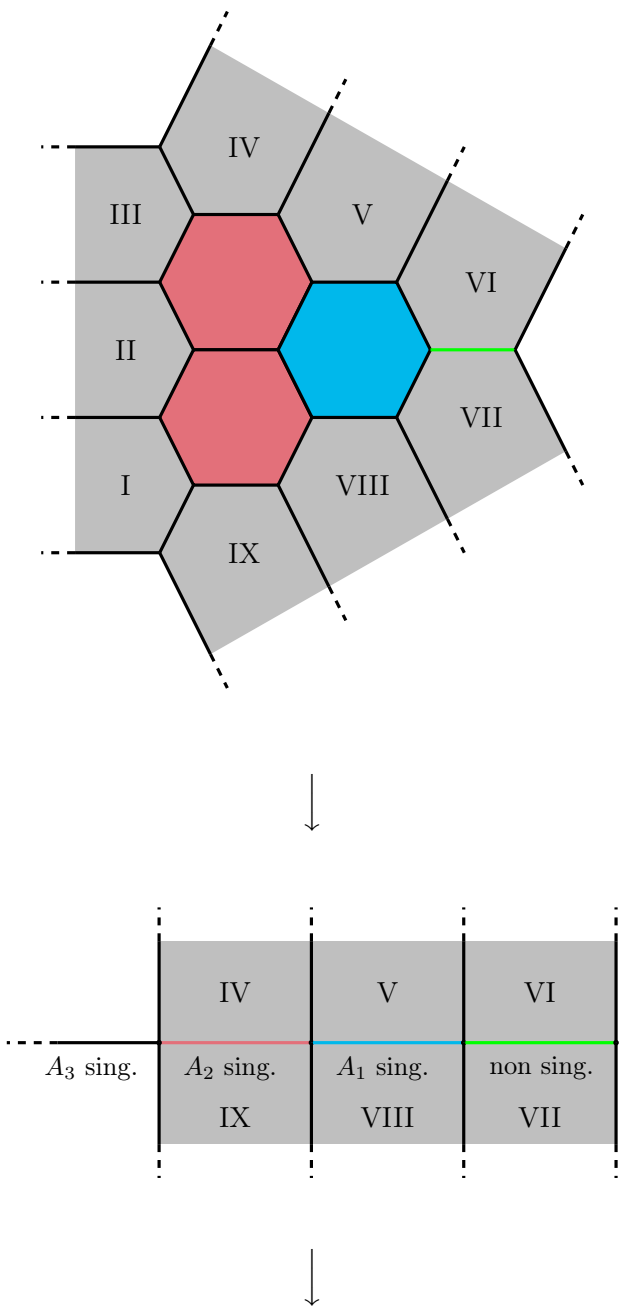
\begin{center}
\begin{tabular}{c}
\begin{tikzpicture}[scale=0.45]
    \draw[fill=lightgray,draw=none] (3,-2)--(5.5,-2)--(7,1)--(3.5,3)--(2,0)--(3,-2);
    \node at (4.5,0) {\small VI};
    \draw[fill=lightgray,draw=none] (-0.5,0)--(2,0)--(3.5,3)--(0,5)--(-1.5,2)--(-0.5,0);
    \node at (1,2) {\small V};
    \draw[fill=lightgray,draw=none] (-4,2)--(-1.5,2)--(0,5)--(-3.5,7)--(-5,4)--(-4,2);
    \node at (-2.5,4) {\small IV};
    \draw[fill=lightgray,draw=none] (3,-2)--(5.5,-2)--(7,-5)--(3.5,-7)--(2,-4)--(3,-2);
    \node at (4.5,-4) {\small VII};
    \draw[fill=lightgray,draw=none] (-0.5,-4)--(2,-4)--(3.5,-7)--(0,-9)--(-1.5,-6)--(-0.5,-4);
    \node at (1,-6) {\small VIII};
    \draw[fill=lightgray,draw=none] (-4,-6)--(-1.5,-6)--(0,-9)--(-3.5,-11)--(-5,-8)--(-4,-6);
    \node at (-2.5,-8) {\small IX};
    \draw[fill=lightgray,draw=none] (-5,-8)--(-7.5,-8)--(-7.5,-4)--(-5,-4)--(-4,-6)--(-5,-8);
    \node at (-6,-6) {\small I};
    \draw[fill=lightgray,draw=none] (-5,-4)--(-7.5,-4)--(-7.5,0)--(-5,0)--(-4,-2)--(-5,-4);
    \node at (-6,-2) {\small II};
    \draw[fill=lightgray,draw=none] (-5,0)--(-7.5,0)--(-7.5,4)--(-5,4)--(-4,2)--(-5,0);
    \node at (-6,2) {\small III};
    \draw[very thick,black,fill=candypink] (-1.5,2)--(-4,2)--(-5,0)--(-4,-2)--(-1.5,-2)--(-0.5,0)--(-1.5,2);
    \draw[very thick,black,fill=candypink] (-1.5,-2)--(-4,-2)--(-5,-4)--(-4,-6)--(-1.5,-6)--(-0.5,-4)--(-1.5,-2);
    \draw[very thick,black,fill=cyan(process)] (-0.5,0)--(2,0)--(3,-2)--(2,-4)--(-0.5,-4)--(-1.5,-2)--(-0.5,0);
    \draw[very thick,green,fill=green] (3,-2)--(5.5,-2);
    \draw[very thick,black,fill=black] (-4,-6)--(-5,-8);
    \draw[very thick,black,fill=black] (-4,2)--(-5,4);
    \draw[very thick,black,fill=black] (2,0)--(3,2)--(3.5,3);
    \draw[very thick, dashed,black,fill=black] (3.5,3)--(4,4);
    \draw[very thick,black,fill=black] (5.5,-2)--(6.5,0)--(7,1);
    \draw[very thick, dashed,black,fill=black] (7,1)--(7.5,2);
    \draw[very thick,black,fill=black] (-1.5,2)--(-0.5,4)--(0,5);
    \draw[very thick, dashed,black,fill=black] (0,5)--(0.5,6);
    \draw[very thick,black,fill=black] (-5,4)--(-4,6)--(-3.5,7);
    \draw[very thick, dashed,black,fill=black] (-3.5,7)--(-3,8);
    \draw[very thick,black,fill=black] (-5,-8)--(-4,-10)--(-3.5,-11);
    \draw[very thick, dashed,black,fill=black] (-3.5,-11)--(-3,-12);
    \draw[very thick,black,fill=black] (-1.5,-6)--(-0.5,-8)--(0,-9);
    \draw[very thick, dashed,black,fill=black] (0,-9)--(0.5,-10);
    \draw[very thick,black,fill=black] (5.5,-2)--(6.5,-4)--(7,-5);
    \draw[very thick, dashed,black,fill=black] (7,-5)--(7.5,-6);
    \draw[very thick,black,fill=black] (2,-4)--(3,-6)--(3.5,-7);
    \draw[very thick, dashed,black,fill=black] (3.5,-7)--(4,-8);
    \draw[very thick,black,fill=black] (-5,-8)--(-7.5,-8);
    \draw[very thick, dashed,black,fill=black] (-7.5,-8)--(-8.5,-8);
    \draw[very thick,black,fill=black] (-5,-4)--(-7.5,-4);
    \draw[very thick, dashed,black,fill=black] (-7.5,-4)--(-8.5,-4);
    \draw[very thick,black,fill=black] (-5,0)--(-7.5,0);
    \draw[very thick, dashed,black,fill=black] (-7.5,0)--(-8.5,0);
    \draw[very thick,black,fill=black] (-5,4)--(-7.5,4);
    \draw[very thick, dashed,black,fill=black] (-7.5,4)--(-8.5,4);
\end{tikzpicture}
\end{tabular}
\end{center}

\begin{center}
\vspace{0.2cm}
 \begin{tabular}{c}
    $\boldsymbol{\xdownarrow{0.5cm}}$ 
    \end{tabular}
\end{center}

    \begin{center}
    \vspace{0.5cm}
     \begin{tabular}{c}
 \begin{tikzpicture}[scale=0.45]
       \draw[fill=lightgray,draw=none] (-4,0.5)--(0.5,0.5)--(0.5,3.5)--(-4,3.5)--(-4,0.5);
       \node at (-1.75,2) {\small IV};
       \draw[fill=lightgray,draw=none] (0.5,0.5)--(5,0.5)--(5,3.5)--(0.5,3.5)--(0.5,0.5);
       \node at (2.75,2) {\small V};
       \draw[fill=lightgray,draw=none] (5,0.5)--(9.5,0.5)--(9.5,3.5)--(5,3.5)--(5,0.5);
       \node at (7.25,2) {\small VI};
       \draw[fill=lightgray,draw=none] (-4,0.5)--(0.5,0.5)--(0.5,-2.5)--(-4,-2.5)--(-4,0.5);
       \node at (-1.75,-1.7) {\small IX};
       \draw[fill=lightgray,draw=none] (0.5,0.5)--(5,0.5)--(5,-2.5)--(0.5,-2.5)--(0.5,0.5);
       \node at (2.75,-1.7) {\small VIII};
       \draw[fill=lightgray,draw=none] (5,0.5)--(9.5,0.5)--(9.5,-2.5)--(5,-2.5)--(5,0.5);
       \node at (7.25,-1.7) {\small VII};
     \draw[very thick,candypink,fill=candypink] (-4,0.5)--(0.5,0.5);
     \node[below] at (-1.9,0.4) {\footnotesize $A_2$ sing.};
     \draw[thick] (0.5,0.5) circle (0.05);
     \draw[very thick,cyan(process),fill=cyan(process)] (0.5,0.5)--(5,0.5);
     \node[below] at (2.6,0.4) {\footnotesize $A_1$ sing.};
      \draw[thick] (5,0.5) circle (0.05);
      \draw[very thick,green,fill=green] (5,0.5)--(9.5,0.5);
       \node[below] at (7.3,0.4) {\footnotesize non sing.};
       \draw[thick] (-4,0.5) circle (0.05);
       \draw[very thick,black,fill=black] (-4,0.5)--(-7,0.5);
       \node[below] at (-6,0.4) {\footnotesize $A_3$ sing.};
       \draw[very thick, dashed,black,fill=black] (-7,0.5)--(-8.5,0.5);
       \draw[thick] (9.5,0.5) circle (0.05);
       \draw[very thick,black,fill=black] (-4,0.5)--(-4,3.5);
       \draw[very thick, dashed,black,fill=black] (-4,3.5)--(-4,4.5);
       \draw[very thick,black,fill=black] (-4,0.5)--(-4,-2.5);
       \draw[very thick, dashed,black,fill=black] (-4,-2.5)--(-4,-3.5);
       \draw[very thick,black,fill=black] (0.5,0.5)--(0.5,3.5);
       \draw[very thick, dashed,black,fill=black] (0.5,3.5)--(0.5,4.5);
        \draw[very thick,black,fill=black] (0.5,0.5)--(0.5,-2.5);
       \draw[very thick, dashed,black,fill=black] (0.5,-2.5)--(0.5,-3.5);
       \draw[very thick,black,fill=black] (5,0.5)--(5,3.5);
       \draw[very thick, dashed,black,fill=black] (5,3.5)--(5,4.5);
        \draw[very thick,black,fill=black] (5,0.5)--(5,-2.5);
       \draw[very thick, dashed,black,fill=black] (5,-2.5)--(5,-3.5);
       \draw[very thick,black,fill=black] (9.5,0.5)--(9.5,3.5);
       \draw[very thick, dashed,black,fill=black] (9.5,3.5)--(9.5,4.5);
        \draw[very thick,black,fill=black] (9.5,0.5)--(9.5,-2.5);
       \draw[very thick, dashed,black,fill=black] (9.5,-2.5)--(9.5,-3.5);
 \end{tikzpicture} 
 \end{tabular}
   \end{center}

\begin{center}
  \begin{tabular}{c}
    $\boldsymbol{\xdownarrow{0.5cm}}$ 
    \end{tabular}
\end{center}

    \begin{figure}[H]
    \begin{center}
     \begin{tabular}{c}

\begin{tikzpicture}[scale=0.45]
   \draw[thick,fill=black] (0,0) circle (0.1);
   \node[below] at (0,-0.2) {\footnotesize $\mathbb{C}^3/(\mathbb{Z}_4\times\mathbb{Z}_4)$};
   \node[below] at (0,-1.2) {\footnotesize singularity};
   \end{tikzpicture}
   \end{tabular}
      \caption{\label{fig: partial resolution} At the top, the fully resolved phase of the $\mathbb{C}^3/(\mathbb{Z}_4 \times \mathbb{Z}_4)$ singularity. In the center, the partially resolved phase corresponding to Figure \ref{fig:trinionquiver} in the main text. At the bottom, the SCFT phase, with no length scale remaining.}
      \label{fig:pqwebtrinion}
      \end{center}
   \end{figure}
   
\section{Higgsing of the flavor symmetry}\label{Appendix B}
In this section, we quantitatively study the Higgsing of the flavor symmetry $F_{\text{rest}}$ when we turn on the mass parameter needed to reach the generalized linear quiver phase of \cref{sec:CBdimensionmolecules}. We are interested in the Higgsing of the flavor group that preserves the $\mathfrak g \times \mathfrak g$ factor. Consequently, we will always omit the $\mathfrak g \times \mathfrak g$ factor in what follows, as it is unaffected by the Higgsing procedure. \\ \indent 
The flavor group unhiggsed by the mass vev coincides with the flavor group of the 4d class-$\mathcal{S}$ molecule, as the Higgs branch is preserved by toroidal reduction. We then want to study the Higgsing 
\begin{equation}
    \label{eq:Higgsingflavormolecules}
    F_{\text{rest}} \to F_{III,4d}, 
\end{equation}
where the 4d flavor group of the class S molecule reads $\mathfrak g \times \mathfrak g \times F_{III,4d}$.\\ \indent
Summing up the result, for all the type of $\mathfrak g$, we are going to show that
\begin{itemize}
    \item the rank of the 4d flavor group is the one of the 5d conformal matter molecule, minus the number of mass parameters that we turned on to reach the generalized quiver phase. 
    \item The Higgsing always happens as $F_{rest} \to F_{III,4d}$;
    \item We will use that the Higgsing of a $
\mathfrak{su}(c_1 n_1 + c_2 n_2)$, with $c_1, c_2 \in \mathbb N$, factors as
\begin{equation}
    \mathfrak{su}(c_1 n_1 + c_2 n_2) \to \mathfrak{su}(c_1 n_1)\times \mathfrak{su}(c_2 n_2)\to 
\ldots,
\end{equation}
where we denoted with the dots the corresponding factor of $F_{III,4d}$.
\item The Higging of a factor $\mathfrak{su}(n_i)^{m_i}$ is always along the diagonal subgroup. One can explicitely check that a generic Higgsing along the diagonal combination of $\mathfrak{su}(n_i)^{m_i}$ breaks the group to a generic combination of the Cartan of the non-diagonal flavor transformations. 
\item We can guess the Higgsing of the abelian $\mathfrak u(1)$ factors in $F_{rest}$ with a case-by-case analysis. 
\end{itemize}
In general, we can geometrically understand this kind of diagonal Higgsing by turning off simultaneously two out of the three $n_i$s. In this case, the $X_{\mathfrak g}^{(i)}$ becomes \begin{eqnarray}
\label{eq:sketchgeom}
    \begin{cases}
    P(x_1,x_2,x_3) = 0,\\ \nonumber
    uv=t^{n_k}, \\ \nonumber 
    t = x_k, \nonumber 
    \end{cases}
\end{eqnarray}
It then makes sense to talk about a ``diagonal'' Higgsing whenever the we have a subfactor $\mathfrak{su}(...)^f$ in $F_{rest}$. Switching on a mass factor along the diagonal corresponds, in  \eqref{eq:sketchgeom}, to perform a diagonal resolution of the  $f$ lines of A-type singularities that is easy to check appear in  \eqref{eq:sketchgeom}, as it corresponds to perform a base-change type resolution now thinking of  \eqref{eq:sketchgeom} as a base-change of the $A_{k-1}$ singularity. Let us now consider the various cases in more details.\\ \indent
First, we consider the $A_{2j+1}$ case. In this case we have just one type of atom, $X_{A_{2j+1}}^{(1)}$, that we can combine in a molecule with $n_1$ junctions. From \cref{sec:CBdimensionmolecules} we get
\begin{equation}
    \label{eq:flavorgroupIIIA2j+1}
    F_{III,4d} = \mathfrak u(1)^{n_1+1} \Rightarrow \text{rank}(F_{III,4d}) = n_1+1. 
\end{equation}
On the other hand, in this case, the 5d flavor group is \cite{DeMarco:2023irn}
\begin{equation}
\label{eq:5dflavorgroupA2j+1molecule}
F_{rest} = \mathfrak{u}(2n_{1}), \quad \Rightarrow \quad \text{rank}(F_{rest}) =  \text{rank}(F_{III,4d}) + (n_1 - 1), 
\end{equation}
This result is expected, as, to reach the generalized linear quiver phase, we gave a vev to $n_1-1$ mass parameters, represented by the K\"ahler volumes of the $\mathbb P^1$s supporting  compact curves of $A_{2j+1}$ singularities. To understand quantitatively the flavor breaking, we can decompose the $\mathfrak su(n_1) \times \mathfrak{su}(n_1)$ factor into its diagonal and antidiagonal part. The decomposition does not split the group as a direct product, and in particular all the elements not in the Cartan of the antidiagonal part gets higgsed once we turn on a diagonal Cartan vev. Hence, we get
\begin{equation}
    \re{\mathfrak su(n_1) \times \mathfrak{su}(n_1)} \times \mathfrak{u}(1) \to \re{\mathfrak{u}(1)^{n_1-1}} \times \mathfrak{u}(1) = \mathfrak{u}(1)^{n_1}. 
\end{equation}

\indent 
Let us now consider the $A_{2j}$ case; we again have just one kind of atom, $X_{A_{2j}}^{(1)}$ and any molecule with $n_{1}$ junctions have a trivial $F_{III,4d}$, hence 
\begin{equation}
    \label{eq:flavorgroupIIIA2j}
\text{rank}(F_{III,4d}) = 0. 
\end{equation}
The 5d flavor group is \cite{DeMarco:2023irn}
\begin{equation}
\label{eq:5dflavorgroupA2j}
F_{rest} = \mathfrak{su}(n_{1}), \quad \Rightarrow \quad \text{rank}(F_{rest}) = n_{1}-1, 
\end{equation}
and hence again the rank of the class-$\mathcal{S}$ flavor group is the rank of the 5d $F_{rest}$ minus the number of masses that we turned on to get the generalized quiver phase. In this case, to completely break $F_{rest}$, we have turned on a generic element on the Cartan $\mathfrak h_{rest} \subset F_{rest}$:
\begin{equation}
    \label{eq:massveva2j}
    M = \sum_{i=1}^{2j} m_{i}\alpha_{i}^*,
\end{equation}
with $\alpha_i$ the $i$-th dual root.\footnote{
Let us define the ``dual of the root $\alpha_j$'', an element of the Cartan $\mathfrak{h}_{rest}$ defined as 
\begin{equation}
    \label{eq:dualroot}
    \comm{\alpha_{j}^*}{e_{\alpha_l}} = \delta_{jl}e_{\alpha_l},
\end{equation}
with $e_{\alpha_l}$ the root vector associated to the simple root $\alpha_l$. Then, given the breaking 
\begin{equation}
    \label{eq:breakingA2j+1}
    \mathfrak{u}(2n_1) \to \mathfrak{u}(1)^{n_1}.
\end{equation}
}
We can then consider the $D_{2j}$ case. For this simple algebra, we have three different kind of atoms, $X_{D_{2j}}^{(1)},X_{D_{2j}}^{(1)}$
and $X_{D_{2j}}^{(3)}$. We have \cite{DeMarco:2023irn}
\begin{eqnarray}
   \label{eq:flavordatad2j}
   &&F_{rest} = \mathfrak{su}(n_1)^2 \times \mathfrak{su}(n_2) \times \mathfrak{su}(n_1 + 2n_3) \times \mathfrak{u}(1)^3, \quad \text{rank}(F_{rest}) = 3 n_1 + n_2 + 2 n_3-1, \nonumber \\
   &&F_{III,4d} = \mathfrak{u}(1)^{2 n_1} \times \mathfrak{su}(2)^{n_3}, \qquad  \text{rank}(F_{III,4d}) = 2n_1 +  n_3. 
\end{eqnarray}
We can again verify that from \eqref{eq:flavordatad2j} that 
\begin{equation}
\text{rank}(F_{III,4d}) =  \text{rank}(F_{rest})- (n_{1} + n_{2} + n_{3} -1) =\text{rank}(F_{rest}) - n_{masses}. 
\end{equation}
The breaking pattern is first 
\begin{equation}
    \mathfrak{su}(n_1 + 2 n_3) \to \mathfrak{su}(n_1) \times \mathfrak{su}(2n_2),
\end{equation}
and then
\begin{eqnarray}
\mathfrak{su}(n_1)^3 \times \mathfrak{u}(1) \to     \mathfrak{u}(1)^{2n_1}, \nonumber \\
\mathfrak{su}(n_2) \to \left\{\mathrm 1\right\}, \nonumber \\
\mathfrak{su}(2n_3) \to \mathfrak{su}(2)^{n_3},\\
\end{eqnarray}
where in the last lines we higgsed turning on the following vev inside $\mathfrak{su}(2n_3)$:
\begin{equation}
\label{eq:Higgsingtosu2}
    M = \sum_{i=1}^{n_3-1}m_i\alpha_{2i}^{*}.
\end{equation}
Finally, to match the class $\mathcal{S}$ molecule flavor group, we give mass to the leftover $\mathfrak{u}(1)^2$ factors.
\\ \indent 
For the $D_{2j+1}$ case, we can proceed similarly, obtaining the following ``accidental'' 5d and 4d flavor groups
\begin{eqnarray}
   \label{eq:flavordatad2j+1}
   &&F_{rest} = \mathfrak{su}(n_1) \times \mathfrak{su}(n_2)^2 \times \mathfrak{su}(n_1 + 2n_3) \times \mathfrak{u}(1)^3, \quad \text{rank}(F_{rest}) = 2 n_1 +2 n_2 + 2 n_3-1, \nonumber \\
   &&F_{III,4d} = \mathfrak{u}(1)^{n_1} \times \mathfrak{u}(1)^{n_2} \times  \mathfrak{su}(2)^{n_3}, \qquad  \text{rank}(F_{III,4d}) = n_1 +  n_2 + n_3 =  \text{rank}(F_{III,4d}) - n_{masses}. \nonumber \\
\end{eqnarray}
The flavor symmetry breaking pattern is first
\begin{equation}
    \mathfrak{su}(n_1 + 2 n_3) \to \mathfrak{su}(n_1) \times \mathfrak{su}(2n_2),
\end{equation}
followed by 
\begin{eqnarray}
\label{eq:eqprev}
\mathfrak{su}(n_1)^2 \times \mathfrak{u}(1) \to     \mathfrak{u}(1)^{n_1}, \nonumber \\
\mathfrak{su}(n_2)^2 \times \mathfrak{u}(1) \to  \mathfrak{u}(1)^{n_2}, \nonumber \\
\mathfrak{su}(2n_3) \to \mathfrak{su}(2)^{n_3}, \nonumber \\
\end{eqnarray}
with the Higgsing in the last line being again \eqref{eq:Higgsingtosu2}. To match the 4d flavor group, we then give a mass to the $\mathfrak{u}(1)$ factor not appearing in \eqref{eq:eqprev}.
\\ \indent
We can then pass to the $E_6$ case. For the most general molecule, we have 
\begin{eqnarray}
   \label{eq:flavordatade6}
   &&F_{rest} = \mathfrak{su}(n_1) \times \mathfrak{su}(n_2)^2 \times \mathfrak{su}(n_3) \times \mathfrak{u}(1)^3, \quad \text{rank}(F_{rest}) = n_1 +2 n_2 + n_3-1, \nonumber \\
   &&F_{III,4d} =  \mathfrak{u}(1)^{n_2}, \qquad  \text{rank}(F_{III,4d}) =  n_2 =  \text{rank}(F_{rest}) - n_{masses}. 
\end{eqnarray}
The symmetry breaking $F_{rest} \to F_{III,4d}$ can be reached by turning on the generic Cartan element along the $\mathfrak{su}(n_1),\mathfrak{su}(n_3)$ factors, turning on a mass along two out of the three $\mathfrak{u}(1)$'s, and Higgsing
\begin{equation}
    \mathfrak{su}(n_2)^2 \times \mathfrak{u}(1) \to \mathfrak{u}(1)^{n_2} 
\end{equation}
via turning on a diagonal combination along the two  $\mathfrak{su}(n_2)$ factors.\\ \indent
For the $E_7$ case, we have 
\begin{eqnarray}
   \label{eq:flavordatae7}
   &&F_{rest} = \mathfrak{su}(n_1) \times \mathfrak{su}(2n_2+n1) \times \mathfrak{su}(n_3) \times \mathfrak{u}(1)^2, \quad \text{rank}(F_{rest}) = 2n_1 +2 n_2 + n_3-1, \nonumber \\
   &&F_{III,4d} =  \mathfrak{u}(1)^{n_1} \times \mathfrak{su}(2)^{n_2}, \qquad  \text{rank}(F_{III,4d}) =  n_1+n_2 =  \text{rank}(F_{rest}) - n_{masses}. \nonumber \\ 
\end{eqnarray}
Again, the Higgsing goes as follows: first we higgs 
\begin{equation}
    \mathfrak{su}(2n_2 +  n_1) \to \mathfrak{su}(n_1) \times \mathfrak{su}(2n_2),
\end{equation}
we give mass to one of the two $\mathfrak{u}(1)$ factors, and finally we higgs
\begin{eqnarray}
\mathfrak{su}(n_1)^2 \times \mathfrak{u}(1) \to     \mathfrak{u}(1)^{n_1}, \nonumber \\
\mathfrak{su}(2n_2) \to  \mathfrak{su}(2)^{n_2}.
\end{eqnarray}
Lastly, we have the $E_8$ case: 
\begin{eqnarray}
   \label{eq:flavordatae8}
   &&F_{rest} = \mathfrak{su}(n_1) \times \mathfrak{su}(n_2) \times \mathfrak{su}(n_3) \times \mathfrak{u}(1)^2, \quad \text{rank}(F_{rest}) = n_1 + n_2 + n_3-1, \nonumber \\
\end{eqnarray}
with trivial $F_{III,4d}$, as expected being the rank of $F_{rest}$ equal to the mass parameters needed to get to the generalized linear quiver phase.  In this case, the Higgsing is compatible with the mass vev being a generic element in the Cartan subalgebra of $F_{rest}$.

\section{Notation for Quivers and Brane Webs}
\label{app:QuiverBWNotation}
In this appendix we briefly summarise the notation we use for quivers and brane webs.

\subsection{Quivers}
\label{app:QuiverNotation}
As is customary, we use square nodes for flavour groups, round nodes for gauge groups, and edges for bifundamental hypermultiplets (or half-hypermultiplets if the bifundamental representation is quaternionic). We colour the nodes white for unitary groups, yellow for special unitary groups (possibly with Chern-Simons level $k$), red for special orthogonal groups, and blue for compact symplectic groups:
\begin{equation}
    \begin{tikzpicture}
        \node[gauge,label=below:{$n$}] at (0,0) {};
        \node[gaugey,label=below:{$(n)_k$}] at (0,-1.5) {};
        \node[gaugey,label=below:{$n$}] at (0,-3) {};
        \node[gauger,label=below:{$n$}] at (4,-0.75) {};
        \node[gaugeb,label=below:{$n$}] at (4,-2.25) {};
        \node at (1,0) {$=$};
        \node at (1,-1.5) {$=$};
        \node at (1,-3) {$=$};
        \node at (5,-0.75) {$=$};
        \node at (5,-2.25) {$=$};
        \node[gauge,label=below:{U($n$)}] at (2,0) {};
        \node[gaugey,label=below:{SU$(n)_k$}] at (2,-1.5) {};
        \node[gaugey,label=below:{SU$(n)_0$}] at (2,-3) {};
        \node[gauger,label=below:{SO($n$)}] at (6,-0.75) {};
        \node[gaugeb,label=below:{USp($n$)}] at (6,-2.25) {};
    \end{tikzpicture}\;.
\end{equation}

\paragraph{Unframed Quivers.}
If the quiver is unframed, i.e.\ there are no flavour nodes, then there can be a 1-form symmetry which can be gauged. (see e.g.\ \cite{Bourget:2020xdz}, where the 1-form symmetry is referred to as '$H$'.).

For an unframed quiver with only unitary gauge nodes there is a U$(1)^{(1)}$ 1-form symmetry which we always take to be gauged, and hence the true gauge group of the quiver is $\prod_i\mathrm{U}(r_i)/\mathrm{U}(1)$, where $r_i$ are the ranks of the individual gauge nodes.

For an unframed quiver $\mathsf{Q}$ with only compact symplectic gauge nodes, and possibly even special orthogonal and/or unitary gauge nodes, there is a $\mathbb{Z}_2^{(1)}$ 1-form symmetry which may be gauged (or not). We denote the gauging of this 1-form symmetry by writing
\begin{equation}
    \begin{tikzpicture}
        \node at (2.5,0) {$\mathsf{Q}$};
        \draw (2.5,-0.5)--(3.5,0.5);
        \node at (3.5,0) {$\mathbb{Z}_2^{(1)}$};
    \end{tikzpicture}\;.
\end{equation}

\subsection{Brane Webs}
\label{app:BWNotation}
We consider webs of $(p,q)5$-branes and $[p,q]7$-branes in Type IIB String Theory \cite{Aharony:1997ju,Aharony:1997bh,DeWolfe:1999hj}. The occupied spacetime directions of the branes are summarised in Table \ref{tab:spacetime}.

\begin{table}[h]
\begin{center}
\begin{tabular}{|c|c|c|c|c|c|c|c|c|c|c|}
\hline
Type IIB & $x^0$ & $x^1$ & $x^2$ & $x^3$ & $x^4$ & $x^5$ & $x^6$ & $x^7$ & $x^8$ & $x^9$\\
\hline
$(p,q)5$-brane & $\times$ & $\times$ & $\times$ & $\times$ & $\times$ & \multicolumn{2}{c|}{angle $\alpha$} & & & \\
\hline
$[p,q]7$-brane & $\times$ & $\times$ & $\times$ & $\times$ & $\times$ & & & $\times$ & $\times$ & $\times$ \\ \hline
\end{tabular}
\caption{Occupation of space-time directions of the $(p,q)5$-branes and $[p,q]7$-branes in Type IIB are denoted by $\times$. The angle $\alpha$ depends on the $(p,q)$ charges and the axio-dilaton $\tau$; $\alpha=\arg(p+\tau q)$. We set $\tau=i$ in the rest of the paper, s.t.\ $\tan(\alpha)=q/p$. Our 5d $\mathcal{N}=1$ theories exist as effective field theories living on fivebranes suspended between sevenbranes.}
\label{tab:spacetime}
\end{center}
\end{table}

We depict brane webs by drawing the $(x^5,x^6)$ plane:
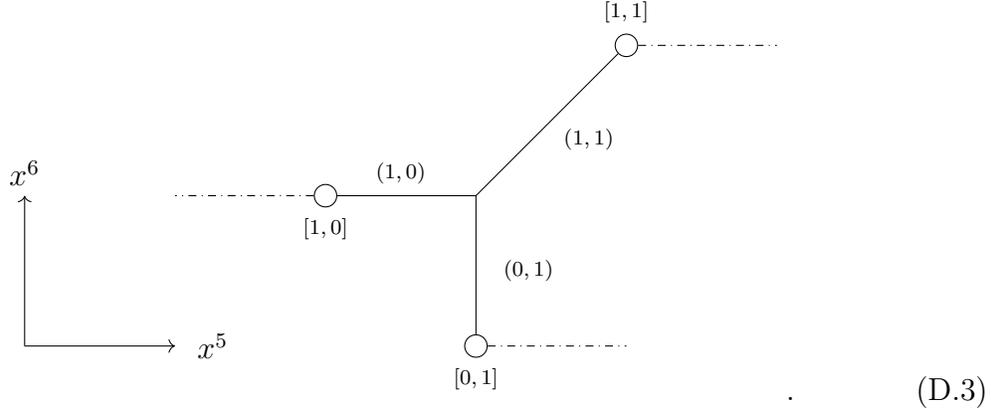
\begin{equation}
    \begin{tikzpicture}
        \node[seven,label=above:{\scriptsize$[1,1]$}] (1) at (2,2) {};
        \node[seven,label=below:{\scriptsize$[1,0]$}] (2) at (-2,0) {};
        \node[seven,label=below:{\scriptsize$[0,1]$}] (3) at (0,-2) {};
        \draw (2)--(0,0)--(3) (0,0)--(1);
        \draw[dash dot] (1)--(4,2) (2)--(-4,0) (3)--(2,-2);
        \node at (-1,0.3) {\scriptsize$(1,0)$};
        \node at (0.7,-1) {\scriptsize$(0,1)$};
        \node at (1.5,0.75) {\scriptsize$(1,1)$};

        \draw[->] (-6,-2)--(-4,-2);
        \draw[->] (-6,-2)--(-6,0);
        \node at (-6,0.3) {$x^6$};
        \node at (-3.5,-2) {$x^5$};
    \end{tikzpicture}\;.
\end{equation}
A $(p,q)5$-brane ends on a $[p,q]7$-brane, or on a fivebrane vertex which preserves $(p,q)$ charges. Multiple $(p,q)5$-branes can end on the same $[p,q]7$-brane as long as the s-rule \cite{Hanany:1996ie,Benini:2009gi,vanBeest:2020kou,Bergman:2020myx} is not violated.

Each $[p,q]7$-brane induces an SL$(2,\mathbb{Z})$ monodromy cut (depicted by a dot-dash line) with associated monodromy matrix
\begin{equation}
    M_{[p,q]}=\begin{pmatrix}
                1-pq & p^2\\
                -q^2 & 1+pq
            \end{pmatrix}\;.
\end{equation}
We usually don't draw the monodromy cuts of sevenbranes if they are oriented in such a way that they do not cross any fivebranes.

An $(r,s)5$-brane crossing the monodromy cut of a $[p,q]7$-brane is affected in the following way:
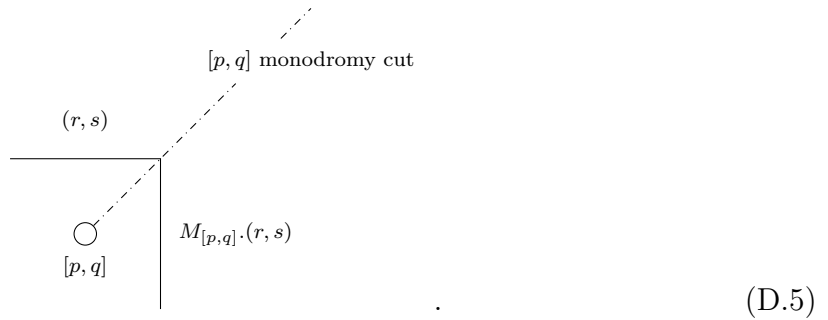
\begin{equation}
    \begin{tikzpicture}
            \draw (0,0)--(2,0)--(2,-2);
            \node[seven,label=below:{\scriptsize$[p,q]$}] (1) at (1,-1) {};
            \draw[dash dot] (1)--(3,1);
            \node at (1,0.5) {\scriptsize$(r,s)$};
            \node at (3,-1) {\scriptsize$M_{[p,q]}.(r,s)$};
            \node at (4,1.3) {\scriptsize$[p,q]$ monodromy cut};
            \draw[dash dot] (3.6,1.6)--(4,2);
        \end{tikzpicture}\;.
\end{equation}
When the $(r,s)5$-brane is pulled through the $[p,q]7$-brane, $|ps-qr|$ extra $(r,s)5$-branes are created via the Hanany-Witten effect:
\begin{equation}
    \begin{tikzpicture}
            \draw (0,0)--(2,0)--(2,-2);
            \node[seven,label=right:{\scriptsize$[p,q]$}] (1) at (4,2) {};
            \node at (1,0.5) {\scriptsize$(r,s)$};
            \node at (3,-1) {\scriptsize$M_{[p,q]}.(r,s)$};
            \draw[thick,double] (2,0)--(1);
            \node at (4,1) {\scriptsize$|ps-qr|(p,q)$};
        \end{tikzpicture}\;.
\end{equation}

We also study brane webs in the presence of orientifold 5-planes. In particular $\mathrm{ON}5^-$ which we draw as a green line, and $\widetilde{\mathrm{ON}}5^-$ (which corresponds to an ON$5^-$ with a stuck $\frac{1}{2}(0,1)5$ on top) which we draw as an orange line. An $\mathrm{ON}5^-$ crossing a $\frac{1}{2}[0,1]7$-brane turns into an $\widetilde{\mathrm{ON}}5^-$:
\begin{equation}
    \begin{tikzpicture}
        \node[seven,label=right:{\scriptsize$[0,1]$}] (1) at (0,1) {};
        \draw[green] (0,0)--(1);
        \draw[orange] (1)--(0,2);
        \node at (-0.5,0.5) {\scriptsize$\mathrm{ON}5^-$};
        \node at (-0.5,1.5) {\scriptsize$\widetilde{\mathrm{ON}}5^-$};
    \end{tikzpicture}
\end{equation}
(suppressing the monodromy cut of the half sevenbrane, which is is pointing upwards along the $\widetilde{\mathrm{ON}}5^-$).

\bibliographystyle{at}
\bibliography{bibliography.bib}

\providecommand{\href}[2]{#2}\begin{thebibliography}{100}

\bibitem{Witten:1995ex}
E.~Witten, ``{String theory dynamics in various dimensions},'' {\em Nucl. Phys. B} {\bf 443} (1995) 85--126, \href{http://arXiv.org/abs/hep-th/9503124}{{\tt hep-th/9503124}}.

\bibitem{Strominger:1995ac}
A.~Strominger, ``{Open p-branes},'' {\em Phys. Lett. B} {\bf 383} (1996) 44--47, \href{http://arXiv.org/abs/hep-th/9512059}{{\tt hep-th/9512059}}.

\bibitem{Witten:1995em}
E.~Witten, ``{Five-branes and M theory on an orbifold},'' {\em Nucl. Phys. B} {\bf 463} (1996) 383--397, \href{http://arXiv.org/abs/hep-th/9512219}{{\tt hep-th/9512219}}.

\bibitem{Ganor:1996mu}
O.~J. Ganor and A.~Hanany, ``{Small E(8) instantons and tensionless noncritical strings},'' {\em Nucl. Phys. B} {\bf 474} (1996) 122--140, \href{http://arXiv.org/abs/hep-th/9602120}{{\tt hep-th/9602120}}.

\bibitem{Seiberg:1996qx}
N.~Seiberg, ``{Nontrivial fixed points of the renormalization group in six-dimensions},'' {\em Phys. Lett. B} {\bf 390} (1997) 169--171, \href{http://arXiv.org/abs/hep-th/9609161}{{\tt hep-th/9609161}}.

\bibitem{Seiberg1996}
N.~Seiberg, ``Five dimensional susy field theories, non-trivial fixed points and string dynamics,'' {\em Physics Letters B} {\bf 388} (Nov, 1996) 753–760.

\bibitem{Morrison_1997}
D.~R. Morrison and N.~Seiberg, ``Extremal transitions and five-dimensional supersymmetric field theories,'' {\em Nuclear Physics B} {\bf 483} (Jan, 1997) 229–247.

\bibitem{Douglas:1996xp}
M.~R. Douglas, S.~H. Katz, and C.~Vafa, ``{Small instantons, Del Pezzo surfaces and type I-prime theory},'' {\em Nucl. Phys. B} {\bf 497} (1997) 155--172, \href{http://arXiv.org/abs/hep-th/9609071}{{\tt hep-th/9609071}}.

\bibitem{DeMarco:2023irn}
M.~De~Marco, M.~Del~Zotto, M.~Graffeo, and A.~Sangiovanni, ``{5d Conformal Matter},'' \href{http://arXiv.org/abs/2311.04984}{{\tt 2311.04984}}.

\bibitem{DelZotto:2014hpa}
M.~Del~Zotto, J.~J. Heckman, A.~Tomasiello, and C.~Vafa, ``{6d Conformal Matter},'' {\em JHEP} {\bf 02} (2015) 054, \href{http://arXiv.org/abs/1407.6359}{{\tt 1407.6359}}.

\bibitem{Heckman:2015bfa}
J.~J. Heckman, D.~R. Morrison, T.~Rudelius, and C.~Vafa, ``{Atomic Classification of 6D SCFTs},'' {\em Fortsch. Phys.} {\bf 63} (2015) 468--530, \href{http://arXiv.org/abs/1502.05405}{{\tt 1502.05405}}.

\bibitem{Bhardwaj:2015oru}
L.~Bhardwaj, M.~Del~Zotto, J.~J. Heckman, D.~R. Morrison, T.~Rudelius, and C.~Vafa, ``{F-theory and the Classification of Little Strings},'' {\em Phys. Rev. D} {\bf 93} (2016), no.~8, 086002, \href{http://arXiv.org/abs/1511.05565}{{\tt 1511.05565}}. [Erratum: Phys.Rev.D 100, 029901 (2019)].

\bibitem{Bhardwaj:2019hhd}
L.~Bhardwaj, ``{Revisiting the classifications of 6d SCFTs and LSTs},'' {\em JHEP} {\bf 03} (2020) 171, \href{http://arXiv.org/abs/1903.10503}{{\tt 1903.10503}}.

\bibitem{Ganor:1996pc}
O.~J. Ganor, D.~R. Morrison, and N.~Seiberg, ``{Branes, Calabi-Yau spaces, and toroidal compactification of the N=1 six-dimensional E(8) theory},'' {\em Nucl. Phys. B} {\bf 487} (1997) 93--127, \href{http://arXiv.org/abs/hep-th/9610251}{{\tt hep-th/9610251}}.

\bibitem{Aharony:1997ju}
O.~Aharony and A.~Hanany, ``{Branes, superpotentials and superconformal fixed points},'' {\em Nucl. Phys. B} {\bf 504} (1997) 239--271, \href{http://arXiv.org/abs/hep-th/9704170}{{\tt hep-th/9704170}}.

\bibitem{Aharony:1997bh}
O.~Aharony, A.~Hanany, and B.~Kol, ``{Webs of (p,q) five-branes, five-dimensional field theories and grid diagrams},'' {\em JHEP} {\bf 01} (1998) 002, \href{http://arXiv.org/abs/hep-th/9710116}{{\tt hep-th/9710116}}.

\bibitem{DeWolfe:1999hj}
O.~DeWolfe, A.~Hanany, A.~Iqbal, and E.~Katz, ``{Five-branes, seven-branes and five-dimensional E(n) field theories},'' {\em JHEP} {\bf 03} (1999) 006, \href{http://arXiv.org/abs/hep-th/9902179}{{\tt hep-th/9902179}}.

\bibitem{Intriligator_1997}
K.~Intriligator, D.~R. Morrison, and N.~Seiberg, ``Five-dimensional supersymmetric gauge theories and degenerations of calabi-yau spaces,'' {\em Nuclear Physics B} {\bf 497} (Jul, 1997) 56–100.

\bibitem{leung1997branes}
N.~C. Leung and C.~Vafa, ``Branes and toric geometry,'' 1997.

\bibitem{Acharya:2024bnt}
B.~S. Acharya, ``{Confinement in Five Dimensions},'' \href{http://arXiv.org/abs/2407.03171}{{\tt 2407.03171}}.

\bibitem{Xie:2017pfl}
D.~Xie and S.-T. Yau, ``{Three dimensional canonical singularity and five dimensional $ \mathcal{N} $ = 1 SCFT},'' {\em JHEP} {\bf 06} (2017) 134, \href{http://arXiv.org/abs/1704.00799}{{\tt 1704.00799}}.

\bibitem{Jefferson:2017ahm}
P.~Jefferson, H.-C. Kim, C.~Vafa, and G.~Zafrir, ``{Towards Classification of 5d SCFTs: Single Gauge Node},'' {\em SciPost Phys.} {\bf 14} (2023) 122, \href{http://arXiv.org/abs/1705.05836}{{\tt 1705.05836}}.

\bibitem{Jefferson:2018irk}
P.~Jefferson, S.~Katz, H.-C. Kim, and C.~Vafa, ``{On Geometric Classification of 5d SCFTs},'' {\em JHEP} {\bf 04} (2018) 103, \href{http://arXiv.org/abs/1801.04036}{{\tt 1801.04036}}.

\bibitem{Apruzzi:2019opn}
F.~Apruzzi, C.~Lawrie, L.~Lin, S.~Sch\"afer-Nameki, and Y.-N. Wang, ``{Fibers add Flavor, Part I: Classification of 5d SCFTs, Flavor Symmetries and BPS States},'' {\em JHEP} {\bf 11} (2019) 068, \href{http://arXiv.org/abs/1907.05404}{{\tt 1907.05404}}.

\bibitem{Apruzzi_2020}
F.~Apruzzi, C.~Lawrie, L.~Lin, S.~Schäfer-Nameki, and Y.-N. Wang, ``Fibers add flavor. part {II}. 5d {SCFTs}, gauge theories, and dualities,'' {\em Journal of High Energy Physics} {\bf 2020} (mar, 2020).

\bibitem{Closset:2020scj}
C.~Closset, S.~Schafer-Nameki, and Y.-N. Wang, ``{Coulomb and Higgs Branches from Canonical Singularities: Part 0},'' {\em JHEP} {\bf 02} (2021) 003, \href{http://arXiv.org/abs/2007.15600}{{\tt 2007.15600}}.

\bibitem{Closset:2020afy}
C.~Closset, S.~Giacomelli, S.~Schafer-Nameki, and Y.-N. Wang, ``{5d and 4d SCFTs: Canonical Singularities, Trinions and S-Dualities},'' {\em JHEP} {\bf 05} (2021) 274, \href{http://arXiv.org/abs/2012.12827}{{\tt 2012.12827}}.

\bibitem{Closset:2021lwy}
C.~Closset, S.~Sch\"afer-Nameki, and Y.-N. Wang, ``{Coulomb and Higgs branches from canonical singularities. Part I. Hypersurfaces with smooth Calabi-Yau resolutions},'' {\em JHEP} {\bf 04} (2022) 061, \href{http://arXiv.org/abs/2111.13564}{{\tt 2111.13564}}.

\bibitem{Collinucci:2021ofd}
A.~Collinucci, M.~De~Marco, A.~Sangiovanni, and R.~Valandro, ``{Higgs branches of 5d rank-zero theories from geometry},'' {\em JHEP} {\bf 10} (2021), no.~18, 018, \href{http://arXiv.org/abs/2105.12177}{{\tt 2105.12177}}.

\bibitem{Collinucci:2021wty}
A.~Collinucci, A.~Sangiovanni, and R.~Valandro, ``{Genus zero Gopakumar-Vafa invariants from open strings},'' {\em JHEP} {\bf 09} (2021) 059, \href{http://arXiv.org/abs/2104.14493}{{\tt 2104.14493}}.

\bibitem{DeMarco:2021try}
M.~De~Marco and A.~Sangiovanni, ``{Higgs Branches of rank-0 5d theories from M-theory on (A$_{j}$, A$_{l}$) and (A$_{k}$, D$_{n}$) singularities},'' {\em JHEP} {\bf 03} (2022) 099, \href{http://arXiv.org/abs/2111.05875}{{\tt 2111.05875}}.

\bibitem{Tian:2021cif}
J.~Tian and Y.-N. Wang, ``{5D and 6D SCFTs from $\mathbb{C}^3$ orbifolds},'' {\em SciPost Phys.} {\bf 12} (2022), no.~4, 127, \href{http://arXiv.org/abs/2110.15129}{{\tt 2110.15129}}.

\bibitem{Closset:2022vjj}
C.~Closset and H.~Magureanu, ``{Partition functions and fibering operators on the Coulomb branch of 5d SCFTs},'' {\em JHEP} {\bf 01} (2023) 035, \href{http://arXiv.org/abs/2209.13564}{{\tt 2209.13564}}.

\bibitem{Collinucci:2022rii}
A.~Collinucci, M.~De~Marco, A.~Sangiovanni, and R.~Valandro, ``{Flops of any length, Gopakumar-Vafa invariants and 5d Higgs branches},'' {\em JHEP} {\bf 08} (2022) 292, \href{http://arXiv.org/abs/2204.10366}{{\tt 2204.10366}}.

\bibitem{DeMarco:2022dgh}
M.~De~Marco, A.~Sangiovanni, and R.~Valandro, ``{5d Higgs branches from M-theory on quasi-homogeneous cDV threefold singularities},'' {\em JHEP} {\bf 10} (2022) 124, \href{http://arXiv.org/abs/2205.01125}{{\tt 2205.01125}}.

\bibitem{Closset:2023pmc}
C.~Closset and H.~Magureanu, ``{Reading between the rational sections: Global structures of 4d $\mathcal{N}=2$ KK theories},'' {\em SciPost Phys.} {\bf 16} (2024), no.~5, 137, \href{http://arXiv.org/abs/2308.10225}{{\tt 2308.10225}}.

\bibitem{Mu:2023uws}
J.~Mu, Y.-N. Wang, and H.~N. Zhang, ``{5d SCFTs from isolated complete intersection singularities},'' {\em JHEP} {\bf 02} (2024) 155, \href{http://arXiv.org/abs/2311.05441}{{\tt 2311.05441}}.

\bibitem{Bourget:2023wlb}
A.~Bourget, A.~Collinucci, and S.~Schafer-Nameki, ``{Generalized Toric Polygons, T-branes, and 5d SCFTs},'' \href{http://arXiv.org/abs/2301.05239}{{\tt 2301.05239}}.

\bibitem{Benini:2009gi}
F.~Benini, S.~Benvenuti, and Y.~Tachikawa, ``{Webs of five-branes and N=2 superconformal field theories},'' {\em JHEP} {\bf 09} (2009) 052, \href{http://arXiv.org/abs/0906.0359}{{\tt 0906.0359}}.

\bibitem{Bergman:2013aca}
O.~Bergman, D.~Rodr\'\i{}guez-G\'omez, and G.~Zafrir, ``{5-Brane Webs, Symmetry Enhancement, and Duality in 5d Supersymmetric Gauge Theory},'' {\em JHEP} {\bf 03} (2014) 112, \href{http://arXiv.org/abs/1311.4199}{{\tt 1311.4199}}.

\bibitem{Zafrir:2014ywa}
G.~Zafrir, ``{Duality and enhancement of symmetry in 5d gauge theories},'' {\em JHEP} {\bf 12} (2014) 116, \href{http://arXiv.org/abs/1408.4040}{{\tt 1408.4040}}.

\bibitem{Hayashi:2015zka}
H.~Hayashi, S.-S. Kim, K.~Lee, and F.~Yagi, ``{6d SCFTs, 5d Dualities and Tao Web Diagrams},'' {\em JHEP} {\bf 05} (2019) 203, \href{http://arXiv.org/abs/1509.03300}{{\tt 1509.03300}}.

\bibitem{Hayashi:2015fsa}
H.~Hayashi, S.-S. Kim, K.~Lee, M.~Taki, and F.~Yagi, ``{A new 5d description of 6d D-type minimal conformal matter},'' {\em JHEP} {\bf 08} (2015) 097, \href{http://arXiv.org/abs/1505.04439}{{\tt 1505.04439}}.

\bibitem{Bergman:2015dpa}
O.~Bergman and G.~Zafrir, ``{5d fixed points from brane webs and O7-planes},'' {\em JHEP} {\bf 12} (2015) 163, \href{http://arXiv.org/abs/1507.03860}{{\tt 1507.03860}}.

\bibitem{Hayashi:2018lyv}
H.~Hayashi, S.-S. Kim, K.~Lee, and F.~Yagi, ``{Dualities and 5-brane webs for 5d rank 2 SCFTs},'' {\em JHEP} {\bf 12} (2018) 016, \href{http://arXiv.org/abs/1806.10569}{{\tt 1806.10569}}.

\bibitem{Hayashi:2018bkd}
H.~Hayashi, S.-S. Kim, K.~Lee, and F.~Yagi, ``{5-brane webs for 5d $ \mathcal{N} $ = 1 G$_{2}$ gauge theories},'' {\em JHEP} {\bf 03} (2018) 125, \href{http://arXiv.org/abs/1801.03916}{{\tt 1801.03916}}.

\bibitem{Hayashi:2019yxj}
H.~Hayashi, S.-S. Kim, K.~Lee, and F.~Yagi, ``{Rank-3 antisymmetric matter on 5-brane webs},'' {\em JHEP} {\bf 05} (2019) 133, \href{http://arXiv.org/abs/1902.04754}{{\tt 1902.04754}}.

\bibitem{Hayashi_2020}
H.~Hayashi, S.-S. Kim, K.~Lee, and F.~Yagi, ``Complete prepotential for 5d $ \mathcal{N} $ = 1 superconformal field theories,'' {\em Journal of High Energy Physics} {\bf 2020} (Feb, 2020).

\bibitem{Bergman:2020myx}
O.~Bergman and D.~Rodr\'\i{}guez-G\'omez, ``{The Cat\textquoteright{}s Cradle: deforming the higher rank E$_{1}$ and $ {\tilde{E}}_1 $ theories},'' {\em JHEP} {\bf 02} (2021) 122, \href{http://arXiv.org/abs/2011.05125}{{\tt 2011.05125}}.

\bibitem{DelZotto:2017pti}
M.~Del~Zotto, J.~J. Heckman, and D.~R. Morrison, ``{6D SCFTs and Phases of 5D Theories},'' {\em JHEP} {\bf 09} (2017) 147, \href{http://arXiv.org/abs/1703.02981}{{\tt 1703.02981}}.

\bibitem{Bhardwaj:2018yhy}
L.~Bhardwaj and P.~Jefferson, ``{Classifying $5d$ SCFTs via $6d$ SCFTs: Rank one},'' {\em JHEP} {\bf 07} (2019) 178, \href{http://arXiv.org/abs/1809.01650}{{\tt 1809.01650}}. [Addendum: JHEP 01, 153 (2020)].

\bibitem{Bhardwaj:2018vuu}
L.~Bhardwaj and P.~Jefferson, ``{Classifying 5d SCFTs via 6d SCFTs: Arbitrary rank},'' {\em JHEP} {\bf 10} (2019) 282, \href{http://arXiv.org/abs/1811.10616}{{\tt 1811.10616}}.

\bibitem{Bhardwaj:2019xeg}
L.~Bhardwaj, ``{Do all 5d SCFTs descend from 6d SCFTs?},'' {\em JHEP} {\bf 04} (2021) 085, \href{http://arXiv.org/abs/1912.00025}{{\tt 1912.00025}}.

\bibitem{Bhardwaj:2020gyu}
L.~Bhardwaj and G.~Zafrir, ``{Classification of 5d $ \mathcal{N} $ = 1 gauge theories},'' {\em JHEP} {\bf 12} (2020) 099, \href{http://arXiv.org/abs/2003.04333}{{\tt 2003.04333}}.

\bibitem{Cremonesi_2014}
S.~Cremonesi, A.~Hanany, and A.~Zaffaroni, ``Monopole operators and hilbert series of coulomb branches of 3d $ \mathcal{N} $ = 4 gauge theories,'' {\em Journal of High Energy Physics} {\bf 2014} (Jan, 2014).

\bibitem{Cremonesi:2015lsa}
S.~Cremonesi, G.~Ferlito, A.~Hanany, and N.~Mekareeya, ``{Instanton Operators and the Higgs Branch at Infinite Coupling},'' {\em JHEP} {\bf 04} (2017) 042, \href{http://arXiv.org/abs/1505.06302}{{\tt 1505.06302}}.

\bibitem{Ferlito:2016grh}
G.~Ferlito and A.~Hanany, ``{A tale of two cones: the Higgs Branch of Sp(n) theories with 2n flavours},'' \href{http://arXiv.org/abs/1609.06724}{{\tt 1609.06724}}.

\bibitem{Ferlito:2017xdq}
G.~Ferlito, A.~Hanany, N.~Mekareeya, and G.~Zafrir, ``{3d Coulomb branch and 5d Higgs branch at infinite coupling},'' {\em JHEP} {\bf 07} (2018) 061, \href{http://arXiv.org/abs/1712.06604}{{\tt 1712.06604}}.

\bibitem{Cabrera:2018ann}
S.~Cabrera and A.~Hanany, ``{Quiver Subtractions},'' {\em JHEP} {\bf 09} (2018) 008, \href{http://arXiv.org/abs/1803.11205}{{\tt 1803.11205}}.

\bibitem{Cabrera:2018jxt}
S.~Cabrera, A.~Hanany, and F.~Yagi, ``{Tropical Geometry and Five Dimensional Higgs Branches at Infinite Coupling},'' {\em JHEP} {\bf 01} (2019) 068, \href{http://arXiv.org/abs/1810.01379}{{\tt 1810.01379}}.

\bibitem{Cabrera:2019izd}
S.~Cabrera, A.~Hanany, and M.~Sperling, ``{Magnetic quivers, Higgs branches, and 6d $N$=(1,0) theories},'' {\em JHEP} {\bf 06} (2019) 071, \href{http://arXiv.org/abs/1904.12293}{{\tt 1904.12293}}. [Erratum: JHEP 07, 137 (2019)].

\bibitem{Bourget:2019aer}
A.~Bourget, S.~Cabrera, J.~F. Grimminger, A.~Hanany, M.~Sperling, A.~Zajac, and Z.~Zhong, ``{The Higgs mechanism \textemdash{} Hasse diagrams for symplectic singularities},'' {\em JHEP} {\bf 01} (2020) 157, \href{http://arXiv.org/abs/1908.04245}{{\tt 1908.04245}}.

\bibitem{Bourget:2019rtl}
A.~Bourget, S.~Cabrera, J.~F. Grimminger, A.~Hanany, and Z.~Zhong, ``{Brane Webs and Magnetic Quivers for SQCD},'' {\em JHEP} {\bf 03} (2020) 176, \href{http://arXiv.org/abs/1909.00667}{{\tt 1909.00667}}.

\bibitem{Cabrera:2019dob}
S.~Cabrera, A.~Hanany, and M.~Sperling, ``{Magnetic quivers, Higgs branches, and 6d $ \mathcal{N} $ = (1, 0) theories \textemdash{} orthogonal and symplectic gauge groups},'' {\em JHEP} {\bf 02} (2020) 184, \href{http://arXiv.org/abs/1912.02773}{{\tt 1912.02773}}.

\bibitem{Grimminger:2020dmg}
J.~F. Grimminger and A.~Hanany, ``{Hasse diagrams for 3d $ \mathcal{N} $ = 4 quiver gauge theories \textemdash{} Inversion and the full moduli space},'' {\em JHEP} {\bf 09} (2020) 159, \href{http://arXiv.org/abs/2004.01675}{{\tt 2004.01675}}.

\bibitem{Bourget:2020gzi}
A.~Bourget, J.~F. Grimminger, A.~Hanany, M.~Sperling, and Z.~Zhong, ``{Magnetic Quivers from Brane Webs with O5 Planes},'' {\em JHEP} {\bf 07} (2020) 204, \href{http://arXiv.org/abs/2004.04082}{{\tt 2004.04082}}.

\bibitem{Bourget:2020asf}
A.~Bourget, J.~F. Grimminger, A.~Hanany, M.~Sperling, G.~Zafrir, and Z.~Zhong, ``{Magnetic quivers for rank 1 theories},'' {\em JHEP} {\bf 09} (2020) 189, \href{http://arXiv.org/abs/2006.16994}{{\tt 2006.16994}}.

\bibitem{Akhond:2020vhc}
M.~Akhond, F.~Carta, S.~Dwivedi, H.~Hayashi, S.-S. Kim, and F.~Yagi, ``{Five-brane webs, Higgs branches and unitary/orthosymplectic magnetic quivers},'' {\em JHEP} {\bf 12} (2020) 164, \href{http://arXiv.org/abs/2008.01027}{{\tt 2008.01027}}.

\bibitem{Bourget:2020mez}
A.~Bourget, S.~Giacomelli, J.~F. Grimminger, A.~Hanany, M.~Sperling, and Z.~Zhong, ``{S-fold magnetic quivers},'' {\em JHEP} {\bf 02} (2021) 054, \href{http://arXiv.org/abs/2010.05889}{{\tt 2010.05889}}.

\bibitem{VanBeest:2020kxw}
M.~Van~Beest, A.~Bourget, J.~Eckhard, and S.~Sch\"afer-Nameki, ``{(5d RG-flow) Trees in the Tropical Rain Forest},'' {\em JHEP} {\bf 03} (2021) 241, \href{http://arXiv.org/abs/2011.07033}{{\tt 2011.07033}}.

\bibitem{Heckman:2018pqx}
J.~J. Heckman, T.~Rudelius, and A.~Tomasiello, ``{Fission, Fusion, and 6D RG Flows},'' {\em JHEP} {\bf 02} (2019) 167, \href{http://arXiv.org/abs/1807.10274}{{\tt 1807.10274}}.

\bibitem{Argyres:1995jj}
P.~C. Argyres and M.~R. Douglas, ``{New phenomena in SU(3) supersymmetric gauge theory},'' {\em Nucl. Phys. B} {\bf 448} (1995) 93--126, \href{http://arXiv.org/abs/hep-th/9505062}{{\tt hep-th/9505062}}.

\bibitem{Chacaltana:2010ks}
O.~Chacaltana and J.~Distler, ``{Tinkertoys for Gaiotto Duality},'' {\em JHEP} {\bf 11} (2010) 099, \href{http://arXiv.org/abs/1008.5203}{{\tt 1008.5203}}.

\bibitem{Chacaltana:2011ze}
O.~Chacaltana and J.~Distler, ``{Tinkertoys for the $D_N$ series},'' {\em JHEP} {\bf 02} (2013) 110, \href{http://arXiv.org/abs/1106.5410}{{\tt 1106.5410}}.

\bibitem{Chacaltana:2012zy}
O.~Chacaltana, J.~Distler, and Y.~Tachikawa, ``{Nilpotent orbits and codimension-two defects of 6d N=(2,0) theories},'' {\em Int. J. Mod. Phys. A} {\bf 28} (2013) 1340006, \href{http://arXiv.org/abs/1203.2930}{{\tt 1203.2930}}.

\bibitem{Chacaltana:2012ch}
O.~Chacaltana, J.~Distler, and Y.~Tachikawa, ``{Gaiotto duality for the twisted A$_{2N-1}$ series},'' {\em JHEP} {\bf 05} (2015) 075, \href{http://arXiv.org/abs/1212.3952}{{\tt 1212.3952}}.

\bibitem{Chacaltana:2013oka}
O.~Chacaltana, J.~Distler, and A.~Trimm, ``{Tinkertoys for the Twisted D-Series},'' {\em JHEP} {\bf 04} (2015) 173, \href{http://arXiv.org/abs/1309.2299}{{\tt 1309.2299}}.

\bibitem{Chacaltana:2014jba}
O.~Chacaltana, J.~Distler, and A.~Trimm, ``{Tinkertoys for the E$_{6}$ theory},'' {\em JHEP} {\bf 09} (2015) 007, \href{http://arXiv.org/abs/1403.4604}{{\tt 1403.4604}}.

\bibitem{Chacaltana:2015bna}
O.~Chacaltana, J.~Distler, and A.~Trimm, ``{Tinkertoys for the Twisted $E_6$ Theory},'' \href{http://arXiv.org/abs/1501.00357}{{\tt 1501.00357}}.

\bibitem{Chacaltana:2016shw}
O.~Chacaltana, J.~Distler, and A.~Trimm, ``{Tinkertoys for the Z3-twisted D4 Theory},'' \href{http://arXiv.org/abs/1601.02077}{{\tt 1601.02077}}.

\bibitem{Chacaltana:2017boe}
O.~Chacaltana, J.~Distler, A.~Trimm, and Y.~Zhu, ``{Tinkertoys for the E$_{7}$ theory},'' {\em JHEP} {\bf 05} (2018) 031, \href{http://arXiv.org/abs/1704.07890}{{\tt 1704.07890}}.

\bibitem{Chacaltana:2018vhp}
O.~Chacaltana, J.~Distler, A.~Trimm, and Y.~Zhu, ``{Tinkertoys for the $E_8$ Theory},'' \href{http://arXiv.org/abs/1802.09626}{{\tt 1802.09626}}.

\bibitem{Yonekura}
K.~Yonekura, ``{Instanton operators and symmetry enhancement in 5d supersymmetric quiver gauge theories},'' {\em JHEP} {\bf 07} (2015) 167, \href{http://arXiv.org/abs/1505.04743}{{\tt 1505.04743}}.

\bibitem{AtomMoleculeHybridupcoming}
A.~Bourget, M.~De~Marco, M.~Del~Zotto, J.~F. Grimminger, and A.~Sangiovanni, ``{upcoming},''.

\bibitem{Ohmori:2015pia}
K.~Ohmori, H.~Shimizu, Y.~Tachikawa, and K.~Yonekura, ``{6d $\mathcal{N}=\left(1,\;0\right) $ theories on S$^{1}$ /T$^{2}$ and class S theories: part II},'' {\em JHEP} {\bf 12} (2015) 131, \href{http://arXiv.org/abs/1508.00915}{{\tt 1508.00915}}.

\bibitem{Martone:2021drm}
M.~Martone and G.~Zafrir, ``{On the compactification of 5d theories to 4d},'' {\em JHEP} {\bf 08} (2021) 017, \href{http://arXiv.org/abs/2106.00686}{{\tt 2106.00686}}.

\bibitem{collingwood1993}
D.~Collingwood and W.~McGovern, ``Nilpotent orbits in semisimple lie algebras,''.

\bibitem{Bhardwaj:2021pfz}
L.~Bhardwaj, M.~Hubner, and S.~Schafer-Nameki, ``{1-form Symmetries of 4d N=2 Class S Theories},'' {\em SciPost Phys.} {\bf 11} (2021) 096, \href{http://arXiv.org/abs/2102.01693}{{\tt 2102.01693}}.

\bibitem{Witten:1996bn}
E.~Witten, ``{Nonperturbative superpotentials in string theory},'' {\em Nucl. Phys. B} {\bf 474} (1996) 343--360, \href{http://arXiv.org/abs/hep-th/9604030}{{\tt hep-th/9604030}}.

\bibitem{Sen:1997kz}
A.~Sen, ``{A Note on enhanced gauge symmetries in M and string theory},'' {\em JHEP} {\bf 09} (1997) 001, \href{http://arXiv.org/abs/hep-th/9707123}{{\tt hep-th/9707123}}.

\bibitem{Sen:1997js}
A.~Sen, ``{Dynamics of multiple Kaluza-Klein monopoles in M and string theory},'' {\em Adv. Theor. Math. Phys.} {\bf 1} (1998) 115--126, \href{http://arXiv.org/abs/hep-th/9707042}{{\tt hep-th/9707042}}.

\bibitem{Acharya:2002gu}
B.~S. Acharya, ``{M theory, G(2)-manifolds and four-dimensional physics},'' {\em Class. Quant. Grav.} {\bf 19} (2002) 5619--5653.

\bibitem{Hanany:1999sj}
A.~Hanany and A.~Zaffaroni, ``{Issues on orientifolds: On the brane construction of gauge theories with SO(2n) global symmetry},'' {\em JHEP} {\bf 07} (1999) 009, \href{http://arXiv.org/abs/hep-th/9903242}{{\tt hep-th/9903242}}.

\bibitem{Bourget:2020xdz}
A.~Bourget, J.~F. Grimminger, A.~Hanany, R.~Kalveks, M.~Sperling, and Z.~Zhong, ``{Magnetic Lattices for Orthosymplectic Quivers},'' {\em JHEP} {\bf 12} (2020) 092, \href{http://arXiv.org/abs/2007.04667}{{\tt 2007.04667}}.

\bibitem{Benini:2010uu}
F.~Benini, Y.~Tachikawa, and D.~Xie, ``{Mirrors of 3d Sicilian theories},'' {\em JHEP} {\bf 09} (2010) 063, \href{http://arXiv.org/abs/1007.0992}{{\tt 1007.0992}}.

\bibitem{Dancer:2024lra}
A.~Dancer, J.~F. Grimminger, J.~Martens, and Z.~Zhong, ``{Complex Symplectic Contractions and 3d Mirrors},'' \href{http://arXiv.org/abs/2406.09626}{{\tt 2406.09626}}.

\bibitem{Ohmori:2015pua}
K.~Ohmori, H.~Shimizu, Y.~Tachikawa, and K.~Yonekura, ``{6d $\mathcal{N}=(1,0)$ theories on $T^2$ and class S theories: Part I},'' {\em JHEP} {\bf 07} (2015) 014, \href{http://arXiv.org/abs/1503.06217}{{\tt 1503.06217}}.

\bibitem{HiggsRGupcoming}
M.~De~Marco, M.~Del~Zotto, J.~F. Grimminger, and A.~Sangiovanni, ``{upcoming},''.

\bibitem{Closset_2019}
C.~Closset, M.~Del~Zotto, and V.~Saxena, ``Five-dimensional scfts and gauge theory phases: an m-theory/type iia perspective,'' {\em SciPost Physics} {\bf 6} (May, 2019).

\bibitem{Hanany:1996ie}
A.~Hanany and E.~Witten, ``{Type IIB superstrings, BPS monopoles, and three-dimensional gauge dynamics},'' {\em Nucl. Phys. B} {\bf 492} (1997) 152--190, \href{http://arXiv.org/abs/hep-th/9611230}{{\tt hep-th/9611230}}.

\bibitem{vanBeest:2020kou}
M.~van Beest, A.~Bourget, J.~Eckhard, and S.~Schafer-Nameki, ``{(Symplectic) Leaves and (5d Higgs) Branches in the Poly(go)nesian Tropical Rain Forest},'' {\em JHEP} {\bf 11} (2020) 124, \href{http://arXiv.org/abs/2008.05577}{{\tt 2008.05577}}.

\end{thebibliography}

\end{document}